\documentclass{article}

\usepackage[preprint,nonatbib]{neurips_2025}

\usepackage[utf8]{inputenc} % allow utf-8 input
\usepackage[T1]{fontenc}    % use 8-bit T1 fonts
\usepackage{hyperref}       % hyperlinks
\usepackage{url}            % simple URL typesetting
\usepackage{booktabs}       % professional-quality tables
\usepackage{amsfonts}       % blackboard math symbols
\usepackage{nicefrac}       % compact symbols for 1/2, etc.
\usepackage{microtype}      % microtypography
\usepackage{xcolor}         % colors
\usepackage{subcaption,graphicx}
\usepackage[numbers]{natbib}

\usepackage{amsmath,tikz,amsthm,makecell,mathdots}
\DeclareMathOperator{\sign}{sign}
\usetikzlibrary{automata,arrows}

\newcommand{\ext}{\texttt{ext}}

\newtheorem{innercustomthrm}{Theorem}

\newenvironment{customthrm}[1]
  {\renewcommand\theinnercustomthrm{#1}\innercustomthrm}
  {\endinnercustomthrm}

\newtheorem{innercustomproposition}{Proposition}

\newenvironment{customprp}[1]
  {\renewcommand\theinnercustomproposition{#1}\innercustomproposition}
  {\endinnercustomproposition}

\title{Last-Iterate Convergence of No-Regret Learning for Equilibria in Bargaining Games}

\author{%
  Serafina Kamp \\
 Computer Science Engineering\\
  University of Michigan\\
  Ann Arbor, MI \\
  \texttt{serafibk@umich.edu} \\
  \And
  Reese Liebman \\
  Computer Science Engineering\\
  University of Michigan\\
  Ann Arbor, MI \\
  \And
  Benjamin Fish \\
 Computer Science Engineering\\
  University of Michigan\\
  Ann Arbor, MI \\
}

\begin{document}

\maketitle

\begin{abstract}
Bargaining games, where agents attempt to agree on how to split utility, are an important class of games used to study economic behavior, which motivates a study of online learning algorithms in these games.  In this work, we tackle when no-regret learning algorithms converge to Nash equilibria in bargaining games. Recent results have shown that online algorithms related to Follow the Regularized Leader (FTRL) converge to Nash equilibria (NE) in the last iterate in a wide variety of games, including zero-sum games. However, bargaining games do not have the properties used previously to established convergence guarantees, even in the simplest case of the ultimatum game, which features a single take-it-or-leave-it offer. Nonetheless, we establish that FTRL (without the modifications necessary for zero-sum games) achieves last-iterate convergence to an approximate NE in the ultimatum game along with a bound on convergence time under mild assumptions. Further, we provide experimental results to demonstrate that convergence to NE, including NE with asymmetric payoffs, occurs under a broad range of initial conditions, both in the ultimatum game and in bargaining games with multiple rounds.  This work demonstrates how complex economic behavior (e.g. learning to use threats and the existence of many possible equilibrium outcomes) can result from using a simple learning algorithm, and that FTRL can converge to equilibria in a more diverse set of games than previously known.
\end{abstract}
% Because of the possibility of take-it-or-leave-it offers, bargaining games are not zero-sum or convex. This includes the ultimatum game, which features a single take-it-or-leave-it offer. 

\section{Introduction}
% start with why bargaining and why it is not obvious to know what strategy to play (e.g., infintiely many equilibria)
%What should players offer in negotiations, and what deals should they agree to? 
How to play in bargaining games is vitally important.  Bargaining games describe wage \& price setting, union negotiations, trade among partners, and many other kinds of economic behavior in markets~\citep{korenok2021wage,feri2011bargaining,prasad2019impact}. The classical Nash bargaining solution only focuses on the desirable properties that the solution should have, without considering how self-interested strategic players should or will actually play in such games (see e.g.~\citet{osborne1990bargaining}).  The primary alternative \cite{osborne1990bargaining,roth1985game,tadelis} has been to compute equilibria in strategic-form bargaining games, since they are often easy to compute. However, there are typically infinitely many equilibrium outcomes where it is not clear how to select between them. 
% And for a variety of solution types and settings, including the ultimatum game, 
Even if there is a unique equilibrium in the one-stage game, there will not be a unique equilibrium when the game is repeated~\citep{fudenberg1986folk}.
% even in the simplest bargaining games.  
% This makes it difficult to know how to select among the many equilibria, given uncertainty about the opponent's play. 

% even if you can compute a unique and strong equilibria, empirical results still diverge and we don't expect repeated play to have the same solution
This includes the \emph{ultimatum game}, the simplest bargaining game, a two-player game which consists of a single take-it-or-leave-it offer.  One player offers a split of the surplus (the total benefit possible to the players as a result of agreement), and the other player accepts the offer, resulting in a deal, or rejects, and neither player receives anything. While any split of the surplus can be supported by a Nash equilibrium, the responder has a weakly dominant strategy of accepting any offer which results in a unique subgame-perfect equilibrium, a stronger notion than Nash.
% -- accepting all offers achieves a payoff at least as large as any other strategy. 
Yet empirical human studies have found that people often deviate from this unique equilibrium strategy~\citep{camerer2011behavioral,debove2016models,falk2006theory,nowak2000fairness,rand2013evolution,thaler1988anomalies}.  

% why online learning for games what are the goals of learning here
Given many possible equilibria strategies, bargainers need to \emph{learn what to play over time}. For example, firms may repeatedly negotiate prices, say for products or advertising revenue, via an online digital market~\citep{beam1999negotiations,juda2009pricing,parvaneh2021show,zhang2020price}. In this work, we consider online learning algorithms for repeated bargaining games where strategies are updated based on feedback from your opponent over time.
% An algorithm in this setting needs to select a strategy to play at each iteration of the game.  The algorithm observes the past play of the opponent, but that opponent may not play at an equilibrium, may play using a learning algorithm of their own, or play entirely arbitrarily.  
% Here, an algorithm selects a strategy to play at each iteration of the game based on how your opponent has played in the past. 
% To ensure an agent would use such an algorithm regardless of how their opponent plays, 
The goal in online learning is to guarantee an algorithm is \emph{no-regret}: The difference between the cumulative utility it receives and the utility using the single best strategy in hindsight up until time $T$ is sublinear in $T$, even against arbitrary play by the opponent.  
% This ensures that an agent may want to use this algorithm regardless of how their opponent plays.
This guarantee may lead to multiple players using the same algorithm and, if they do,
% Given that the opponent may then want to use the same algorithm, if both agents play the same algorithm,
the goal is for the played strategies to converge to a Nash equilibrium (NE), i.e.\ \emph{last-iterate convergence}.  Then, at least in the long run, neither player has incentive to deviate from the algorithm's selected strategies.

% intuition for why bargaining games do not have the nice learning properties of previous work
There is a large and growing literature on finding algorithms that simultaneously achieve no-regret and last-iterate convergence to Nash equilibria in a wide variety of games, including zero-sum games, auction games, non-negative regret games, and variationally stable games~\citep{anagnostides2022last,cai2024uncoupled,daskalakis2018last,deng2022nash,farina2022near,hart2000simple,hsieh2021adaptive,mertikopoulos2019online}. 
Unfortunately, bargaining games do not belong to any of these categories.  
% The central challenge is showing that there is a no-regret algorithm that converges (even before considering whether it converges to an equilibrium or not). %do not have the properties necessary in this literature to prove last-iterate convergence in any online learning setting: 
A key feature of bargaining games is that an agent's utility is discontinuous:  rejecting an offer below a threshold results in zero utility, whereas accepting an offer just above a threshold results in much more utility.  This discontinuity means that the agents do not have concave utility functions, a notably difficult setting for learning. 
% , and, further, this structure implies bargaining games are not equivalent to auction games, zero-sum games, and other games previously examined in the literature.  
% We detail differences between bargaining games and these other categories of games below in Section~\ref{sec:related_works}.
%Since there are also many deals that are in NE, bargaining games are not guaranteed to have non-negative regret.
%Finally, since bargaining strategies generally include an acceptance threshold of one or both agents, they generally do not have strict Nash equilibrium solution concepts which implies they do not satisfy strict variational stability. 
However, recent work~\citep{kamp2025equal} has pointed out that there is an algorithm that achieves no-regret in at least some bargaining games, including those examined in this work:  Follow-the-Regularized Leader (FTRL)~\citep{shalev2006convex}. 
% In fact, FTRL and its variants are commonly used to achieve simultaneous no-regret and last-iterate convergence in a
% remarkably wide 
% variety of games~\citep{anagnostides2022last,deng2022nash,giannou2021survival}. 
Moreover, previous work has shown that there are variants of FTRL that converge in the last iterate in the ultimatum game, though these variants are not no-regret here~\citep{kamp2025equal,marden2007regret}.  This suggests the possibility of establishing no-regret and last-iterate NE convergence guarantees simultaneously in at least 
% some kinds of bargaining games, and the bargaining game that may be 
%to achieve this with is
the simplest bargaining game: the ultimatum game.  
% In this work, we focus on the \emph{full-feedback} model 
% (where the learning algorithm observes the entire strategy of the opponent from previous iterations of the game) 
% since this is the setting where FTRL has proven successful in previous work.
% though there is success with other feedback models, so not necessarily needed to point this out and just state it in our model later.

%However, previous work shows the possibility of last-iterate convergence to NE for weakly acyclic games, a class that includes the ultimatum game with acceptance threshold response strategies, using a regret-based (though, not no-regret) algorithm~\citep{marden2007regret} as well as convergence of the ultimatum game to NE under a variant of FTRL that is not no-regret~\citep{kamp2025equal}. 

In this work, our main theoretical result is that \textbf{FTRL with the Euclidean regularizer and full feedback converges to an approximate NE in the last iterate in the ultimatum game}.  This is a no-regret algorithm, establishing our desired simultaneous no-regret and last-iterate NE convergence.  We also provide an upper bound on the convergence rate under mild assumptions on the learning rate and discuss the limitations in establishing a tighter bound. 
These results might be surprising, since we do not require optimism or any other modifications of FTRL to establish last-iterate convergence, as is required even for zero-sum games~\citep{anagnostides2022last,daskalakis2018last}.  
In non-convex games, no-regret algorithms only converge to coarse correlated equilibria (strictly weaker than NE)~\citep{hannan1957approximation,hart2000simple}, and only games with strict NE can be stable and attracting under FTRL dynamics~\citep{giannou2021survival}.  However, as we show, FTRL in bargaining games turn out to be a special case where it is possible to avoid those worst-case dynamics.
As such, FTRL enjoys even stronger guarantees of last-iterate convergence than previous results suggest, and 
% suggests that these kinds of 
strong learning guarantees may be possible in other bargaining games. 
%However, they assume generic games in their result, and, notably, the fact that multiple acceptance thresholds can be a best response to a given offer implies that the ultimatum game is \textit{degenerate} in a non-trivial way which allows us to get around the impossibility result.

% why is it not trivial to solve, yet we still expect success 
% most important thing to explain -- bargaining games are not variationally stable when they are degenerate (think, acceptance threshold in your strategy), and this covers a wide range of games that have convergence guarantees. also not non-negative regret because of pure NE, but different, at each time step gives more utility than the best in hindsight strategy which only gets some of the payoffs. These differences show there is not a unique equilibrium that necessarily acts like an attractor -- each NE has the chance a priori to be an attractor, but determining which one depends on the initial conditions. Then, convergence still occurs because the best NE does not cycle (i.e., $w_{\max}$ can only decrease).

% now, why do we care about which outcome the algorithm converges and how do we explore this
% Our results raise another important question:
% % Will the algorithms that people are incentivized to use result in an equal split, or will it result in asymmetric outcomes? 
Our theoretical results establish convergence in the ultimatum game, but they do not specify \textit{which} NE is converged to, and 
% there are many possible such NE, including ones
it could be one with asymmetric payoffs. This is an important detail since the ultimatum game has long been studied to understand how people value and act on fairness concerns~\citep{camerer2011behavioral,debove2016models,rand2013evolution}. Here, an equal split of the surplus 
% (in the absence of outside options) 
is equated with fairness, with increasingly asymmetric outcomes increasingly unfair. There is now an expansive literature on how learning algorithms learn to play at asymmetric outcomes as a result of biased training data (see e.g.~\citet{barocas-hardt-narayanan}), including at equilibrium~\citep{elzayn2020effects,milli2019social}.  But the full range of conditions, beyond asymmetries in training data, that are possible to cause algorithms to result in unfairness or asymmetric outcomes remains an open question.  Bargaining games have been shown to be a promising setting to demonstrate some alternative possibilities, including in markets where all agents are identical except in their chosen strategies ~\citep{fish2022s,kamp2025equal}.  Thus exploring how and when FTRL converges to asymmetric NE in the ultimatum game helps address an underexplored source of unfairness in learning: the dynamics of learning and the initial parameters used by the learning algorithm.

%In some settings, such as wage or price setting, there are important fairness concerns for which NE agents play at. In particular, any asymmetric Nash equilibrium outcome where equal-merit agents are getting different payoffs can potentially be considered an unfair, yet stable outcome~\citep{fish2022s, kamp2025equal}. To address this point and explore bargaining games and learning settings beyond our assumptions in our theoretical results,
To do so, we conduct numerical experiments to simulate agents learning bargaining strategies for the ultimatum game via FTRL under a variety of initial conditions, including varying the reference point used in the regularized and the initial strategies used by the agents.  We find that the algorithm does converge under all observed conditions, including parameter settings that are not included in our theoretical results, and to a wide variety of approximate NE outcomes where many are asymmetric.

However, the ultimatum game has only one round of bargaining, so players are not able to execute \emph{threats} -- e.g.\ rejecting a worse offer off the equilibrium path to ensure a better offer on the equilibrium path.  Because of the importance of threats to many kinds of bargaining, we explore how FTRL can learn to play threats in a two-round extensive-form alternating bargaining game. As in the ultimatum game, we show empirically that FTRL converges to approximate NE, providing evidence that our theoretical results can be extended to this game.
%Then, we explore learning in the 2 round extensive form alternating bargaining game with no outside options. In all of our results, we observe both convergence of strategies and  a variety of approximate NE outcomes in terms of the payoffs each agent gets. 
% what are the takaways 
%Not only do our experiments support our hypothesis that stronger learning guarantees may be possible in a variety of bargaining games, they also demonstrate that many NE outcomes are possible, depending on the initial conditions and learning setting, and
We also see the emergence of threats, both credible (if the opponent deviates, the agent would still want to execute the threat) and non-credible.  These results also provide a novel way to potentially understand how complex economic behavior emerges, including fairness attitudes and deviance from some equilibria (e.g.\ the unique subgame-perfect equilibrium in the ultimatum game), when agents learn to bargain.

In Section~\ref{sec:related_works}, we review related work and highlight how our setting is different from previous last-iterate convergence guarantees. In Section~\ref{sec:preliminaries} we introduce our setting, the bargaining games we consider, and the learning setting for these games. In Section~\ref{sec:main_theorems}, we present our theoretical contribution and in Section~\ref{sec:experimental} we discuss experimental results. 
% Finally, we conclude with a discussion and future work in Section~\ref{sec:discussion}.

\section{Related Work}
\label{sec:related_works}
There is extensive literature on online learning in games, and we provide comparisons to a select few papers to highlight the relevant differences between bargaining games and previous classes of games that have last-iterate NE convergence guarantees. This includes zero-sum games, auction games, non-negative regret games, potential games, and variationally stable games (i.e., games with strict NE)~\citep{anagnostides2022last,cai2024uncoupled,daskalakis2018last,deng2022nash,farina2022near,hart2000simple,hsieh2021adaptive,mertikopoulos2019online} and some of these works use FTRL and its variants~\citep{anagnostides2022last,deng2022nash,giannou2021survival}. Because of the multiplicity of NE, the ultimatum game can get negative regret and, by its utility functions, it is not equivalent to zero-sum games, potential games, or auction games. Additionally, the NE in our game are not strict because there are generally multiple acceptance thresholds that are best responses to a single offer. Further, since we prove last-iterate convergence to a mixed NE, non-strict equilibria allow us to get around the impossibility result of~\citet{vlatakis2020no} where they prove that, in generic games, only quasi-strict and pure NE can be stochastically stable and attracting.

Many multi-agent no-regret guarantees and last-iterate NE convergence results, including for zreo-sum games, rely on the use of optimism which adds a duplicate of the previous utility feedback to the agent's update step~\citep{anagnostides2022last,cai2022finite,daskalakis2018last,farina2022near,golowich2020tight,hsieh2021adaptive}. However, we are able to establish last-iterate convergence to a NE in the ultimatum game using FTRL \textit{without optimism}.

Finally, previous work shows convergence of weakly acyclic games (which includes our bargaining game) to NE~\citep{marden2007regret} and convergence of the normal form ultimatum game to NE under FTRL with an $\ell_1$ regularizer and particular learning rates\citep{kamp2025equal}. However, both works use algorithms that are notably not no-regret, so this leave open the question of whether no-regret and last-iterate convergence can be simultaneously guaranteed. In this work, we demonstrate that FTRL with the Euclidean regularizer, a no-regret algorithm, achieves last-iterate convergence to a NE in the ultimatum game.

\section{Preliminaries}
\label{sec:preliminaries}
\paragraph{Setting} We consider a bargaining game between a single firm $f$ and a single worker $w$. We use this notation to keep the first proposer and first responder consistent between single and multi-round bargaining, though our results hold for any two kinds agents bargaining. We assume the agents are bargaining over the split of a surplus normalized to 1. We assume throughout that the firm is always the first to propose a surplus split and the worker is always the first to respond. The action set of the proposing agent is to make an offer to the responding agent from the set $\mathcal{A} = [0,1]$. The action set of the responding agent is to specify whether they would accept or reject each possible offer. The payoff to the agents is given as a tuple $(u_f,u_w)$ where $u_f$ is the payoff to the firm and $u_w$ is the payoff to the worker. We consider two versions of the bargaining game: the \emph{normal form ultimatum game} and the \emph{2-round alternating extensive form bargaining game}. 

% In this section, we introduce both games with continuous action sets, and in Section~\ref{sec:learning_strategies} we describe the convex version of each game used for learning.

\paragraph{Normal form ultimatum game}
In the ultimatum game~\footnote{See~\citet{tadelis} for an overview of variations of the ultimatum game as well as other game theoretic concepts.}, the firm makes an offer $a \in \mathcal{A}$ and the worker can either accept $a$ or reject $a$. If the worker accepts, the payoff to the agents is $(1-a,a)$, and if the worker rejects, the payoff to the agents is $(0,0)$. In the normal form version of the game, the agents specify their actions simultaneously where the firm chooses an offer $a_f \in \mathcal{A}$ and the worker chooses an acceptance threshold $a_w \in \mathcal{A}$, i.e., specifies the lower bound on offers they are willing to accept. We will refer to the strategy profile of the agents as a tuple specifying each agent's action: $(a_f,a_w)$. Finally, the utility functions of the agents are 
\begin{align*}
    u_f(a_f,a_w)&=(1-a_f)\cdot \mathbf{1}\{a_w \le a_f\}, & \tag{1}\label{firm-utility-continuous}\\
    u_w(a_f,a_w)&= a_f \cdot \mathbf{1}\{a_w \le a_f\}.&\tag{2}\label{worker-utiltiy-continuous}
\end{align*}
There are infinitely many pure Nash equilibria for this game. For each $a\in\mathcal{A}$, the strategy profile $(a,a)$ is in NE as well as the strategy profile $(0,1)$ where both agents get 0. Given the worker's acceptance threshold, the firm gets the most utility by making the lowest possible offer that will get accepted, and, given the firm's offer, the worker gets equal utility from any acceptance threshold at or below this offer. As a result, the Nash equilibria are not \textit{strict}, i.e., for an offer $a_f >0$, \[u_w(a_f, a_w) = a_f, \forall a_w \le a_f, a_w\in\mathcal{A}.\]
We prove convergence to a \textit{mixed} NE, and we discuss its structure below when we introduce the learning setting.
% mixed NE here 
% The mixed Nash equilibria in this game follow the structure of the firm making a pure offer $a_f \in \mathcal{A}$ and the worker mixing over acceptance thresholds $a_w\in\mathcal{A}$ where the largest acceptance threshold the worker plays with non-zero probability is $a_f$. Notably, in order to be in NE, the worker must be playing the acceptance threshold $a_f$ with sufficiently high probability to prevent the firm from preferring to deviate to a lower offer. We define this mixed NE in detail in Section~\ref{sec:learning_strategies}. 
% Finally, it is of note that in the sequential version of the game there is a unique subgame perfect equilibrium where the worker would accept any offer greater than 0, so the firm proposes the lowest possible non-zero offer. However, we are interested in the conditions that lead to convergence to different equilibria, especially given the divergence from the subgame perfect outcome in real-world experiments of the ultimatum game~\citep{debove2016models}, so our results focus on the normal form version of the game,
Finally, the ultimatum game can also be represented in the extensive form, and we discuss the implications of this in Section~\ref{sec:nfg-experiments}. 

\paragraph{2-round alternating extensive form bargaining game}
To explore multi-round bargaining, we consider a 2-round alternating extensive form bargaining game with complete information and perfect recall. 
% Actions are now performed sequentially instead of simultaneously and the agents take turns making offers and responding to offers. 
Here, the agents take actions sequentially: The firm makes the first offer $a_f \in \mathcal{A}$, then the worker either accepts or rejects the offer. If the offer is accepted, the agents receive the payoff $(1-a_f, a_f)$. Otherwise, the agents switch roles and the worker now makes a counter-offer $a_w \in \mathcal{A}$ to which the firm can either accept or reject. As is typical, a time discount factor $0<\delta<1$ is applied to payoffs in the second round. So, if the firm accepts the counter offer the agents receive the payoff $\delta(a_w,1-a_w)$, and if the firm rejects the counter offer the agents receive $\delta(0,0)$. In the sequential setting, it is possible for agents to make \textit{threats} in their strategy by specifying actions they would take if their opponent deviates from any given strategy they use. We discuss threats in Section~\ref{sec:efg-experiments}.

\paragraph{Learning bargaining strategies online}
% \label{sec:learning_strategies}
% In this work, we consider agents learning bargaining strategies online
% , in line with previous literature of learning in games~\citep{cesa2006prediction,mertikopoulos2019online}.
In online learning in games with multiple agents, agents update their strategies over time based on the utility feedback they see given the previous strategies of their opponent(s). The performance of an online algorithm is judged by the regret the learner incurs for any possible sequence of utility feedback they could see.
% Further, there exist \textit{no-regret} algorithms where the strategy an agent learn online gets as much utility, on average, as the best-in-hindsight strategy. 
Formally, let $a_i^{(t)} \in \mathcal{A}$ be the action agent $i$ took at time $t$ and let $u_i^{(t)}(a)$ be the utility agent $i$ receives at time step $t$ when playing action $a$. Then, the regret of an algorithm after $T$ time steps is $$\texttt{Regret}_T = \arg\max_{a\in \mathcal{A}}\sum_{t=1}^Tu_i^{(t)}(a) - \sum_{t=1}^Tu_i^{(t)}(a_i^{(t)}).$$
An algorithm is said to be \textit{no-regret} if $\texttt{Regret}_T$ is sublinear in $T$ for any arbitrary sequence of utility feedback functions $u_i^{(1)},\ldots,u_i^{(T)}$ drawn from a class of utility feedback functions. Online learning algorithms are particularly well-suited for convex optimization problems~\footnote{See~\citet{hazan2016introduction} for an overview of online convex optimization.}. However, given the non-convexity of our games,  we discretize $\mathcal{A}$ to use the convex expected utility function instead. 

\paragraph{Action discretization}
For any integer $D>1$, let $\mathcal{A} =\{0,\frac{1}{D}, \ldots, \frac{D-1}{D},1\}$. Then, let $\mathcal{G}^{(1)}$ be the normal form ultimatum game with the discretized action set $\mathcal{A}$ and let $\mathcal{G}^{(2)}$ be the 2-round alternating bargaining game with the discretized offer action set $\mathcal{A}$. In $\mathcal{G}^{(1)}$, let $x_i^{(t)} \in \Delta(\mathcal{A})$ be the mixed strategy of agent $i$ at time $t$ for $i \in \{f,w\}$ and let $x_{i,a}^{(t)}$ be the probability mass agent $i$ puts on action $a \in \mathcal{A}$ at time $t$. In $\mathcal{G}^{(2)}$, an extensive form game, we represent strategies as realization plans from the space of the sequence-form polytope associated with the game tree in line with previous work\citep{farina2022near}. We provide an explanation of this representation in Appendix~\ref{appendix:sequence-form}.

Here, we are assuming both agents discretize the action space in the same way since their utility functions in equations~\ref{firm-utility-continuous} and~\ref{worker-utiltiy-continuous} only depend on whether an offer is above or below an acceptance threshold, not their distance. We expand on how learning over the discretized action set is still meaningful in the original ultimatum game with a continuous action set in Appendix~\ref{appendix-proofs}. 
 
\paragraph{Mixed NE in the normal form ultimatum game} 

For $x_f, x_w \in\Delta(\mathcal{A})$, the strategy profile $(x_f,x_w)$ is in a mixed NE if the firm uses the pure strategy $x_f = a_f$ for $a_f\in \mathcal{A}$, $\max\{a_w|x_{w,a_w} >0\} = a_f$, and \begin{align*}
    (1-a_f) \ge (1-a)\cdot \sum_{a_w \le a}x_{w,a_w}, \forall a < a_f. &\tag{3}\label{mixed_NE_condition}
\end{align*}
Here, since $\max\{a_w|x_{w,a_w} >0\} = a_f$, then the worker accepts an offer of $a_f$ with probability 1, so the expected utility to the firm for an offer of $a_f$ is $(1-a_f)$. Any higher offer would also be accepted with probability 1, so the firm would get strictly worse utility from making an offer higher than $a_f$. Further, the condition~\ref{mixed_NE_condition} ensures the firm does not get more expected utility by lowering their offer. When, $x_f = a_f$ the expected utility of the worker is $a_f$ for any distribution over acceptance thresholds $a \le a_f$ is $a_f$ and all other mixed strategies get strictly less than $a_f$. 

\paragraph{Follow-the-Regularized-Leader}
The online algorithm we consider is Follow-the-Regularized-Leader (FTRL)~\citep{shalev2006convex}. We use the vanilla Euclidean regularizer throughout and let $\eta>0$ be the learning rate. We assume all algorithm parameters and initial strategies are rational numbers. For each $i \in \{f,w\}$, let $\mathcal{X}_i$ be the strategy space of agent $i$. We use the full feedback setting where agent $i$ has access to a vector $U_i^{(t)}$ specifying their utility feedback for any action in the support of their strategy space. The FTRL update step of agent $i$'s strategy at time $t$ for either game $\mathcal{G}^{(1)}$ or $\mathcal{G}^{(2)}$ is: 
\begin{align*}
    \arg\max_{x \in\mathcal{X}_i}\eta  \langle U_i^{(t)}, x\rangle - \frac{1}{2}\|x-\alpha_i\|_2^2. & \tag{1}\label{alg:ftrl}
\end{align*}
The term $\alpha_i$ is the \textit{reference point} of the regularizer. We let $\alpha_i = a$ for $a\in\mathcal{A}$ represent a pure strategy while $\alpha_i =\mathbf{0}$ is the vector of all 0's.
We will assume a reference point of $\alpha_i=\mathbf{0}$ in Section~\ref{sec:main_theorems}, but in Section~\ref{sec:experimental} we experiment with a variety of pure strategy reference points for learning in $\mathcal{G}^{(1)}$. 

\section{Last-Iterate Convergence to $\epsilon$-NE}
\label{sec:main_theorems}
We are now ready to state our main theoretical results.  All proofs for this section can be found in Appendix~\ref{appendix-proofs}.  First, in Theorem~\ref{thrm:ftrl_convergence_n=1-main-paper}, we show that agents using Algorithm~\ref{alg:ftrl} in the ultimatum game ($\mathcal{G}^{(1)}$) converge to an approximate equilibrium, regardless of the initial strategies that agents choose.  This result shows that FTRL, even without optimism~\citep{anagnostides2022last,daskalakis2018last,rakhlin2013optimization}, is sufficient to achieve simultaneous no-regret learning and convergence to equilibrium in the ultimatum game.

\begin{customthrm}{1}
    \label{thrm:ftrl_convergence_n=1-main-paper}
    Suppose agents learn strategies for $\mathcal{G}^{(1)}$ using Algorithm~\ref{alg:ftrl} with $\alpha_i=\mathbf{0}$, any $\eta >0, D>2$, and arbitrary initial conditions $x_w^{(1)}, x_f^{(1)} \in \Delta(\mathcal{A})$. Then, for any $\epsilon>0$, there exists a finite time $t_\epsilon$ where $(x_f^{(\tau)},x_w^{(\tau)})$ is in $\epsilon$-Nash equilibrium for all $\tau \ge t_\epsilon$. 
\end{customthrm}
We consider the setting $\alpha_f=\alpha_w=\mathbf{0}$  to align with regularization functions used in previous literature on last-iterate convergence in regularized learning~\citep{giannou2021survival,lee2021last,vlatakis2020no}. These reference points imply that the regularizer for both agents is minimized by the uniform mixed strategy. Note that both agents using a uniform mixed strategy is not an equilibrium point of $\mathcal{G}^{(1)}$, however we are still able to establish convergence for any strictly positive value of $\eta$ because our proof of Theorem~\ref{thrm:ftrl_convergence_n=1-main-paper} only depends on $\eta$ insofar as the value $\frac{1}{\eta}$ is finite. Thus, it may be possible to extend our theoretical results to other reference points, even those that do not imply the regularizer minimizer strategies are in equilibrium, and we leave additional theoretical results to future work. We outline the proof of Theorem~\ref{thrm:ftrl_convergence_n=1-main-paper} below.
To establish convergence to approximate NE, we take advantage of the structure of the game.  First,
% Consider the case where the firm uses a pure strategy, and the worker uses a non-pure mixed strategy, where 
let $w_{\max}$ be the largest acceptance with non-zero probability mass threshold in the worker's strategy.  Note this action is time-dependent, but for the sake of this discussion we will suppress this dependence in the notation.  Fix $w_{\max}=\frac{k}{D}$ for some $k\in[D]$. The core of the proof relies on showing that $w_{\max}$ can never increase and that the the firm either prefers the offer $w_{\max}$ long enough for convergence to occur there or $w_{\max}$ decreases in value. To see this, note that  
% If we consider only strategies for the worker where $w_{\max}=\frac{k}{D}$,
 the offer $w_{\max} = \frac{k}{D}$ is strictly dominant for the firm when 
% the worker's probability mass on $w_{\max}$, $x_{w,w_{\max}}$, satisfies 
$x_{w,w_{\max}} > \frac{1}{D-k+1}$, weakly dominant for the firm when $x_{w,w_{\max}} = \frac{1}{D-k+1}$, and the offer $\frac{k-1}{D}$ strictly dominates the offer $w_{\max}$ when $x_{w,w_{\max}} < \frac{1}{D-k+1}$.  This threshold is derived from Equation~\ref{mixed_NE_condition}.  Then if $x_{w,w_{\max}} \ge \frac{1}{D-k+1}$ for sufficiently long, the firm will converge to the strategy $w_{\max}$, which is an equilibrium.  However, $x_{w,w_{\max}}$ decreases proportionally to the firm's probability mass on offers less than $w_{\max}$. So, if $x_{w,w_{\max}}$ falls below the threshold $\frac{1}{D-k+1}$ before the firm converges, 
% it may decrease until it's below it, at which point 
then the firm starts receiving more utility for offers below $w_{\max}$.  
% If this happens, 
Eventually the firm prefers smaller offers, at which point so does the worker, and $w_{\max}$ decreases.  Since there only $D+1$ 
% possible strategies and therefore $D+1$ 
possible values of $w_{\max}$, eventually this process must terminate. 
% The core of the proof relies on showing that $w_{\max}$ can never increase, and then enumerating over the following possibilities:  whether or not the firm has non-zero probability mass above, at, or below $w_{\max}$, and whether or not $x_{w,w_{\max}}$ is smaller or larger than $\frac{1}{D-k+1}$.  
% This produces five distinct cases. 
% (it can also matter whether the firm's smallest action with non-zero mass is exactly equal to $w_{\max}$ or strictly above or below).  
% For each case, 
%(see Figure~\ref{fig:thrm_1_state_machine}), 
% we show either that it only lasts finitely long (therefore eliminating that case or decreasing $w_{\max}$) or ends in an approximate equilibrium.

This is sufficient to establish convergence to an approximate equilibrium, but does not a priori establish a convergence rate.  The challenge is establishing a convergence rate in the case when $x_{w,w_{\max}} > \frac{1}{D-k+1}$, where the firm's mass on $w_{\max}$ increases at a decreasing rate while the worker's mass on $w_{\max}$ decreases at a decreasing rate. 
%Eventually, either convergence is reached at that $w_{\max}$ value or $x_{w,w_{\max}}$ passes the threshold $\frac{1}{D-k+1}$ and $w_{\max}$ must decrease in value.  
The time it takes for $w_{\max}$ to decrease below the threshold $\frac{1}{D-k+1}$, if it does so, turns out to be inversely proportional to $x_{w,w_{\max}} - \frac{1}{D-k+1}$.  Thus we need to know how close $x_{w,w_{\max}}$ is to $\frac{1}{D-k+1}$ at the instant that $w_{\max}$ decreased to $\frac{k}{D}$ so we can upper bound the time it takes for $w_{\max}$ to decrease again in the worst case.
% but this depends on the exact value of $x_{w,w_{\max}}$ before $w_{\max}$ decreased.  

Exactly determining dynamics here appears difficult, so we avoid this problem entirely by simply counting bits.  Note that $x_{w,w_{\max}}$ is algorithmically determined in polynomial time, with polynomially many additions and multiplications.  This means that there's only an exponential number of bits in $x_{w,w_{\max}}$, as a function of the number of time steps so far, which in turn controls the size of $x_{w,w_{\max}} - \frac{1}{D-k+1}$.  Unfortunately, this means that after the first time that $w_{\max}$ decreases, the maximum number of time steps is already exponential, and the second times that $w_{\max}$ decreases is exponential in the time taken so far, which is already exponential.  Since this can happen up to $D$ times, this results in an upper bound that is an exponential iterated $D$ times.

\begin{customthrm}{2}
\label{thrm:ftrl_convergence_rate-main-paper}
    Suppose agents learn strategies for $\mathcal{G}^{(1)}$ using Algorithm~\ref{alg:ftrl} with $\alpha_i=\mathbf{0}$, any $0<\eta \le 1, D>2$, and arbitrary initial conditions $x_w^{(1)}, x_f^{(1)} \in \Delta(\mathcal{A})$. Then, for any $\epsilon>0$, the time to convergence of Algorithm~\ref{alg:ftrl} to an $\epsilon$-Nash equilibrium is 
    % \[O\left(\frac{D}{\eta}\uparrow \uparrow 4\cdot D\right) +\log\left(\frac{2\epsilon}{3D^2}\right)\cdot \log\left(1-\frac{\eta}{2}\right)^{-1}.\]
    % \[O\left(\underbrace{\frac{D}{\eta}^{\frac{D}{\eta}^{\iddots^{\frac{D}{\eta}}}}}_{4\cdot D}\right)+\log\left(\frac{2\epsilon}{3D^2}\right)\cdot \log\left(1-\frac{\eta}{2}\right)^{-1}.\]
    \[O\left(\underbrace{\frac{D}{\eta}^{\frac{D}{\eta}^{\iddots^{\frac{D}{\eta}}}}}_{3\cdot D}+\log\left(\frac{D^3}{\epsilon}\right)\cdot \frac{D^2}{\eta}\right).\]
    % \[O\left(\log\left(\frac{D^2}{\epsilon}\right)\cdot \log\left(\frac{2}{2-\eta}\right)^{-1}\cdot \underbrace{\frac{D}{\eta}^{\frac{D}{\eta}^{\iddots^{\frac{D}{\eta}}}}}_{4\cdot D}\right).\]
\end{customthrm}

We add the assumption $\eta \le 1$ since this is sufficient to ensure monotonicity of the dynamics of the firm's mass increasing on $w_{\max}$ while the worker's mass decreases on $w_{\max}$, and we leave extending the convergence time to all $\eta$ values for future work since convergence is guaranteed for any positive $\eta$ value by Theorem~\ref{thrm:ftrl_convergence_n=1-main-paper}. However, $\eta \le 1$ is sufficient to ensure FTRL is no-regret because $\eta = \frac{1}{T^{1/k}}$ for any $k>1$ is sufficient to establish sublinear regret of FTRL for our setting~\citep{hazan2016introduction}.  

The proof of Theorem~\ref{thrm:ftrl_convergence_rate-main-paper} shows how we can find a time where we can describe the dynamics of the firm's mass increasing on $w_{\max}$ while the worker's mass decreases on $w_{\max}$ as two non-homogeneous linear recurrence relations which can be solved in the closed form. Then, we can use these closed form equations and possible initial conditions of the recurrence relations to determine whether convergence occurs at $w_{\max}$ and, if so, whether exact convergence is possible. When exact convergence is not possible, the firm's mass on $w_{\max}$ approaches 1 at a rate proportional to, in the worst case, the rate at which $D^2(1-\frac{\eta}{2})^t$ approaches 0. This results in the leading term of Theorem~\ref{thrm:ftrl_convergence_rate-main-paper} to bound the time to approximate equilibrium 
convergence while the iterated exponential term is needed to guarantee that convergence will happen at the current $w_{\max}$. We do not believe this bound is tight, and we leave this to future work along with extending the convergence time to all $\eta$ values. We turn to Section~\ref{sec:experimental} to empirically explore possible convergence outcomes and their dependence on algorithm parameters.

\section{Experimental Results}
\label{sec:experimental}
% technical settings and convergence definition
We implemented Algorithm~\ref{alg:ftrl} to simulate the agents learning strategies for $\mathcal{G}^{(1)}$ and $\mathcal{G}^{(2)}$ using CVXPY~\citep{diamond2016cvxpy} to approximate the update step with the Clarabel solver. The Clarabel solver is accurate up to a threshold of $10^{-8}$, so we define convergence at time $T$ when, for both agents, no strategy dimension (i.e., the mass on a certain action) has changed by more than $10^{-7}$. That is, for $i\in\{f,w\}$, 
% and $\tau\in\{1,2\}$\ben{why 2 and not 1?}for at least 2 time steps
\[\|x_i^{(T)}- x_i^{(T-1)}\|_{\infty} \le 10^{-7}.\]We found greater sensitivity to small probability changes in $\mathcal{G}^{(2)}$ compared to $\mathcal{G}^{(1)}$, so we set the convergence threshold to $10^{-6}$ in those experiments and we verified that this threshold does not qualitatively change the strategy profiles compared to running the algorithm for more iterations. 

% Then, if the agents' strategy profile $(x_f^{(T)}, x_w^{(T)})$ converges at time $T$, we check to see if the strategy profile is in an $\epsilon$-Nash equilibrium for $\epsilon = 10^{-7}$.

% goal of experimental section 
The first goal of our experiments is to check whether convergence occurs in $\mathcal{G}^{(1)}$ with reference points besides the $\bf{0}$ reference point ($\alpha_f=\alpha_w=\bf{0}$) as well as in $\mathcal{G}^{(2)}$. Across all of our experiments, 8000 time steps was sufficient for strategy convergence in $\mathcal{G}^{(1)}$ for non-zero and zero reference points while 15000 steps was sufficient for convergence in $\mathcal{G}^{(2)}$, though under most parameters convergence time was much faster. Both bounds are much less than the bound given in Theorem~\ref{thrm:ftrl_convergence_rate} indicating that the bound is either not tight or at least the worst-case was too rare to appear in our experiments. Further, all strategy profiles at convergence were in $\epsilon$-NE for $\epsilon = 10^{-7}$. This strongly indicates that convergence is likely for many kinds of reference points, not only the $\mathbf{0}$ reference point considered in Theorem~\ref{thrm:ftrl_convergence_n=1-main-paper}, and that convergence may be possible for multi-round bargaining games.

In Sections~\ref{sec:nfg-experiments} and~\ref{sec:efg-experiments}, we explore the dependence of outcomes (the payoffs of the NE that the algorithm empirically converges to) on the initial conditions of the algorithm (i.e., initial strategy, $D, \eta$, reference points) for $\mathcal{G}^{(1)}$ and $\mathcal{G}^{(2)}$ respectively.  For $\mathcal{G}^{(1)}$, we also investigate one possibility for how agents can strategically select the initial strategies: by treating the map from initial strategies to predicted payoffs at convergence as a zero-sum game.
%Additionally, the set of possible payoffs at equilibrium given each pair of initial strategies is a zero-sum game, so we also investigate how agents could strategically choose initial strategies by leveraging this information. 
Finally, in Section~\ref{sec:efg-experiments}, we also describe the emergence of both credible and non-credible threats in the two-round strategy profiles at convergence. 

\subsection{Normal form ultimatum game}
\label{sec:nfg-experiments}
\begin{figure}[h]
    \centering
    \begin{subfigure}[t]{0.32\textwidth}
     \centering
        \includegraphics[height=4.5cm]{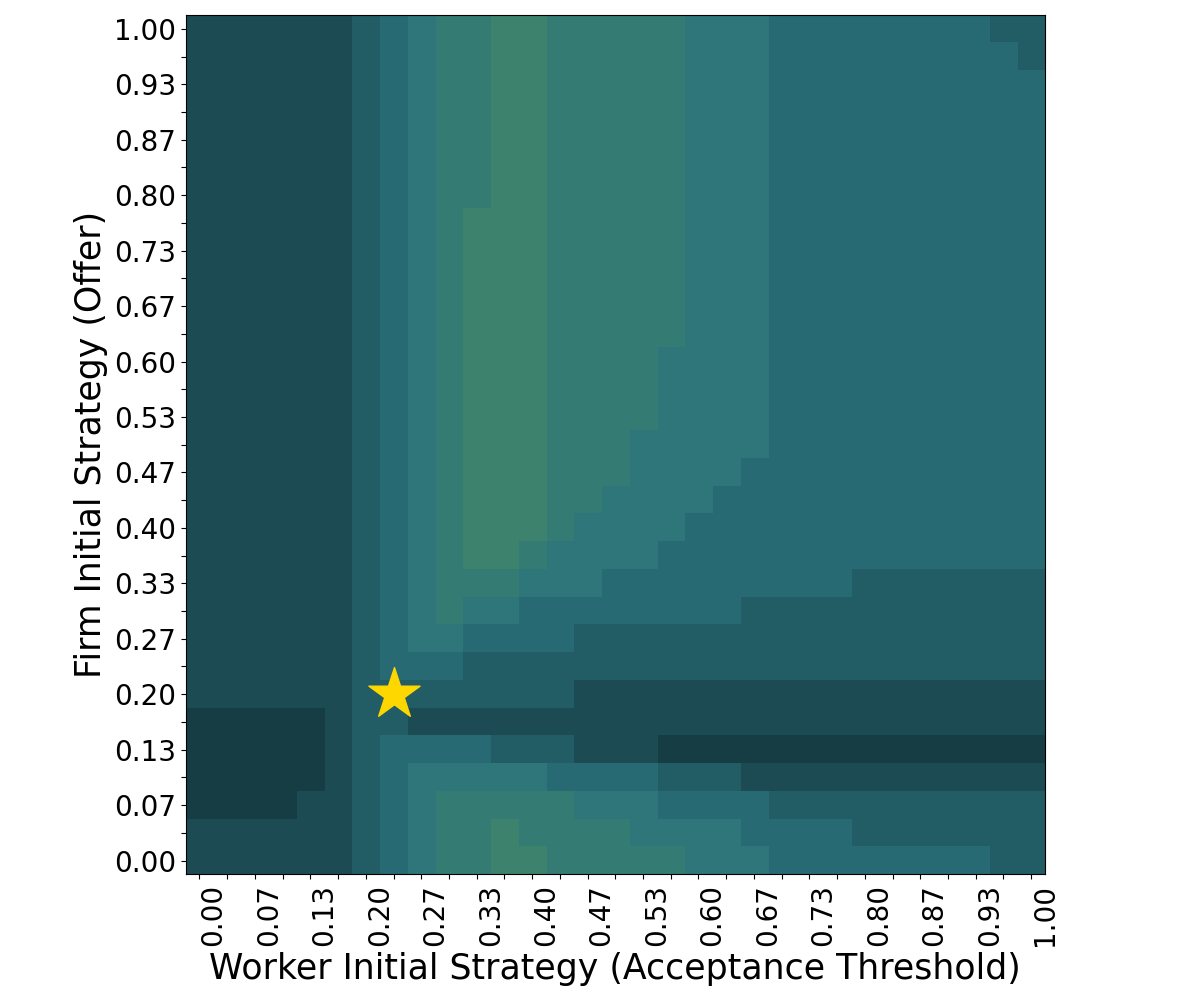}
        \caption{$\alpha_f = \alpha_w = \mathbf{0}$.}
        \label{Figure:nfg-no-reference}
    \end{subfigure}\hfill
    \begin{subfigure}[t]{0.33\textwidth}
    \centering
        \includegraphics[height=4.5cm]{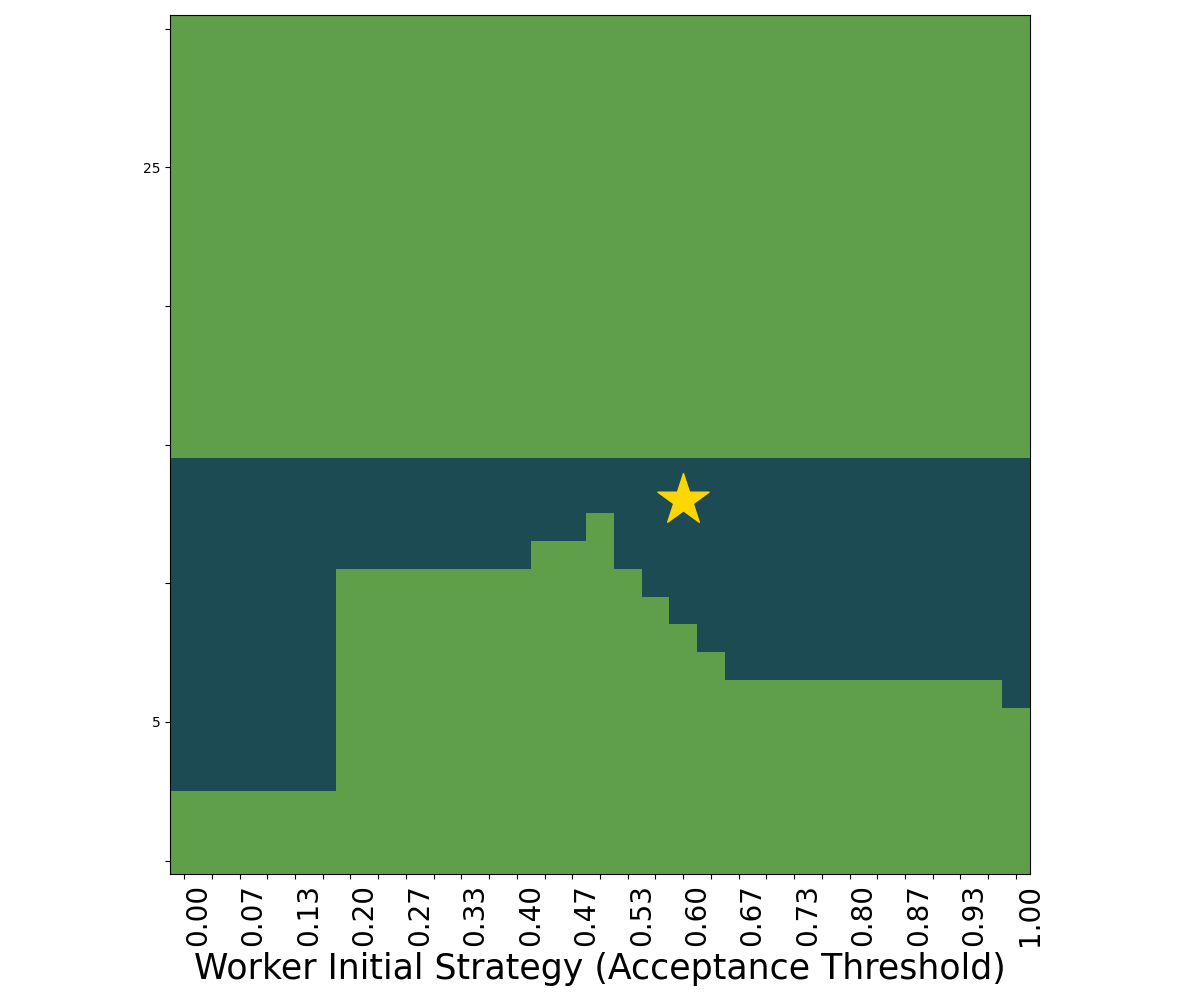}
        \caption{$\alpha_f = \frac{1}{6}$,  $\alpha_w = \frac{1}{2}$. }
        \label{Figure:nfg-reference-2}
    \end{subfigure}\hfill
    \begin{subfigure}[t]{0.32\textwidth}
    \centering
        \includegraphics[height=4.5cm]{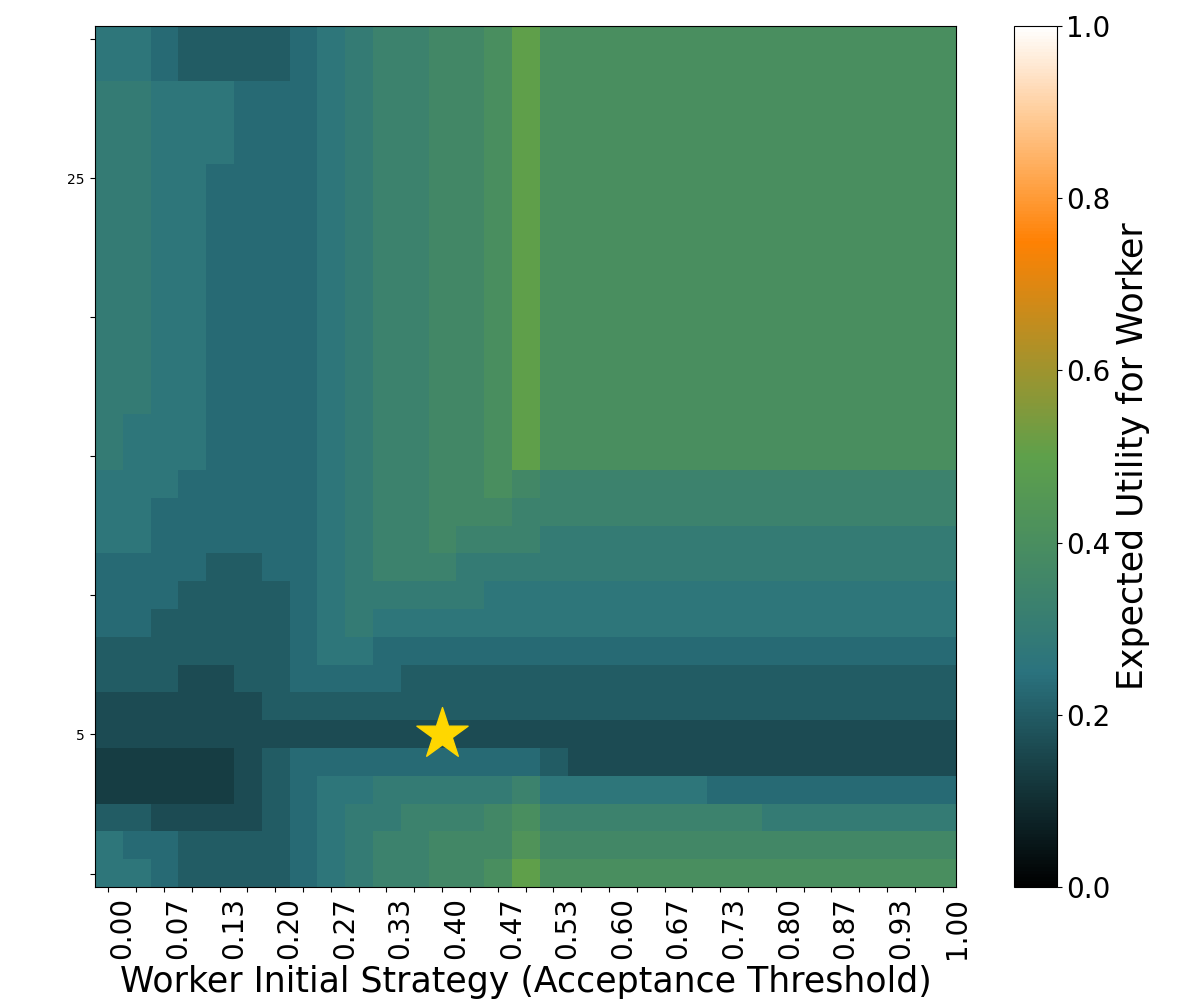}
        \caption{$\alpha_f = \frac{1}{2}$,  $\alpha_w =\frac{29}{30}$. }
        \label{Figure:nfg-reference-1}
    \end{subfigure}
    \caption{Comparison of Nash equilibrium payoff outcomes at convergence for the worker when agents are learning strategies for $\mathcal{G}^{(1)}$ using Algorithm~\ref{alg:ftrl} with $D=30,\eta=0.5$ and a variety of reference points. The y-axis represents the range of possible pure initial strategies for the firm while the x-axis represents the range of possible pure initial strategies for the worker. The gold star on each plot represents an equilibrium point in the zero-sum game of choosing an initial strategy.}
    \label{fig:nfg-plots}
\end{figure}

In our first set of experiments for $\mathcal{G}^{(1)}$, we run Algorithm~\ref{alg:ftrl} on a variety of initial parameter settings, including
% A single run of the algorithm takes a few minutes, and this justifies why it is practical for agents to hypothetically use this algorithm to learn strategies. However, this becomes computationally intractable to test an extensive set of initial parameter conditions, but
  a range of $D$ values between 10 and 50, $\eta$ values in $(0,1)$, and $\alpha_i$ values where $\alpha_f >\alpha_w$, $\alpha_w> \alpha_f$,  $\alpha_f = \alpha_w = 1$, and $\alpha_f=\alpha_w = \bf{0}$. For a given combination of $D,\eta,$ and $\alpha_i$, we run Algorithm~\ref{alg:ftrl} on each possible pure initial strategies for the agents. 
% \ben{summarize the range of values tested here}
% In this section, we abuse notation slightly and let $\alpha_i = a$ for $a\in\mathcal{A}$ represent a pure strategy while $\alpha_i =\mathbf{0}$ is the vector of all 0's.
% reference point convergence took longer, within 2000 time steps as opposed to 500, but note that both are much smaller than bound on no reference point convergence rate.
% characterize results broadly, then go into specifics 
% To display our results, for fixed $D,\eta$, and $\alpha_i$ values, we display the equilibrium payoff value to the worker (recall that the payoff to the firm is then $1-$ this value) for each possible pair of initial pure strategies for the agents. 
For each initial condition setting, we display the worker's payoff at the converged NE outcomes, denoted as $u_w$ (the firm's payoff is $1-u_w$). Across all of our experiments, there were many possible outcomes for the worker: The lowest $u_w$ value was $0.05$ while the highest was $0.8$ and, further, the range of $u_w$ values per experiment varied from $0$ to $0.75$ where there were generally more outcomes when $\eta$ was larger. All of the figures from our experimental results can be found in Appendix~\ref{appendix:nfg-graphs}, but we highlight three examples here, in Figure~\ref{fig:nfg-plots} and Table~\ref{tab:nfg-summary-statistics}, which demonstrate some common trends we observed in the experiments.  In these examples, we varied the reference, and fixed $D=30$ and $\eta = 0.5$. Additionally, we ran experiments for uniform mixed initial strategies and the ultimatum game represented as an extensive-form game. These graphs and further discussion can be found in Appendix~\ref{appendix:nfg-graphs}. 
% We now turn to $\mathcal{G}^{(2)}$ to discuss learning in extensive-form bargaining games further.
%In these experiments, we used $D=30, \eta = 0.5$, and compare between agents using a variety of reference points, including $\alpha_f=\alpha_w = \mathbf{0}$. The results for these experiments are in Figure~\ref{fig:nfg-plots} and Table~\ref{tab:nfg-summary-statistics}.
\begin{table}[h]
    \centering
    \begin{tabular}{|c|c|c|c|c|}
    \hline
        $(\alpha_f,\alpha_w)$  & Prop. $u_w\ge$ $x_w^{(1)}$ & Prop. $u_w \ge \alpha_w$& Min. $u_w$ & Max. $u_w$\\\hline
        & & & & \\
        ($\frac{1}{2}$, $\frac{29}{30}$) & $0.2997$ & $0.0$ & $0.1333$ & $0.5$\\
        & & & & \\
        ($\frac{1}{6}$, $\frac{1}{2}$) & $0.4662$ & $0.7721$ & $0.1667$ & $0.5$\\
        & & & &\\
        ($\mathbf{0}$, $\mathbf{0}$) & $0.2258$ & N/A & $0.1333$ & $0.3333$\\
        \hline       
    \end{tabular}
    \caption{Attributes of $u_w$ for each plot in Figure~\ref{fig:nfg-plots}. The second column gives the proportion of outcomes where $u_w \ge x_w^{(1)}$, the worker's initial acceptance threshold. The third column gives the proportion of outcomes where $u_w \ge \alpha_w$, the worker's reference point. The last two columns give the minimum and maximum $u_w$ value in each plot.}
    \label{tab:nfg-summary-statistics}
\end{table}

\paragraph{Payoffs at convergence} The plots in Figure~\ref{fig:nfg-plots} show that the payoffs are non-linear in the initial strategies. We noticed similar trends to these plots in our other experiments except when $\eta$ was small where we saw convergence to only one outcome: $\min\{\alpha_w,\alpha_f\}$. This indicates that the choice of reference point can be very influential in the converged outcome, but plot~\ref{Figure:nfg-reference-1} shows that this is not always the case. Notably, in Figure~\ref{fig:nfg-plots}, all plots have a minimum $u_w$ value of at least $\frac{1}{15}$ which means the worker is guaranteed more than the smallest split of the surplus, $\frac{1}{30}$, even when they use an initial strategy of accepting any offer. However, the maximum $u_w$ value here is $\frac{1}{2}$ with a majority of outcomes smaller than $\frac{1}{2}$ in plots~\ref{Figure:nfg-no-reference} and~\ref{Figure:nfg-reference-1} indicating that asymmetric outcomes where the worker is worse off are likely. In all of our experiments, the only time the worker had a majority of outcomes above $\frac{1}{2}$ was when the firm had $\alpha_f >\frac{1}{2}$, indicating their explicit preference for the worker to have a better outcome. We leave formalizing the dependence between reference points, initial strategies, and payoff outcomes to future work.

% To explore outcomes when agents start with mixed strategies, we also experimented with one agent using a uniform mixed strategy while the other sweeps over all possible pure initial strategies in the setting of $\alpha_f=\alpha_w=\mathbf{0}$. However, we found in both cases that this slightly decreased the minimum value of $u_w$ to $\frac{1}{10}$ while leaving the maximum value unchanged and there were generally lower payoffs to the worker with larger values of $D$. We have added these Figures in Appendix~\ref{appendix:nfg-graphs} for completeness. 

% \ben{this paragraph should have less repeating what's in the table, and more conclusions:  worker never gets above .5, as expected, most are asymmetric outcomes, etc.}
% \ben{also want a brief summary of the effect of reference point, learning rate, and $D$, beyond just these three examples.}

% grid over payoffs creates a constant sum game
\paragraph{Strategic choice of initial strategy}
Given that the outcomes at convergence are dependent on the parameters, the strategic agent may want to choose the parameters to maximize their outcomes at convergence. 
% If an agent doesn't have any information on how their opponent will play, it's not feasible to strategically choose parameters (beyond guaranteeing no regret), so let's assume that both agents are going to use FTRL, and this is common knowledge.  Computational constraints may limit an agent's ability to choose $D$ or $\eta$, so 
For simplicity, we focus on the case when $D,\eta$ are fixed and shared, and  $\alpha_f,\alpha_w$ are fixed as well so that the only parameter the agents choose strategically is their initial strategy.   

For each pair of initial strategies, we can define the payoff of each player with their utility under the equilibrium at convergence, as displayed in Figure~\ref{fig:nfg-plots}.  This results in a constant-sum payoff matrix, where the agent's payoffs always sum to one. Thus,
% ~\footnote{Note that we can always increase the precision of the computation of these utilities by running the algorithm for more iterations.}
% That is, given a value of $\eta$, $D$, and a reference point for each player, this is
we can compute the minimax values
% i.e., the expected utility of any NE~\citep{v1928theorie} \ben{always fun to cite ancient literature, but since we use lots of basic concepts in game theory, I usually point to a textbook for everything, and not have individual citations for each concept} 
for each game.  The NE that achieve the minimax values are not necessarily unique, but in Figure~\ref{fig:nfg-plots}, we show one such NE (depicted as a gold star). 
%Note that there are many possible NE since there are multiple initial strategies that lead to the same payoff, i.e., the zero-sum game is degenerate. Plots~\ref{Figure:nfg-reference-1} and~\ref{Figure:nfg-no-reference} 
For example, for both $\alpha_f=1/2,\alpha_w=29/30$ and $\alpha_f=\alpha_w=\bf{0}$ (Figures~\ref{Figure:nfg-reference-1} and~\ref{Figure:nfg-no-reference}), the minimax value is $0.2$ for the worker and $0.8$ for the firm.  While the firm does leverage their first-mover advantage, the worker's value is more than the minimum value of $u_w$ across the possible initial strategies, which means the firm does not have unilateral power to force the worker to the worst outcome among those possible at convergence. However, this is not always the case, as when $\alpha_f=\frac{1}{6}$ and $\alpha_w=\frac{1}{2}$ (Figure~\ref{Figure:nfg-reference-2}), 
where the minimax value for the worker is exactly the minimum $u_w$, $\frac{1}{6}$, despite the majority of outcomes across the possible pure initial strategies resulting in $u_w = \frac{1}{2}$. 

Across all of our $\mathcal{G}^{(1)}$ experiments, we computed the minimax value and a NE of their associated constant-sum games, there was no unique minimax strategy at the pure strategy $0$, nor are the strategies used at convergence.  That is, they do not start nor end at the unique subgame-perfect equilibrium where the worker has no choice but to accept any offer, and consequently receives zero payoff.  Intriguingly, this same phenomenon occurs in the literature~\citep{camerer2011behavioral,tadelis} featuring experiments where people play the ultimatum game in a large variety of settings. They find that people generally offer the responder (i.e.\ the worker) between $20-50\%$ of the surplus, even though the game theoretic prediction is that the 
responder should only see the minimum offer~\citep{camerer2011behavioral,cooper2011dynamics,tadelis}. Some explanations for this behavior in this literature include people's preference for fairness or social norm influence~\citep{camerer2011behavioral} while others point to the evolution of fair outcomes when proposers have access to some previous deals responders have accepted or uncertainty exists in which strategy is best~\citep{nowak2000fairness,rand2013evolution}. While our goal was not to match any experimental value present in this literature, our experiments offer another approach for explaining for why responders can demand more and proposers offer more than the smallest possible offer: If both agents aim to ensure their play is robust against arbitrary play while still valuing play close to best responses, then each of the agents can use their knowledge that the other player is trying to do that to maximize their own payoffs.  At least in this setting, this results in the equivalent of fairness norms or the like, without requiring agents do anything but maximize their own utility. 
%that the about an algorithm that could be used to learn bargaining strategies, especially one that has rich outcomes as seen in Figure~\ref{fig:nfg-plots}, creates new information that agents can use to rationally determine an initial strategy that would lead to the highest utility they could get, i.e., via strategy choice in a zero-sum game.

% just a sentence about how this is not the only representation, see appendix for efg 1-round, make this a connection to the 2 round case. 

% \bigskip
% Finally, the ultimatum game can also be represented as an extensive-form game rather than a normal-form game.  The extensive-form game has the same equilibria, but this does not imply that learning in the extensive-form game is equivalent.  We observed convergence in this setting as well, though the worker tended to get smaller $u_w$ values than when using the normal-form representation

\subsection{2-Round Alternating Extensive-Form Bargaining Game}
\label{sec:efg-experiments}
In our second set of experiments, we consider agents learning bargaining strategies for $\mathcal{G}^{(2)}$. The space of strategies for each agent in $\mathcal{G}^{(2)}$ is larger (now each agent makes both an offer and an acceptance threshold), so we use a fixed $D=5$ and $\alpha_f=\alpha_w=\mathbf{0}$ throughout our experiments. Then, we vary $\delta\in\{0.1,0.55,0.9\}$ and $\eta\in\{0.1,0.5,0.8\}$. We chose to explore a variety of $\delta$ values since this value changes the second round utilities for the agents which could influence an agent's ability to make threats. For our purposes, we will define a \textit{credible} threat by the worker to be one where their strategy is to reject an offer lower than the equilibrium offer and their counter offer is a best-response, given the firm's strategy. Next, we define one kind of \textit{non-credible} threat the firm can make: If the worker is accepting the equilibrium offer, and $\delta \cdot \frac{D-1}{D}$ is greater than the equilibrium offer, then the firm is making a non-credible threat if their strategy is to reject $\frac{1}{D}$ with non-zero probability. 

For each initial condition setting, we run Algorithm~\ref{alg:ftrl} on pairs of the possible pure initial strategies for the agents characterized by the offer the agent makes as a proposer and the acceptance threshold it sets as a responder. For each outcome, we display the value $u_w$ as well as whether a non-credible (circles) and / or a credible threat (diagonal hatching)  occurred in the converged strategies. We show the results for $D=5, \eta=0.5$, and $\delta=0.1,0.55,0.9$ in Figure~\ref{fig:efg-plots} and additional plots for $\eta = 0.25$ and $\eta=0.8$ are in Appendix~\ref{appendix:efg-graphs}.

\begin{figure}[t]
    \centering
    \begin{subfigure}{0.32\textwidth}
        \includegraphics[height=4.75cm]{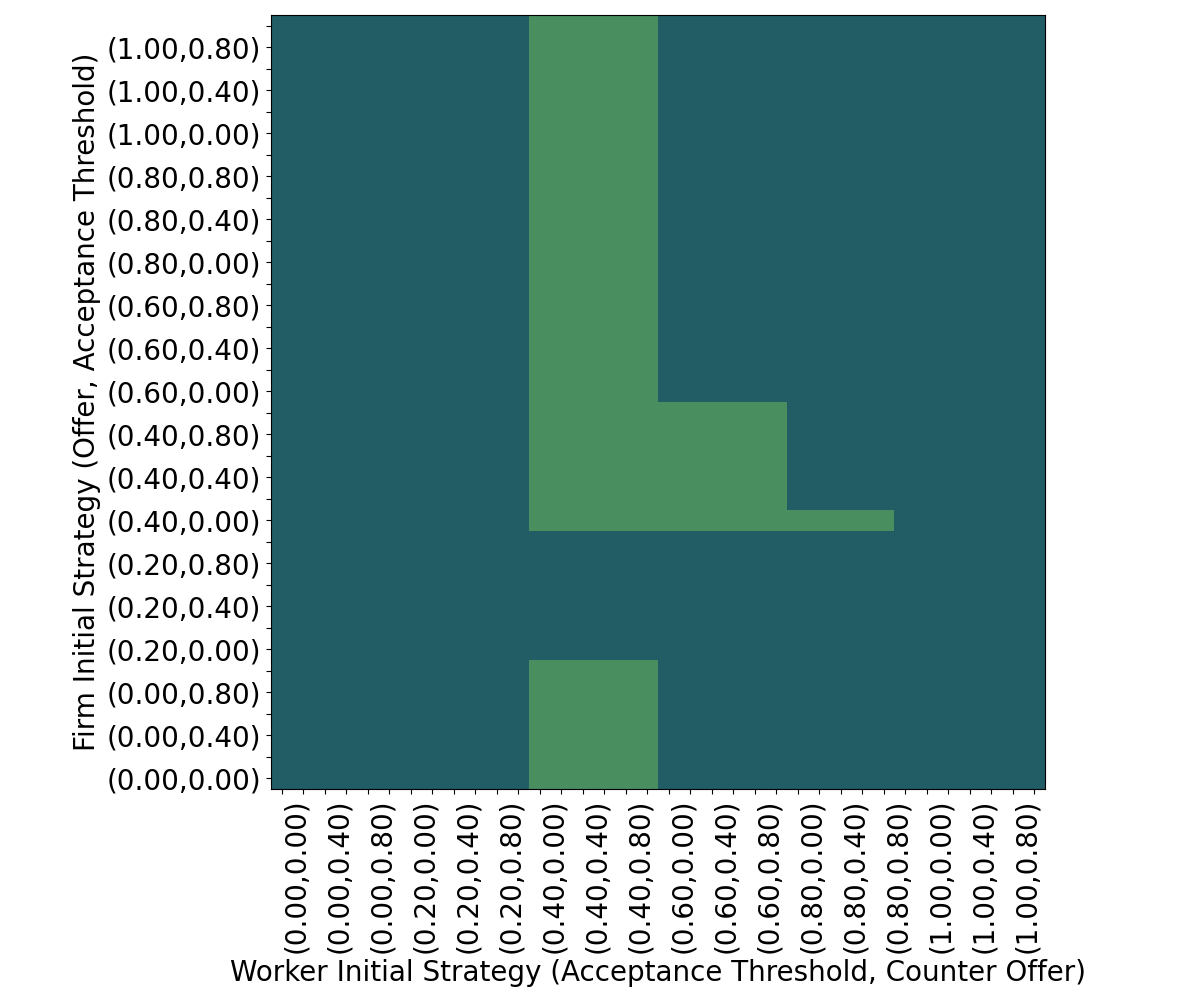}
        \caption{$\delta = 0.1$.}
        \label{Figure:efg-delta-1}
    \end{subfigure}\hfill
    \begin{subfigure}{0.32\textwidth}
    \centering
        \includegraphics[height=4.75cm]{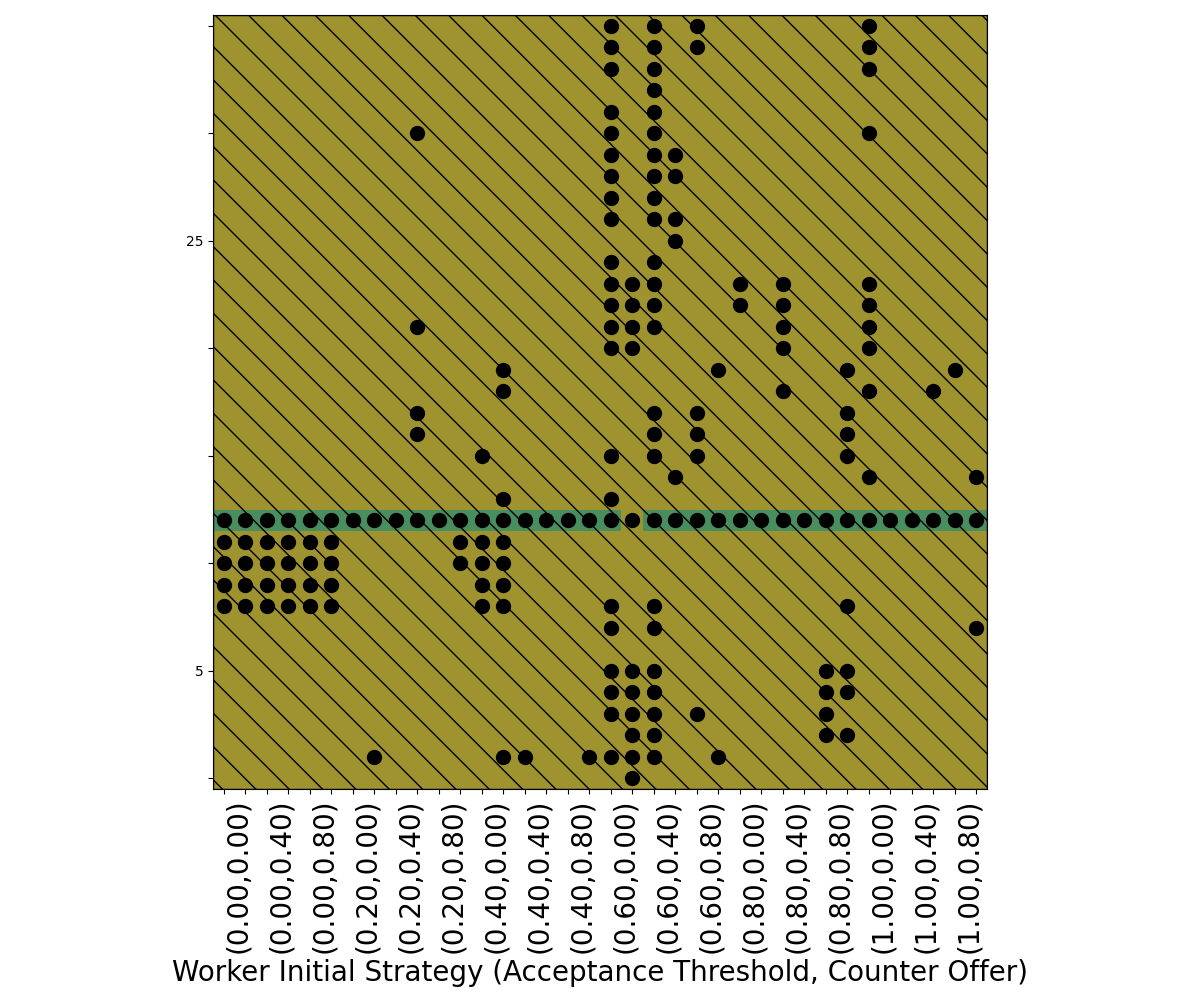}
        \caption{$\delta=0.55$. }
        \label{Figure:efg-delta-2}
    \end{subfigure}\hfill
     \begin{subfigure}{0.32\textwidth}
     \centering
        \includegraphics[height=4.75cm]{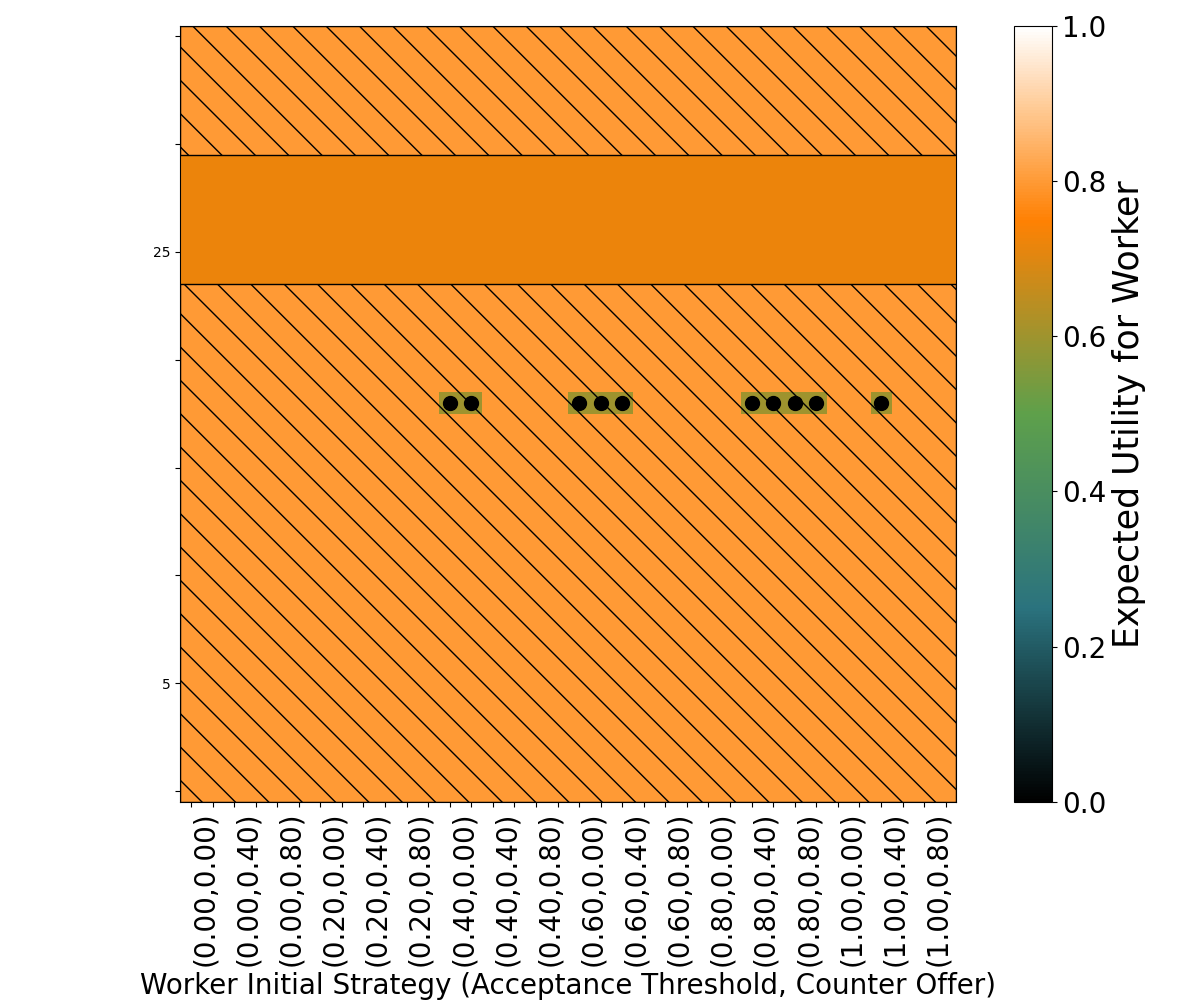}
        \caption{$\delta=0.9$.}
        \label{Figure:efg-delta-3}
    \end{subfigure}
    \caption{Comparison of NE payoff outcomes at convergence for the worker when agents are learning strategies for $\mathcal{G}^{(2)}$ using Algorithm~\ref{alg:ftrl} with $D=5,\eta = 0.5, \alpha_i=\bf{0}$, and a variety of time discount factors. The y-axis represents the range of possible pure initial strategies for the firm while the x-axis represents the range of possible pure initial strategies for the worker. The black circles indicate the existence of \textit{non-credible} firm threats. The diagonal hatching indicates \textit{credible} worker threats.}
    \label{fig:efg-plots}
\end{figure}

\paragraph{Payoffs at convergence} The plots in Figure~\ref{fig:efg-plots} not only show convergence to NE, but they also show a variety of NE outcomes and generally fewer possible $u_w$ values than those in Figure~\ref{fig:nfg-plots}.  Here, the worker's maximum value of $u_w$ shrinks with $\delta$. This is because smaller values of $\delta$ create smaller payoffs in the second round for both agents, and this causes the worker to prefer accepting smaller first round offers, leading to lower payoffs for the worker. Even when only using the zero reference point, we still see a non-trivial relationship between the initial strategies and the final NE outcome. We leave formalizing this dependence for future work, but our results so far already suggest complex economic behavior at convergence to explain the possibility of different NE outcomes, namely, \textit{threats}. 

\paragraph{Threats} In extensive-form games, agents specify strategies that say what action they will take at all possible nodes in the game, including strategies off the equilibrium path that are never realized. In Algorithm~\ref{alg:ftrl}, agents commit to a strategy and their opponent receives full feedback about this strategy, so each explicitly take into account what their opponent does at all nodes in the tree when updating their strategy. In fact, in Figure~\ref{fig:efg-plots}, we see the emergence of the both \textit{credible} and \textit{non-credible} threats as we described above. We know empirically that bargaining can involve non-credible threats~\citep{croson2003cheap}, so the fact that non-credible threats emerged through learning shows that our approach potentially offers some insight for how people choose bargaining strategies rationally in ways other than traditional game theory would predict, e.g., via backwards induction which only includes credible threats.

\bibliographystyle{ACM-Reference-Format}
\bibliography{bibliography}


\begin{thebibliography}{47}

%%% the Web address.

\ifx \showCODEN    \undefined \def \showCODEN     #1{\unskip}     \fi
\ifx \showDOI      \undefined \def \showDOI       #1{#1}\fi
\ifx \showISBNx    \undefined \def \showISBNx     #1{\unskip}     \fi
\ifx \showISBNxiii \undefined \def \showISBNxiii  #1{\unskip}     \fi
\ifx \showISSN     \undefined \def \showISSN      #1{\unskip}     \fi
\ifx \showLCCN     \undefined \def \showLCCN      #1{\unskip}     \fi
\ifx \shownote     \undefined \def \shownote      #1{#1}          \fi
\ifx \showarticletitle \undefined \def \showarticletitle #1{#1}   \fi
\ifx \showURL      \undefined \def \showURL       {\relax}        \fi
% invisible to TeX
\providecommand\bibfield[2]{#2}
\providecommand\bibinfo[2]{#2}
\providecommand\natexlab[1]{#1}
\providecommand\showeprint[2][]{arXiv:#2}

\bibitem[Anagnostides et~al\mbox{.}(2022)]%
        {anagnostides2022last}
\bibfield{author}{\bibinfo{person}{Ioannis Anagnostides}, \bibinfo{person}{Ioannis Panageas}, \bibinfo{person}{Gabriele Farina}, {and} \bibinfo{person}{Tuomas Sandholm}.} \bibinfo{year}{2022}\natexlab{}.
\newblock \showarticletitle{On last-iterate convergence beyond zero-sum games}. In \bibinfo{booktitle}{\emph{International Conference on Machine Learning}}. PMLR, \bibinfo{pages}{536--581}.
\newblock


\bibitem[Baker(1984)]%
        {baker1984concise}
\bibfield{author}{\bibinfo{person}{Alan Baker}.} \bibinfo{year}{1984}\natexlab{}.
\newblock \bibinfo{booktitle}{\emph{A concise introduction to the theory of numbers}}.
\newblock \bibinfo{publisher}{Cambridge University Press}.
\newblock


\bibitem[Barocas et~al\mbox{.}(2019)]%
        {barocas-hardt-narayanan}
\bibfield{author}{\bibinfo{person}{Solon Barocas}, \bibinfo{person}{Moritz Hardt}, {and} \bibinfo{person}{Arvind Narayanan}.} \bibinfo{year}{2019}\natexlab{}.
\newblock \bibinfo{booktitle}{\emph{Fairness and Machine Learning: Limitations and Opportunities}}.
\newblock \bibinfo{publisher}{fairmlbook.org}.
\newblock
\newblock
\shownote{\url{http://www.fairmlbook.org}}.


\bibitem[Beam et~al\mbox{.}(1999)]%
        {beam1999negotiations}
\bibfield{author}{\bibinfo{person}{Carrie Beam}, \bibinfo{person}{Arie Segev}, \bibinfo{person}{Martin Bichler}, {and} \bibinfo{person}{Ramayya Krishnan}.} \bibinfo{year}{1999}\natexlab{}.
\newblock \showarticletitle{On negotiations and deal making in electronic markets}.
\newblock \bibinfo{journal}{\emph{Information Systems Frontiers}}  \bibinfo{volume}{1} (\bibinfo{year}{1999}), \bibinfo{pages}{241--258}.
\newblock


\bibitem[Cai et~al\mbox{.}(2024)]%
        {cai2024uncoupled}
\bibfield{author}{\bibinfo{person}{Yang Cai}, \bibinfo{person}{Haipeng Luo}, \bibinfo{person}{Chen-Yu Wei}, {and} \bibinfo{person}{Weiqiang Zheng}.} \bibinfo{year}{2024}\natexlab{}.
\newblock \showarticletitle{Uncoupled and convergent learning in two-player zero-sum Markov games with bandit feedback}.
\newblock \bibinfo{journal}{\emph{Advances in Neural Information Processing Systems}}  \bibinfo{volume}{36} (\bibinfo{year}{2024}).
\newblock


\bibitem[Cai et~al\mbox{.}(2022)]%
        {cai2022finite}
\bibfield{author}{\bibinfo{person}{Yang Cai}, \bibinfo{person}{Argyris Oikonomou}, {and} \bibinfo{person}{Weiqiang Zheng}.} \bibinfo{year}{2022}\natexlab{}.
\newblock \showarticletitle{Finite-time last-iterate convergence for learning in multi-player games}.
\newblock \bibinfo{journal}{\emph{Advances in Neural Information Processing Systems}}  \bibinfo{volume}{35} (\bibinfo{year}{2022}), \bibinfo{pages}{33904--33919}.
\newblock


\bibitem[Camerer(2011)]%
        {camerer2011behavioral}
\bibfield{author}{\bibinfo{person}{Colin~F Camerer}.} \bibinfo{year}{2011}\natexlab{}.
\newblock \bibinfo{booktitle}{\emph{Behavioral game theory: Experiments in strategic interaction}}.
\newblock \bibinfo{publisher}{Princeton university press}.
\newblock


\bibitem[Cooper and Dutcher(2011)]%
        {cooper2011dynamics}
\bibfield{author}{\bibinfo{person}{David~J Cooper} {and} \bibinfo{person}{E~Glenn Dutcher}.} \bibinfo{year}{2011}\natexlab{}.
\newblock \showarticletitle{The dynamics of responder behavior in ultimatum games: a meta-study}.
\newblock \bibinfo{journal}{\emph{Experimental Economics}}  \bibinfo{volume}{14} (\bibinfo{year}{2011}), \bibinfo{pages}{519--546}.
\newblock


\bibitem[Croson et~al\mbox{.}(2003)]%
        {croson2003cheap}
\bibfield{author}{\bibinfo{person}{Rachel Croson}, \bibinfo{person}{Terry Boles}, {and} \bibinfo{person}{J~Keith Murnighan}.} \bibinfo{year}{2003}\natexlab{}.
\newblock \showarticletitle{Cheap talk in bargaining experiments: lying and threats in ultimatum games}.
\newblock \bibinfo{journal}{\emph{Journal of Economic Behavior \& Organization}} \bibinfo{volume}{51}, \bibinfo{number}{2} (\bibinfo{year}{2003}), \bibinfo{pages}{143--159}.
\newblock


\bibitem[Daskalakis and Panageas(2018)]%
        {daskalakis2018last}
\bibfield{author}{\bibinfo{person}{Constantinos Daskalakis} {and} \bibinfo{person}{Ioannis Panageas}.} \bibinfo{year}{2018}\natexlab{}.
\newblock \showarticletitle{Last-iterate convergence: Zero-sum games and constrained min-max optimization}.
\newblock \bibinfo{journal}{\emph{arXiv preprint arXiv:1807.04252}} (\bibinfo{year}{2018}).
\newblock


\bibitem[Debove et~al\mbox{.}(2016)]%
        {debove2016models}
\bibfield{author}{\bibinfo{person}{St{\'e}phane Debove}, \bibinfo{person}{Nicolas Baumard}, {and} \bibinfo{person}{Jean-Baptiste Andr{\'e}}.} \bibinfo{year}{2016}\natexlab{}.
\newblock \showarticletitle{Models of the evolution of fairness in the ultimatum game: a review and classification}.
\newblock \bibinfo{journal}{\emph{Evolution and Human Behavior}} \bibinfo{volume}{37}, \bibinfo{number}{3} (\bibinfo{year}{2016}), \bibinfo{pages}{245--254}.
\newblock


\bibitem[Deng et~al\mbox{.}(2022)]%
        {deng2022nash}
\bibfield{author}{\bibinfo{person}{Xiaotie Deng}, \bibinfo{person}{Xinyan Hu}, \bibinfo{person}{Tao Lin}, {and} \bibinfo{person}{Weiqiang Zheng}.} \bibinfo{year}{2022}\natexlab{}.
\newblock \showarticletitle{Nash convergence of mean-based learning algorithms in first price auctions}. In \bibinfo{booktitle}{\emph{Proceedings of the ACM Web Conference 2022}}. \bibinfo{pages}{141--150}.
\newblock


\bibitem[Diamond and Boyd(2016)]%
        {diamond2016cvxpy}
\bibfield{author}{\bibinfo{person}{Steven Diamond} {and} \bibinfo{person}{Stephen Boyd}.} \bibinfo{year}{2016}\natexlab{}.
\newblock \showarticletitle{{CVXPY}: {A} {P}ython-embedded modeling language for convex optimization}.
\newblock \bibinfo{journal}{\emph{Journal of Machine Learning Research}} \bibinfo{volume}{17}, \bibinfo{number}{83} (\bibinfo{year}{2016}), \bibinfo{pages}{1--5}.
\newblock


\bibitem[Elaydi et~al\mbox{.}({[n.\,d.]})]%
        {elaydiintroduction}
\bibfield{author}{\bibinfo{person}{Saber~N Elaydi} {et~al\mbox{.}}} \bibinfo{year}{[n.\,d.]}\natexlab{}.
\newblock \showarticletitle{An Introduction to Difference Equations [electronic resource]}.
\newblock  (\bibinfo{year}{[n.\,d.]}).
\newblock


\bibitem[Elzayn and Fish(2020)]%
        {elzayn2020effects}
\bibfield{author}{\bibinfo{person}{Hadi Elzayn} {and} \bibinfo{person}{Benjamin Fish}.} \bibinfo{year}{2020}\natexlab{}.
\newblock \showarticletitle{The effects of competition and regulation on error inequality in data-driven markets}. In \bibinfo{booktitle}{\emph{Proceedings of the 2020 conference on fairness, accountability, and transparency}}. \bibinfo{pages}{669--679}.
\newblock


\bibitem[Falk and Fischbacher(2006)]%
        {falk2006theory}
\bibfield{author}{\bibinfo{person}{Armin Falk} {and} \bibinfo{person}{Urs Fischbacher}.} \bibinfo{year}{2006}\natexlab{}.
\newblock \showarticletitle{A theory of reciprocity}.
\newblock \bibinfo{journal}{\emph{Games and economic behavior}} \bibinfo{volume}{54}, \bibinfo{number}{2} (\bibinfo{year}{2006}), \bibinfo{pages}{293--315}.
\newblock


\bibitem[Farina et~al\mbox{.}(2022)]%
        {farina2022near}
\bibfield{author}{\bibinfo{person}{Gabriele Farina}, \bibinfo{person}{Ioannis Anagnostides}, \bibinfo{person}{Haipeng Luo}, \bibinfo{person}{Chung-Wei Lee}, \bibinfo{person}{Christian Kroer}, {and} \bibinfo{person}{Tuomas Sandholm}.} \bibinfo{year}{2022}\natexlab{}.
\newblock \showarticletitle{Near-optimal no-regret learning dynamics for general convex games}.
\newblock \bibinfo{journal}{\emph{Advances in Neural Information Processing Systems}}  \bibinfo{volume}{35} (\bibinfo{year}{2022}), \bibinfo{pages}{39076--39089}.
\newblock


\bibitem[Feri and Gantner(2011)]%
        {feri2011bargaining}
\bibfield{author}{\bibinfo{person}{Francesco Feri} {and} \bibinfo{person}{Anita Gantner}.} \bibinfo{year}{2011}\natexlab{}.
\newblock \showarticletitle{Bargaining or searching for a better price?--an experimental study}.
\newblock \bibinfo{journal}{\emph{Games and Economic Behavior}} \bibinfo{volume}{72}, \bibinfo{number}{2} (\bibinfo{year}{2011}), \bibinfo{pages}{376--399}.
\newblock


\bibitem[Fish and Stark(2022)]%
        {fish2022s}
\bibfield{author}{\bibinfo{person}{Benjamin Fish} {and} \bibinfo{person}{Luke Stark}.} \bibinfo{year}{2022}\natexlab{}.
\newblock \showarticletitle{It’s not fairness, and it’s not fair: the failure of distributional equality and the promise of relational equality in complete-information hiring games}. In \bibinfo{booktitle}{\emph{Proceedings of the 2nd ACM Conference on Equity and Access in Algorithms, Mechanisms, and Optimization}}. \bibinfo{pages}{1--15}.
\newblock


\bibitem[Fudenberg and Maskin(1986)]%
        {fudenberg1986folk}
\bibfield{author}{\bibinfo{person}{Drew Fudenberg} {and} \bibinfo{person}{Eric Maskin}.} \bibinfo{year}{1986}\natexlab{}.
\newblock \showarticletitle{The folk theorem in repeated games with discounting or with incomplete information}.
\newblock \bibinfo{journal}{\emph{Econometrica}} \bibinfo{volume}{54}, \bibinfo{number}{3} (\bibinfo{year}{1986}), \bibinfo{pages}{533--554}.
\newblock


\bibitem[Giannou et~al\mbox{.}(2021)]%
        {giannou2021survival}
\bibfield{author}{\bibinfo{person}{Angeliki Giannou}, \bibinfo{person}{Emmanouil~Vasileios Vlatakis-Gkaragkounis}, {and} \bibinfo{person}{Panayotis Mertikopoulos}.} \bibinfo{year}{2021}\natexlab{}.
\newblock \showarticletitle{Survival of the strictest: Stable and unstable equilibria under regularized learning with partial information}. In \bibinfo{booktitle}{\emph{Conference on Learning Theory}}. PMLR, \bibinfo{pages}{2147--2148}.
\newblock


\bibitem[Golowich et~al\mbox{.}(2020)]%
        {golowich2020tight}
\bibfield{author}{\bibinfo{person}{Noah Golowich}, \bibinfo{person}{Sarath Pattathil}, {and} \bibinfo{person}{Constantinos Daskalakis}.} \bibinfo{year}{2020}\natexlab{}.
\newblock \showarticletitle{Tight last-iterate convergence rates for no-regret learning in multi-player games}.
\newblock \bibinfo{journal}{\emph{Advances in neural information processing systems}}  \bibinfo{volume}{33} (\bibinfo{year}{2020}), \bibinfo{pages}{20766--20778}.
\newblock


\bibitem[Hannan(1957)]%
        {hannan1957approximation}
\bibfield{author}{\bibinfo{person}{James Hannan}.} \bibinfo{year}{1957}\natexlab{}.
\newblock \showarticletitle{Approximation to Bayes risk in repeated play}.
\newblock \bibinfo{journal}{\emph{Contributions to the Theory of Games}} \bibinfo{volume}{3}, \bibinfo{number}{2} (\bibinfo{year}{1957}), \bibinfo{pages}{97--139}.
\newblock


\bibitem[Hart and Mas-Colell(2000)]%
        {hart2000simple}
\bibfield{author}{\bibinfo{person}{Sergiu Hart} {and} \bibinfo{person}{Andreu Mas-Colell}.} \bibinfo{year}{2000}\natexlab{}.
\newblock \showarticletitle{A simple adaptive procedure leading to correlated equilibrium}.
\newblock \bibinfo{journal}{\emph{Econometrica}} \bibinfo{volume}{68}, \bibinfo{number}{5} (\bibinfo{year}{2000}), \bibinfo{pages}{1127--1150}.
\newblock


\bibitem[Hazan et~al\mbox{.}(2016)]%
        {hazan2016introduction}
\bibfield{author}{\bibinfo{person}{Elad Hazan} {et~al\mbox{.}}} \bibinfo{year}{2016}\natexlab{}.
\newblock \showarticletitle{Introduction to online convex optimization}.
\newblock \bibinfo{journal}{\emph{Foundations and Trends{\textregistered} in Optimization}} \bibinfo{volume}{2}, \bibinfo{number}{3-4} (\bibinfo{year}{2016}), \bibinfo{pages}{157--325}.
\newblock


\bibitem[Hsieh et~al\mbox{.}(2021)]%
        {hsieh2021adaptive}
\bibfield{author}{\bibinfo{person}{Yu-Guan Hsieh}, \bibinfo{person}{Kimon Antonakopoulos}, {and} \bibinfo{person}{Panayotis Mertikopoulos}.} \bibinfo{year}{2021}\natexlab{}.
\newblock \showarticletitle{Adaptive learning in continuous games: Optimal regret bounds and convergence to nash equilibrium}. In \bibinfo{booktitle}{\emph{Conference on Learning Theory}}. PMLR, \bibinfo{pages}{2388--2422}.
\newblock


\bibitem[Juda et~al\mbox{.}(2009)]%
        {juda2009pricing}
\bibfield{author}{\bibinfo{person}{Adam~Isaac Juda}, \bibinfo{person}{S Muthukrishnan}, {and} \bibinfo{person}{Ashish Rastogi}.} \bibinfo{year}{2009}\natexlab{}.
\newblock \showarticletitle{Pricing guidance in ad sale negotiations: the PrintAds example}. In \bibinfo{booktitle}{\emph{Proceedings of the Third International Workshop on Data Mining and Audience Intelligence for Advertising}}. \bibinfo{pages}{61--68}.
\newblock


\bibitem[Kamp and Fish(2025)]%
        {kamp2025equal}
\bibfield{author}{\bibinfo{person}{Serafina Kamp} {and} \bibinfo{person}{Benjamin Fish}.} \bibinfo{year}{2025}\natexlab{}.
\newblock \showarticletitle{Equal Merit Does Not Imply Equality: Discrimination at Equilibrium in a Hiring Market with Symmetric Agents}. In \bibinfo{booktitle}{\emph{Proceedings of the AAAI Conference on Artificial Intelligence}}, Vol.~\bibinfo{volume}{39}. \bibinfo{pages}{17724--17732}.
\newblock


\bibitem[Knuth(1976)]%
        {knuth1976mathematics}
\bibfield{author}{\bibinfo{person}{Donald~E Knuth}.} \bibinfo{year}{1976}\natexlab{}.
\newblock \showarticletitle{Mathematics and Computer Science: Coping with Finiteness: Advances in our ability to compute are bringing us substantially closer to ultimate limitations.}
\newblock \bibinfo{journal}{\emph{Science}} \bibinfo{volume}{194}, \bibinfo{number}{4271} (\bibinfo{year}{1976}), \bibinfo{pages}{1235--1242}.
\newblock


\bibitem[Korenok and Munro(2021)]%
        {korenok2021wage}
\bibfield{author}{\bibinfo{person}{Oleg Korenok} {and} \bibinfo{person}{David Munro}.} \bibinfo{year}{2021}\natexlab{}.
\newblock \showarticletitle{Wage bargaining in a matching market: Experimental evidence}.
\newblock \bibinfo{journal}{\emph{Labour Economics}}  \bibinfo{volume}{73} (\bibinfo{year}{2021}), \bibinfo{pages}{102078}.
\newblock


\bibitem[Lee et~al\mbox{.}(2021)]%
        {lee2021last}
\bibfield{author}{\bibinfo{person}{Chung-Wei Lee}, \bibinfo{person}{Christian Kroer}, {and} \bibinfo{person}{Haipeng Luo}.} \bibinfo{year}{2021}\natexlab{}.
\newblock \showarticletitle{Last-iterate convergence in extensive-form games}.
\newblock \bibinfo{journal}{\emph{Advances in Neural Information Processing Systems}}  \bibinfo{volume}{34} (\bibinfo{year}{2021}), \bibinfo{pages}{14293--14305}.
\newblock


\bibitem[LeVeque(1996)]%
        {leveque1996fundamentals}
\bibfield{author}{\bibinfo{person}{William~Judson LeVeque}.} \bibinfo{year}{1996}\natexlab{}.
\newblock \bibinfo{booktitle}{\emph{Fundamentals of number theory}}.
\newblock \bibinfo{publisher}{Courier Corporation}.
\newblock


\bibitem[Marden et~al\mbox{.}(2007)]%
        {marden2007regret}
\bibfield{author}{\bibinfo{person}{Jason~R Marden}, \bibinfo{person}{G{\"u}rdal Arslan}, {and} \bibinfo{person}{Jeff~S Shamma}.} \bibinfo{year}{2007}\natexlab{}.
\newblock \showarticletitle{Regret based dynamics: convergence in weakly acyclic games}. In \bibinfo{booktitle}{\emph{Proceedings of the 6th international joint conference on Autonomous agents and multiagent systems}}. \bibinfo{pages}{1--8}.
\newblock


\bibitem[Mertikopoulos(2019)]%
        {mertikopoulos2019online}
\bibfield{author}{\bibinfo{person}{Panayotis Mertikopoulos}.} \bibinfo{year}{2019}\natexlab{}.
\newblock \emph{\bibinfo{title}{Online optimization and learning in games: Theory and applications}}.
\newblock \bibinfo{thesistype}{Ph.\,D. Dissertation}. \bibinfo{school}{Grenoble 1 UGA-Universit{\'e} Grenoble Alpes}.
\newblock


\bibitem[Milli et~al\mbox{.}(2019)]%
        {milli2019social}
\bibfield{author}{\bibinfo{person}{Smitha Milli}, \bibinfo{person}{John Miller}, \bibinfo{person}{Anca~D Dragan}, {and} \bibinfo{person}{Moritz Hardt}.} \bibinfo{year}{2019}\natexlab{}.
\newblock \showarticletitle{The social cost of strategic classification}. In \bibinfo{booktitle}{\emph{Proceedings of the conference on fairness, accountability, and transparency}}. \bibinfo{pages}{230--239}.
\newblock


\bibitem[Nowak et~al\mbox{.}(2000)]%
        {nowak2000fairness}
\bibfield{author}{\bibinfo{person}{Martin~A Nowak}, \bibinfo{person}{Karen~M Page}, {and} \bibinfo{person}{Karl Sigmund}.} \bibinfo{year}{2000}\natexlab{}.
\newblock \showarticletitle{Fairness versus reason in the ultimatum game}.
\newblock \bibinfo{journal}{\emph{Science}} \bibinfo{volume}{289}, \bibinfo{number}{5485} (\bibinfo{year}{2000}), \bibinfo{pages}{1773--1775}.
\newblock


\bibitem[Osborne and Rubinstein(1990)]%
        {osborne1990bargaining}
\bibfield{author}{\bibinfo{person}{Martin~J Osborne} {and} \bibinfo{person}{Ariel Rubinstein}.} \bibinfo{year}{1990}\natexlab{}.
\newblock \bibinfo{booktitle}{\emph{Bargaining and Markets}}.
\newblock \bibinfo{publisher}{Academic Press Limited}.
\newblock


\bibitem[Parvaneh et~al\mbox{.}(2021)]%
        {parvaneh2021show}
\bibfield{author}{\bibinfo{person}{Amin Parvaneh}, \bibinfo{person}{Ehsan Abbasnejad}, \bibinfo{person}{Qi Wu}, \bibinfo{person}{Javen~Qinfeng Shi}, {and} \bibinfo{person}{Anton Van Den~Hengel}.} \bibinfo{year}{2021}\natexlab{}.
\newblock \showarticletitle{Show, price and negotiate: A negotiator with online value look-ahead}.
\newblock \bibinfo{journal}{\emph{IEEE Transactions on Multimedia}}  \bibinfo{volume}{24} (\bibinfo{year}{2021}), \bibinfo{pages}{1426--1434}.
\newblock


\bibitem[Prasad et~al\mbox{.}(2019)]%
        {prasad2019impact}
\bibfield{author}{\bibinfo{person}{Sanjay Prasad}, \bibinfo{person}{Ravi Shankar}, {and} \bibinfo{person}{Sreejit Roy}.} \bibinfo{year}{2019}\natexlab{}.
\newblock \showarticletitle{Impact of bargaining power on supply chain profit allocation: a game-theoretic study}.
\newblock \bibinfo{journal}{\emph{Journal of Advances in Management Research}} \bibinfo{volume}{16}, \bibinfo{number}{3} (\bibinfo{year}{2019}), \bibinfo{pages}{398--416}.
\newblock


\bibitem[Rakhlin and Sridharan(2013)]%
        {rakhlin2013optimization}
\bibfield{author}{\bibinfo{person}{Sasha Rakhlin} {and} \bibinfo{person}{Karthik Sridharan}.} \bibinfo{year}{2013}\natexlab{}.
\newblock \showarticletitle{Optimization, learning, and games with predictable sequences}.
\newblock \bibinfo{journal}{\emph{Advances in Neural Information Processing Systems}}  \bibinfo{volume}{26} (\bibinfo{year}{2013}).
\newblock


\bibitem[Rand et~al\mbox{.}(2013)]%
        {rand2013evolution}
\bibfield{author}{\bibinfo{person}{David~G Rand}, \bibinfo{person}{Corina~E Tarnita}, \bibinfo{person}{Hisashi Ohtsuki}, {and} \bibinfo{person}{Martin~A Nowak}.} \bibinfo{year}{2013}\natexlab{}.
\newblock \showarticletitle{Evolution of fairness in the one-shot anonymous ultimatum game}.
\newblock \bibinfo{journal}{\emph{Proceedings of the National Academy of Sciences}} \bibinfo{volume}{110}, \bibinfo{number}{7} (\bibinfo{year}{2013}), \bibinfo{pages}{2581--2586}.
\newblock


\bibitem[Roth(1985)]%
        {roth1985game}
\bibfield{author}{\bibinfo{person}{Alvin~E Roth}.} \bibinfo{year}{1985}\natexlab{}.
\newblock \bibinfo{booktitle}{\emph{Game-theoretic models of bargaining}}.
\newblock \bibinfo{publisher}{Cambridge University Press}.
\newblock


\bibitem[Shalev-Shwartz and Singer(2006)]%
        {shalev2006convex}
\bibfield{author}{\bibinfo{person}{Shai Shalev-Shwartz} {and} \bibinfo{person}{Yoram Singer}.} \bibinfo{year}{2006}\natexlab{}.
\newblock \showarticletitle{Convex repeated games and Fenchel duality}.
\newblock \bibinfo{journal}{\emph{Advances in neural information processing systems}}  \bibinfo{volume}{19} (\bibinfo{year}{2006}).
\newblock


\bibitem[Tadelis(2013)]%
        {tadelis}
\bibfield{author}{\bibinfo{person}{Steven Tadelis}.} \bibinfo{year}{2013}\natexlab{}.
\newblock \showarticletitle{Game Theory; an Introduction}.
\newblock  (\bibinfo{year}{2013}).
\newblock


\bibitem[Thaler(1988)]%
        {thaler1988anomalies}
\bibfield{author}{\bibinfo{person}{Richard~H Thaler}.} \bibinfo{year}{1988}\natexlab{}.
\newblock \showarticletitle{Anomalies: The ultimatum game}.
\newblock \bibinfo{journal}{\emph{Journal of economic perspectives}} \bibinfo{volume}{2}, \bibinfo{number}{4} (\bibinfo{year}{1988}), \bibinfo{pages}{195--206}.
\newblock


\bibitem[Vlatakis-Gkaragkounis et~al\mbox{.}(2020)]%
        {vlatakis2020no}
\bibfield{author}{\bibinfo{person}{Emmanouil-Vasileios Vlatakis-Gkaragkounis}, \bibinfo{person}{Lampros Flokas}, \bibinfo{person}{Thanasis Lianeas}, \bibinfo{person}{Panayotis Mertikopoulos}, {and} \bibinfo{person}{Georgios Piliouras}.} \bibinfo{year}{2020}\natexlab{}.
\newblock \showarticletitle{No-regret learning and mixed nash equilibria: They do not mix}.
\newblock \bibinfo{journal}{\emph{Advances in Neural Information Processing Systems}}  \bibinfo{volume}{33} (\bibinfo{year}{2020}), \bibinfo{pages}{1380--1391}.
\newblock


\bibitem[Zhang and Chung(2020)]%
        {zhang2020price}
\bibfield{author}{\bibinfo{person}{Lingling Zhang} {and} \bibinfo{person}{Doug~J Chung}.} \bibinfo{year}{2020}\natexlab{}.
\newblock \showarticletitle{Price bargaining and competition in online platforms: An empirical analysis of the daily deal market}.
\newblock \bibinfo{journal}{\emph{Marketing Science}} \bibinfo{volume}{39}, \bibinfo{number}{4} (\bibinfo{year}{2020}), \bibinfo{pages}{687--706}.
\newblock


\end{thebibliography}

\appendix
% \documentclass{article}
% \usepackage{graphicx, amsmath, enumitem,amssymb,amsthm} % Required for inserting images
% \usepackage{ulem, amsfonts, tikz}
% \usetikzlibrary{positioning, calc}
% \usetikzlibrary {decorations.pathreplacing}
% \usepackage{algorithm,xcolor,algorithmic}
% \DeclareMathOperator*{\argmax}{arg\,max}
% \DeclareMathOperator*{\argmin}{arg\,min}
% \usepackage{hyperref}
% \usepackage{amstext} % for \text macro
% \usepackage{array, makecell,tabularx}   % for \newcolumntype macro
% \usepackage{chngpage}
% \usepackage{bbm}

% \newcolumntype{L}{>{$}l<{$}} % math-mode version of "l" column type
\setlength\parindent{0pt}

\newtheorem{innercustomlem}{Lemma}

\newenvironment{customlem}[1]
  {\renewcommand\theinnercustomlem{#1}\innercustomlem}
  {\endinnercustomlem}

  \newtheorem{innercustomclm}{Claim}

\newenvironment{customclm}[1]
  {\renewcommand\theinnercustomclm{#1}\innercustomclm}
  {\endinnercustomclm}

\newtheorem{innercustomdef}{Definition}

\newenvironment{customdef}[1]
  {\renewcommand\theinnercustomdef{#1}\innercustomdef}
  {\endinnercustomdef}

%% notation key
% firm = first firm in the game
% worker = first worker in the game

% This document contains proofs of convergence to NE when agents use FTRL with the Euclidean regularizer.

\section{$\mathcal{G}^{(1)}$ Under Algorithm~\ref{alg:ftrl} Always Converges to $\epsilon$-Mixed NE}
\label{appendix-proofs}
\subsection{Discretization Justification}
Previous work has shown that no-regret strategies for the discretized normal form ultimatum game get sublinear regret with respect to the best action in hindsight from the original continuous space~\citep{kamp2025equal}. Additionally, in the ultimatum game, any strategy profile that is in $\epsilon$-NE with respect to mixed strategies in the discretized space is also in an $\epsilon$-NE with respect to pure strategies in the action set $[0,1]$ which we formalize in Proposition~\ref{prop:ftrl_discrete_ne_implies_continuous-main-paper}.
\begin{customprp}{1}
\label{prop:ftrl_discrete_ne_implies_continuous-main-paper}
    Let $\epsilon>0$ and suppose $(x_f^{(T)},x_w^{(T)})$, is an $\epsilon$-Nash Equilibrium for some $\epsilon>0$ with respect to mixed strategies in $\Delta(\mathcal{A})$. Then, $(x_f^{(T)},x_w^{(T)})$ is an $\epsilon$-Nash Equilibrium with respect to pure strategies from the action set $[0,1]$.
\end{customprp}
\begin{proof}
    Let $\epsilon>0$ and suppose $(x_f^{(T)},x_w^{(T)})$ is in $\epsilon$-Nash Equilibrium for $x_f^{(T)}, x_w^{(T)} \in \Delta(\mathcal{A})$. 
    % Then, by Theorem~\ref{thrm:ftrl_convergence_n=1}, there exists $k \in \{1,\ldots,D\}$ such that $w_{\max}^{(T)} =\frac{k}{D}$ where $x_{w,w_{\max}^{(T)}}^{(T)} \ge \frac{1}{D-k+1}$ and $x_{f,w_{\max}^{(T)}}^{(T)} \ge 1-\epsilon$. 
    We will now show that 
    % $(x_f^{(T)},x_w^{(T)})$ is an $\epsilon$-best response for the firm and worker, respectively, in the continuous game.That is, 
    there is no action in $[0,1]$ that gets at least $\epsilon$ more utility than $x_w^{(T)}$ for the worker and similarly for the firm. Since $(x_f^{(T)},x_w^{(T)})$ is in an $\epsilon$-NE and $\mathcal{A} \subset [0,1]$, then it suffices to argue that the best response for the worker to $x_f^{(T)}$ is an action $\mathcal{A}$ and similarly the best response for the firm to $x_w^{(T)}$ is also an action in $\mathcal{A}$.
    
    First, for the worker, their best response to the firm's strategy $x_f^{(T)}$ from the action set $[0,1]$ is the acceptance threshold, say $a_w^*$, that accepts every offer that has positive probability in $x_f^{(T)}$. Since $x_f^{(T)} \in \Delta(\mathcal{A})$, then clearly $a_w^* \in \mathcal{A}$.
    
    % but $(x_f^{(T)},x_w^{(T)})$ being an $\epsilon$-NE and $f_{\min}^{(T)} \in \mathcal{A}$ implies
    % % $x_{f,w_{\max}^{(T)}}^{(T)} \ge 1-\epsilon$ implies
    % $$u_w(x_f^{(T)},x_w^{(T)}) \ge u_w(x_f^{(T)}, f_{\min}^{(T)})-\epsilon.$$
    %  Therefore, no action in the continuous set of actions that gets at least $\epsilon$ more utility than $x_w^{(T)}$.

     Next, for the firm, their best response to the worker's strategy $x_w^{(T)}$ is the offer \[a_f^* = \arg\max_{a\in[0,1]} \left(\sum_{a' \le a}x_{w,a}^{(T)}\right) \cdot (1-a).\]
     That is, if two offers are accepted by the worker with the same probability, then the firm prefers the lower offer. Thus, $a_f^*$ will be exactly equal to an acceptance threshold that the worker plays with positive probability in $x_w^{(T)}$. Again, since $x_w^{(T)} \in \Delta(\mathcal{A})$, then clearly $a_f^* \in \mathcal{A}$.
     
    %  for all $\ell \in\{1,\ldots,k\}$, consider an offer $a \in [0,1]$ from the continuous game where $w_{\max}^{(T)} - \frac{\ell}{D} \le a < w_{\max}^{(T)}-\frac{\ell-1}{D}$. Then, \begin{align*}
    % u_f(a,x_r^{(T)}) &= (1-\sum_{i=0}^{\ell-1} x_{w,w_{\max}^{(T)}-\frac{i}{D}}^{(T)})\cdot (1-a) \\&\le (1-\sum_{i=0}^{\ell-1} x_{w,w_{\max}^{(T)}-\frac{i}{D}}^{(T)})\cdot (1-(w_{\max}^{(T)} - \frac{\ell}{D}))\\
    % &= u_f(w_{\max}^{(T)} - \frac{\ell}{D},x_r^{(T)})
    % \end{align*}

    % So, if $w_{\max}^{(T)}$ is a best response offer with respect to the offer set $\{\frac{1}{D},\ldots,1\}$, then, it must also be a best response with respect to the offer set $[0,1]$. Therefore, if $x_{f,w_{\max}^{(T)}}^{(T)} \ge 1-\epsilon$, then $x_f^{(T)}$ is an $\epsilon$-best response for the firm in the continuous game as well. 
\end{proof}
\subsection{Set-up and the Lagrangian}
The Lagrangian of this quadratic program for Algorithm~\ref{alg:ftrl} is 
\begin{align*}
    \mathcal{L}_i(x_i,\lambda_i, \mu_i) = \frac{1}{2}\|x_i\|_2^2 - \eta \langle U_i^t, x_i\rangle +\lambda_i \left(\sum_{a \in \mathcal{A}} x_{i,a}  - 1\right) - \sum_{a \in \mathcal{A}}\mu_{i,a} x_{i,a}
\end{align*}

The dual of this Lagrangian is 
\begin{align*}
    \max_{\lambda_i, \mu_i} \min_{x \in \mathbb{R}^{|\mathcal{A}|}}& \mathcal{L}_i(x,\lambda_i, \mu_i) \\
    \text{subject to}&\\
     & \mu_i \ge 0
\end{align*}

The quadratic program has strong duality by Slater's condition since the objective function is convex, the inequality constraint is convex, the equality constraint is affine, and there exists a point $x \in \Delta(\mathcal{A})$ where the equality constraint is satisfied and the inequality is strictly satisfied.

Then, by the KKT theorem, any problem that satisfies strong duality also satisfies the following KKT conditions:
\begin{itemize}
    \item \textbf{Stationarity:}  $0 \in \nabla \mathcal{L}(x,\lambda_i,\mu_i)|_{x=x^*}$ for the primal optimal $x^*$.
    \item \textbf{Primal Feasibility:} The primal constraints  are satisfied for the primal optimal $x^*$.
    \item \textbf{Dual Feasibility:}  $\mu_a \ge 0, \forall a \in \mathcal{A}$ for the dual optimal variables.
    \item \textbf{Complementary Slackness:} $\mu_ax_a^*=0$. 
\end{itemize}
% Stationarity implies that we can find the optimal $x$ value by taking the gradient of $\mathcal{L}_i(x,\lambda_i,\mu_i)$ and set it equal to 0 to solve for the conditions that each $x_{i,a}$ must satisfy in the chosen optimal mixed strategy $x_i^{(t+1)}$. 
Notably, by stationarity, for each $i \in\{f,w\}$  and for each $a \in \mathcal{A}$, 
$$x_{i,a}^{(t+1)} = \eta  U_{i,a}^{(t)}-\lambda_i + \mu_{i,a}.$$

\begin{customclm}{1}
\label{claim:stationarity_reqs}
    If $x_{i,a}^{(t+1)} >0$ and $x_{i,a'}^{(t+1)} >0$, then \begin{align*}
    x_{i,a}^{(t+1)}-x_{i,a'}^{(t+1)} = \eta U_{i,a}^{(t)}-\eta U_{i,a'}^{(t)} &\label{lagrangian_1}\tag{1}
\end{align*}
\end{customclm} 
\begin{proof}
    If $x_{i,a}^{(t+1)} >0, x_{i,a'}^{(t+1)} >0$, then by complementary slackness, we have $\mu_{i,a} = \mu_{i,a'} = 0$.

    By stationarity, this implies that $$x_{i,a}^{(t+1)} = \eta U_{i,a}^{(t)} - \lambda_i,$$ and 
    $$x_{i,a'}^{(t+1)} = \eta U_{i,a'}^{(t)} - \lambda_i.$$ 

    The claim immediately follows.
\end{proof}

\begin{customclm}{2}
\label{claim:kkt_ordering}
    Consider agent $i$ and two possible strategies of $i$: $a,a' \in \mathcal{A}$. If at least one of $x_{i,a}^{(t+1)}, x_{i,a'}^{(t+1)}$ has non-zero probability mass, then $x_{i,a}^{(t+1)} \ge x_{i,a'}^{(t+1)} $ if and only if $U_{i,a}^{(t)} \ge U_{i,a'}^{(t)}$ with equality if and only if $x_{i,a}^{(t+1)} = x_{i,a'}^{(t+1)}$.
\end{customclm}
\begin{proof}
    To begin, the KKT conditions imply 
    $$x_{i,a}^{(t+1)}-\eta U_{i,a}^{(t)}+\lambda_i \ge 0,$$
    and 
    $$x_{i,a}^{(t+1)}(x_{i,a}^{(t+1)}-\eta U_{i,a}^{(t)}+\lambda_i) = 0.$$
    This implies $$x_{i,a}^{(t)} = \begin{cases}
        \eta U_{i,a}^{(t)}-\lambda_i & \lambda_i < \eta U_{i,a}^{(t)}\\
        0 & \lambda_i \ge \eta U_{i,a}^{(t)}
    \end{cases}$$

    First, suppose $x_{i,a}^{(t+1)}\ge x_{i,a'}^{(t+1)}$. If $x_{i,a}^{(t+1)}>0$ and $x_{i,a'}^{(t+1)} = 0$, then from above we have $ \eta U_{i,a'}^{(t)}\le \lambda_i <\eta U_{i,a}^{(t)}$ and immediately we have $U_{i,a'}^{(t)}< U_{i,a}^{(t)}$. If both $x_{i,a}^{(t+1)}>0$ and $x_{i,a'}^{(t+1)} > 0$, then from Claim~\ref{claim:stationarity_reqs}, we have $$x_{i,a}^{(t+1)} - x_{i,a'}^{(t+1)}  = \eta\left(U_{i,a}^{(t)}-U_{i,a'}^{(t)}\right)\ge 0,$$
    so we have $U_{i,a'}^{(t)}\le U_{i,a}^{(t)}$ with equality if and only if $x_{i,a}^{(t+1)} = x_{i,a'}^{(t+1)}$.

    Next, suppose $U_{i,a'}^{(t)}\le U_{i,a}^{(t)}$. If both $x_{i,a}^{(t+1)}>0$ and $x_{i,a'}^{(t+1)} > 0$, then, $x_{i,a}^{(t+1)} \ge x_{i,a'}^{(t+1)}$ with equality if and only if $U_{i,a'}^{(t)}= U_{i,a}^{(t)}$ immediately follows from Claim~\ref{claim:stationarity_reqs}. Next, if $x_{i,a}^{(t+1)}>0$ and $x_{i,a'}^{(t+1)} = 0$, then we immediately have $x_{i,a}^{(t+1)}>x_{i,a'}^{(t+1)}$. Further, it cannot be the case that $U_{i,a'}^{(t)} =  U_{i,a}^{(t)}$ because this case implies $ \eta U_{i,a'}^{(t)}\le \lambda_i <\eta U_{i,a}^{(t)}$. Finally, if $x_{i,a}^{(t+1)}=0$ and $x_{i,a'}^{(t+1)} >0$, then we must have $ \eta U_{i,a}^{(t)}\le \lambda_i <\eta U_{i,a'}^{(t)}$ which contradicts our original assumption $U_{i,a'}^{(t)}\le U_{i,a}^{(t)}$ and we can conclude that such a probability assignment in $x_i^{(t+1)}$ is not possible.

\end{proof}
\newpage

\subsection{Convergence to $\epsilon$-Mixed NE}

First we state the main theorem to prove in this section. We will then prove several lemmas that will be necessary to prove this theorem. We will close by proving the main theorem. Throughout this section, we assume a firm and worker agent are learning strategies for $\mathcal{G}^{(1)}$ using Algorithm~\ref{alg:ftrl} parameterized by any $\eta >0, D>2$.

\begin{customthrm}{1}
    \label{thrm:ftrl_convergence_n=1}
    Suppose agents learn strategies for $\mathcal{G}^{(1)}$ using Algorithm~\ref{alg:ftrl} with $\alpha_i=\mathbf{0}$, any $\eta >0, D>2$, and arbitrary initial conditions $x_w^{(1)}, x_f^{(1)} \in \Delta(\mathcal{A})$. Then, for any $\epsilon>0$, there exists a finite time $t_\epsilon$ where $(x_f^{(\tau)},x_w^{(\tau)})$ is in $\epsilon$-Nash Equilibrium for all $\tau \ge t_\epsilon$. 
\end{customthrm}

\begin{customlem}{1}
\label{lem:worker_pmf_structure}
The sequence $x_{w,0}^{(t)}, \ldots, x_{w,1}^{(t)}$ is non-increasing at all time steps $t > 1$ for any arbitrary sequence of firm mixed strategies $x_{f}^{(1)}, \ldots, x_{f}^{(t-1)}$. Further, the sequence $u_w(x_f^{(t)},0), \ldots, u_w(x_f^{(t)},1)$ is non-increasing at all time steps $t > 1$ for any arbitrary firm mixed strategy $x_f^{(t)}$.
\end{customlem}
\begin{proof}
For any arbitrary sequence of firm mixed strategies $x_{f}^{(1)}, \ldots, x_{f}^{(t-1)}$, the cumulative utility the worker gets through time $t-1$ of an acceptance threshold $a \in \mathcal{A}$ is 
$$U_{w,a}^{(t-1)} = \sum_{\tau=1}^{t-1}\sum_{a_p \ge a}x_{f,a_p}^{(\tau)} \cdot a_p.$$
This implies the following cumulative utility relation between subsequent strategies $a_k < a_{k+1}$:
$$U_{w,a_k}^{(t-1)} = U_{w,a_{k+1}}^{(t-1)} + \sum_{\tau=1}^{t-1}x_{f,a_k}^{(\tau)} \cdot a_k.$$
Since $x_f$ is a probability distribution and each $a_k$ is non-negative, we can conclude $$U_{w,a_0}^{(t-1)} \ge \ldots \ge U_{w,a_D}^{(t-1)}.$$ 

By Claim~\ref{claim:kkt_ordering},  $x_{w,a_k}^{(t)} \ge x_{w,a_{k+1}}^{(t)}$ if and only if $U_{w,a_k}^{(t-1)} \ge U_{w,a_{k+1}}^{(t-1)}$ with equality if and only if $x_{w,a_k}^{(t)} = x_{w,a_{k+1}}^{(t)}$. Therefore, the sequence $x_{w,0}^{(t)},\ldots,x_{w,1}^{(t)}$ is non-increasing. 

The result above holds for the expected utility to the worker at any time step $t$ as well: $$u_w(x_f^{(t)},a_k) = u_w(x_f^{(t)},a_{k+1})  +x_{f,a_k}^{(t)} \cdot a_k,$$
so we may conclude
$$u_w(x_f^{(t)},a_0) \ge \ldots \ge u_w(x_f^{(t)},a_D).$$
\end{proof}

\begin{customlem}{2}
\label{lem:firm_pmf_structure}
     The sequence $x_{f,0}^{(t)},\ldots,x_{f,1}^{(t)}$ is unimodal at all time steps $t > 1$ for any arbitrary sequence of worker mixed strategies $x_{w}^{(1)}, \ldots, x_{w}^{(t-1)}$ that satisfy Lemma~\ref{lem:worker_pmf_structure}. Further, the sequence $u_f(0,x_w^{(t)}), \ldots, u_f(1,x_w^{(t)})$ is unimodal at all time steps $t > 1$ for any worker mixed strategy $x_w^{(t)}$ that satisfies Lemma~\ref{lem:worker_pmf_structure}.
\end{customlem}
\begin{proof}
    For any arbitrary sequence of worker mixed strategies $x_{w}^{(1)}, \ldots, x_{w}^{(t-1)}$, the cumulative utility the firm gets through time $t-1$ for an offer of $a \in \mathcal{A}$ is
    $$U_{f,a}^{(t-1)} = \sum_{\tau=1}^{t-1}\sum_{a_r\le a}x_{w,a_r}^{(\tau)} \cdot (1-a).$$
    We begin by showing the sequence $U_{f,0}^{(t-1)},\ldots,U_{f,1}^{(t-1)}$ is unimodal when the sequence $x_{w}^{(1)}, \ldots, x_{w}^{(t-1)}$ satisfies Lemma~\ref{lem:worker_pmf_structure}.

    Consider subsequent strategies $a_\ell = \frac{\ell}{D}, a_{\ell+1} = \frac{\ell+1}{D}$, then we have 
    \begin{align*}
        U_{f,a_{\ell+1}}^{(t-1)} - U_{f,a_{\ell}}^{(t-1)} = \sum_{\tau = 1}^{t-1} \left[x_{w,a_{\ell+1}}^{(\tau)}\left(\frac{D-\ell-1}{D}\right)- \sum_{a\le a_\ell} x_{w,a}^{(\tau)} \frac{1}{D}\right].\tag{1}\label{utility-diff}
    \end{align*}

    From expression~\ref{utility-diff}, note that
    \begin{align*}
        U_{f,a_{\ell}}^{(t-1)}\le U_{f,a_{\ell+1}}^{(t-1)} \iff \frac{\sum_{\tau=1}^{t-1}x_{w,a_{\ell+1}}^{(\tau)}}{\sum_{\tau=1}^{t-1}\sum_{a\le a_{\ell}}x_{w,a}^{(\tau)}} \ge \frac{1}{D-\ell-1}.\tag{2}\label{l+1>=l}
    \end{align*}

    Suppose $a_k = \frac{k}{D}$ is a cumulative utility maximizer, i.e., \begin{align*}
        U_{f,a_k}^{(t-1)} - U_{f,a}^{(t-1)} \ge 0,\forall a \neq a_k \in \mathcal{A}.\tag{3}\label{utility-max}
    \end{align*}

    So, to establish that $U_{f,a_1}^{(t-1)},\ldots,U_{f,a_{D-1}}^{(t-1)}$ is unimodal, it suffices to show
    \begin{align*}
        U_{f,a_{\ell-1}}^{(t-1)} - U_{f,a_{\ell}}^{(t-1)}, &\le 0&\forall \ell \in \{1,\ldots,k\}\\
        U_{f,a_\ell}^{(t-1)} - U_{f,a_{\ell+1}}^{(t-1)} &\ge 0. &\forall \ell \in \{k,\ldots,D-1\}
    \end{align*}

    By Lemma~\ref{lem:worker_pmf_structure}, $x_w^{(\tau)}$ is non-increasing as the acceptance thresholds $a \to 1$ at every time step $\tau$. Further, each $x_{w,a}^{(\tau)}\ge 0$ at every time step $\tau$, so whenever $i \ge j$, 
    $$\sum_{a\le a_i}x_{w,a}^{(\tau)} \ge \sum_{a\le a_j}x_{w,a}^{(\tau)}.$$
    Therefore, $\forall \ell \in\{1,\ldots,k\}$, 
    $$\frac{\sum_{\tau=1}^{t-1}x_{w,a_{\ell}}^{(\tau)}}{\sum_{\tau=1}^{t-1}\sum_{a\le a_{\ell-1}}x_{w,a}^{(\tau)}} \ge \frac{\sum_{\tau=1}^{t-1}x_{w,a_{k}}^{(\tau)}}{\sum_{\tau=1}^{t-1}\sum_{a\le a_{k-1}}x_{w,a}^{(\tau)}}\ge \frac{1}{D-k},$$
    and $\forall \ell \in\{k,\ldots,D-1\}$, 
    $$\frac{\sum_{\tau=1}^{t-1}x_{w,a_{\ell+1}}^{(\tau)}}{\sum_{\tau=1}^{t-1}\sum_{a\le a_\ell}x_{w,a}^{(\tau)}} \le \frac{\sum_{\tau=1}^{t-1}x_{w,a_{k+1}}^{(\tau)}}{\sum_{\tau=1}^{t-1}\sum_{a\le a_k}x_{w,a}^{(\tau)}}\le \frac{1}{D-k-1},$$
    where the last inequality in each expression follows from combining expressions~\ref{l+1>=l} and~\ref{utility-max}.
    Therefore,
    \begin{align*}
        \frac{\sum_{\tau=1}^{t-1}x_{w,a_{\ell}}^{(\tau)}}{\sum_{\tau=1}^{t-1}\sum_{a\le a_{\ell-1}}x_{w,a}^{(\tau)}} &\ge \frac{1}{D-\ell} &\forall \ell \in \{1,\ldots,k\},\\
        \frac{\sum_{\tau=1}^{t-1}x_{w,a_{\ell+1}}^{(\tau)}}{\sum_{\tau=1}^{t-1}\sum_{a\le a_\ell}x_{w,a}^{(\tau)}} &\le \frac{1}{D-\ell-1} & \forall \ell\in\{k,\dots,D-1\}.
    \end{align*}

    Finally, Claim~\ref{claim:kkt_ordering} implies that if the sequence $U_{f,0}^{(t-1)},\ldots,U_{f,1}^{(t-1)}$ is unimodal, then the sequence $x_{f,0}^{(t)},\ldots,x_{f,1}^{(t)}$ is unimodal as well.

    Further, the above logic holds for any time step $t$ where $x_{w}^{(t)}$ satisfies Lemma~\ref{lem:worker_pmf_structure}, so we can conclude $u_f(0,x_w^{(t)}), \ldots, u_f(1,x_w^{(t)})$ is unimodal as well. 
    
\end{proof}
Recall the following notation which will be used throughout the subsequent lemmas and main theorems.  
\begin{align*}
    w_{\max}^{(t)} &=  \max \{a | x_{w,a}^{(t)} >0 \},\\
    f_{\min}^{(t)} &= \min \{a | x_{f,a}^{(t)} >0 \}.
\end{align*}
\begin{customlem}{3}
    \label{lem:stationary_x_w}
    If agents play strategies at time $t$ such that $w_{\max}^{(t)} \le f_{\min}^{(t)}$, then $x_w^{(t+1)} = x_w^{(t)}$.
\end{customlem}
\begin{proof}
    Notice that, for any acceptance threshold $ a'\le f_{\min}^{(t)}$, $$u_w(x_f^{(t)},a') = \sum_{a \ge f_{\min}^{(t)}}x_{f,a}^{(t)} \cdot a$$
    because $x_{f,a}^{(t)} =0$ for all $a < f_{\min}^{(t)}$ by definition. This implies, for any $a,a' \le f_{\min}^{(t)}$, 
     \begin{align*}
         U_{w,a}^{(t)} - U_{w,a'}^{(t)} = U_{w,a}^{(t-1)} - U_{w,a'}^{(t-1)}. &\tag{1}\label{cumulative-utility-diff-same}
     \end{align*}
    % Next, for $a,a' \in \mathcal{A}$ where $a \le f_{\min}^{(t)} < a'$, $$u_w(x_f^{(t)},a) - u_w(x_f^{(t)},a') \ge x_{f,f_{\min}^{(t)}}^{(t)} \cdot f_{\min}^{(t)} >0.$$ 

    First, we show that any acceptance threshold that gets some mass at time $t$ and time $t+1$ must have the same probability mass difference with other such acceptance thresholds. Since $w_{\max}^{(t)} \le f_{\min}^{(t)}$, then for any $a,a' \in \mathcal{A}$ where $x_{w,a}^{(t)}>0, x_{w,a'}^{(t)}>0$, by Claim~\ref{claim:stationarity_reqs} and equation~\ref{cumulative-utility-diff-same}, $$x_{w,a}^{(t)} -x_{w,a'}^{(t)} = \eta(U_{w,a}^{(t-1)} - U_{w,a'}^{(t-1)}) = \eta(U_{w,a}^{(t)} - U_{w,a'}^{(t)} ).$$
    This implies that, if $x_{w,a}^{(t+1)}>0, x_{w,a'}^{(t+1)}>0$, then \begin{align*}
        x_{w,a}^{(t+1)} -x_{w,a'}^{(t+1)} = x_{w,a}^{(t)} -x_{w,a'}^{(t)} & \tag{2}\label{mass_diff}
    \end{align*}

    Next, suppose for some $ a,a' \le w_{\max}^{(t)}$, we have $x_{w,a}^{(t)}>0, x_{w,a'}^{(t)}>0$, but  $x_{w,a}^{(t+1)}>0, x_{w,a'}^{(t+1)}=0$. By Claim~\ref{claim:kkt_ordering}, the only way for this case to occur is for at least one acceptance threshold $a' \le w_{\max}^{(t)}$ to get 0 mass at time $t+1$ and no acceptance threshold greater than $w_{\max}^{(t)}$, which gets 0 mass at time $t$ by definition, to have non-zero mass at time $t+1$. Therefore, this is the only case to consider for an acceptance threshold getting 0 mass at time $t+1$ after having non-zero mass at time $t$.
    Here, the number of acceptance thresholds that get mass must be strictly less than those that do at time $t$, so by the primal constraints and equation~\ref{mass_diff} we must have $$ x_{w,a}^{(t+1)} > x_{w,a}^{(t)}.$$
    Then, by the KKT conditions, $$\lambda_w^{(t+1)} = \eta(U_{w,a}^{(t-1)}+u_w(x_f^{(t)},a))-x_{w,a}^{(t+1)} \ge \eta(U_{w,a'}^{(t-1)}+u_w(x_f^{(t)},a')).$$ 
    Since $w_{\max}^{(t)} \le f_{\min}^{(t)}$, then $u_w(x_f^{(t)},a)=u_w(x_f^{(t)},a')$, so $$ \eta(U_{w,a}^{(t-1)}-U_{w,a'}^{(t-1)}) \ge x_{w,a}^{(t+1)}.$$ 
    By Claim~\ref{claim:stationarity_reqs}, this implies 
    $$ x_{w,a}^{(t)} - x_{w,a'}^{(t)} \ge x_{w,a}^{(t+1)},$$ 
    however, since $x_{w,a'}^{(t)} >0$, this implies the contradiction
    $$ x_{w,a}^{(t)} > x_{w,a}^{(t+1)}.$$
    Therefore, it is impossible for some $a,a' \le w_{\max}^{(t)}$ to satisfy $x_{w,a}^{(t)}>0, x_{w,a'}^{(t)}>0$, but  $x_{w,a}^{(t+1)}>0, x_{w,a'}^{(t+1)}=0$.

    Finally, suppose for some $ a \le w_{\max}^{(t)} < a'$, we have $x_{w,a}^{(t)}>0, x_{w,a'}^{(t)}=0$, but  $x_{w,a}^{(t+1)}>0, x_{w,a'}^{(t+1)}>0$. Note by Claim~\ref{claim:kkt_ordering}, it is impossible for $x_{w,a}^{(t)}=0, x_{w,a'}^{(t)}>0$ since the cumulative utility functions are non-increasing as $a$ increases, so this is the only case to consider for an acceptance threshold gaining mass at time $t+1$ after having 0 mass at time $t$. Further, by the primal constraints and equation~\ref{mass_diff}, this implies $$x_{w,a}^{(t+1)} < x_{w,a}^{(t)}.$$
    First, by the KKT conditions, $$\eta U_{w,a'}^{(t-1)}\le\lambda_w^{(t)} = \eta U_{w,a}^{(t-1)}- x_{w,a}^{(t)},$$
    but if $x_{w,a'}^{(t+1)}>0$ $$\lambda_w^{(t+1)}< \eta  U_{w,a'}^{(t)}.$$
    Since $x_{w,a}^{(t+1)} < x_{w,a}^{(t)}$, $$\lambda_w^{(t+1)} = \eta(U_{w,a}^{(t-1)}+u_w(x_f^{(t)},a))- x_{w,a}^{(t+1)} \ge \lambda_w^{(t)} + \eta u_w(x_f^{(t)},a).$$
    However, since $u_w(x_f^{(t)},a) \ge u_w(x_f^{(t)},a')$ by Lemma~\ref{lem:worker_pmf_structure}, $$\lambda_w^{(t)}+\eta u_w(x_f^{(t)},a) \ge \eta( U_{w,a'}^{(t-1)}+u_w(x_f^{(t)},a')),$$
    which implies the contradiction $$ \lambda_w^{(t+1)} \ge \eta U_{w,a'}^{(t)}.$$
    Therefore, the same acceptance thresholds get non-zero probability mass at time $t$ and $t+1$ and their probability mass differences must remain the same, thus, $x_w^{(t+1)} = x_w^{(t)}$.
\end{proof}

\begin{customlem}{4}
    \label{lem:x_w_a_r,+_decreases}
    Suppose at time $t$ that $f_{\min}^{(t)} < w_{\max}^{(t)}$. Then, $x_{w,w_{\max}^{(t)}}^{(t+1)} < x_{w,w_{\max}^{(t)}}^{(t)}$.
\end{customlem}
\begin{proof}
    To begin, if $x_{w,w_{\max}^{(t)}}^{(t+1)} = 0$, then immediately by definition of $w_{\max}^{(t)}$, $$x_{w,w_{\max}^{(t)}}^{(t+1)} < x_{w,w_{\max}^{(t)}}^{(t)}.$$
    
    Next suppose $x_{w,w_{\max}^{(t)}}^{(t+1)} > 0$. First, by Lemma~\ref{lem:worker_pmf_structure} and the definition of $w_{\max}^{(t)}$, for all $a < w_{\max}^{(t)}$, $$x_{w,a}^{(t)} \ge x_{w,w_{\max}^{(t)}}^{(t)}>0,$$
    and Claim~\ref{claim:kkt_ordering} implies $$U_{w,a}^{(t-1)} \ge U_{w,w_{\max}^{(t)}}^{(t-1)}.$$
    Next, since $f_{\min}^{(t)} < w_{\max}^{(t)}$ it must be the case that, for all $a < w_{\max}^{(t)}$, $$u_w(x_f^{(t)},a) - u_w(x_f^{(t)},w_{\max}^{(t)})>0.$$
    Therefore, \begin{align*}
        U_{w,a}^{(t)} - U_{w,w_{\max}^{(t)}}^{(t)}  > U_{w,a}^{(t-1)} - U_{w,w_{\max}^{(t)}}^{(t-1)} &\tag{1}\label{growing_diff}
    \end{align*}
    Since $x_{w,w_{\max}^{(t)}}^{(t+1)} > 0$, then by Lemma~\ref{lem:worker_pmf_structure}, $x_{w,a}^{(t+1)} > 0$ for all $a < w_{\max}^{(t)}$. Then, Claim~\ref{claim:stationarity_reqs} and inequality~\ref{growing_diff} implies for all $a < w_{\max}^{(t)}$, $$x_{w,a}^{(t+1)} - x_{w,w_{\max}^{(t)}}^{(t+1)}>x_{w,a}^{(t)} - x_{w,w_{\max}^{(t)}}^{(t)}.$$
    If $x_{w,w_{\max}^{(t)}}^{(t+1)} \ge x_{w,w_{\max}^{(t)}}^{(t)}$, then this implies for all $a < w_{\max}^{(t)}$, 
    \begin{align*}
        x_{w,a}^{(t+1)} > x_{w,a}^{(t)} & \tag{2}\label{larger_a}
    \end{align*}
    However, by the primal constraint $$\sum_{a \le w_{\max}^{(t)}}x_{w,a}^{(t)} = 1,$$
    so the assumption $x_{w,w_{\max}^{(t)}}^{(t+1)} \ge x_{w,w_{\max}^{(t)}}^{(t)}$ along with inequality~\ref{larger_a} implies that  $$\sum_{a \le w_{\max}^{(t)}}x_{w,a}^{(t+1)} > 1,$$
    which violates the primal constraint at time $t+1$. Therefore, $$x_{w,w_{\max}^{(t)}}^{(t+1)} < x_{w,w_{\max}^{(t)}}^{(t)}.$$

\end{proof}
\begin{customlem}{5}
    \label{lem:a_r,+_never_increases}
    At any time step $t$, $w_{\max}^{(\tau)} \le w_{\max}^{(t)}$ for all $\tau \ge t$.
\end{customlem}
\begin{proof}
    First, if $f_{\min}^{(t)} \ge w_{\max}^{(t)}$, then by Lemma~\ref{lem:stationary_x_w}, $$x_{w}^{(t+1)} = x_w^{(t)} \implies x_{w,w_{\max}^{(t)}}^{(t+1)} = x_{w,w_{\max}^{(t)}}^{(t)}.$$
    
    Otherwise if $f_{\min}^{(t)} < w_{\max}^{(t)}$, then we will show it's impossible to have $a>w_{\max}^{(t)}$ with $x_{w,a}^{(t)}=0$ but $x_{w,a}^{(t+1)}>0$. 
    First, by Lemma~\ref{lem:x_w_a_r,+_decreases}, $$x_{w,w_{\max}^{(t)}}^{(t+1)} < x_{w,w_{\max}^{(t)}}^{(t)}.$$
    If $x_{w,w_{\max}^{(t)}}^{(t+1)} = 0$, then $x_{w,a}^{(t+1)} =0$ necessarily by Lemma~\ref{lem:worker_pmf_structure}. 

    Otherwise suppose $x_{w,w_{\max}^{(t)}}^{(t+1)}>0$.
    Note that the KKT conditions at time $t$ imply $$\eta U_{w,w_{\max}^{(t)}}^{(t-1)} -   \eta U_{w,a}^{(t-1)} \ge x_{w,w_{\max}^{(t)}}^{(t)}.$$
    Then, if $x_{w,a}^{(t+1)}>0$ and $x_{w,w_{\max}^{(t+1)}}^{(t+1)}>0$, by Claim~\ref{claim:stationarity_reqs}, it must be true that $$x_{w,w_{\max}^{(t)}}^{(t+1)} - x_{w,a}^{(t+1)} = \eta U_{w,w_{\max}^{(t)}}^{(t)} -   \eta U_{w,a}^{(t)},$$
    which implies $$x_{w,w_{\max}^{(t)}}^{(t+1)} > \eta U_{w,w_{\max}^{(t)}}^{(t)} -   \eta U_{w,a}^{(t)}.$$
    However, since $x_{w,w_{\max}^{(t)}}^{(t+1)} < x_{w,w_{\max}^{(t)}}^{(t)}$, then this implies $$U_{w,w_{\max}^{(t)}}^{(t-1)} -   \eta U_{w,a}^{(t-1)} > \eta U_{w,w_{\max}^{(t)}}^{(t)} -   \eta U_{w,a}^{(t)},$$ or $$u_w(x_f^{(t)},a) > u_w(x_f^{(t)},w_{\max}^{(t)})$$
    which is impossible by Lemma~\ref{lem:worker_pmf_structure}. Therefore, $w_{\max}^{(\tau)} \le w_{\max}^{(t)}$ for all $\tau \ge t$. 
\end{proof}

\begin{customlem}{6}
\label{lem:a_p,-_<=_a_r,+}
    Suppose at time $t$ that $w_{\max}^{(t)} < f_{\min}^{(t)}$. Then, there is always a time $t' >t$ where $f_{\min}^{(t')} \le w_{\max}^{(t')}$. 
\end{customlem}
\begin{proof}

     Suppose instead it is the case that for all $\tau \ge t$, $w_{\max}^{(\tau)} < f_{\min}^{(\tau)}$. First, notice by Lemma~\ref{lem:stationary_x_w} that for all $\tau \ge t$ where $w_{\max}^{(\tau)} < f_{\min}^{(\tau)}$,  $$x_w^{(\tau+1)} = x_w^{(\tau)},$$ thus,  $w_{\max}^{(\tau)} = w_{\max}^{(t)}$ for all $\tau \ge t$.
     
     Next, by definition of the firm's utility function, 
     \begin{align*}
     u_f(w_{\max}^{(\tau)}, x_w^{(\tau)}) \ge u_f(a, x_w^{(\tau)})+\frac{1}{D}, \forall a >w_{\max}^{(\tau)} 
     \end{align*}
    Therefore, since $w_{\max}^{(\tau)}$ is fixed for all $\tau \ge t$, there exists a time $t' > t$ where 
    \begin{align*}
        U_{f,w_{\max}^{(t')}}^{(t')} \ge  U_{f,a}^{(t')}, \forall a >w_{\max}^{(t')}
    \end{align*}
    So, if there exists $a> w_{\max}^{(t')}$ where $x_{f,a}^{(t')} >0$, then by Claim~\ref{claim:kkt_ordering}, it must be the case that $x_{f,w_{\max}^{(t')}}^{(t')} \ge x_{f,a}^{(t')} >0$ which implies 
    $$f_{\min}^{(t')} \le w_{\max}^{(t')}.$$
    Otherwise, if no such $a> w_{\max}^{(t')}$ where $x_{f,a}^{(t')} >0$ exists, then by definition of $f_{\min}^{(t')}$ and the primal constraints we again have
    $$f_{\min}^{(t')} \le w_{\max}^{(t')}.$$
    Therefore, by contradiction, there always exists a time $t' > t$ where $f_{\min}^{(t')} \le w_{\max}^{(t')}$.

\end{proof}

\begin{customlem}{7}
\label{lem:a_p,-_<=_a_r,+_for_all_t}
    Suppose at time $t$, $f_{\min}^{(t)} \le w_{\max}^{(t)}$. Then, there exists a finite time $t' \ge t$ where $f_{\min}^{(\tau)} \le w_{\max}^{(t)}$ for all $\tau \ge t'$. 
\end{customlem}
\begin{proof}
    Suppose at time $t$, $f_{\min}^{(t)} \le w_{\max}^{(t)}$. First, by Lemma~\ref{lem:a_r,+_never_increases}, $w_{\max}^{(\tau)} \le w_{\max}^{(t)}$ for all $\tau \ge t$. As a result, for all $\tau \ge t$, $$u_f(w_{\max}^{(t)}, x_w^{(\tau)}) \ge u_f(a, x_w^{(\tau)})+\frac{1}{D}, \forall a >w_{\max}^{(t)}.$$
    % Suppose there exists $a > w_{\max}^{(t)}$ where $x_{f,a}^{(t)}$
   which implies there exists a time $t'\ge t$ where for all $a >w_{\max}^{(t)}$ and all $\tau \ge t'$, $$U_{f,w_{\max}^{(t)}}^{(\tau)} \ge U_{f,a}^{(\tau)}.$$
    Therefore, by Claim~\ref{claim:kkt_ordering}, for all $\tau \ge t'$, it is impossible for at least one $a >w_{\max}^{(t)}$ to have $x_{f,a}^{(\tau)} >0$, but  $x_{f,w_{\max}^{(t)}}^{(\tau)} =0$. Thus, $f_{\min}^{(\tau)} \le w_{\max}^{(t)}$ for all $\tau \ge t'$.
\end{proof}

\begin{customlem}{8}
    \label{lem:a_p,-<a_r,+}
    Suppose at time $t$, $w_{\max}^{(t)} = \frac{k}{D}$ for some $k \in\{2,\ldots, D\}$,  $x_{w,w_{\max}^{(t)}}^{(t)} < \frac{1}{D-k+1}$, and $f_{\min}^{(t)} = w_{\max}^{(t)}$. Then, there is a finite time $t' > t$ where $f_{\min}^{(\tau)} < w_{\max}^{(t')}$ for all $\tau \ge t'$.
\end{customlem}
\begin{proof}
    Suppose at time $t$, $f_{\min}^{(t)} = w_{\max}^{(t)}$. First, by Lemma~\ref{lem:a_p,-_<=_a_r,+_for_all_t}, there exists a finite time $t' \ge t$ where $f_{\min}^{(\tau)} \le w_{\max}^{(t)}$ for all $\tau \ge t'$, so suppose $f_{\min}^{(\tau)} = w_{\max}^{(t)}$ for all $\tau \ge t'$. Then,  by Lemma~\ref{lem:stationary_x_w}, $x_w^{(\tau)} = x_w^{(t)}$ for all $\tau \ge t'$. Note that since $k \ge 2$, there exists a smaller action than $w_{\max}^{(t)}$: $w_{\max}^{(t)} - \frac{1}{D} \in\mathcal{A}$. Then $x_{w_{\max}^{(t)}}^{(\tau)}< \frac{1}{D-k+1}$ implies for all $\tau \ge t'$, 
    \begin{align*}
        u_f(w_{\max}^{(t)} - \frac{1}{D},x_w^{(\tau)}) > \left(1-\frac{1}{D-k+1}\right)\cdot \left(1-w_{\max}^{(t)} + \frac{1}{D}\right) = u_f(w_{\max}^{(t)},x_w^{(\tau)})
    \end{align*}
    which implies $$u_f(w_{\max}^{(t)} - \frac{1}{D},x_w^{(\tau)}) - u_f(w_{\max}^{(t)},x_w^{(\tau)})$$ is a constant, positive value for all $\tau \ge t'$. Therefore, there exists another time $t^*$ where 
    \begin{align*}
        U_{f,w_{\max}^{(t)} - \frac{1}{D}}^{(t^*)} \ge U_{f,w_{\max}^{(t)}}^{(t^*)} 
    \end{align*}
    Since $f_{\min}^{(\tau)} = w_{\max}^{(t)}$ for all $\tau \ge t'$, then we must have $x_{f,w_{\max}^{(t)}}^{(t^*)} >0$, but by Claim~\ref{claim:kkt_ordering},$$x_{f,w_{\max}^{(t)}-\frac{1}{D}}^{(t^*)} \ge x_{f,w_{\max}^{(t)}}^{(t^*} >0,$$
    which immediately implies $f_{\min}^{(t^*)} < w_{\max}^{(t)}$.

    Next, suppose at time $t$, $f_{\min}^{(t)} < w_{\max}^{(t)}$ and $x_{w,w_{\max}^{(t)}}^{(t)} < \frac{1}{D-k+1}$. Then, we will show it is impossible at time step $t+1$ to have $x_{f,a}^{(t+1)} = 0$ for all $a < w_{\max}^{(t)}$. 

    To begin, since  $x_{w,w_{\max}^{(t)}}^{(t)} < \frac{1}{D-k+1}$ and $u_f(w_{\max}^{(t)},x_w^{(t)}) \ge u_f(a,x_w^{(t)}) +\frac{1}{D}$ for all $a> w_{\max}^{(t)}$,
    \begin{align*}
        u_f(w_{\max}^{(t)} - \frac{1}{D},x_w^{(t)}) >  u_f(a,x_w^{(t)}),\forall a\ge w_{\max}^{(t)} &\tag{1}\label{inequality}
    \end{align*}
    First, suppose $x_{f,w_{\max}^{(t)}-\frac{1}{D}}^{(t)} >0.$ By Claim~\ref{claim:stationarity_reqs}, this implies for all $a \ge w_{\max}^{(t)}$ where $x_{f,a}^{(t)} >0$,
    \begin{align*}
        x_{f,w_{\max}^{(t)}-\frac{1}{D}}^{(t)} - x_{f,a}^{(t)}=\eta U_{f,w_{\max}^{(t)}-\frac{1}{D}}^{(t-1)} -\eta U_{f,a}^{(t-1)}, &\tag{2}\label{stationarity}
    \end{align*}and for all $a \ge w_{\max}^{(t)}$ where $x_{f,a}^{(t)} =0$,
    \begin{align*}
        U_{f,w_{\max}^{(t)}-\frac{1}{D}}^{(t-1)} >U_{f,a}^{(t-1)}. &\tag{3}\label{ordering}
    \end{align*}     
    Then, suppose $x_{f,w_{\max}^{(t)}-\frac{1}{D}}^{(t+1)} = 0$ and there exists $a \ge w_{\max}^{(t)}$ where $x_{f,a}^{(t+1)} >0$. This implies 
    $$U_{f,a}^{(t-1)}>U_{f,w_{\max}^{(t)}-\frac{1}{D}}^{(t-1)},$$
    so by inequalities~\ref{inequality} and~\ref{ordering}, it is impossible for such an $a$ to have $x_{f,a}^{(t)} =0$. So, it must be the case that $x_{f,a}^{(t)} >0$. Then by the KKT conditions, 
    \begin{align*}
        x_{f,a}^{(t+1)} &\le \eta  U_{f,a}^{(t)} - \eta U_{f,w_{\max}^{(t)}-\frac{1}{D}}^{(t)}
    \end{align*}
    Subtracting both sides by equation~\ref{stationarity} and applying inequality~\ref{inequality} implies
    $$x_{f,a}^{(t+1)} < x_{f,a}^{(t)}.$$
    Since this is true for any $a \ge w_{\max}^{(t)}$, then $$\sum_{a\ge w_{\max}^{(t)}}x_{f,a}^{(t+1)} < \sum_{a\ge w_{\max}^{(t)}}x_{f,a}^{(t)} \le 1,$$
    which implies by the primal constraints that it is impossible for all $a < w_{\max}^{(t)}$ to have $x_{f,a}^{(t+1)} = 0$ and this contradicts $x_{f,w_{\max}^{(t)}-\frac{1}{D}}^{(t+1)} = 0$.

    Next, suppose $x_{f,w_{\max}^{(t)}-\frac{1}{D}}^{(t)} =0.$ Then by Lemma~\ref{lem:firm_pmf_structure} there exists $f_{\min}^{(t)} \le a^* < w_{\max}^{(t)}-\frac{1}{D}$ such that 
    \begin{align*}
        U_{f,a^*}^{(t-1)} \ge U_{f,a}^{(t-1)}, \forall a \neq a^* & \tag{4}\label{mode_point}
    \end{align*}
    % Further, by Lemma~\ref{lem:firm_pmf_structure}, it is impossible for any $a \ge w_{\max}^{(t)}$ to have $x_{f,a}^{(t)} > 0$, so 
    % $$U_{f,a^*}^{(t-1)} > U_{f,a}^{(t-1)}, \forall a \neq a^*,$$.
    This implies, by Lemma~\ref{lem:firm_pmf_structure}, 
    \begin{align*}
        U_{f,w_{\max}^{(t)}-\frac{1}{D}}^{(t-1)} \ge U_{f,a}^{(t-1)}, \forall a \ge w_{\max}^{(t)} &\tag{5}\label{sequence}
    \end{align*}
    Then, if there exists $a \ge w_{\max}^{(t)}$ where $x_{f,a}^{(t+1)} > 0$, but all $a' < w_{\max}^{(t)}$ have $x_{f,a'}^{(t+1)} = 0$, then by Claim~\ref{claim:kkt_ordering}, $$U_{f,a}^{(t)} > U_{f,a^*}^{(t)},$$
    which along with inequality~\ref{mode_point} implies $$u_f(a,x_w^{(t)}) > u_f(a^*,x_w^{(t)}).$$
    However, by inequalities~\ref{inequality} and~\ref{sequence},  $$U_{f,w_{\max}^{(t)}-\frac{1}{D}}^{(t)} \ge U_{f,a}^{(t)},$$
    so by Claim~\ref{claim:kkt_ordering}, if $x_{f,a}^{(t+1)}> 0$, then $x_{f,w_{\max}^{(t)}-\frac{1}{D}}^{(t)} >0$ which contradicts all $a' < w_{\max}^{(t)}$ have $x_{f,a'}^{(t+1)} = 0$.
    
     Therefore, in all possible cases, $f_{\min}^{(t+1)} < w_{\max}^{(t)}$. Further, by Lemma~\ref{lem:x_w_a_r,+_decreases}, it is also the case that $$x_{w,w_{\max}^{(t)}}^{(t+1)}<x_{w,w_{\max}^{(t)}}^{(t)} < \frac{1}{D-k+1},$$
     so we can conclude $f_{\min}^{(\tau)} < w_{\max}^{(t)}$ for all $\tau \ge t$.

\end{proof}

\begin{customlem}{9}
\label{lem:a_r,+_decreasing_conditions}
    Suppose at time $t$,  $w_{\max}^{(t)} = \frac{k}{D}$ for some $k \in\{2,\ldots, D\}$,  $x_{w,w_{\max}^{(t)}}^{(t)} < \frac{1}{D-k+1}$, and $f_{\min}^{(\tau)} < w_{\max}^{(t)}$ for all $\tau \ge t$. Then, there is a finite time $t' > t$ where $w_{\max}^{(t')} < w_{\max}^{(t)}$.
\end{customlem}
\begin{proof}
    Suppose $f_{\min}^{(\tau)} < w_{\max}^{(t)}$ for all $\tau \ge t$. Then, we will show there must be another finite time $t' > t$ where $x_{w,w_{\max}^{(t)}}^{(t')} = 0$. Then by Lemma~\ref{lem:worker_pmf_structure} this implies it is also the case that $x_{w,a}^{(t')} = 0$ for all $a \ge w_{\max}^{(t)}$, and  we can conclude $w_{\max}^{(t')} < w_{\max}^{(t)}$. 
    
    First, note that since $k \ge 2$, there exists a smaller action than $w_{\max}^{(t)}$: $w_{\max}^{(t)} - \frac{1}{D} \in\mathcal{A}$.
    % \begin{align*}
    %     u_f(w_{\max}^{(t)}-\frac{1}{D},x_w^{(t)}) &= (1-x_{w,w_{\max}^{(t)}}^{(t)}) (1-(w_{\max}^{(t)}-\frac{1}{D})),\\
    %     u_f(w_{\max}^{(t)},x_w^{(t)}) &=  (1-w_{\max}^{(t)}).
    % \end{align*}
    Then, since $x_{w,w_{\max}^{(t)}}^{(t)} < \frac{1}{D-k+1}$, $$u_f(w_{\max}^{(t)}-\frac{1}{D},x_w^{(t)}) >u_f(w_{\max}^{(t)},x_w^{(t)}).$$
    Further, since $f_{\min}^{(\tau)} < w_{\max}^{(t)}$ for all $\tau \ge t$, by Lemma~\ref{lem:x_w_a_r,+_decreases} either $x_{w,w_{\max}^{(t)}}^{(\tau)} = 0$ or  $x_{w,w_{\max}^{(t)}}^{(\tau+1)}<x_{w,w_{\max}^{(t)}}^{(\tau)}$ for all $\tau \ge t$. Then, this implies by the definition of the firm's utility function that for all $\tau \ge t$,
    $$u_f(w_{\max}^{(t)}- \frac{1}{D}, x_w^{(\tau)}) -  u_f(w_{\max}^{(t)}, x_w^{(\tau)})> u_f(w_{\max}^{(t)}- \frac{1}{D}, x_w^{(t)}) -u_f(w_{\max}^{(t)}, x_w^{(t)})>0.$$
   %  Therefore,
   % $$U_{f,w_{\max}^{(t)} - \frac{1}{D}}^{(\tau+1)} - U_{f,w_{\max}^{(t)} - \frac{1}{D}}^{(\tau)}  > U_{f,w_{\max}^{(t)}}^{(\tau+1)}- U_{f,w_{\max}^{(t)}}^{(\tau)}.$$
   This implies there exists a time $t^*$ where  for all $\tau \ge t^*$, $$U_{f,w_{\max}^{(t)} - \frac{1}{D}}^{(\tau)} - U_{f,w_{\max}^{(t)}}^{(\tau)} \ge \frac{1}{\eta},$$
    and by Claim~\ref{claim:stationarity_reqs} and the primal constraints, it cannot be the case that both$$x_{f,w_{\max}^{(t)} - \frac{1}{D}}^{(\tau)}>0\text{ and }  
    x_{f,w_{\max}^{(t)}}^{(\tau)}>0,$$ and since $$U_{f,w_{\max}^{(t)} - \frac{1}{D}}^{(\tau)} > U_{f,w_{\max}^{(t)}}^{(\tau)},$$
    by Claim~\ref{claim:kkt_ordering}, we can conclude 
    $$x_{f,w_{\max}^{(t)}}^{(\tau)}=0, \forall \tau \ge t^*.$$
    
    So, there must always be a time $t^*> t$ where either $x_{w,w_{\max}^{(t)}}^{(t^*)} = 0$ or  $x_{w,w_{\max}^{(t)}}^{(t^*)}>0$ and $x_{f,w_{\max}^{(t)}}^{(\tau)} = 0$ for all $\tau \ge t^*$.
    The latter case implies that for all $\tau \ge t^*$, $$U_{w,w_{\max}^{(t)}}^{(\tau)} = U_{w,w_{\max}^{(t)}}^{(t^*-1)}.$$ 
    However, since $k\ge 2$, $\frac{1}{D}$ is a lower acceptance threshold for the worker than $w_{\max}^{(t)}$. Then, by the worker's utility function
    \begin{align*}
        u_w(x_f^{(t)},\frac{1}{D}) = \sum_{a \ge \frac{1}{D}}x_{f,a}^{(t)}\cdot a \ge (1-x_{f,0}^{(t)})\cdot \frac{1}{D}. &\tag{1}\label{1/D_bound}
    \end{align*}
    Note that by the fact that an acceptance threshold of $0$ for the worker cannot get more utility than an acceptance threshold of $\frac{1}{D}$, then by Lemma~\ref{lem:worker_pmf_structure} $$U_{w,0}^{(t)} = U_{w,\frac{1}{D}}^{(t)} \ge U_{w,a}^{(t)}, \forall a > \frac{1}{D}$$ for all time steps $t$, so by Claim~\ref{claim:stationarity_reqs} it is always the case that $x_{w,0}^{(t)} = x_{w,\frac{1}{D}}^{(t)} >0$. This implies that when $D>2$,
    $$u_f(\frac{1}{D},x_w^{(t)})  > u_f(0,x_w^{(t)}), \forall t.$$
    Therefore, $$U_{f,\frac{1}{D}}^{(t)} > U_{f,0}^{(t)}, \forall t$$
    So, by Claim~\ref{claim:kkt_ordering} $$x_{f,\frac{1}{D}}^{(t)}> x_{f,0}^{(t)}, \forall t. $$
    By the primal constraints, this implies $x_{f,0}^{(t)} < \frac{1}{2}$ for all $t$, so combining this fact with the lower bound~\ref{1/D_bound}, then for any time step $t$,
    $$U_{w,\frac{1}{D}}^{(t)} \ge U_{w,\frac{1}{D}}^{(t-1)}+ \frac{1}{2D}.$$ Therefore, since the cumulative utility of the offer $w_{\max}^{(t)}$ stops growing after time $t^*$, there must be a finite time $t' \ge t^*$ where $$U_{w,\frac{1}{D}}^{(t')} - U_{w,w_{\max}^{(t)}}^{(t')} \ge \frac{1}{\eta},$$
    and by Claim~\ref{claim:stationarity_reqs} and the primal constraints, it must be the case that $x_{w,w_{\max}^{(t)}}^{(t')}=0$ and we can conclude $w_{\max}^{(t')} < w_{\max}^{(t)}$.

\end{proof}
\begin{customlem}{10}
\label{lem:final_NE_convergence}
    Suppose at time $t$, $w_{\max}^{(t)} = \frac{k}{D}$ for some $k \in \{1,\ldots,D\}$, $f_{\min}^{(\tau)} \le w_{\max}^{(t)}$, and $x_{w,w_{\max}^{(t)}}^{(\tau)} \ge \frac{1}{D-k+1}$ for all $\tau \ge t$. Then, for any $\epsilon>0$, there exists a time $t_\epsilon \ge t$ where, for all $\tau \ge t_{\epsilon}$, $(x_f^{(\tau)}, x_w^{(\tau)})$ is in an $\epsilon$-mixed Nash Equilibrium. 
\end{customlem}
\begin{proof}
    % finite time until all greater offers have 0 mass
    Suppose at time $t$, $f_{\min}^{(\tau)} \le w_{\max}^{(t)}$ and $x_{w,w_{\max}^{(t)}}^{(\tau)} \ge \frac{1}{D-k+1}$ for all $\tau \ge t$. First, by definition of $w_{\max}^{(t)}$ and the firm's expected utility function, \begin{align*}
        u_f(w_{\max}^{(t)}, x_w^{(\tau)}) \ge u_f(a, x_w^{(\tau)})+\frac{1}{D}, \forall a > w_{\max}^{(t)} & \tag{1}\label{best-response-firm-greater-offers}
    \end{align*}
    This implies there exists a time $t' \ge t$ where for all $\tau \ge t'$ $$U_{f,w_{\max}^{(t)}}^{(\tau)} - U_{f,a}^{(\tau)} \ge \frac{1}{\eta}, \forall a> w_{\max}^{(t)}.$$
    Then, by Claim~\ref{claim:stationarity_reqs}, for each $a> w_{\max}^{(t)}$ and all time steps $\tau \ge t'$, it is impossible for $$x_{f,w_{\max}^{(t)}}^{(\tau)}>0, x_{f,a}^{(\tau)}>0.$$
    By expression~\ref{best-response-firm-greater-offers}, it must be the case that for all $\tau \ge t'$,
    \begin{align*}
        x_{f,a}^{(\tau)}= 0, \forall a> w_{\max}^{(t)} &\tag{2}\label{greater-offers-0}
    \end{align*}
    % next, a_r,+ is a utility maximizer
    Next, since, for all $\tau \ge t$,
    $$x_{w,w_{\max}^{(t)}}^{(\tau)} \ge \frac{1}{D-k+1},$$
    then for all $\tau \ge t$,
    \begin{align*}
        u_f(w_{\max}^{(t)}-\frac{1}{D},x_w^{(\tau)}) &\le (1-\frac{1}{D-k+1}) \frac{D-k+1}{D}\\
        &= \frac{D-k}{D} \\
        &= u_f(w_{\max}^{(t)},x_w^{(\tau)})
    \end{align*}
    By Lemma~\ref{lem:firm_pmf_structure}, if $u_f(w_{\max}^{(t)}-\frac{1}{D},x_w^{(\tau)}) \le u_f(w_{\max}^{(t)},x_w^{(\tau)})$, then it must also be the case that for all $a < w_{\max}^{(t)}$ and all $\tau \ge t$, $$u_f(a,x_w^{(\tau)}) \le u_f(w_{\max}^{(t)},x_w^{(\tau)}).$$
    Combining this with expression~\ref{best-response-firm-greater-offers}, we can conclude for all $\tau \ge t$ and all $a\neq w_{\max}^{(t)}$, 
    \begin{align*}
        u_f(w_{\max}^{(t)},x_w^{(\tau)}) \ge u_f(a,x_w^{(\tau)}). &\tag{3}\label{best-response-firm}
    \end{align*}

    We now break into the individual cases of $f_{\min}^{(\tau)} = w_{\max}^{(t)}$ for all $\tau \ge t$ and $f_{\min}^{(\tau)} < w_{\max}^{(t)}$ for all $\tau \ge t$ to finish the proof. It is sufficient to consider these two cases because after time $t'$ where expression~\ref{greater-offers-0} becomes true, then if $f_{\min}^{(t')} < w_{\max}^{(t)}$, but there exists a time $t^* > t'$ where $f_{\min}^{(t^*)} = w_{\max}^{(t)}$, then we immediately have $x_{f, w_{\max}^{(t)}}^{(t^*)} = 1$ and the first case below shows this implies convergence.
    
    % state 4 coverged to mixed-NE
    In the first case, suppose $f_{\min}^{(\tau)} = w_{\max}^{(t)}$ for all $\tau \ge t$. This implies that for all $a < w_{\max}^{(t)}$ and all $\tau \ge t$, $$x_{f,a}^{(\tau)} = 0.$$ Combining this with expression~\ref{greater-offers-0}, we can conclude that it must be the case that, for all $\tau \ge t'$ $$x_{f,w_{\max}^{(t)}}^{(\tau)} = 1.$$
    Thus, after time $t'$, the firm purely offers $w_{\max}^{(t)}$ for all future time steps and by expression~\ref{best-response-firm}, this offer will always be a best response to the worker's strategy. Further, $w_{\max}^{(t)}$ is the largest acceptance threshold with non-zero probability by definition, it is impossible for the worker switch acceptance thresholds to get more utility than $w_{\max}^{(t)}$. So, any mixture over acceptance thresholds $a\le w_{\max}^{(t)}$ is a best response to $x_{f,w_{\max}^{(t)}}^{(t')} = 1$. Therefore, for any $\epsilon\ge 0$, then for all $\tau \ge t'$,  $(x_w^{(\tau)}, x_f^{(\tau)})$ is in an $\epsilon$-mixed Nash Equilibrium.

    % state 5 converges to epsilon-NE for all epsilon>0 (only way for this case to make sense is if it approaches threshold but is never equal to it)
    In the second case, suppose $f_{\min}^{(\tau)} < w_{\max}^{(t)}$.  
    % $x_{w,w_{\max}^{(t)}}^{(\tau)} \ge \frac{1}{D-k+1}$ for all $\tau \ge t$. Let $t'\ge t$ be the time where expression~\ref{greater-offers-0} guarantees offers greater than $w_{\max}^{(t)}$ get 0 probability mass in all future time steps.
    The following two properties must hold in this case
    \begin{align*}
        x_{w,w_{\max}^{(t)}}^{(\tau)}> \frac{1}{D-k+1}, \forall \tau \ge t, &\tag{4}\label{state-5-prop-1}
    \end{align*}
    and there exists a $t^* \ge t'$ where
    \begin{align*}
        x_{f,w_{\max}^{(t)}}^{(t^*)} \ge x_{f,a}^{(t^*)}, \forall a \neq w_{\max}^{(t)}. &\tag{5}\label{state-5-prop-2}
    \end{align*}
    By Lemma~\ref{lem:x_w_a_r,+_decreases} $x_{w,w_{\max}^{(t)}}^{(\tau+1)} < x_{w,w_{\max}^{(t)}}^{(\tau)}$ when $f_{\min}^{(\tau)} < w_{\max}^{(t)}$, so $x_{w,w_{\max}^{(t)}}^{(\tau+1)} < \frac{1}{D-k+1}$ if $x_{w,w_{\max}^{(t)}}^{(\tau)} = \frac{1}{D-k+1}$, so property~\ref{state-5-prop-1} must hold. Next, if for all $\tau \ge t'$ there exists $f_{\min}^{(\tau)} \le a < w_{\max}^{(t)}$ where $$x_{f,a}^{(\tau)} > x_{f,w_{\max}^{(t)}}^{(\tau)}, $$
    then by the primal constraints this implies 
    $$\sum_{f_{\min}^{(\tau)} \le a < w_{\max}^{(t)}}x_{f,a}^{(\tau)} > \frac{1}{2},$$ 
    so by the worker's utility function $$u_w(x_f^{(\tau)},\frac{1}{D}) \ge u_w(x_f^{(\tau)},w_{\max}^{(t)}) + \frac{1}{2D},$$
    since a lower acceptance threshold, $\frac{1}{D}$ gets at least $\frac{1}{D}$ more utility than the acceptance threshold $w_{\max}^{(t)}$ with probability at least $\frac{1}{2}$. This implies there exists a time $t^* > t'$ where $$U_{w,\frac{1}{D}}^{(t^*)} - U_{w,w_{\max}^{(t)}}^{(t^*)} \ge \frac{1}{\eta},$$
    which implies $x_{w,w_{\max}^{(t)}}^{(t^*)} = 0$. Therefore, property~\ref{state-5-prop-2} must be true as well.
    
    Now, by property~\ref{state-5-prop-1}, then by the definition of the firm's utility function, for all $\tau \ge t$  $$u_f(w_{\max}^{(t)}, x_w^{(\tau)}) > u_f(w_{\max}^{(t)}-\frac{1}{D}, x_w^{(\tau)}),$$
    so by Lemma~\ref{lem:firm_pmf_structure},$$u_f(w_{\max}^{(t)}, x_w^{(\tau)}) > u_f(a, x_w^{(\tau)}), \forall a <w_{\max}^{(t)},$$
    and combining this with expression~\ref{best-response-firm-greater-offers},
    \begin{align*}
        u_f(w_{\max}^{(t)}, x_w^{(\tau)}) > u_f(a, x_w^{(\tau)}), \forall a \neq w_{\max}^{(t)} &\tag{4}\label{strict-best-response-state-5}
    \end{align*}
    
    % This implies for any $\epsilon>0$, there exists a $t_\epsilon$ where $$x_{w,w_{\max}^{(t)}}^{(t_{\epsilon})} - \frac{1}{D-k+1}<\epsilon.$$
    %  Then, if $x_{f,w_{\max}^{(\tau)}} = 0$ for all $\tau \ge t'$, then by examining the proof of Lemma~\ref{lem:a_r,+_decreasing_conditions}, it is necessarily the case that $w_{\max}$ decreases and the agents move out of state 5. So, there must be a time $t^* \ge t'$ where $$x_{f,w_{\max}^{(t)}}^{(t^*)} >0.$$
    % Next, given expression~\ref{strict-best-response-state-5}, we will show it is impossible for $x_{f,w_{\max}^{(t)}}^{(t^*+1)}<x_{f,w_{\max}^{(t)}}^{(t^*)}.$ Suppose $x_{f,w_{\max}^{(t)}}^{(t'+1)}<x_{f,w_{\max}^{(t)}}^{(t')},$ then by the primal constraints there must be $a \neq w_{\max}^{(t)}$ where $x_{f,a}^{(t'+1)}>x_{f,a}^{(t')}.$ If $x_{f,a}^{(t')} >0$, then since $x_{f,w_{\max}^{(t)}}^{(t')} >0$ as well, by Claim~\ref{claim:stationarity_reqs}, $$x_{f,a}^{(t')} - x_{f,w_{\max}^{(t)}}^{(t')} = \eta  U_{f,a}^{(t'-1)} - \eta  U_{f,w_{\max}^{(t)}}^{(t'-1)},$$
    % and if $x_{f,a}^{(t'+1)} >0$ as well, 
    % $$x_{f,a}^{(t'+1)} - x_{f,w_{\max}^{(t)}}^{(t'+1)} = \eta  U_{f,a}^{(t')} - \eta  U_{f,w_{\max}^{(t)}}^{(t')}.$$
    % However, since we assume $x_{f,w_{\max}^{(t)}}^{(t'+1)}<x_{f,w_{\max}^{(t)}}^{(t')}$ and $x_{f,a}^{(t'+1)}>x_{f,a}^{(t')},$ then this implies $$u_f(a,x_w^{(t')}) >u_f(w_{\max}^{(t)},x_w^{(t')}),$$
    % which contradicts expression~\ref{strict-best-response-state-5}. 
    % need 3 other cases of 0,>0 >0,0 and 0,0 (make this a lemma?) -- feels like there's an easier argument.
    Next, by property~\ref{state-5-prop-2}, Claim~\ref{claim:kkt_ordering},
    and expression~\ref{strict-best-response-state-5}, then for all $\tau \ge t^*$
    $$U_{f,w_{\max}^{(t)}}^{(\tau+1)} - U_{f,a}^{(\tau+1)} >U_{f,w_{\max}^{(t)}}^{(\tau)} - U_{f,a}^{(\tau)} \ge 0, \forall a \neq w_{\max}^{(t)}.$$
    Then, by Claim~\ref{claim:stationarity_reqs}, this implies for all $\tau \ge t^*$ and for all $a \neq w_{\max}^{(t)}$ where $x_{f,a}^{(\tau)}>0$ and $x_{f,a}^{(\tau+1)} >0$,
    $$x_{f,w_{\max}^{(t)}}^{(\tau+1)} - x_{f,a}^{(\tau+1)} > x_{f,w_{\max}^{(t)}}^{(\tau)} - x_{f,a}^{(\tau)}.$$
    Further, it cannot be the case that $x_{f,w_{\max}^{(t)}}^{(\tau+1)} = x_{f,w_{\max}^{(t)}}^{(\tau)}$ while $x_{f,a}^{(\tau+1)} < x_{f,a}^{(\tau)}$ for all such $a\neq w_{\max}^{(t)}$ because this implies $$\sum_{a \in \mathcal{A}}x_{f,a}^{(\tau+1)} < \sum_{a \in \mathcal{A}}x_{f,a}^{(\tau)}=1,$$
    which would violate the primal constraint at time $t+1$.

    So, we can conclude 
    $x_{f,w_{\max}^{(t)}}^{(\tau+1)} >x_{f,w_{\max}^{(t)}}^{(\tau)}$ for all $\tau \ge t^*$. Since for all $a >w_{\max}^{(t)}$, $x_{f,a}^{(\tau)} = 0$ for all $\tau \ge t'$ and $t^* \ge t'$, then this immediately implies
    $$\sum_{f_{\min}^{(\tau+1)}\le a<w_{\max}^{(t)}}x_{f,a}^{(\tau+1)} < \sum_{f_{\min}^{(\tau)}\le a<w_{\max}^{(t)}}x_{f,a}^{(\tau)}, \forall \tau \ge t^*$$
    So, by the primal constraints, we can conclude $$\underset{\tau\to\infty}{\lim}\sum_{f_{\min}^{(\tau)}\le a<w_{\max}^{(t)}}x_{f,a}^{(\tau)} = 0,$$
    and 
    $$\underset{\tau\to\infty}{\lim} x_{f,w_{\max}^{(t)}}^{(\tau)}= 1.$$
    Therefore, for any $\epsilon>0$ there exists a time $t_\epsilon$ where, for all $\tau \ge t_{\epsilon}$, $$x_{f,w_{\max}^{(t)}}^{(\tau)} > 1- \epsilon.$$
    By expression~\ref{strict-best-response-state-5}, the offer $w_{\max}^{(t)}$ is a best-response for the firm for all $\tau \ge t$, so $x_{f,w_{\max}^{(t)}}^{(\tau)} > 1- \epsilon$ implies $$u_f(x_f^{(\tau)}, x_w^{(\tau)}) \ge u_f(x_f', x_w^{(\tau)}) - \epsilon, \forall x_f' \in \Delta(\mathcal{A}).$$
    Further, $x_{f,w_{\max}^{(t)}}^{(\tau)} > 1- \epsilon$ implies the worker gets at most $\epsilon$ more utility by lowering their acceptance threshold from $w_{\max}^{(t)}$, so we also have
    $$u_w(x_f^{(\tau)}, x_w^{(\tau)}) \ge u_w(x_f^{(\tau)}, x_w') - \epsilon, \forall x_w' \in \Delta(\mathcal{A}).$$
    Therefore, the strategy profile $(x_f^{(t_\epsilon)}, x_w^{(t_\epsilon)})$ is an $\epsilon$-mixed NE.
\end{proof}

\newpage

Finally, we prove the main theorem of this section.

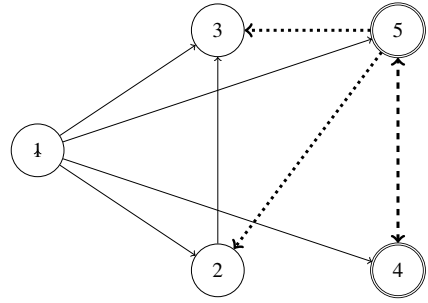
\begin{figure}[h]
\centering
\begin{subfigure}[t]{0.45\textwidth}
    \centering
    \vfill
    \begin{tabular}{|c|c|}
    \hline
            1 & $f_{\min} > w_{\max}$  \\\hline
             2 & \thead{$f_{\min} = w_{\max},$\\$x_{w,w_{\max}} <\frac{1}{D-k+1}$} \\\hline
             3 & \thead{$f_{\min} < w_{\max},$\\$x_{w,w_{\max}} <\frac{1}{D-k+1}$} \\\hline
             4 & \thead{$f_{\min} = w_{\max},$\\$x_{w,w_{\max}} \ge\frac{1}{D-k+1}$} \\\hline
             5 & \thead{$f_{\min} < w_{\max},$\\$x_{w,w_{\max}} \ge \frac{1}{D-k+1}$} \\\hline
    \end{tabular}
    \caption{Complete list of possible conditions Algorithm~\ref{alg:ftrl} satisfies at time $t$. }
    \label{thrm1-condition-table}
\end{subfigure}
\hfill
\begin{subfigure}[t]{0.45\textwidth}
    \vfill
    \resizebox{0.9\textwidth}{!}{
    \begin{tikzpicture}
    \node[state] (1) at (-5,-1) {1};
    \node[state] (2) at (-2,-3) {2};
    \node[state] (3) at (-2,1) {3};
    \node[state, accepting] (4) at (1,-3) {4};
    \node[state, accepting] (5) at (1,1) {5};

    \draw[->, solid,black] (1) to (2);
    \draw[->, solid,black] (1) to (1);
    \draw[->, solid,black] (1) to (3);
    \draw[->, solid,black] (1) to (4);
    \draw[->, solid,black] (1) to (5);
    \draw[<->, dashed, black,line width=0.5mm] (4) to (5);
    \draw[->, dotted,black,line width=0.5mm] (5) to (2);
    \draw[->, dotted,black,line width=0.5mm] (5) to (3);
    \draw[->, solid, black] (2) to (3);

    \end{tikzpicture}
    }
    \caption{Possible condition transitions per $w_{\max}$ value.}
    \label{thrm1-condition-transitions}
\end{subfigure}
    \caption{Figure~\ref{thrm1-condition-table} gives the complete list of possible conditions that the agents' strategies can satisfy at any time. Figure~\ref{thrm1-condition-transitions} demonstrates how the strategies can change between conditions for a given $w_{\max}$ value. The \textbf{solid lines} denote that there exists a time where the algorithm will transition from one condition to another and never transition back. The \textbf{dashed line} indicates that only finitely many transitions may happen between two conditions per $w_{\max}$ value before the algorithm remains in one condition or the other. The \textbf{dotted lines} indicate that the transition can only occur if convergence does not happen at $w_{\max}$. Finally, conditions 4 and 5 have double circles to indicate that those are the only conditions where convergence can occur.}
    \label{fig:thrm_1_state_machine}
\end{figure}

\begin{customthrm}{1}
    \label{thrm:ftrl_convergence_n=1}
    Suppose agents learn strategies for $\mathcal{G}^{(1)}$ using Algorithm~\ref{alg:ftrl} with $\alpha_i=\mathbf{0}$, any $\eta >0, D>2$, and arbitrary initial conditions $x_w^{(1)}, x_f^{(1)} \in \Delta(\mathcal{A})$. Then, for any $\epsilon>0$, there exists a finite time $t_\epsilon$ where $(x_f^{(\tau)},x_w^{(\tau)})$ is in $\epsilon$-Nash Equilibrium for all $\tau \ge t_\epsilon$. 
\end{customthrm}
\begin{proof}

    % new structure, like a state machine, one lemma -> another lemma, a_r,+ decreases a finite amount of time and each decrease takes a finite amount of time OR we are in epsilon-NE
    
    %% scaffolding
    To prove the theorem, we will show that, regardless of the initial conditions, the agents must always reach or approach a mixed Nash Equilibrium (NE) asymptotically, such that we can conclude the agents end in an $\epsilon$-NE at the last iterate.

    % map possible changes at time $t+1$
    We begin by describing all the possible conditions the agents' strategy profile, $(x_f^{(t)}, x_w^{(t)})$, could satisfy at any time $t$. Then, we use induction to show there is always a finite time where the agents are in one of two conditions for all future time steps. We conclude by showing that this implies the agents have converged to an $\epsilon$-NE for any $\epsilon>0$. See Figure~\ref{fig:thrm_1_state_machine} for a visual of how the algorithm moves through the possible conditions.

    % conditions
    To begin, at any time step $t$, $$w_{\max}^{(t)} = \frac{k}{D}, $$ for some $k \in \{1,\ldots,D\}$. Then, exactly one of the following conditions is satisfied by the agents' strategy profile at time $t$.
    \begin{enumerate}
        \item $w_{\max}^{(t)} < f_{\min}^{(t)}$
        \item $w_{\max}^{(t)} = f_{\min}^{(t)}$ and $x_{w,w_{\max}^{(t)}}^{(t)} < \frac{1}{D-k+1}$
        \item $f_{\min}^{(t)}<w_{\max}^{(t)} $ and $x_{w,w_{\max}^{(t)}}^{(t)} < \frac{1}{D-k+1}$
        \item $w_{\max}^{(t)} = f_{\min}^{(t)}$ and $x_{w,w_{\max}^{(t)}}^{(t)} \ge \frac{1}{D-k+1}$
        \item $f_{\min}^{(t)}< w_{\max}^{(t)} $ and $x_{w,w_{\max}^{(t)}}^{(t)} \ge \frac{1}{D-k+1}$
    \end{enumerate}
    
    % induction proof to reduce conditions
    Now, we consider each condition separately, and show the possible conditions that can be satisfied in time step $t+1$, given the condition satisfied at time step $t$. We say the agents move to condition $i$ if $(x_f^{(t+1)}, x_w^{(t+1)})$ satisfies condition $i$ for $i\in\{1,2,3,4,5\}$.\\

    First, suppose $(x_f^{(t)}, x_w^{(t)})$ is in \textbf{condition 1}. Then, in the next time step, either $f_{\min}^{(t+1)} > w_{\max}^{(t+1)}$ and the agents remain in condition 1 or $f_{\min}^{(t+1)} \le w_{\max}^{(t+1)}$ and the agents move to condition 2, 3, 4, or 5. \\

    Next, suppose $(x_f^{(t)}, x_w^{(t)})$ is in \textbf{condition 2}. First, by Lemma~\ref{lem:stationary_x_w}, $$x_w^{(t+1)} = x_w^{(t)},$$ so it cannot be the case that the agents move to condition 4 or 5. If, $f_{\min}^{(t+1)} >  w_{\max}^{(t+1)}$, then the agents move to condition 1. Next, if $f_{\min}^{(t+1)} = f_{\min}^{(t)}$, then the agents remain in condition 2. Finally, if $f_{\min}^{(t+1)} < f_{\min}^{(t)}$, then this implies $f_{\min}^{(t+1)} < w_{\max}^{(t+1)}$ since $f_{\min}^{(t)} = w_{\max}^{(t)}=w_{\max}^{(t+1)}$ and the agents move to condition 3.\\

    Next, suppose $(x_f^{(t)}, x_w^{(t)})$ is in \textbf{condition 3}. First, by Lemma~\ref{lem:x_w_a_r,+_decreases}, $$x_{w,w_{\max}^{(t)}}^{(t+1)} < x_{w,w_{\max}^{(t)}}^{(t)}.$$
    If $x_{w,w_{\max}^{(t)}}^{(t+1)} =0$, then by definition $w_{\max}^{(t+1)}\neq w_{\max}^{(t)}$, so by Lemma~\ref{lem:a_r,+_never_increases},
    $w_{\max}^{(t+1)} < w_{\max}^{(t)}$. Now, the agents can move to condition 1, 2, 3, 4, or 5, with the new $w_{\max}$ value.
    Otherwise, if $x_{w,w_{\max}^{(t)}}^{(t+1)}>0$, then by Lemma~\ref{lem:a_r,+_never_increases}, $w_{\max}^{(t+1)} = w_{\max}^{(t)}$. Since $x_{w,w_{\max}^{(t)}}^{(t+1)} < x_{w,w_{\max}^{(t)}}^{(t)}$, then the agents cannot move to condition 4 or 5. Further, since $w_{\max}^{(t+1)} = w_{\max}^{(t)}$, then by Lemma~\ref{lem:a_p,-<a_r,+}, $f_{\min}^{(t+1)} < w_{\max}^{(t+1)}$, and the agents cannot move back to condition 1 or 2. So, the agents remain in condition 3 in this case.\\

    Next, suppose $(x_f^{(t)}, x_w^{(t)})$ is in \textbf{condition 4}. First, by Lemma~\ref{lem:stationary_x_w} $$x_w^{(t+1)} = x_w^{(t)},$$  so the agents cannot move to condition 2 or 3. Next, if $f_{\min}^{(t+1)} > w_{\max}^{(t+1)}$, then the agents move to condition 1. If $f_{\min}^{(t+1)} = f_{\min}^{(t)}$, then the agents remain in condition 4. Otherwise, if $f_{\min}^{(t+1)} < f_{\min}^{(t)}$, then since $f_{\min}^{(t)} = w_{\max}^{(t)}=w_{\max}^{(t+1)}$, this implies $f_{\min}^{(t+1)} < w_{\max}^{(t+1)}$ and the agents move to condition 5.\\
    
    Next, suppose $(x_f^{(t)}, x_w^{(t)})$ is in \textbf{condition 5}. First,  by Lemma~\ref{lem:x_w_a_r,+_decreases}, $$x_{w,w_{\max}^{(t)}}^{(t+1)} < x_{w,w_{\max}^{(t)}}^{(t)}.$$ If $f_{\min}^{(t+1)} > w_{\max}^{(t+1)}$, then the agents move to condition 1.
    Next, if $x_{w,w_{\max}^{(t)}}^{(t+1)} \ge \frac{1}{D-k+1}$ and $f_{\min}^{(t+1)} < w_{\max}^{(t+1)}$, then the agents remain in condition 5. Next, if $x_{w,w_{\max}^{(t)}}^{(t+1)} \ge \frac{1}{D-k+1}$ but $f_{\min}^{(t+1)} = w_{\max}^{(t+1)}$, then the agents move to condition 4. Next, if $0< x_{w,w_{\max}^{(t)}}^{(t+1)} < \frac{1}{D-k+1}$, then the agents move to condition 2 or 3. Finally, if $x_{w,w_{\max}^{(t)}}^{(t+1)} =0 $ then by Lemma~\ref{lem:a_r,+_never_increases}, $$w_{\max}^{(t+1)} < w_{\max}^{(t)},$$ and the agents move to condition 1, 2, 3, 4, or 5.\\

    % finite time until only condition 4 or only condition 5
    Next, we show there exists a finite time where the agents remain in condition 4 or condition 5 for all future time steps. First, Lemma~\ref{lem:a_r,+_never_increases} shows that the value of $w_{\max}$ is non-increasing in all time steps, and further the number of $w_{\max}$ values is $|\mathcal{A}|$. So, it suffices to show for each unique $w_{\max}$ value that either the agents must remain in condition 4 or condition 5 for all future time steps or the value of $w_{\max}$ must decrease in finite time.\\ 
    
    Suppose $(x_f^{(t)}, x_w^{(t)})$ is in condition 1 at time $t$. From the cases above, the agents either remain in condition 1 or move to one of the other conditions, and Lemma~\ref{lem:a_p,-_<=_a_r,+} shows that it takes finite time for the agents to move to condition 2, 3, 4, or 5. Further,  Lemma~\ref{lem:a_p,-_<=_a_r,+_for_all_t} shows that once agents leave condition 1 for the first time per unique value of $w_{\max}$, it takes finite time to ensure the agents never enter condition 1 again for that $w_{\max}$ value. So, we may assume that the agents never enter condition 1 for the remainder of the cases. Next, if the agents are in condition 4 or 5, but don't stay there for all future time steps and  $w_{\max}$ does not decrease, then there must be a finite time where the agents move to condition 2 or 3. Next, if the agents are in condition 2, then from the cases above, they either remain there or move to condition 3, and Lemma~\ref{lem:a_p,-<a_r,+} shows that it takes finite time for the agents to move to condition 3. Then, from the cases above, agents must stay in condition 3 until $w_{\max}$ decreases in value, and Lemma~\ref{lem:a_r,+_decreasing_conditions} shows it takes finite time for $w_{\max}$ to decrease. Therefore, for each unique $w_{\max}$ value, either it takes a finite amount of time for its value to decrease, or the agents never leave condition 4 or 5.\\
    
    Finally, in the base case, suppose at time $t$, $w_{\max}^{(t)} =\frac{1}{D}$. Note that since $$U_{w,0}^{(t)} = U_{w,\frac{1}{D}}^{(t)}, \forall t$$ then by Claim~\ref{claim:kkt_ordering}, this is the smallest value in $\mathcal{A}$ that $w_{\max}^{(t)}$ can be. By definition of $w_{\max}$, this implies $x_{w,\frac{1}{D}}^{(t)} \ge \frac{1}{2}$ which satisfies the probability mass lower bound of condition 4 and 5 for $D>2$. Then, Lemmas~\ref{lem:a_p,-_<=_a_r,+} and ~\ref{lem:a_p,-_<=_a_r,+_for_all_t}, along with the fact that the lower bound of $f_{\min}^{(t)}$ is $\frac{0}{D}$, shows that it takes finite time for $f_{\min}^{(t)} \le  w_{\max}^{(t)}$ for all future time steps. Therefore, the conditions for the agents being in condition 4 or condition 5 are satisfied at the lowest value of $w_{\max}$. So, we can conclude there is always a finite time where the agents are in condition 4 or condition 5 for all future time steps.\\

    % todo -- special considerations for 0/1?
    % condition 4 or 5 => convergence to epsilon^*-NE if we remain in one for all future time steps
     Once agents are either in condition 4 or condition 5 for all future time steps, then the conclusion follows from Lemma~\ref{lem:final_NE_convergence}.
\end{proof}

 % theorem set up 
\paragraph{Calculating a convergence rate} Theorem~\ref{thrm:ftrl_convergence_n=1} shows that approximate convergence to $\epsilon$-NE is guaranteed under broad conditions. In our second main result, Theorem~\ref{thrm:ftrl_convergence_rate}, we give conditions under which we can derive an upper bound on the number of iterations until convergence to $\epsilon$-NE when $\eta \le 1$. We do not believe our bound is tight, but our proof highlights the phases of the algorithm that make the convergence rate large in the worst case. In particular, as the conditions of Theorem~\ref{thrm:ftrl_convergence_n=1} show, the agents' strategies evolve differently depending on whether $x_{w,w_{\max}^{(t)}}^{(t)}$ is above or below the threshold $\frac{1}{D-k+1}$, and the \textit{rate} at which they change depends on how far $x_{w,w_{\max}^{(t)}}^{(t)}$ is from this threshold. However, since we assume all initial strategy masses and algorithm parameters are reduced rational numbers, then all strategy masses will be rational numbers which ensures, at any time $t$, $x_{w,w_{\max}^{(t)}}^{(t)}$ cannot be arbitrarily close to the threshold $\frac{1}{D-k+1}$. 

So, to calculate the convergence rate, we bound the largest bit size of the reduced form denominator of the strategy masses using standard bit arithmetic and then we use a bit counting argument to lower bound the distance of a strategy mass from a threshold to find the maximum number of iterations until the agents' strategies satisfy another condition. That is, if we first find the largest value of the denominator of one rational term in reduced form is  $D_1$ and the largest value of the denominator of a second rational term in reduced form, say $D_2$, finally, we can conclude the absolute difference between these two rational terms is at least \[2^{-(\log(D_1) +\log(D_2))}.\]
This logic is formalized in Lemma~\ref{lem:bits-to-bound}.

\paragraph{Assumptions}
% assumption recall and set up to be used throughout
Throughout these results, we assume that all initial strategy masses and algorithm parameters are rational numbers and we will represent the numbers in their reduced form, i.e., $\frac{m}{n}$ for $m,n \in\mathbb{Z}, n\neq 0$ where $m$ and $n$ are coprime. We also assume that for each $i \in \{f,w\}$ and all $a\in \mathcal{A}$ the numerator and denominator of both $x_{i,a}^{(1)}$ and the learning rate $\eta$ (in their reduced forms) are each represented in at most $c$ bits for some integer $c>0$. For any $t$, if $x_{i,a}^{(t)} =0$, then we use the convention that its numerator and denominator are represented in 1 bit, i.e., it is represented as $0 = \frac{0}{1}$. Finally, note that for all $k \in \{0,\ldots,D+1\}$, $k$ is at most $\log(D+1)$ bits. 

% Thus, there exists a constant integer $d>0$ such that the bits to represent the numerator and denominator of $\eta$ as well as all $k \in  \{0,\ldots,D+1\}$ can be upper bounded by $d \log(\frac{D}{\eta})$, and we will use $d$ as such throughout this section.

First, we state our second main result, prove several lemmas, and finally prove the main theorem.
\begin{customthrm}{2}
\label{thrm:ftrl_convergence_rate}
    Suppose agents learn strategies for $\mathcal{G}^{(1)}$ using Algorithm~\ref{alg:ftrl} with $\alpha_i=\mathbf{0}$, any $0<\eta \le 1, D>2$, and arbitrary initial conditions $x_w^{(1)}, x_f^{(1)} \in \Delta(\mathcal{A})$. Then, for any $\epsilon>0$, the time to convergence of Algorithm~\ref{alg:ftrl} to an $\epsilon$-Nash equilibrium is  
    \[O\left(\left[\left(2^{\frac{D}{\eta}^{{\frac{D^{31}}{\eta^5}}}}\right)\uparrow \uparrow D\right] +\log\left(\frac{D^3}{\epsilon}\right)\cdot \frac{D^2}{\eta}\right).\]
    % \[\left(\left(2^{D^{{M\cdot \frac{D^{27}}{\eta^5}}}}\right)\uparrow \uparrow D \right)+2\log\left(\frac{3D^3}{\epsilon}\right)\cdot \log\left(\frac{4D^2}{4D^2-\eta}\right)^{-1}.\]
    % \[\left(2^{D^{D^{12M\frac{D^{18}}{\eta}}}}\right)\uparrow \uparrow D +\log\left(\frac{3D^2}{2\epsilon}\right)\cdot \log\left(\frac{2}{2-\eta}\right)^{-1}\]
    % where \[M= 3483 c\cdot m\cdot d,\]
    % for a constant $M>0$.
\end{customthrm}

\begin{customlem}{12}
    \label{lem:max_bits_of_x_i,a^t}
    Let $D>2$ be an integer and $0<\eta\le 1$ be a rational number. 
    % and suppose $x_{i,a}^{(1)}$ is a  rational number where the numerator and denominator are each at most $c>0$ bits. 
    Then, for all $t> 1$, $x_{i,a}^{(t)}$ is a rational number and the maximum number of bits to represent the numerator and the denominator of $x_{i,a}^{(t)}$ are each upper bounded by $20c^2\log\left(D+1\right)(25D^8)^t$ bits.
\end{customlem}
\begin{proof}

    To begin, note that at time $t$, if $x_{i,a}^{(t)}>0$, then the primal constraints along with Claim~\ref{claim:stationarity_reqs} imply \[x_{i,a}^{(t)} = \frac{1}{n}\cdot \left(1+\eta \sum_{a'\in \mathcal{A}, x_{i,a'}^{(t)} >0, a' \neq a}\left(U_{i,a}^{(t-1)} - U_{i,a'}^{(t-1)}\right)\right).\tag{1}\label{x-i-t}\]
    where  
    \[n = |\{a : a\in \mathcal{A}, x_{i,a}^{(t)} >0\}|,\]
    and 
    \begin{align*}
        U_{f,a}^{(t)} - U_{f,a'}^{(t)} &= \sum_{\tau=1}^t\left[\sum_{a'' \le a}x_{w,a''}^{(\tau)}\cdot (1-a) - \sum_{a'' \le a'}x_{w,a''}^{(\tau)}\cdot (1-a')\right],\\
        U_{w,a}^{(t)} - U_{w,a'}^{(t)} &= \sum_{\tau=1}^t\left[\sum_{a'' \ge a}x_{f,a''}^{(\tau)}\cdot a'' - \sum_{a'' \ge a'}x_{f,a''}^{(\tau)}\cdot a''\right].
    \end{align*}
    Thus, $x_{i,a}^{(t)}$ is the result of a finite number of addition and multiplication operations of rational numbers, so $x_{i,a}^{(t)}$ will be rational as well.
    
    Next, finding the maximum number of bits the denominator and numerator of $x_{i,a}^{(t)}$ could have is equivalent to counting bits that result from the maximum number of multiplication and addition operations that occur between rational numbers in Equation~\ref{x-i-t}. Note that if there exists $i\in\{f,w\}, a \in \mathcal{A}, 1\le \tau \le t$ such that $x_{i,a}^{(\tau)} = 0$, then this necessarily reduces the number of operations in the term \[\sum_{a'\in \mathcal{A}, x_{i,a'}^{(t)} >0, a' \neq a}\left(U_{i,a}^{(t-1)} - U_{i,a'}^{(t-1)}\right).\] Thus, for a bit upper bound, it suffices to assume all variables have non-zero mass for all time steps $t$. Under this assumption, Equation~\ref{x-i-t} can be written as the recurrence relation
    \[x_{i,a}^{(t)} = x_{i,a}^{(t-1)}+\frac{1}{D+1}\cdot \left(1+\eta \sum_{a'\in \mathcal{A}, a' \neq a}\left(u_{i}(a,x_{-i}^{(t-1)}) - u_{i}(a',x_{-i}^{(t-1)})\right)\right),\tag{2}\label{x-i-t-reccurrence}\]
    where 
    \begin{align*}
        u_{f}(a,x_{w}^{(t-1)}) - u_{f}(a',x_{w}^{(t-1)}) &=\sum_{a'' \le a}x_{w,a''}^{(t-1)}\cdot (1-a) - \sum_{a'' \le a'}x_{w,a''}^{(t-1)}\cdot(1-a'),  \tag{3}\label{firm-bits}\\
        u_{w}(x_{f}^{(t-1)},a) - u_{w}(x_{f}^{(t-1)},a') &=\sum_{a'' \ge a}x_{f,a''}^{(t-1)}\cdot a'' - \sum_{a'' \ge a'}x_{f,a''}^{(t-1)}\cdot a''.\tag{4}\label{worker-bits}
    \end{align*}
    % setting up rational number bit calculations of multiplication and addition operations
    % Note that for  $r,s \in \mathbb{Q}$ where $r=\frac{r_1}{r_2},s=\frac{s_1}{s_2}$ for $r_1,r_2,s_1,s_2 \in \mathbb{Z}$ and $r_2,s_2 \neq 0$, \[r\cdot s = \frac{r_1 \cdot s_1}{r_2 \cdot s_2}, r + s = \frac{r_1\cdot s_2 + s_1 \cdot r_2}{r_2\cdot s_2}.\tag{5}\label{rational-math}\] Therefore, if $p_1,p_2$ are the number of bits to represent $r_1,r_2$, respectively, and $q_1,q_2$ are the number of bits to represent $s_1,s_2$, respectively, then the denominator of $r\cdot s$ and $r+s$ needs at most $p_2+q_2$ bits to be represented and the numerator of $r\cdot s$ needs at most $p_1 + q_1$ bits to be represented while the numerator of $r+s$ needs at most $\max\{p_1 + q_2, s_1+p_2\} +1$ bits to be represented.
    Then, we can bound the number of bits in $x_{i,a}^{(t)}$ as a function of the number of bits in $x_{i,a}^{(t-1)}$. This will create a recurrence relation on the maximum number of bits in the numerator and denominator of $x_{i,a}^{(t)}$ which we can solve to get our final bound.
    
    % calculating max bits
    % 1. assumption on bits at t-1
    Suppose for each $i \in \{f,w\}$ and all $a \in \mathcal{A}$, the numerator and denominator of the rational number $x_{i,a}^{(t-1)}$ can each be represented in at most $c_{t-1}$ bits. Recall our assumption that $\eta$ is rational  such that the bits to represent the numerator and denominator of $\eta$ in its reduced form is $c$ and that $D+1$ can be represented in $\log(D+1)$ bits. 
    Then, using Equation~\ref{x-i-t-reccurrence} (and, consequently, Equations~\ref{firm-bits} and~\ref{worker-bits}) and standard rational bit arithmetic, we can conclude $x_{i,a}^{(t)}$ has an numerator with at most \[2D(D+1)\left(c_{t-1} + \log\left(D+1\right)\right)+ \log\left(D+1\right) +c+ c_{t-1}+ 2D + 3\] bits and a denominator with at most \[2D(D+1)\left(c_{t-1} + \log\left(D+1\right)\right)+ \log\left(D+1\right) +c+ c_{t-1}\] bits. 

    Clearly, $D>2$ implies the maximum number of bits of the numerator is larger than the denominator, so to find the maximum number of bits that can represent both the numerator and denominator in $x_{i,a}^{(t)}$, call this value $c_t$, it suffices to solve the linear recurrence 
    \[c_t = (2D(D+1)+1)c_{t-1} + \log\left(D+1\right)(2D(D+1)+1) + 2D+3+c.\]
    Through simple iteration, the closed form solution of this recurrence is
    \begin{align*}c_t = (2D(D+1)+1)^{t-1} c_1 &+ \left(\log\left(D+1\right)(2D(D+1)+1) + 2D+3+c\right)\cdot (\sum_{\tau=1}^{t-2}(2D(D+1)+1)^\tau)\\ &+ \log\left(D+1\right)(2D(D+1)+1) + 2D+3+c.
    \end{align*}
    Using the fact that, for any integers $t>0, x>0$, $\sum_{i=1}^tx^i \le x^{2t}$ and $ D>2$ implies $\log(D+1)>1$, then \[c_t \le 20c_1c \log\left(D+1\right)(25D^8)^t.\]
    % For the denominator, this yields the linear recurrence
    % \[c_t = (2D(D-1)+1)c_{t-1} + 2d(D^2-D+1),\]
    % which has the closed form 
    % \[c_t = (2D(D-1)+1)^{t-1} c_1 + 2d(D^2-D+1)\cdot (\sum_{\tau=1}^{t-2}(2D(D-1)+1)^\tau) + 2d(D^2-D+1).\]
    % % generous upper bound for both
    % This value can be upper bounded by 
    % % \[c_t = (3D^2)^{2(t-1)} c_1 + 2dD^2\cdot(3D^2)^{2(t-1)} + 2dD^2(3D^2)^{2(t-1)}.\]
    % \[c_t = 5c_1d(9D^8)^{t}.\]
    % substitute initial assumption into final bound
    Thus, since $x_{i,a}^{(1)}$ is assumed to have a numerator and a denominator each with at most $c$ bits, then we can conclude $c_1 = c$ and $x_{i,a}^{(t)}$ has a numerator and a denominator each represented by at most $ 20c^2 \log\left(D+1\right)(25D^8)^t$ bits.

\end{proof}

\begin{customlem}{13}
\label{lem:recurrence_relation_dynamics}
Suppose the agents run Algorithm~\ref{alg:ftrl} with $D>2$ and $\eta >0$.  Further, suppose at time $t$, $w_{\max}^{(t)} = \frac{k}{D}$ for some $k\in\{2,\ldots,D\}$, $x_{w,w_{\max}^{(t)}}^{(t)} \ge \frac{1}{D-k+1}$, $f_{\min}^{(t)} < f_{\max}^{(t)}$, and for all $a \in\mathcal{A} \setminus\{\frac{k}{D}, \frac{k-1}{D}\}$,\[U_{f,w_{\max}^{(t)}}^{(t-1)} -U_{f,a}^{(t-1)} \ge \frac{1}{\eta}.\] 
%$x_{w,w_{\max}^{(t)}}^{(t)} >\frac{1}{D-k+1}$, $f_{\min}^{(t)} = \frac{k-1}{D}$,
% Then, for all $\tau \ge t$ where $w_{\max}^{(\tau)}  = \frac{k}{D}$, $x_{w,w_{\max}^{(\tau)}}^{(\tau)} >\frac{1}{D-k+1}$, and $f_{\min}^{(\tau)} = \frac{k-1}{D}$, 
Then, each agent's mass on $w_{\max}^{(t)}$ evolves according to the following recurrence relations:
\[x_{f,w_{\max}^{(t)}}^{(t+1)} = \begin{cases}
    1 & f_{\min}^{(t+1)} = f_{\max}^{(t+1)},\\
    x_{f,w_{\max}^{(t)}}^{(t)} +\frac{\eta(D-k+1)}{2D}\left(x_{w,w_{\max}^{(t)}}^{(t)} - \frac{1}{D-k+1}\right) & f_{\min}^{(t+1)} < f_{\max}^{(t+1)},
\end{cases}\]
and 
\[x_{w,w_{\max}^{(t)}}^{(t+1)} = \begin{cases}
0 & w_{\max}^{(t+1)} < w_{\max}^{(t)},\\
    x_{w,w_{\max}^{(t)}}^{(t)} -\frac{\eta(k-1) k}{(k+1)D}\cdot \left(1-x_{f,w_{\max}^{(t)}}^{(t)}\right) & w_{\max}^{(t+1)} = w_{\max}^{(t)}.
\end{cases}\]
Further, for all $a \in\mathcal{A} \setminus\{\frac{k}{D}, \frac{k-1}{D}\}$,\[U_{f,w_{\max}^{(t)}}^{(t)} -U_{f,a}^{(t)} \ge \frac{1}{\eta}.\]
% Then, there exists a finite time $t_1\ge t$ where either $x_{w,w_{\max}^{(t)}}^{(t_1)} < \frac{1}{D-k+1}$, $x_{f,w_{\max}^{(t)}}^{(t_1)} =  1$, or $x_{f,w_{\max}^{(t)}}^{(\tau)} \ge 1-\epsilon$ for any $\epsilon>0$ and all $\tau \ge t_1$. 
% Further, there exists constants $c>0,d>0, m>0$ and values $\tau_{dec}, \tau_{conv},$ and $\tau_{\epsilon}$ such that, in the first case,  $x_{w,w_{\max}^{(t)}}^{(t+\tau_{dec})} < \frac{1}{D-k+1}$ is guaranteed when \[\tau_{dec} = \frac{\log(\frac{4D^3}{\eta}) + 20 c d(9D^8)^{t} +4d +3md + 5}{\log(1+\frac{\eta}{2D(D+1)})}+2,\]
% in the second case,  $x_{f,w_{\max}^{(t)}}^{(t+\tau_{conv})} =  1$ is guaranteed when \[\tau_{conv} = \frac{\log (\frac{3D^2}{\eta})+20c d(9D^8)^t + 4d + 3md + 5}{\log(1+\frac{\eta}{2D(D+1)})} +2,\]
% and in the third case, for any $\epsilon>0$, $x_{f,w_{\max}^{(t)}}^{(\tau)} \ge 1-\epsilon$ for all $\tau \ge t+\tau_{\epsilon}$ is guaranteed when
% \[\tau_{\epsilon} = \log\left(\frac{3D^2}{2\epsilon}\right)\cdot \log\left(\frac{2}{2-\eta}\right)^{-1} +2.\]
    
\end{customlem}
\begin{proof}
    First, we derive the expression for $x_{f,w_{\max}^{(t)}}^{(t+1)}$. 
    
    % for $\tau\ge t$.  Note we can assume that for $\tau$ such that $t \le \tau < t+\tau_{dec}$, $x_{w,w_{\max}}^{(\tau)} \ge \frac{1}{D-k+1}$.  Otherwise, there is certainly a time $\tau<t+\tau_{dec}$ where $x_{w,w_{\max}}^{(\tau)} < \frac{1}{D-k+1}$, as desired.  Let $t \le \tau < t+\tau_{dec}$. 
    To begin, the assumption $x_{w,w_{\max}^{(t)}}^{(t)} \ge \frac{1}{D-k+1}$ along with Lemma~\ref{lem:firm_pmf_structure} implies that for all $a \neq \frac{k}{D}$, \[u_f(w_{\max}^{(t)},x_w^{(t)})\ge u_f(a,x_w^{(t)}).\tag{1}\label{w-max-dominates}\]
    
    Since by assumption, for $ a\in\mathcal{A} \setminus\{\frac{k}{D}, \frac{k-1}{D}\}$,\[U_{f,w_{\max}^{(t)}}^{(t-1)} -U_{f,a}^{(t-1)} \ge \frac{1}{\eta},\]
    then Equation~\ref{w-max-dominates} implies that for all $a  \in\mathcal{A} \setminus\{\frac{k}{D}, \frac{k-1}{D}\}$,
    \[U_{f,w_{\max}^{(t)}}^{(t)} -U_{f,a}^{(t)} \ge \frac{1}{\eta}.\]

    Thus, by Claims~\ref{claim:stationarity_reqs} and~\ref{claim:kkt_ordering}, for $a  \in\mathcal{A} \setminus\{\frac{k}{D}, \frac{k-1}{D}\}$, $x_{f,a}^{(t)} = 0$ and $x_{f,a}^{(t+1)} = 0$.  
    % Moreover, if $x_{f,w_{\max}-\frac{1}{D}}^{(\tau)} = 0$, then $x_{f,w_{\max}}^{(\tau)} = 1$, as desired.  So we can assume that for $\tau < t + \tau_{conv}$, $x_{f,w_{\max}-\frac{1}{D}}^{(\tau)} > 0$.
    % Further, for all $t$ where $x_{w,w_{\max}}^{(t)} < \frac{1}{D-k+1}$, it holds that 
    % \[u_f(w_{\max}-\frac{1}{D},x_w^{(t)})> u_f(w_{\max},x_w^{(t)}).\tag{2}\label{w-max-1/D-dominates-w-max}\]
    This fact along with the assumption $f_{\min}^{(t)} < f_{\max}^{(t)}$ and $w_{\max}^{(t)} = \frac{k}{D}$ imply   \[x_{f,w_{\max}^{(t)}-\frac{1}{D}}^{(t)}>0 \text{ and } x_{f,w_{\max}^{(t)}}^{(t)}>0,\]
    so by Claim~\ref{claim:stationarity_reqs}, 
    \[x_{f,w_{\max}^{(t)}}^{(t)}- x_{f,w_{\max}^{(t)}-\frac{1}{D}}^{(t)} = \eta U_{f,w_{\max}^{(t)}}^{(t-1)} - \eta U_{f,w_{\max}^{(t)}-\frac{1}{D}}^{(t-1)}.\tag{2}\label{firm-mass-t}\]
    Note that by definition, $f_{\min}^{(t+1)} \le f_{\max}^{(t+1)}$. First, suppose $f_{\min}^{(t+1)} = f_{\max}^{(t+1)}$, then this implies $x_{f,f_{\max}^{(t+1)}}^{(t+1)} = 1$. Since $x_{f,a}^{(t+1)}=0$ for all $a  \in\mathcal{A} \setminus\{\frac{k}{D},\frac{k-1}{D}\}$ and $u_f(w_{\max}^{(t)},x_w^{(t)})\ge u_f(w_{\max}^{(t)}-\frac{1}{D},x_w^{(t)})$, then $f_{\max}^{(t+1)} = w_{\max}^{(t)}$, thus, $x_{f,w_{\max}^{(t)}}^{(t+1)} =1$.

    Otherwise, suppose $f_{\min}^{(t+1)} < f_{\max}^{(t+1)}$. Since $x_{f,a}^{(t+1)}=0$ for all $a  \in\mathcal{A} \setminus\{\frac{k}{D},\frac{k-1}{D}\}$, then $f_{\min}^{(t+1)} < f_{\max}^{(t+1)}$ implies $x_{f,w_{\max}^{(t)}-\frac{1}{D}}^{(t+1)} > 0$ and $x_{f,w_{\max}^{(t)}}^{(t+1)} > 0$. So, by Claim~\ref{claim:stationarity_reqs}, \[x_{f,w_{\max}^{(t)}}^{(t+1)}- x_{f,w_{\max}^{(t)}-\frac{1}{D}}^{(t+1)} = \eta U_{f,w_{\max}^{(t)}}^{(t)} - \eta U_{f,w_{\max}^{(t)}-\frac{1}{D}}^{(t)}.\]  
    This fact along with Equation~\ref{firm-mass-t} and the primal constraint that the probability mass must sum to 1 implies
    \[x_{f,w_{\max}^{(t)}}^{(t+1)} = x_{f,w_{\max}^{(t)}}^{(t)} +\frac{\eta}{2}\left(u_f(w_{\max}^{(t)},x_w^{(t)}) - u_f(w_{\max}^{(t)}-\frac{1}{D},x_w^{(t)})\right).\]
    Since \[u_f(w_{\max}^{(t)},x_w^{(t)}) = \left(1-\frac{k}{D}\right)\cdot 1,\]
    and \[u_f(w_{\max}^{(t)}-\frac{1}{D},x_w^{(t)}) = \left(1-\frac{k-1}{D}\right)\cdot (1-x_{w,w_{\max}^{(t)}}^{(t)}),\]
    then, 
    \[x_{f,w_{\max}^{(t)}}^{(t+1)} = x_{f,w_{\max}^{(t)}}^{(t)} +\frac{\eta(D-k+1)}{2D}\left(x_{w,w_{\max}^{(t)}}^{(t)} - \frac{1}{D-k+1}\right).\]

    Next, we write the expression for $x_{w,w_{\max}^{(t)}}^{(t+1)}$.
    
    % As before, we can assume that for $\tau$ such that $t \le \tau < t+\tau_{dec}$, $x_{w,w_{\max}}^{(\tau)} > 0$.  
    To begin, since $w_{\max}^{(t)}$ is defined as the largest acceptance threshold with mass at time $t$, the assumption $w_{\max}^{(t)} = \frac{k}{D}$ implies $x_{w,w_{\max}^{(t)}}^{(t)} >0$ and for all $a > w_{\max}^{(t)}$, $x_{w,a}^{(t)} =0$. Moreover, by Lemma~\ref{lem:worker_pmf_structure}, for all $a < w_{\max}^{(t)}$, $x_{w,a}^{(t)} >0$.  Then, by Claim~\ref{claim:stationarity_reqs}, for all $a < w_{\max}^{(t)}$,
    \[x_{w,w_{\max}^{(t)}}^{(t)} - x_{w,a}^{(t)} = \eta U_{w,w_{\max}^{(t)}}^{(t-1)} - \eta U_{w,a}^{(t-1)}.\tag{3}\label{worker-mass-t}\]
    
    Next, by Lemma~\ref{lem:a_r,+_never_increases}, $w_{\max}^{(t+1)} \le w_{\max}^{(t)}$. First, suppose $w_{\max}^{(t+1)} < w_{\max}^{(t)}$, then this implies by definition of $w_{\max}$, $x_{w,w_{\max}^{(t)}}^{(t+1)} = 0$. 
    
    Otherwise, suppose $w_{\max}^{(t+1)} = w_{\max}^{(t)}$. Then, as above, for all $a > w_{\max}^{(t)}$, $x_{w,a}^{(t)} =0$, and for all $a < w_{\max}^{(t)}$, 
    \[x_{w,w_{\max}^{(t)}}^{(t+1)} - x_{w,a}^{(t+1)} = \eta U_{w,w_{\max}^{(t)}}^{(t)} - \eta U_{w,a}^{(t)}.\]
    This fact along with Equation~\ref{worker-mass-t} and the primal constraint that the probability mass must sum to 1 implies 
    \[x_{w,w_{\max}^{(t)}}^{(t+1)} = x_{w,w_{\max}^{(t)}}^{(t)} +\frac{\eta}{k+1}\left(\sum_{a < w_{\max}}\left( u_w(x_f^{(t)}, w_{\max}^{(t)}) - u_w(x_f^{(t)}, a)\right)\right).\]
    
    From the above derivation of $x_{f,w_{\max}^{(t)}}^{(t+1)}$, the assumptions of this lemma imply $x_{f,a}^{(t)} =0$ for all $a \in\mathcal{A}\setminus\{\frac{k}{D}, \frac{k-1}{D}\}$ and $x_{f,w_{\max}^{(t)}}^{(t)} >0,x_{f,w_{\max}^{(t)}-\frac{1}{D}}^{(t)} >0$. Thus, \[u_w(x_f^{(t)}, w_{\max}^{(t)}) = \frac{k}{D}\cdot x_{f,w_{\max}^{(t)}}^{(t)},\]
    and for all $a < \frac{k}{D}$,
    \[u_w(x_f^{(t)}, a) = \frac{k}{D}\cdot x_{f,w_{\max}^{(t)}}^{(t)} + \frac{k-1}{D}\cdot (1-x_{f,w_{\max}^{(t)}}^{(t)}).\]
    Therefore,
    \[x_{w,w_{\max}^{(t)}}^{(t+1)} = x_{w,w_{\max}^{(t)}}^{(t)} -\frac{\eta(k-1) k}{(k+1)D}\cdot \left(1-x_{f,w_{\max}^{(t)}}^{(t)}\right).\]
    \end{proof}

    \begin{customlem}{14}
    \label{lem:bounding_recurrence_relation_time}
        Suppose the agents run Algorithm~\ref{alg:ftrl} with $D>2$ and $0< \eta \le 1$.  Further, suppose at time $t$, $w_{\max}^{(t)} = \frac{k}{D}$ for some $k\in\{2,\ldots,D\}$, $x_{w,w_{\max}^{(t)}}^{(t)} \ge \frac{1}{D-k+1}$, $f_{\min}^{(t)} < f_{\max}^{(t)}$, and for all $a \in\mathcal{A} \setminus\{\frac{k}{D}, \frac{k-1}{D}\}$,\[U_{f,w_{\max}^{(t)}}^{(t-1)} -U_{f,a}^{(t-1)} \ge \frac{1}{\eta}.\]
        
       Then, there exists an integer $\tau \ge 0$ where either 1) $x_{w,w_{\max}^{(t)}}^{(t+\tau)} < \frac{1}{D-k+1}$, 2) $x_{f,w_{\max}^{(t)}}^{(t+\tau)} = 1$, or 3) for any $\epsilon>0$, $x_{f,w_{\max}^{(t)}}^{(\tau')} \ge 1-\epsilon$ for all $\tau' \ge t+\tau$.\\
       In case 1, $\tau \le \tau_{dec}$ where
       \[\tau_{dec} = \frac{519c^2\left(25\frac{D^{10}}{\eta}\right)^t}{\log(1+\frac{\eta}{2D(D+1)})}.\]
       % \frac{\log(\frac{4D^3}{\eta}) + 20 c d(9D^8)^{t} +4d +3md + 5}{\log(1+\frac{\eta}{2D(D+1)})}+2,\]
       In case 2, $\tau \le \tau_{conv}$ where 
       \[\tau_{conv} = \frac{519c^2\left(25\frac{D^{10}}{\eta}\right)^t}{\log(1+\frac{\eta}{2D(D+1)})}.\]
       % \frac{\log (\frac{3D^2}{\eta})+20c d(9D^8)^t + 4d + 3md + 5}{\log(1+\frac{\eta}{2D(D+1)})} +2,\]
       In case 3, for any $\epsilon>0$, $\tau \le \tau_{\epsilon}$ where 
      \[\tau_{\epsilon} = 2\log\left(\frac{3D^3}{\epsilon}\right)\cdot \frac{4D^2}{\eta}.\]
    \end{customlem}
    \begin{proof}
    The assumptions at time $t$ satisfy those of  Lemma~\ref{lem:recurrence_relation_dynamics}, so there are three possible outcomes at time $t+1$. First, $w_{\max}^{(t+1)}< w_{\max}^{(t)}$ implies $x_{w,w_{\max}^{(t)}}^{(t+1)} = 0$ which implies the outcome $x_{w,w_{\max}^{(t)}}^{(t+1)} < \frac{1}{D-k+1}$ is reached at time $t+1$. Second, $f_{\min}^{(t+1)} = f_{\max}^{(t+1)}$ implies $x_{f,w_{\max}^{(t)}}^{(t+1)} = 1$ which implies the outcome $x_{f,w_{\max}^{(t)}}^{(t+1)}=1$ is reached at time $t+1$. Finally, $f_{\min}^{(t+1)} < f_{\max}^{(t+1)}$ and $w_{\max}^{(t+1)}= w_{\max}^{(t)}$ imply
    \[x_{w,w_{\max}^{(t)}}^{(t+1)} = x_{w,w_{\max}^{(t)}}^{(t)} -\frac{\eta(k-1) k}{(k+1)D}\cdot \left(1-x_{f,w_{\max}^{(t)}}^{(t)}\right),\tag{1}\label{worker-recurrence}\]
    and
    \[x_{f,w_{\max}^{(t)}}^{(t+1)} = x_{f,w_{\max}^{(t)}}^{(t)} +\frac{\eta(D-k+1)}{2D}\left(x_{w,w_{\max}^{(t)}}^{(t)} - \frac{1}{D-k+1}\right).\tag{2}\label{firm-recurrence}\]
    Further, for all $a \in\mathcal{A} \setminus\{\frac{k}{D}, \frac{k-1}{D}\}$,\[U_{f,w_{\max}^{(t)}}^{(t)} -U_{f,a}^{(t)} \ge \frac{1}{\eta}.\tag{3}\label{w-max-still-max}\]
    % Note that Equation~\ref{worker-recurrence} along with $f_{\min}^{(t)} < f_{\max}^{(t)}$ implies $x_{w,w_{\max}^{(t)}}^{(t+1)} < x_{w,w_{\max}^{(t)}}^{(t)}$ while Equation~\ref{firm-recurrence} along with $x_{w,w_{\max}^{(t)}}^{(t)} \ge \frac{1}{D-k+1}$ implies $x_{f,w_{\max}^{(t)}}^{(t+1)} > x_{f,w_{\max}^{(t)}}^{(t)}$ if $x_{w,w_{\max}^{(t)}}^{(t)} > \frac{1}{D-k+1}$ and otherwise $x_{f,w_{\max}^{(t)}}^{(t+1)} = x_{f,w_{\max}^{(t)}}^{(t)}$. 
    If either $x_{w,w_{\max}^{(t)}}^{(t+1)} < \frac{1}{D-k+1}$ or $x_{f,w_{\max}^{(t)}}^{(t+1)}=1$, then the conclusion again holds at time $t+1$, otherwise $x_{w,w_{\max}^{(t)}}^{(t+1)} \ge \frac{1}{D-k+1}$ and $x_{f,w_{\max}^{(t)}}^{(t+1)}<1$  and the assumptions of Lemma~\ref{lem:recurrence_relation_dynamics} are satisfied again. Therefore, $x_{w,w_{\max}^{(t)}}^{(t)}$ strictly decreases in mass while $x_{f,w_{\max}^{(t)}}^{(t)}$ weakly increases in mass according to Equations~\ref{worker-recurrence} and~\ref{firm-recurrence} until there exists a value $\tau \ge 0$  where either $x_{w,w_{\max}^{(t)}}^{(t+\tau)} < \frac{1}{D-k+1}$, $x_{f,w_{\max}^{(t)}}^{(t+\tau)} = 1$, or, for any $\epsilon>0$, $x_{f,w_{\max}^{(t)}}^{(\tau')} \ge 1-\epsilon$ for all $\tau' \ge t+\tau$. 
    
    To bound the time it takes for one of these cases to occur, we next find a closed form solution for the strategy mass recurrence relations. Note that Equations~\ref{worker-recurrence} and~\ref{firm-recurrence} only give a valid mass of $x_{w,w_{\max}^{(t)}}^{(\tau+1)}$ and  $x_{f,w_{\max}^{(t)}}^{(\tau+1)}$, respectively, if both $x_{w,w_{\max}^{(t)}}^{(\tau)} \ge \frac{1}{D-k+1}$ and $x_{f,w_{\max}^{(t)}}^{(\tau)} < 1$. So, we introduce new recurrent variables $w$ and $f$ where \[w^{(0)} = x_{w,w_{\max}^{(t)}}^{(t)}, f^{(0)} = x_{f,w_{\max}^{(t)}}^{(t)},\]
    and
    \begin{align*}
    w^{(\tau+1)} &= w^{(\tau)} -\frac{\eta(k-1) k}{(k+1)D}\cdot \left(1-f^{(\tau)}\right), &\tag{4}\label{worker-recurrence-proxy}\\
    f^{(\tau+1)} &= f^{(\tau)} +\frac{\eta(D-k+1)}{2D}\left(w^{(\tau)} - \frac{1}{D-k+1}\right),&\tag{5}\label{firm-recurrence-proxy}\
    \end{align*}
    such that we can conclude, for all $\tau \ge 0$, \[\begin{cases}
        x_{w,w_{\max}^{(t)}}^{(t+\tau+1)} = w^{(\tau+1)} & w^{(\tau)} \ge \frac{1}{D-k+1}, f^{(\tau)} < 1\\
        x_{w,w_{\max}^{(t)}}^{(t+\tau+1)} < \frac{1}{D-k+1} & w^{(\tau+1)} < \frac{1}{D-k+1}
    \end{cases}\]
     and 
     \[\begin{cases}
        x_{f,w_{\max}^{(t)}}^{(t+\tau+1)} = f^{(\tau+1)} & w^{(\tau)} \ge \frac{1}{D-k+1}, f^{(\tau)} < 1\\
        x_{f,w_{\max}^{(t)}}^{(t+\tau+1)} = 1 & f^{(\tau+1)} \ge 1
    \end{cases}\]
    Thus, we can analyze the limiting behavior of the closed form of Equations~\ref{worker-recurrence-proxy} and~\ref{firm-recurrence-proxy} to determine the outcomes of the variables $x_{w,w_{\max}^{(t)}}^{(t+\tau)} $ and $x_{f,w_{\max}^{(t)}}^{(t+\tau)}$ as a function of the conditions at time $t$.
    
    For ease of notation, let \[A:= \frac{\eta(k-1) k}{(k+1)D},B:= \frac{\eta(D-k+1)}{2D},C:= \frac{\eta}{2D}.\tag{6}\label{notation-1}\] Since $\eta >0$ and $D>2$, then $A,B,C$ are all positive reals.
    
    % Then, recurrence relations~\ref{worker-recurrence} and~\ref{firm-recurrence} simplify to 
    % \begin{align*}
    %     w^{(t+1)} &= w^{(t)} + A \cdot f^{(t)} - A,\\
    %     f^{(t+1)} &= f^{(t)} + B \cdot w^{(t)} - C.
    % \end{align*}
    Then, by applying Lemma~\ref{lem:solving_recurrence_relations} to Equations~\ref{worker-recurrence-proxy} and~\ref{firm-recurrence-proxy},
    \begin{align*}
        w^{(\tau)} &= \alpha_1^w\sqrt{AB}(1+\sqrt{AB})^{\tau-1} - \alpha_2^w\sqrt{AB}(1-\sqrt{AB})^{\tau-1} + \frac{1}{D-k+1},&\tag{7}\label{worker-closed}\\
        f^{(\tau)} &= \alpha_1^f\sqrt{AB}(1+\sqrt{AB})^{\tau-1} - \alpha_2^f\sqrt{AB}(1-\sqrt{AB})^{\tau-1} + 1,&\tag{8}\label{firm-closed}
    \end{align*} 
    where, in Expressions~\ref{a_1_w} through~\ref{a_2_f} in the statement of Lemma~\ref{lem:solving_recurrence_relations},
    \[c_w := w^{(0)} - \frac{1}{D-k+1},c_f := 1-f^{(0)}\tag{9}\label{notation-2}\] and either $\sign(\alpha_1^f)= \sign(\alpha_1^w)$ or $\alpha_1^f=\alpha_1^w = 0$. So, there are three possible cases to consider and we will show that these cases correspond to the three possible outcomes: 1) $\alpha_1^w,\alpha_1^f<0$ imply there exists $\tau \ge 0$ where $x_{w,w_{\max}^{(t)}}^{(t+\tau)} < \frac{1}{D-k+1}$, 2) $\alpha_1^w,\alpha_1^f>0$ imply  there exists $\tau \ge 0$ where $x_{f,w_{\max}^{(t)}}^{(t+\tau)} = 1$, or 3) $\alpha_1^w=\alpha_1^f=0$ imply for any $\epsilon>0$, there exists $\tau \ge 0$ where $x_{f,w_{\max}^{(t)}}^{(\tau')} \ge 1-\epsilon$ for all $\tau' \ge t+\tau$.
    
    %Bounding $\tau_{dec}$
    \textbf{Case 1:} $\alpha_1^w,\alpha_1^f <0$.\\
    Suppose $\alpha_1^w <0$ and note that $0<\eta \le 1$ implies $0<A < 1$ and $0<B < 1$ for all $k \in \{2,\ldots,D\}$. Then, $\alpha_1^w<0$ and $0<AB < 1$ imply \[\lim_{\tau \to \infty}w^{(\tau)} = -\infty.\] Thus, there exists $\tau_{dec} \ge 0$ where $w^{(\tau_{dec})} < \frac{1}{D-k+1}$, thereby implying $x_{w,w_{\max}^{(t)}}^{(t+\tau_{dec})} < \frac{1}{D-k+1}$. To upper bound the value of $\tau_{dec}$, we upper bound Equation~\ref{worker-closed} while maintaining its limiting behavior. That is, we first lower bound $\alpha_2^w$, and then we upper bound $\alpha_1^w$ with a negative number to ensure the final upper bound on $w^{(\tau)}$ also limits to $-\infty$.
      
    \textbf{Bounding $\alpha_2^w$ and $\alpha_1^w$.}\\
    First, we lower bound $\alpha_2^w$.
    \begin{align*}
        \alpha_2^w &\ge -\frac{1}{2B}  - \frac{1}{2\sqrt{AB}}\tag{10}\label{line1-alpha-2}\\
        &\ge -\frac{D + D(D+1)}{\eta}\tag{11}\label{line2-alpha-2}\\
        & \ge -\frac{3D^2}{\eta} 
    \end{align*}
    where Line~\ref{line1-alpha-2} follows from Equation~\ref{a_2_w} and that fact that $c_w,c_f \in [0,1]$ and Line~\ref{line2-alpha-2} follows from $B \ge \frac{\eta}{2D}$ and $\sqrt{AB} \ge \frac{\eta}{2D(D+1)}$ for all $k \in \{2,\ldots,D\}$.
    
    Next, we upper bound $\alpha_1^w$ with a negative value. First, note that $\alpha_1^w <0$ and $AB < 1$ imply $c_w-\frac{1}{B}c_f <0$, so $|\alpha_1^w|$ is minimized when $c_w-A\cdot c_f > 0$ and, thus, it suffices to upper bound $\alpha_1^w$ under the assumption $c_w-A\cdot c_f > 0$. 
    Then, the fact that $\sqrt{AB} <1$ along with Equation~\ref{a_1_w} and the assumption $c_w-A\cdot c_f > 0$ imply that any $x \in \mathbb{R}$ that satisfies \[\frac{c_w-Ac_f}{\frac{1}{B}c_f-c_w}<x \le \sqrt{AB},\tag{13}\label{case-1-bounds}\]
    also satisfies
     \[\alpha_1^w \le \frac{1}{2}(c_w - \frac{1}{B}c_f)+\frac{1}{2x}(c_w-A\cdot c_f)<0. \tag{14}\label{case-1-x-expression}\]
     Finally, since $\eta, D, c_w,c_f, A,$ and $B$ are all rational, we want to find an $x \in \mathbb{Q}$ that satisfies the bounds in~\ref{case-1-bounds} in order to count bits from a series of multiplication and addition operations of rational numbers.
     \paragraph{Constructing $x$.} To construct such a rational $x$, we lower bound the difference $\sqrt{AB} -\frac{c_w-Ac_f}{\frac{1}{B}c_f-c_w}$ by a rational number $r$ such that it suffices to set $x = \frac{c_w-Ac_f}{\frac{1}{B}c_f-c_w} + r$.
   
    To begin, using the definition of $A$ and $B$ from~\ref{notation-1},\[\sqrt{AB} -\frac{c_w-Ac_f}{\frac{1}{B}c_f-c_w} =\frac{\eta}{2D(k-1)}\left(\sqrt{2k(k-1)(k+1)(D-k+1)} - \frac{\frac{2D(k-1)}{\eta}c_w - \frac{2k(k-1)^2}{(k+1)}c_f}{\frac{2D}{\eta(D-k+1)}c_f-c_w}\right).\]
    % \[\frac{D_k^1\cdot c_w^N\cdot c_f^D\cdot \eta^D - D_k^2\cdot c_f^N\cdot \eta^N\cdot c_w^N}{D_k^3\cdot c_f^N\cdot c_w^D\cdot \eta^D - D_k^4\cdot c_w^N\cdot \eta^N \cdot c_f^D}\]
    Note that the first term in the parentheses is of the form $\sqrt{n}$ for $n \in \mathbb{Z}_{>0}$ and the second term in the parentheses is a rational number and let $\frac{R}{S}$ denote the reduced form of the second term in the parentheses. Further, note that the largest possible square in the radicand of the first term is $D^4$. Since $\sqrt{s^2 r} - a = s(\sqrt{r}-\frac{a}{s})$ for any $s,a \in \mathbb{Z}_{>0}$ and square-free $r \in \mathbb{Z}_{\ge 0}$, then pulling out all square terms to ensure the radicand is square-free increases the denominator of the second term by at most a factor of $D^2$.  Thus, by Lemma~\ref{lem:rational-approx-bound}, there exists coprime integers $P,Q$ where $Q \le C_k \cdot S$ for a constant $C_k <128(2k(k-1)(k+1)(D-k+1))^2$ such that \[\sqrt{AB} -\frac{c_w-Ac_f}{\frac{1}{B}c_f-c_w} \ge \frac{\eta}{2D(k-1)}\left(\frac{P}{Q} - \frac{R}{S}\right)>0.\] 
    Note that, for all $k \in \{2,\ldots,D\}$,
    \[C_k < 512D^8.\]
    % which is represented in at most $8\log(2D)$ bits.
    
    Further, let the reduced rational representation of $c_w,c_f,$ and $\eta$ be denoted as \[c_w = \frac{c_w^N}{c_w^D}, c_f = \frac{c_f^N}{c_f^D},\eta = \frac{\eta^N}{\eta^D},\]
    where $c_w^N, c_f^N,\eta^N \in \mathbb{Z} $ and $c_w^D, c_f^D,\eta^D \in \mathbb{Z}_{>0}$.
    Therefore, the largest denominator of $\frac{\eta}{2D(k-1)}\left(\frac{P}{Q} - \frac{R}{S}\right)$ is \[\eta^D\cdot 2\cdot D\cdot (k-1)\cdot 512\cdot D^8\cdot S^2, \tag{15}\label{largest-denominator-of-diff}\]
    where, using standard fraction simplification along with the factor of $D^2$ to ensure the radicand is square-free,
    \[S \le D^2\cdot \left(2D\cdot(k+1)\cdot c_f^N\cdot c_w^D\cdot \eta^D - (D-k+1)\cdot (k+1)\cdot c_w^N\cdot \eta^N \cdot c_f^D\right). \tag{16}\label{S-upper-bound}\]

    Finally, using standard bit arithmetic, we can calculate an upper bound on the maximum number of bits in expression~\ref{largest-denominator-of-diff}, denoted as $\beta$. Recall that the bits to represent the numerator and denominator of the rational $\eta$ in the reduced form is $c$ and for all $k \in \{0,\ldots,D+1\}$, the number of bits to represent $k$ is at most $\log(D+1)$.     
    \begin{align*}
        \beta &\le  2\log\left(D+1\right) + c + 2 + 8\log(2D) + 2 \log(S)\\
        &\le 10\log\left(2D+1\right) + c + 2  + 2\left(2\log(D) + 2\log\left(D+1\right) + c+44c^2\log\left(D+1\right)(25D^8)^t+2\right)&\tag{17}\label{line-2-beta}\\
        &\le 115c^2\log\left(2D+1\right)(25D^8)^t
    \end{align*}
    where Line~\ref{line-2-beta} follows from bound~\ref{S-upper-bound} and, using Definition~\ref{notation-2} and Lemma~\ref{lem:max_bits_of_x_i,a^t}, that, for $i \in \{f,w\}$, $c_i^N$ and $c_i^D$ are both represented in at most $22c^2\log\left(D+1\right)(25D^8)^t$ bits.
    
    Let $\beta'$ be the upper bound on $\beta$, i.e., $\beta' =115c^2\log\left(2D+1\right)(25D^8)^t$ and by Lemma~\ref{lem:bits-to-bound}, \[\sqrt{AB} -\frac{c_w-Ac_f}{\frac{1}{B}c_f-c_w} \ge 2^{-\beta'}=: r,\] and we set $x =\frac{c_w-Ac_f}{\frac{1}{B}c_f-c_w} + 2^{-\beta'} $.
    
    \paragraph{Using $x$ to bound $\alpha_1^w$.} Plugging this value of $x$ into expression~\ref{case-1-x-expression} gives a negative upper bound on $\alpha_1^w$ that is rational and, using similar bit arithmetic as above, the maximum number of bits to represent its denominator is \[\mathcal{B} = 259c^2\log\left(2D+1\right)(25D^8)^t.\tag{18}\label{final-beta-case-1}\]
    Thus, by Lemma~\ref{lem:bits-to-bound}, \[\alpha_{1}^w \le -2^{-\mathcal{B}}.\]

    \textbf{Bounding $\tau_{dec}$.}\\ 
    Since $0<AB\le 1$, then for all $\tau\ge 1$, \[(1-\sqrt{AB})^{\tau-1} < 1.\] 
    Therefore,
    \begin{align*}
        w^{(\tau)} &\le  -2^{-\mathcal{B}}(\sqrt{AB})(1+\sqrt{AB})^{\tau-1} +\frac{3D^2}{\eta}(\sqrt{AB}) + \frac{1}{D-k+1}.
    \end{align*}
    This immediately implies $w^{(\tau)} < \frac{1}{D-k+1}$ when 
    \[\tau > \frac{\log(\frac{3D^2}{\eta}) + \mathcal{B}}{\log(1+\sqrt{AB})}+1.\]

Finally, since $1+\sqrt{AB} \ge 1+\frac{\eta}{2D(D+1)}$, then, substituting in the definition~\ref{final-beta-case-1} and simplifying, the setting 
% \[\tau_{dec} = \frac{514cd\log\left(\frac{D^2}{\eta}\right)(25D^8)^t}{\log(1+\frac{\eta}{2D(D+1)})},\]
\[\tau_{dec} = \frac{519c^2\left(25\frac{D^{10}}{\eta}\right)^t}{\log(1+\frac{\eta}{2D(D+1)})},\]
    is sufficient to conclude $w^{(\tau_{dec})}  < \frac{1}{D-k+1}$, and, therefore, to conclude $x_{w,w_{\max}^{(t)}}^{(t+\tau_{dec})} < \frac{1}{D-k+1}$.

    \textbf{Case 2:} $\alpha_1^w,\alpha_1^f>0$.\\ 
     Suppose $\alpha_1^f>0$. Then, $\alpha_1^f>0$ and $0<AB < 1$ imply \[\lim_{\tau \to \infty}f^{(\tau)} = +\infty.\] Thus, there exists $\tau_{conv} \ge 0$ where $f^{(\tau_{conv})} \ge 1$ which implies $x_{f,w_{\max}^{(t)}}^{(t+\tau_{conv})} =1$. We use Equations~\ref{firm-closed},~\ref{a_1_f}, and~\ref{a_2_f} to bound the value of $\tau_{conv}$ in the same way as we bound $\tau_{dec}$ in Case 1, though with slightly different constants. Thus, we omit the details and state the final bound of 
     % \[\tau_{conv} = \frac{ 515cd\log\left(\frac{D^2}{\eta}\right)(25D^8)^t}{\log(1+\frac{\eta}{2D(D+1)})}.\]
     \[\tau_{conv} = \frac{519c^2\left(25\frac{D^{10}}{\eta}\right)^t}{\log(1+\frac{\eta}{2D(D+1)})},\]

    \textbf{Case 3:} $\alpha_1^w =\alpha_1^f=0$.\\
     Finally, suppose $\alpha_1^f = 0$. Then, by Equations~\ref{a_1_f} and~\ref{a_2_f}, $\alpha_1^f=0$ and $AB < 1$ imply $\alpha_2^f>0$ and, thus, by Equation~\ref{firm-closed}, 
     \[f^{(\tau)} = -\alpha_2^f\sqrt{AB}(1-\sqrt{AB})^{\tau-1} + 1,\]
     and, since $0<\sqrt{AB} <1$,
     \[\lim_{\tau\to \infty} f^{(\tau)} = 1.\] 
     % Further, using equations~\ref{a_1_w_2} and~\ref{a_2_w_2}, $\alpha_1^w=0$ and $AB < 1$ imply $\alpha_2^w<0$ and, thus, by equation~\ref{worker-closed},\[\lim_{\tau\to \infty} w^{(t+\tau)} = \frac{1}{D-k+1},\]
    So, for any $\epsilon>0$, there exists $\tau_{\epsilon}\ge 0$ where $f^{(\tau)} \ge 1-\epsilon$ for all $\tau\ge \tau_{\epsilon}$ which implies $x_{f,w_{\max}^{(t)}}^{(\tau)} \ge 1-\epsilon$ for all $\tau \ge t+\tau_{\epsilon}$.
    
    To find the largest such $\tau_{\epsilon}$, we first find an upper bound on $\alpha_2^f$.
    \begin{align*}
        \alpha_2^f &\le \frac{1}{2A}+\frac{1}{2\sqrt{AB}}\tag{19}\label{line1-alpha-2-f}\\
        &\le \frac{3D^2+3D}{2\eta}\tag{20}\label{line2-alpha-2-f}\\
        & \le \frac{3D^2}{\eta} 
    \end{align*}
    where Line~\ref{line1-alpha-2-f} follows from Equation~\ref{a_2_f_2} and that fact that $c_w,c_f \in [0,1]$ and Line~\ref{line2-alpha-2-f} follows from $A \ge \frac{\eta}{D(D+1)}$ and $\sqrt{AB} \ge \frac{\eta}{2D(D+1)}$ for all $k \in \{2,\ldots,D\}$.
    
    % First, since $c_w,c_f \in[0,1]$,then \[\alpha_{2}^f \le \frac{1}{2A}+\frac{1}{2\sqrt{AB}}.\]
    % Further, $\frac{1}{A} \le \frac{D(D+1)}{\eta}$ and $\frac{1}{\sqrt{AB}} \le \frac{2D(D+1)}{\eta}$ for all $k \in \{2,\ldots,D\}$ imply
    % \[\alpha_{2}^f \le \frac{3D^2+3D}{2\eta}\le \frac{3D^2}{\eta}.\]
    
    Then, since $\sqrt{AB} \le \eta D$ for all $k \in \{2,\ldots,D\}$,  
    \[-\alpha_2^f(\sqrt{AB}) \ge -3D^3.\]
    Finally, since $1-\sqrt{AB} \le 1-\frac{\eta}{2D(D+1)}$ for all $k\in\{2,\ldots,D\}$, then \[f^{(\tau)} \ge -3D^3(1-\frac{\eta}{2D(D+1)})^{\tau-1}+1.\]

    Let $\epsilon>0$. Then, $f^{(\tau)} \ge 1-\epsilon$ for all $\tau \ge \tau_{\epsilon}$ when \[\tau_{\epsilon} \ge \log\left(\frac{3D^3}{\epsilon}\right)\cdot \log\left(\frac{2D(D+1)}{2D(D+1)-\eta}\right)^{-1} +1.\]

    Finally, since $\log(1+x) \ge \frac{x}{1+x}$ for all $x>-1$, it suffices to have \[\tau_{\epsilon} = 2\log\left(\frac{3D^3}{\epsilon}\right)\cdot \frac{4D^2}{\eta}.\]
    
    % \[\tau_{\epsilon} = 2\log\left(\frac{3D^3}{\epsilon}\right)\cdot \log\left(\frac{4D^2}{4D^2-\eta}\right)^{-1}.\]

\end{proof}

\begin{customthrm}{2}
\label{thrm:ftrl_convergence_rate}
    Suppose agents learn strategies for $\mathcal{G}^{(1)}$ using Algorithm~\ref{alg:ftrl} with $\alpha_i=\mathbf{0}$, any $0<\eta \le 1, D>2$, and arbitrary initial conditions $x_w^{(1)}, x_f^{(1)} \in \Delta(\mathcal{A})$. Then, for any $\epsilon>0$, the maximum time to convergence of Algorithm~\ref{alg:ftrl} to an $\epsilon$-Nash equilibrium is
    \[O\left(\left[\left(2^{\frac{D}{\eta}^{{\frac{D^{31}}{\eta^5}}}}\right)\uparrow \uparrow D\right] +\log\left(\frac{D^3}{\epsilon}\right)\cdot \frac{D^2}{\eta}\right).\]
    % \[\left(2^{D^{{ M\cdot \frac{D^{27}}{\eta^5}}}}\right)\uparrow \uparrow D +2\log\left(\frac{3D^3}{\epsilon}\right)\cdot \log\left(\frac{4D^2}{4D^2-\eta}\right)^{-1}.\]
    % \[\left(2^{D^{D^{12M\frac{D^{18}}{\eta}}}}\right)\uparrow \uparrow D +\log\left(\frac{3D^2}{2\epsilon}\right)\cdot \log\left(\frac{2}{2-\eta}\right)^{-1}\]
    % where \[M= 3483 c\cdot m\cdot d,\]
    % for a constant $M>0$.
\end{customthrm}
\begin{proof}

    % start by formalizing the conditions for convergence to eps-NE using theorem 1. reader should know what the goal of the bound is in this paragraph.
     To begin, recall that Lemma~\ref{lem:final_NE_convergence} gives the conditions at some time $t^*$ under which we can conclude that $(x_f^{(\tau)},x_w^{(\tau)})$ is in the same $\epsilon$-Nash equilibrium for all $\tau \ge t^*$, that is, the agents have converged to an $\epsilon$-Nash equilibrium. Thus, to find the maximum time to convergence of Algorithm~\ref{alg:ftrl} to an $\epsilon$-Nash equilibrium, it suffices to bound the time $t^*$ at which we are guaranteed the following conditions hold: Let $w_{\max}^{(t^*)} = \frac{k}{D}$ for some $k = \{1,\ldots,D\}$, then $x_{w,w_{\max}^{(t^*)}}^{(\tau)} \ge \frac{1}{D-k+1}$ for all $\tau \ge t^*$ and either $x_{f,w_{\max}^{(t^*)}}^{(\tau)} = 1$ for all $\tau \ge t^*$ or $x_{f,w_{\max}^{(t^*)}}^{(\tau)} \ge 1-\epsilon$ for all $\tau \ge t^*$. 
     
     % better scaffolding
    To do so, we first bound the time $t_1$ until, for $w_{\max}^{(t_1)} = \frac{k}{D}$ for some $k \in \{1,\ldots,D\}$, $x_{w,w_{\max}^{(t_1)}}^{(\tau)} \ge \frac{1}{D-k+1}$ for all $\tau \ge t_1$. Then, for any $\epsilon \ge 0$, we bound the maximum time $t_2$ until $x_{f,w_{\max}^{(t_1)}}^{(\tau)} \ge 1-\epsilon$ for all $\tau \ge t_1+t_2$. Thus, the maximum time to convergence for both agents will be $t^*= t_1 + t_2$. 

    % step 1 -- maximum time for a w_max to be guaranteed to pass threshold, call this time t_1 (should this be a separate lemma -- idk, this is most of the theorem actually).
    \textbf{Bounding $t_1$.}\\
     Here, we say $w_{\max}$ decreases in value at time $t$ if $w_{\max}^{(t+1)} < w_{\max}^{(t)}$. By Lemma~\ref{lem:a_r,+_never_increases}, $w_{\max}^{(t+1)} \le w_{\max}^{(t)}$ for all $t \ge 1$, so $w_{\max}^{(t)}$ can only ever decrease in value once per $k \in \{2,\ldots,D\}$. Note that in the proof of Theorem~\ref{thrm:ftrl_convergence_n=1}, the smallest possible value of $w_{\max}$ at convergence is $\frac{1}{D}$. Thus, to bound $t_1$, it suffices to first bound the maximum number of time steps for $w_{\max} = \frac{k}{D}$ to decrease in value for any $k \in \{2,\ldots,D\}$ as a function of the first time step where $w_{\max} = \frac{k}{D}$. Then, we can compose such a function $D-1$ times to get the final bound on $t_1$.

    Suppose $t_0$ is the first time step where $w_{\max}^{(t_0)}= \frac{k}{D}$ for some $k \in \{2,\ldots,D\}$ and that there exists a time where $w_{\max}^{(t_0)}$ decreases in value. We begin by bounding the time $\tau_1 \ge t_0$ where $x_{w,w_{\max}^{(t_0)}}^{(\tau_1)} < \frac{1}{D-k+1}$. Note that by Lemmas~\ref{lem:x_w_a_r,+_decreases} and~\ref{lem:stationary_x_w}, $x_{w,w_{\max}^{(t_0)}}^{(t)} \le x_{w,w_{\max}^{(t_0)}}^{(t-1)}$ for all $t >t_0$, i.e., the mass weakly decreases, so $\tau_1$ is the first time $x_{w,w_{\max}^{(t_0}}^{(\tau_1)}$ cross the threshold $\frac{1}{D-k+1}$.
    
    \paragraph{Bounding $\tau_1$} Here, we will show that there exists a time $\tau_1$ where \[t_0 \le \tau_1 \le t_0 + \frac{D^2}{\eta} + D^2\cdot t_0 +  \frac{519c^2\left(25\frac{D^{10}}{\eta}\right)^{t_0}}{\log(1+\frac{\eta}{2D(D+1)})}\]
    and $x_{w,w_{\max}^{(t_0)}}^{(\tau_1)} < \frac{1}{D-k+1}$.

    First, let $\tau_1' = t_0 + \frac{D^2}{\eta} + D^2\cdot t_0 +  \frac{519c^2\left(25\frac{D^{10}}{\eta}\right)^{t_0}}{\log(1+\frac{\eta}{2D(D+1)})}$ and suppose $x_{w,w_{\max}^{(t_0)}}^{(t)} \ge \frac{1}{D-k+1}$ for all $ t_0\le t < \tau_1'$ since, otherwise, we have found a $\tau_1$ within the bounds where $x_{w,w_{\max}^{(t_0}}^{(\tau_1)}$ cross the threshold $\frac{1}{D-k+1}$. 
    
     This assumption implies $x_{w,w_{\max}^{(t_0)}}^{(t)} >0$ for $t_0 \le t< \tau_1'$, so \[u_f(w_{\max}^{(t_0)}, x_w^{(t)}) = \left(1-\frac{k}{D}\right)\cdot 1.\] 
    This implies, for all offers $a > w_{\max}^{(t_0)}$ and all $t_0 \le t < \tau_1'$, \begin{align*}
        u_f(w_{\max}^{(t_0)}, x_w^{(t)}) - u_f(a, x_w^{(t)}) \ge \frac{1}{D}. & \tag{1}\label{u_bound_1}
    \end{align*}
    Next, for all offers $a = \frac{\ell}{D} < w_{\max}^{(t_0)}-\frac{1}{D}$ and all $t_0 \le t < \tau_1'$,
    \begin{align*}
        u_f(w_{\max}^{(t_0)},x_w^{(t)}) - u_f(a,x_w^{(t)}) &= \left(1-\frac{k}{D}\right) - \left(1-\frac{\ell}{D}\right)\cdot\left(1-\sum_{\frac{\ell}{D}<a'\le w_{\max}^{(t_0)}}x_{w,a'}^{(t)}\right)\\
        &\ge \left(1-\frac{k}{D}\right) - \left(1-\frac{\ell}{D}\right)\cdot\left(1-(k-\ell)\frac{1}{D-k+1}\right)&\tag{2}\label{u-bound-2}\\
        &= \frac{(k-\ell)^2 - (k-\ell)}{D(D-k+1)}\\
        &\ge \frac{1}{D^2} &\tag{3}\label{u-bound-3}
    \end{align*}
    Line~\ref{u-bound-2} follows from the assumption $x_{w_{\max}^{(t_0)}}^{(t)} \ge \frac{1}{D-k+1}$ and Lemma~\ref{lem:worker_pmf_structure} and Line~\ref{u-bound-3} follows from the bounds $1< k < D$ and $\ell < k-1$ and from the fact that $\ell \in \mathbb{Z}_{\ge 0}$ imply $(k-\ell)^2 - (k-\ell) >1$.
    
    % Then, combining bounds~\ref{u_bound_1} and~\ref{u-bound-3}, then for all $a \neq w_{\max}^{(t_0)}, a\neq w_{\max}-\frac{1}{D}$, for all $\tau \ge t_0$ where $x_{w,w_{\max}^{(t_0)}}^{(\tau)} \ge \frac{1}{D-k+1}$, $$u_f(w_{\max}^{(t_0)},x_w^{(\tau)}) - u_f(a,x_w^{(\tau)}) \ge \frac{1}{D^2}.$$
    Then, since the maximum utility any agent gains in a single time step is 1, then for all $a \neq w_{\max}^{(t_0)}, a\neq w_{\max}^{(t_0)}-\frac{1}{D}$,
    \[U_{f,w_{\max}^{(t_0)}}^{(t_0)} - U_{f,a}^{(t_0)} \ge -t_0. \tag{4}\label{cumulative-lower-bound-1}\]
    So, by bounds~\ref{u_bound_1},~\ref{u-bound-3}, and~\ref{cumulative-lower-bound-1} and the assumption that  $x_{w_{\max}^{(t_0)}}^{(t)} \ge \frac{1}{D-k+1}$ for all $t_0 \le t < \tau_1'$, we are guaranteed, for all $a \neq w_{\max}^{(t_0)}, w_{\max}^{(t_0)}-\frac{1}{D}$ that for all $t_0 + \frac{D^2}{\eta} + D^2\cdot t_0 \le t' < \tau_1'$,
    \[U_{f,w_{\max}^{(t_0)}}^{(t')} - U_{f,a}^{(t')} \ge \frac{1}{\eta}.\tag{5}\label{cumulative-lower-bound-2}\]
    Then, for all $a \neq w_{\max}^{(t_0)},w_{\max}^{(t_0)}-\frac{1}{D}$, Claims~\ref{claim:stationarity_reqs} and~\ref{claim:kkt_ordering} and the probability constraint along with bound~\ref{cumulative-lower-bound-2} imply 
    % it is impossible for both $x_{f,w_{\max}^{(t_0)}}^{(t_0 + \frac{D^2}{\eta} + D^2\cdot t_0)}>0$ and $x_{f,a}^{(t_0 + \frac{D^2}{\eta} + D^2\cdot t_0)}>0$. Further, since $U_{f,w_{\max}^{(t_0)}}^{(t_0 + \frac{D^2}{\eta} + D^2\cdot t_0)} > U_{f,a}^{(t_0 + \frac{D^2}{\eta} + D^2\cdot t_0)}$, then by Claim~\ref{claim:kkt_ordering}, we are guaranteed 
    $x_{f,a}^{(t')}=0$ and, thus, $f_{\min}^{(t')} \ge w_{\max}^{(t_0)}-\frac{1}{D}$ and $f_{\max}^{(t')} \le w_{\max}^{(t_0)}$ for all $ t_0 + \frac{D^2}{\eta} + D^2\cdot t_0 \le t'< \tau_1'$. Note that, by definition, $f_{\min}^{(t)} \le f_{\max}^{(t)}$ for all $t>0$, so we now go through the three possible cases implied by $f_{\min}^{(t')}, f_{\max}^{(t')} \in \{w_{\max}^{(t_0)}-\frac{1}{D}, w_{\max}^{(t_0)}\}$ for all $ t_0 + \frac{D^2}{\eta} + D^2\cdot t_0 \le t'<\tau_1'$.

    % cases to show either we apply lemma 17 or we reach the point earlier. (needs work)
    First, suppose that $f_{\min}^{(t')} = f_{\max}^{(t')} = w_{\max}^{(t_0)}$. Then, this implies $x_{f,w_{\max}^{(t_0)}}^{(t')} = 1$ and we can conclude convergence has occurred given our assumption $x_{w,w_{\max}^{(t_0)}}^{(t')} \ge \frac{1}{D-k+1}$ and we are done. 
    Next, suppose  $f_{\min}^{(t')} = f_{\max}^{(t')} = w_{\max}^{(t_0)}-\frac{1}{D}$ for all $ t_0 + \frac{D^2}{\eta} + D^2\cdot t_0 \le t'< \tau_1'$. Note that if there exists $t_0\le t <\tau_1$ where $f_{\min}^{(t)} \neq w_{\max}^{(t_0)}-\frac{1}{D}$ or $f_{\max}^{(t)} \neq w_{\max}^{(t_0)}-\frac{1}{D}$, then we will be in one of the other two cases. Then, for all $a < w_{\max}^{(t_0)}$ and all $ t_0 + \frac{D^2}{\eta} + D^2\cdot t_0 \le t'< \tau_1'$,
    \[u_w(x_f^{(t')}, a) - u_w(x_f^{(t')},w_{\max}^{(t_0)}) = \frac{k-1}{D}. \tag{6}\label{worker-fmin-utility-bound}\]
    Then, by Claim~\ref{claim:stationarity_reqs} and the probability mass constraint, \[x_{w,w_{\max}^{(t_0)}}^{(t'+1)} = x_{w,w_{\max}^{(t_0)}}^{(t')}+\frac{\eta}{k}\left(\sum_{a<w_{\max}^{(t_0)}}u_w(x_f^{(t')}, w_{\max}) - u_w(x_f^{(t')},a)\right),\]
    so substituting in Equation~\ref{worker-fmin-utility-bound} and using the fact that $k \in \{2,\ldots,D\}$ results in the following bound,
    \[x_{w,w_{\max}^{(t_0)}}^{(t'+1)} \le x_{w,w_{\max}^{(t_0)}}^{(t')}-\frac{\eta}{D^2}.\]
    Therefore, we are guaranteed $x_{w,w_{\max}^{(t_0)}}^{(t'+\frac{D}{\eta})}<\frac{1}{D-k+1}$.  
    
    Finally, suppose $f_{\min}^{(t')}  = w_{\max}^{(t_0)}-\frac{1}{D}$ and $f_{\max}^{(t')} = w_{\max}^{(t_0)}$. We further break this case into two subcases. First, suppose $x_{w,w_{\max}^{(t_0)}}^{(t')} = \frac{1}{D-k+1}$. Since $f_{\min}^{(t')} < w_{\max}^{(t_0)}$, then by Lemma~\ref{lem:x_w_a_r,+_decreases}, $x_{w,w_{\max}^{(t_0)}}^{(t'+1)} < \frac{1}{D-k+1}$. 
    Otherwise, suppose $x_{w,w_{\max}^{(t_0)}}^{(t')} > \frac{1}{D-k+1}$. Then, the assumptions of Lemma~\ref{lem:bounding_recurrence_relation_time} are satisfied and since we have assumed there exists a time where $w_{\max}^{(t_0)}$ decreases in value, we can conclude it will take at most \[ \tau_{dec} = \frac{519c^2\left(25\frac{D^{10}}{\eta}\right)^{t_0}}{\log(1+\frac{\eta}{2D(D+1)})}\] time steps to be guaranteed $x_{w,w_{\max}^{(t_0)}}^{(t'+ \tau_{dec})} < \frac{1}{D-k+1}$. 
    
    Since $\eta \le 1$ implies $\log(1+\frac{\eta}{2D(D+1)})<1,D>2$, and $t_0 \ge 1$, it is clear that $\tau_{dec} > \frac{D}{\eta} > 1$, thus we can conclude
    \[\tau_1 \le t_0 + \frac{D^2}{\eta} + D^2\cdot t_0 + \tau_{dec} = \tau_1'.\] 
    
    Next, we derive the maximum time steps $\tau_2 \ge \tau_1$ until $x_{w,w_{\max}^{(t_0)}}^{(\tau_2)}=0$ given that $x_{w,w_{\max}^{(t_0)}}^{(\tau_1)}< \frac{1}{D-k+1}$. 
    \paragraph{Bounding $\tau_2$.} We begin by bounding the time it takes for the firm's mass on  $w_{\max}^{(t_0)}$ to reach 0. We start by lower bounding \[\frac{1}{D-k+1}- x_{w,w_{\max}^{(t_0)}}^{(t_1)}\] 
    by a positive value since the firm's mass decreases on $w_{\max}^{(t_0)}$ at a rate proportional to this difference. By Lemma~\ref{lem:max_bits_of_x_i,a^t}, we know $x_{w,w_{\max}^{(t_0)}}^{(\tau_1)}$ has a denominator represented by at most $20c^2\log\left(D+1\right)(25D^8)^{\tau_1}$ bits and $\frac{1}{D-k+1}$ has a denominator represented by at most $\log(D)$ bits for any $k \in \{2,\ldots,D\}$, so their difference has a denominator that can be represented by at most $20c^2\log\left(D+1\right)(25D^8)^{\tau_1} + \log(D)$ bits. Then, by Lemma~\ref{lem:bits-to-bound}, \[\frac{1}{D-k+1} - x_{w,w_{\max}^{(t_0)}}^{(\tau_1)} >2^{-(20c^2\log\left(D+1\right)(25D^8)^{\tau_1} + \log(D))}.\tag{7}\label{D-k+1-diff-bound}\]
    
    Since $x_{w,w_{\max}^{(t_0)}}^{(\tau_1)}$ does not increase by Lemmas~\ref{lem:stationary_x_w} and~\ref{lem:x_w_a_r,+_decreases}, then the bound~\ref{D-k+1-diff-bound} implies for all $\tau \ge \tau_1$,
    \[u_f(w_{\max}^{(t_0)}-\frac{1}{D}, x_w^{(\tau)}) - u_f(w_{\max}^{(t_0)},x_w^{(\tau)}) > \frac{1}{D}\cdot 2^{-(20c^2\log\left(D+1\right)(25D^8)^{\tau_1} + \log(D))}.\]

    Next, since $$U_{f,w_{\max}-\frac{1}{D}}^{(\tau_1)} - U_{f,w_{\max}^{(t_0)}}^{(\tau_1)} \ge - \tau_1,$$
    Then it will take at most \[t'=\frac{D}{\eta}2^{20c^2\log\left(D+1\right)(25D^8)^{\tau_1} + \log(D)} + \tau_1 2^{20c^2\log\left(D+1\right)(25D^8)^{\tau_1} + \log(D)}\]
    time steps to be guaranteed, for all $\tau \ge \tau_1+t'$,
    \[U_{f,w_{\max}-\frac{1}{D}}^{(\tau)} - U_{f,w_{\max}^{(t_0)}}^{(\tau)} \ge \frac{1}{\eta}.\]

    Thus, by Claims~\ref{claim:stationarity_reqs} and~\ref{claim:kkt_ordering} and the probability constraint, we can say $x_{f,w_{\max}^{(t_0)}}^{(\tau)} =0$ for all $\tau \ge \tau_1+t'$. 
    
    Finally, we bound the time it takes for the worker to no longer put mass on $w_{\max}^{(t_0)}$. First, by Lemma~\ref{lem:worker_pmf_structure}, \[U_{w,\frac{1}{D}}^{(\tau_1+t')} - U_{w,w_{\max}^{(t_0)}}^{(\tau_1+t')} \ge 0,\] 
    and since $x_{f,w_{\max}^{(t_0)}}^{(\tau)} =0$ implies $f_{\max}^{(\tau)} < w_{\max}^{(t_0)}$ for all $\tau \ge \tau_1 + t'$ and, by the utility functions, $x_{f,0}^{(t)} \le \frac{1}{2}$ for all $t$, then for all $\tau \ge \tau_1+t'$,
    \[u_{w}(x_f^{(\tau)},\frac{1}{D}) - u_{w}(x_f^{(\tau)},w_{\max}^{(t_0)}) \ge \frac{1}{2D}.\]
    Thus, for all $\tau \ge \tau_1+t'+\frac{2D}{\eta}$, \[U_{w,\frac{1}{D}}^{(\tau)} - U_{w,w_{\max}^{(t_0)}}^{(\tau)} \ge \frac{1}{\eta}.\]
    Then, by Claims~\ref{claim:stationarity_reqs} and~\ref{claim:kkt_ordering} and the probability constraint, we can conclude  $x_{w,w_{\max}^{(t_0)}}^{(\tau)}=0$ for all $\tau \ge \tau_1+t'+\frac{2D}{\eta}$, i.e., $w_{\max}^{(t_0)}$ must decrease in value, and we can conclude \[\tau_2 \le \tau_1+t'+\frac{2D}{\eta}.\]

    \paragraph{Deriving the final bound on $t_1$} In total, the maximum number of time steps before $w_{\max}$ decreases as a function of the starting time step $t_0$ is $\tau_2 - t_0$ which can be bounded as follows
    \begin{align*}
       \tau_2 -t_0 &\le  \tau_1 + t' + \frac{2D}{\eta} - t_0\\
        &\le \tau_1+ \frac{2D}{\eta} + \frac{D}{\eta}2^{20c^2\log\left(D+1\right)(25D^8)^{2\tau_1}+\log(D)} - t_0 &\tag{8}\label{line-1-t-1}\\
        &\le 3\frac{D^2}{\eta}t_0\tau_{dec} + \frac{2D}{\eta}+  \frac{D}{\eta}2^{20c^2\log\left(D+1\right)(25D^8)^{8\frac{D^2}{\eta}t_0\tau_{dec}}+\log(D)} &\tag{9}\label{line-2-t-1}\\
        &\le 2^{\frac{D}{\eta}^{\left(\log\left(1+\frac{\eta}{2D(D+1)}\right)\right)^{-1}\cdot M \cdot \left(25^2\frac{D^{28}}{\eta^4}\right)^{t_0}}} 
    \end{align*}
    where $M = 164430 c^2$ and Line~\ref{line-1-t-1} follows from the expansion of $t'$ and Line~\ref{line-2-t-1} follows from the bound on $\tau_1$ and the final line finds a loose upper bound by combining and simplifying terms using the definition of $\tau_{dec}$.

 % \[\frac{D^2}{\eta} + D^2\cdot t_0 + \tau_{dec} + \frac{2D}{\eta} + \frac{D}{\eta}2^{d+1+5c d(9D^8)^{t_0 + \frac{D^2}{\eta} + D^2\cdot t_0 + \tau_{dec}}}\\+ (t_0 + \frac{D^2}{\eta} + D^2\cdot t_0 + \tau_{dec})2^{d+1+5c d(9D^8)^{t_0 + \frac{D^2}{\eta} + D^2\cdot t_0 + \tau_{dec}}}.\]

    % We can upper bound this term by 
    % \[ (\frac{6D^2}{\eta}t_0+ \tau_{dec})2^{(c+2 +(2+3d+2D)\cdot (\frac{3D^2}{\eta} t_0 + \tau_{dec}))}.\]
    % % We can further upper bound by 
    % \[2^{c+2+(3+3d+2D)(\frac{6D^2}{\eta}t_0+\tau_{dec})}.\]
    % Using the definition of $\tau_{dec}$, this term can be upper bounded by 
    % \[2^{c+2+(3+3d+2D)(\frac{6D^2}{\eta}t_0+\frac{c + 4+2m\cdot d  + (3+3d+2D)(\frac{4D^3}{\eta}t_0)}{\log(1+\frac{\eta}{2D(D+1)})})}.\]
    % % This is upper bounded by 
    % \[2^{\frac{1}{\log(1+\frac{\eta}{2D(D+1)})}\cdot \left(2c + 6 +2m\cdot d + (3+3d+2D)\frac{10D^3}{\eta}t_0\right)}.\]
    % \[8^{D^{\log\left(1+\frac{\eta}{2D(D+1)}\right)^{-1}\cdot D^{M\frac{D^{16}}{\eta}t_0}}}\]
    % where \[M= 3483c\cdot m\cdot d.\]
    
    % then the bound on the maximum number of time steps is
    % \[2^{\frac{1}{\log(1+\frac{\eta}{2D(D+1)})}\cdot \left(M\frac{20D^4}{\eta}\right)t_0}.\]

    Finally, since there are $D-1$ values of $w_{\max}$ that can decrease, i.e., $w_{\max} = \frac{k}{D}$ for $k \in \{2,\ldots,D\}$ and the derived bound on time to decrease does not depend on the value of $k$, then the maximum number of time steps to be guaranteed that $w_{\max}$ will no longer decrease in value is
    % \[8^{(9D^8)^{\frac{1}{\log(1+\frac{\eta}{2D(D+1)})}\cdot\left(9\frac{D^8}{\eta}\right)^{M\frac{D^2}{\eta}t_0}}} \uparrow \uparrow (D-1)\]
    % \[t_1=\left(8^{D^{\log\left(1+\frac{\eta}{2D(D+1)}\right)^{-1}\cdot D^{M\frac{D^{16}}{\eta}}}}\right) \uparrow \uparrow (D-1)\]
    \[t_1 = \left(2^{\frac{D}{\eta}^{\left(\log\left(1+\frac{\eta}{2D(D+1)}\right)\right)^{-1}\cdot M \cdot \left(25^2\frac{D^{28}}{\eta^4}\right)}}\right) \uparrow \uparrow (D-1)\]
    time steps. Note that $\uparrow \uparrow$ is the Knuth up arrow notation to denote $D-1$ tetrations~\citep{knuth1976mathematics}.

    % Step 2 -- maximum time to converge to eps-NE or exact NE
    \textbf{Bounding $t_2$.}\\
    From the previous section, we know $x_{w,w_{\max}^{(t_1)}}^{(\tau)} \ge \frac{1}{D-k+1}$ for all $\tau \ge t_1$. First, as in the first step of the $t_1$ bound, for all $a \neq w_{\max}^{(t_1)}$ and $a \neq w_{\max}^{(t_1)}- \frac{1}{D}$, then for all $\tau \ge t_1 + \frac{D^2}{\eta} + D^2 t_1$ \[x_{f,a}^{(\tau)} =0.\tag{10}\label{0-out-a}\]
    Then, if $k \ge 2$, we can apply Lemma~\ref{lem:bounding_recurrence_relation_time} to conclude that for any $\epsilon>0$, it will take at most 
    \[\tau_{eq} = \max\{\tau_{conv}, \tau_{\epsilon}\}\]
    % \[\tau_{conv} = \frac{\log (\frac{3D^2}{\eta})+20c d(9D^8)^{t_1 + \frac{D^2}{\eta} + D^2 t_1} + 4d + 3md + 5}{\log(1+\frac{\eta}{2D(D+1)})} +2\]
    time steps for $x_{f,w_{\max}^{(t_1)}}^{(\tau)} \ge 1-\epsilon$ for all $\tau \ge t_1 + \frac{D^2}{\eta} + D^2 t_1 + \tau_{eq}$.\\

    Otherwise, if $k=1$, then we are guaranteed the algorithm will eventually converge to an exact ($\epsilon=0$) NE. First, note that for all $t >0$ and all $a > \frac{1}{D}$, \[U_{w,0}^{(t)} = U_{w,\frac{1}{D}}^{(t)} \ge U_{w,a}^{(t)}.\]
    So, by Claims~\ref{claim:stationarity_reqs} and~\ref{claim:kkt_ordering}, for all $t>0$ and all $a > \frac{1}{D}$, \[x_{w,0}^{(t)} = x_{w,\frac{1}{D}}^{(t)} \ge x_{w,a}^{(t)}.\]
    Therefore, when $w_{\max}^{(t_1)} = \frac{1}{D}$, then we are guaranteed for all $\tau \ge t_1$
    \[x_{w,0}^{(\tau)} = x_{w,\frac{1}{D}}^{(\tau)} = \frac{1}{2}.\]
    This implies, for all $\tau \ge t_1$, \[u_f(\frac{1}{D},x_w^{(\tau)}) - u_f(0,x_w^{(\tau)}) = \frac{D-1}{D} - \frac{1}{2} = \frac{D-2}{2D}  >\frac{1}{2D}\tag{11}\label{1/d-better-bound}\]
    where the last inequality follows from $D>2$.
    Let $t' = t_1 + \frac{D^2}{\eta} + D^2 t_1$ and note by~\ref{0-out-a}, $f_{\min}^{(\tau)},f_{\max}^{(\tau)} \in\{0,\frac{1}{D}\}$ for all $\tau \ge t'$, so we next bound the time $t'' \ge t'$ where we are guaranteed $f_{\min}^{(\tau)} = f_{\max}^{(\tau)} = \frac{1}{D} = w_{\max}^{(t_1)}$ for all $\tau \ge t''$. Note that, \[U_{f,\frac{1}{D}}^{(t')} - U_{f,0}^{(t')} \ge -t'.\]
    Then, for $t'' = t' + \frac{2D}{\eta} + 2Dt'$ and for all $\tau \ge t''$, by bound~\ref{1/d-better-bound},
    \[U_{f,\frac{1}{D}}^{(\tau)} - U_{f,0}^{(\tau)} \ge \frac{1}{\eta}.\]
     Thus, by Claims~\ref{claim:stationarity_reqs} and~\ref{claim:kkt_ordering} and the probability constraint, we can conclude $f_{\min }^{(\tau)} = f_{\max}^{(\tau)} = w_{\max}^{(t_1)}$ for all $\tau \ge t''$ which immediately implies $x_{f,w_{\max}^{(t_1)}}^{(\tau)} =1$ for all $\tau \ge t''$.

    Therefore, the maximum time to convergence to an $\epsilon$-NE if $k\ge 2$ or exact NE if $k=1$ after time $t_1$ is \[t_2 = \frac{D^2}{\eta} + D^2 t_1 + \max\{ \tau_{eq}, \frac{2D}{\eta} + 2Dt_1 + \frac{2D^3}{\eta} + 2D^3t_1\}.\]
    \textbf{Final bound on time to convergence.}\\
    Using the definition of $\tau_{conv}$ and $\tau_{\epsilon}$ and the constant $M$ defined above, then the maximum time to convergence to $\epsilon$-NE can be bounded by \[t_1+ \max\left\{2^{\frac{D}{\eta}^{\left(\log\left(1+\frac{\eta}{2D(D+1)}\right)\right)^{-1}\cdot M \cdot \left(25^2\frac{D^{28}}{\eta^4}\right)^{t_1}}}, 2\log\left(\frac{3D^3}{\epsilon}\right)\cdot \frac{4D^2}{\eta}\right\}\]
    Since $\max\{x,y\} \le x+y$ for $x,y>0$, then the overall time to convergence is bounded by \[\left[\left(2^{\frac{D}{\eta}^{\left(\log\left(1+\frac{\eta}{2D(D+1)}\right)\right)^{-1}\cdot M \cdot \left(25^2\frac{D^{28}}{\eta^4}\right)}}\right)\uparrow \uparrow D\right] +2\log\left(\frac{3D^3}{\epsilon}\right)\cdot \frac{4D^2}{\eta}.\]
    Since $\log(1+x)^{-1} \le \frac{x+1}{x}$ for $x >0$, then \[\log\left(1+\frac{\eta}{2D(D+1)}\right)^{-1} \le \frac{4D^3}{\eta}.\]
    Thus, the bound can be simplified as 
    \[O\left(\left[\left(2^{\frac{D}{\eta}^{{\frac{D^{31}}{\eta^5}}}}\right)\uparrow \uparrow D\right] +\log\left(\frac{D^3}{\epsilon}\right)\cdot \frac{D^2}{\eta}\right).\]

\end{proof}

\section{Additional Lemmas for Theorem~\ref{thrm:ftrl_convergence_rate}}
    \begin{customlem}{15}
    \label{lem:solving_recurrence_relations}
     Suppose the variables $f$ and $w$ evolve according to the following recurrence relations for all $n \ge 0$:
    \begin{align*}
        w^{(n+1)} &= w^{(n)} + A \cdot f^{(n)} - A,\\
        f^{(n+1)} &= f^{(n)} + B \cdot w^{(n)} - C.
    \end{align*}
    For any positive constants $A,B,C \in \mathbb{R}_{>0}$ and any real initial values at time $t$, $f^{(0)}, w^{(0)} \in \mathbb{R}$, then the closed form solutions for $w^{(n)}$ and $f^{(n)}$ for all $n \ge 0$ are 
    \begin{align*}
        w^{(n)} &= \alpha_1^w\sqrt{AB}(1+\sqrt{AB})^{n-1} - \alpha_2^w\sqrt{AB}(1-\sqrt{AB})^{n-1} + \frac{C}{B},\\
        f^{(n)} &= \alpha_1^f\sqrt{AB}(1+\sqrt{AB})^{n-1} - \alpha_2^f\sqrt{AB}(1-\sqrt{AB})^{n-1} + 1,
    \end{align*} 
    where
    \begin{align*}
        \alpha_1^w &= \frac{1}{2}(c_w-\frac{1}{B}\cdot c_f) + \frac{1}{2\sqrt{AB}}(c_w-A\cdot c_f), &\tag{B.1}\label{a_1_w}\\
        \alpha_2^w &=\frac{1}{2}(c_w-\frac{1}{B}\cdot c_f) - \frac{1}{2\sqrt{AB}}(c_w-A\cdot c_f),&\tag{B.2}\label{a_2_w}\\
        \alpha_1^f &= \frac{1}{2}(\frac{1}{A}\cdot c_w-c_f) + \frac{1}{2\sqrt{AB}}(B\cdot c_w-c_f),&\tag{B.3}\label{a_1_f}\\
        \alpha_2^f &= \frac{1}{2}(\frac{1}{A}\cdot c_w-c_f) - \frac{1}{2\sqrt{AB}}(B\cdot c_w-c_f).&\tag{B.4}\label{a_2_f}
    \end{align*}
    and either $\sign(\alpha_1^f) = \sign(\alpha_1^w)$ or $\alpha_1^f=\alpha_1^w = 0$.
    
    \end{customlem}
    \begin{proof}
         Let \[W^{(n)} := \sum_{i = 0}^{n} w^{(i)}, F^{(n)} := \sum_{i=0}^{n} f^{(i)}.\tag{1}\label{sum-def}\] Then, using the recurrence relations for $w^{(n+1)}$ and $f^{(n+1)}$, we can derive the following non-homogeneous second order linear recurrences
    \begin{align*}
        W^{(n)} &= 2W^{(n-1)} -(1-A\cdot B)W^{(n-2)} - (n-1)\cdot A\cdot C + A\cdot f^{(0)}-A,\\
        F^{(n)} &= 2F^{(n-1)} -(1-A\cdot B)F^{(n-2)} - (n-1)\cdot A\cdot B + B\cdot w^{(0)}-C,\\
    \end{align*}
    with the initial conditions
    \begin{align*}
        W^{(0)} &= w^{(0)} & W^{(1)} = w^{(0)} + w^{(0)} + A\cdot f^{(0)} -A,\\
        F^{(0)} &= f^{(0)} & F^{(1)} = f^{(0)} + f^{(0)} + B\cdot f^{(0)} -C.
    \end{align*}
    It is well-known how to solve these kinds of recurrences, see, e.g., Section 2.4 of~\citet{elaydiintroduction}. Let \[c_w = w^{(0)} - \frac{C}{B}, c_f = 1-f^{(0)}.\]
    Then, the closed form solutions for $W^{(n)}$ and $F^{(n)}$ are given below. 
    \begin{align*}
        W^{(n)} &= \alpha_1^w\left(1+\sqrt{AB}\right)^n+\alpha_2^w\left(1-\sqrt{AB}\right)^n + \frac{C}{B}\cdot n + \frac{C+1-f^{(0)}}{B}\\
        F^{(n)} &= \alpha_1^f\left(1+\sqrt{AB}\right)^n+\alpha_2^f\left(1-\sqrt{AB}\right)^n + n + \frac{AB-Bw^{(0)} +C}{AB}
    \end{align*}
    where 
    \begin{align*}
        \alpha_1^w &= \frac{1}{2}(c_w-\frac{1}{B}\cdot c_f) + \frac{1}{2\sqrt{AB}}(c_w-A\cdot c_f), &\tag{2}\label{a_1_w_2}\\
        \alpha_2^w &=\frac{1}{2}(c_w-\frac{1}{B}\cdot c_f) - \frac{1}{2\sqrt{AB}}(c_w-A\cdot c_f),&\tag{3}\label{a_2_w_2}\\
        \alpha_1^f &= \frac{1}{2}(\frac{1}{A}\cdot c_w-c_f) + \frac{1}{2\sqrt{AB}}(B\cdot c_w-c_f),&\tag{4}\label{a_1_f_2}\\
        \alpha_2^f &= \frac{1}{2}(\frac{1}{A}\cdot c_w-c_f) - \frac{1}{2\sqrt{AB}}(B\cdot c_w-c_f).&\tag{5}\label{a_2_f_2}
    \end{align*}
    Thus, using Definition~\ref{sum-def}, the closed form solution of the original recurrence relations are 
    \begin{align*}
        w^{(n)} &= \alpha_1^w\sqrt{AB}(1+\sqrt{AB})^{n-1} - \alpha_2^w\sqrt{AB}(1-\sqrt{AB})^{n-1} + \frac{C}{B},\\
        f^{(n)} &= \alpha_1^f\sqrt{AB}(1+\sqrt{AB})^{n-1} - \alpha_2^f\sqrt{AB}(1-\sqrt{AB})^{n-1} + 1.
    \end{align*} 
    Note that we assume $AB>0$, so all roots are real.
    
    Finally, Equations~\ref{a_1_w_2} through~\ref{a_2_f_2} imply $\alpha_1^{w}$ and $\alpha_1^{f}$ are either both $0$ or both have the same sign. To see this, note that 
    \begin{align*}
        \alpha_1^w >0 \iff c_w > \frac{\sqrt{AB}+AB}{B(1+\sqrt{AB})}c_f,\\
        \alpha_1^w =0 \iff c_w = \frac{\sqrt{AB}+AB}{B(1+\sqrt{AB})}c_f,\\
        \alpha_1^w <0 \iff c_w < \frac{\sqrt{AB}+AB}{B(1+\sqrt{AB})}c_f.
    \end{align*}
    Similarly,
    \begin{align*}
        \alpha_1^f >0 \iff c_w > \frac{A(1+\sqrt{AB})}{\sqrt{AB}+AB}c_f,\\
        \alpha_1^f =0 \iff c_w = \frac{A(1+\sqrt{AB})}{\sqrt{AB}+AB}c_f,\\
        \alpha_1^f <0 \iff c_w < \frac{A(1+\sqrt{AB})}{\sqrt{AB}+AB}c_f.
    \end{align*}
    Finally, note that \begin{align*}
        \frac{\sqrt{AB}+AB}{B(1+\sqrt{AB})} &= \frac{A(1+\sqrt{AB})}{\sqrt{AB}+AB}\\
        \iff (\sqrt{AB}+AB)^2 &= AB(1+\sqrt{AB})^2\\
        \iff AB + 2AB\sqrt{AB} + A^2B^2 &= AB+2AB\sqrt{AB} + A^2B^2.
    \end{align*}
    \end{proof}

    \begin{customlem}{16}
    \label{lem:rational-approx-bound}
        Let $n,r,s \in \mathbb{Z}$ where $n>0, r \ge 0,s > 0$, and $n$ is square-free. Suppose $\sqrt{n} > \frac{r}{s}$. Then, there exists a rational number $\frac{p}{q}$ where $p$ and $q$ are coprime, \[\frac{r}{s} < \frac{p}{q} <\sqrt{n},\]
        and \[q < 128 n^2s.\]
    \end{customlem}
    \begin{proof}
        To begin, we use the following facts about continued fractions and approximating the quadratic irrational $\sqrt{n}$ from Chapter 6 Sections 2-4 and Chapter 8 Section 1 of~\citet{baker1984concise}. Note that all notation for continued fractions is consistent with these sections as well.
        \begin{itemize}
            \item $n\in \mathbb{Z}_{>0}$ where $n$ is square-free implies $\sqrt{n}$ is a quadratic irrational which implies there exists an  integer $h \ge 2$ such that \[\sqrt{n} = [a_0;\overline{a_{1},\ldots,a_{h-1},2a_0}],\] where $a_0$ is a non-negative integer, $a_1, \ldots, a_{h-1}$ are positive integers, and $a_0,\ldots,a_{h-1}$ are all referred to as partial quotients. Then, denote the largest partial quotient by \[C_n := \max \{a_0,\ldots,a_{h},2a_0\}.\tag{1}\label{max_partial_quotient}\]
            \item For any integer $k>0$, the $k$th rational convergent of $\sqrt{n}$ is $\frac{p_k}{q_k} = [a_0; a_1,\ldots, a_k]$ where \[q_k = a_kq_{k-1}  + q_{k-2},\tag{2}\label{cf_denominator_evolution}\]
            and $p_k$ and $q_k$ are coprime. Further, if $\frac{p_k}{q_k} >\sqrt{n}$, then $\frac{p_{k+1}}{q_{k+1}} < \sqrt{n}$ and vice versa. Finally, each convergent gets strictly closer to $\sqrt{n}$, i.e., \[\left|\sqrt{n} - \frac{p_k}{q_k}\right| >\left|\sqrt{n} - \frac{p_{k+1}}{q_{k+1}}\right|.\]
            \item Suppose $\frac{p_k}{q_k}$ is the $k$th convergent of $\sqrt{n}$. Then, for all integers $q \le q_k$ and all integers $p$,
            \[\left|\sqrt{n} - \frac{p_k}{q_k}\right| \le \left|\sqrt{n} - \frac{p}{q}\right|.\]
        \end{itemize}

        Since the sequence of convergent denominators is strictly increasing, there exists convergent denominators $q_k,q_{k-1}$ where \[q_{k-1} \le s <q_k. \tag{3}\label{denominator_ordering}\]
        Next, since the convergents alternate as under and over-approximations of $\sqrt{n}$ and the convergents get successively strictly closer in terms of absolute distance to $\sqrt{n}$, then there exists a convergent $\frac{p_i}{q_i}\in \{\frac{p_k}{q_k}, \frac{p_{k+1}}{q_{k+1}}, \frac{p_{k+2}}{q_{k+2}},\frac{p_{k+3}} {q_{k+3}}\}$ that satisfies $\frac{r}{s} < \frac{p_i}{q_i} <\sqrt{n}$. 

        By Equation~\ref{cf_denominator_evolution}, $q_i$ strictly increases with $i$ and we can bound the size of $q_i$ as follows:
        \begin{align*}
            q_i &\le q_{k+3}\\
            % &\le C_n q_{k+2} + q_{k+1}\\
            % &\le C_n^2q_{k+1} + C_nq_{k} + C_nq_{k} + q_{k-1}\\
            % & \le C_n^3 q_k + C_n^2q_{k-1} + C_n^2 q_{k-1} + C_nq_{k-2} + C_n^2 q_{k-1} + C_nq_{k-2} + q_{k-1}\\
            % &\le C_n^4q_{k-1} + C_n^3q_{k-2} + C_n^2q_{k-1} + C_n^2 q_{k-1} + C_nq_{k-2} +  C_n^2 q_{k-1} + C_nq_{k-2} + q_{k-1}\\
            &\le (C_n^4 + 3C_n^2 + 1)q_{k-1} + (C_n^3+2C_n)q_{k-2} &\tag{4}\label{line_1}\\
            &\le (C_n^4 + 3C_n^2 + 1 + C_n^3+2C_n) s & \tag{5}\label{line_2}\\
            &\le  8C_n^4s
        \end{align*}
        Line~\ref{line_1} follows from Definition~\ref{max_partial_quotient} and Equation~\ref{cf_denominator_evolution} and Line~\ref{line_2} follows from Inequality~\ref{denominator_ordering}.
        Finally, it is known that $C_n < 2\sqrt{n}$, see page 245, problem 3(f) of~\citet{leveque1996fundamentals}, thus \[q_i < 128n^2s.\]
    \end{proof}

    \begin{customlem}{17}
    \label{lem:bits-to-bound}
        Suppose $\frac{m}{n} \in \mathbb{Q}$ where $m\in \mathbb{Z}$ and $n\in \mathbb{Z}_{>0}$ and $n$ is represented in at most $\beta$ bits. Then, \[\left|\frac{m}{n}\right| \ge 2^{-\beta}.\]
    \end{customlem}
    \begin{proof}
        Suppose $n$ is represented in at most $\beta$ bits, then \[n \le 2^{\beta}.\]
        Further, since $m\in \mathbb{Z}$, then $|m| \ge 1$. Thus,
        \[\left|\frac{m}{n}\right| \ge 2^{-\beta}.\]
    \end{proof}
\section{Sequence Form Representation of 2-Round Alternating Bargaining Game}
\label{appendix:sequence-form}
 A \textit{sequence} $\sigma$ is a sequential string of actions an agent must take to get to some node in the game tree. For example, if agents are at the payoff node $(1-a_i, a_i)$, then the sequence the firm took is $\sigma_f = a_i$ and the sequence the worker took is $\sigma_w = A_{a_i}$. The sequences and associated payoffs for the two round bargaining game parameterized by discount factor $0<\delta <1$ are given in the table below with the firm's sequences on the rows and the worker's sequences on the columns where $a,b \in \mathcal{A}$ are arbitrary offers in $\mathcal{A}$. The line $-$ indicates that the combination of sequences does not result in a terminal node.
\begin{table}[h]
\centering
\begin{tabular}{|c|c|c|}
\hline
    & $A_a$ & $R_ab$ \\\hline
    $a$ & $(1-a,a)$ & $-$  \\\hline
    $aA_b$ & $-$ & $\delta(b,1-b)$ \\\hline
    $aR_b$ & $-$ & $\delta (0,0)$ \\ \hline 
\end{tabular}
\end{table}\\
Let $\Sigma_i$ be the set of all terminal sequences of agent $i$ for $i \in \{f,w\}$ and let $\emptyset$ represent the root node of the extensive form game tree. Then,
\begin{align*}
    \Sigma_f &= \{\emptyset,a,aA_b,aR_b|a,b\in\mathcal{A}\}\\
    \Sigma_w &= \{\emptyset,A_a,R_ab|a,b\in\mathcal{A}\}
\end{align*}
Next, let $I_i$ be the information set of agent $i$. Since our game is complete information and perfect recall, for each $I \in I_i$, $I$ is a singleton set with one node $h$ and, further, there is a unique sequence $\sigma_i \in \Sigma_i$ that leads to $h$. For each $I \in I_i$, let $\ext(I)$ be the set of extensions of the unique sequence $\sigma_h \in \Sigma_i$ that leads to the node $h \in I$ by 1 valid action in $\Sigma_i$. For example, if $h\in I$ is the node corresponding to the firm responding to a counteroffer of $b\in\mathcal{A}$ from the worker after giving an initial offer of $a\in\mathcal{A}$, then $\sigma_h=a$ is the unique sequence leading to the node $h\in I$ and  $$\ext(I) = \{aA_b,aR_b|b\in \mathcal{A}\}.$$

Next, a \textit{realization plan} represents the probability mass an agent puts on reaching each terminal sequence. Formally, $r_i: \Sigma_i \to [0,1]$ such that 
\begin{align*}
    r_i(\emptyset) &= 1 \\
    \sum_{\sigma^+ \in \ext(I)} r_i(\sigma^+) &= r_i(\sigma) &\forall I \in I_i\\
    r_i(\sigma) &\ge 0 &\forall \sigma \in \Sigma_i
\end{align*}
From a realization plan $r_i$, a behavioral strategy of agent $i$ can be recovered. Let $\sigma_i \in \Sigma_i$ where $\sigma_i$ is the unique sequence leading to $I_{\sigma_i} \in I_i$. Then, let $\sigma_ia_i \in \ext(I_{\sigma_i})$, and the behavioral strategy at action $a_i$ is:
$$\beta_i(I_{\sigma_i},\sigma_ia_i) = \frac{r_i(\sigma_ia)}{r_i(\sigma_i)}.$$

Let $U_{i,\sigma}^{(t)}(\{r_{-i}^{(\tau)}\}_{1\le\tau\le t})$ be the cumulative expected utility agent $i$ gets at terminal sequence $\sigma \in \Sigma_i \setminus\{\emptyset\}$ through time $t$:
  $$U_{i,\sigma}^{(t)}(\{r_{-i}^{(\tau)}\}_{1\le\tau\le t}) = \sum_{\tau=1}^t u_i(\sigma,r_{-i}^{(\tau)})$$
where, for all first round offers $a\in \mathcal{A}$  and second round offers $b\in\mathcal{A}$, $$u_f(\sigma, r_{w}^{(\tau)}) = \begin{cases}
    (1-a) \cdot r_w^{(\tau)}(A_a) & \sigma = a\\
    \delta \cdot b \cdot r_w^{(\tau)}(R_ab) & \sigma = aA_b\\
    \delta  \cdot 0  \cdot r_w^{(\tau)}(R_ab) & \sigma=aR_b
\end{cases}$$
and 
$$
u_w(r_{f}^{(\tau)},\sigma) = \begin{cases}
    a \cdot r_f^{(\tau)}(a) & \sigma = A_a\\
    \delta \cdot (1-b) \cdot r_f^{(\tau)}(aA_b) & \sigma = R_ab
\end{cases}$$

For ease of notation, we will shorten the cumulative expected utility of agent $i$ at terminal sequence $\sigma$ to $U_{i,\sigma}^{(t)}$ and we will refer to the realization plan mass that agent $i$ puts on terminal sequence $\sigma$ at time $t$ as $r_{i,\sigma}^{(t)}$. Then, $U_i^{(t)}$ is the cumulative expected utility vector of agent $i$ at time $t$ and $r_i^{(t)}$ is the realization plan of agent $i$ at time $t$. Finally, let $\mathcal{Q}_i$ be the set of valid realization plans of agent $i$. We abuse notation slightly and suppose $r \in \mathcal{Q}_i$ is represented as a vector. Then, the expected utility of a realization plan, given a cumulative expected utility vector $U_i^{(t)}$, can be denoted as $$\langle U_i^{(t)}, r\rangle.$$

 % Let $I_i$ for $i \in \{f,w\}$ be the information set for each agent. Since our game is complete information, note that there is exactly one node for each $I \in I_i$. Let $h_{i,p,\sigma}$ be the node where agent $i$ is making a proposal after their opponent's previous action $\sigma$ and $h_{i,r,\sigma}$ be the node where agent $i$ is responding after their opponent's previous action $\sigma$. Then, an agent's behavioral strategy is $$\beta_i: I \times\mathcal{A} \cup\{\text{Accept},\text{Reject}\} \to [0,1].$$ 
%   Then, the convex version of an extensive form game can be derived from its sequence form representation~\footnote{See Appendix~\ref{appendix:sequence-form} for details of the sequence form of the 2-Round Alternating Bargaining game and see~\citet{shoham2008multiagent} for more details on the sequence form representation of extensive form games in general.}. Let $r_i$ be the realization plan of agent $i$ mapping action sequences of player $i$ to probability masses. Let $\mathcal{Q}_i$ be the set of valid realization plans of agent $i$. We abuse notation slightly and suppose $r \in \mathcal{Q}_i$ is represented as a vector of probability masses on sequences leading to payoff nodes. Then, the expected utility of a realization plan, given a cumulative expected utility vector $U_i^{(t)}$, can be denoted as $\langle U_i^{(t)}, r\rangle$. Finally, note that every realization plan has a one-to-one correspondence with a behavioral strategy

\section{Experimental Results for $\mathcal{G}^{(1)}$}
\label{appendix:nfg-graphs}

% NFG graph parameter sweeps, mutliple gold stars implies a mixed strategy at meta-game equilibrium
In this section, we include additional graphs from our numerical experiments for agents learning strategies for $\mathcal{G}^{(1)}$ via Algorithm~\ref{alg:ftrl}. In each plot, the worker payoff at the converged equilibrium is displayed along with gold star(s) to indicate the (mixed) equilibrium point of the 0-sum game of agents choosing an initial strategy for a given set of algorithm parameters. Each run of the algorithm converged according to the convergence criteria in Section~\ref{sec:experimental} within 5000 time steps and each convergence outcome was verified to be a $\epsilon$-Nash equilibrium for $\epsilon=10^{-7}$. The code we used to run these experiments can be accessed anonymously at the following link: \href{https://anonymous.4open.science/r/learning_bargaining_strategies-D6B7/README.md}{https://anonymous.4open.science/r/learning\_bargaining\_strategies-D6B7/README.md}. All of our experiments were run on standard laptops without any extra compute. The goal of our experiments was not to optimize run time or implement an algorithm for bargaining strategies in general, but rather to explore the space of convergence outcomes that were not covered by our theoretical results. 

\begin{figure}[hp]
    \centering
    \begin{subfigure}{0.32\textwidth}
        \includegraphics[width=\linewidth]{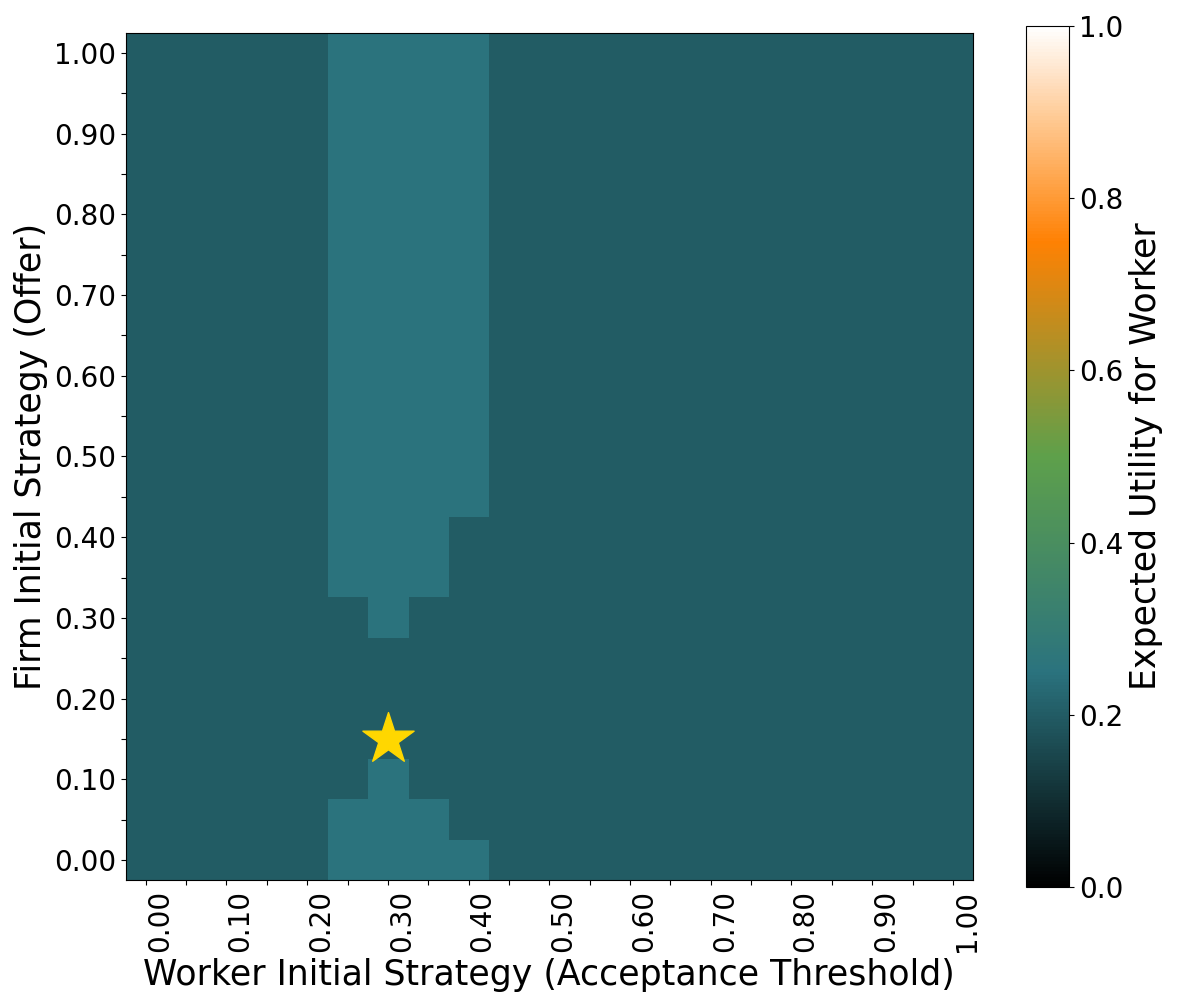}
        \caption{$D=20, \eta = 0.1$,$\alpha_f=\alpha_w=\bf{0}$.}
        \label{fig:enter-label}
    \end{subfigure}\hfill
    \begin{subfigure}{0.32\textwidth}
        \includegraphics[width=\linewidth]{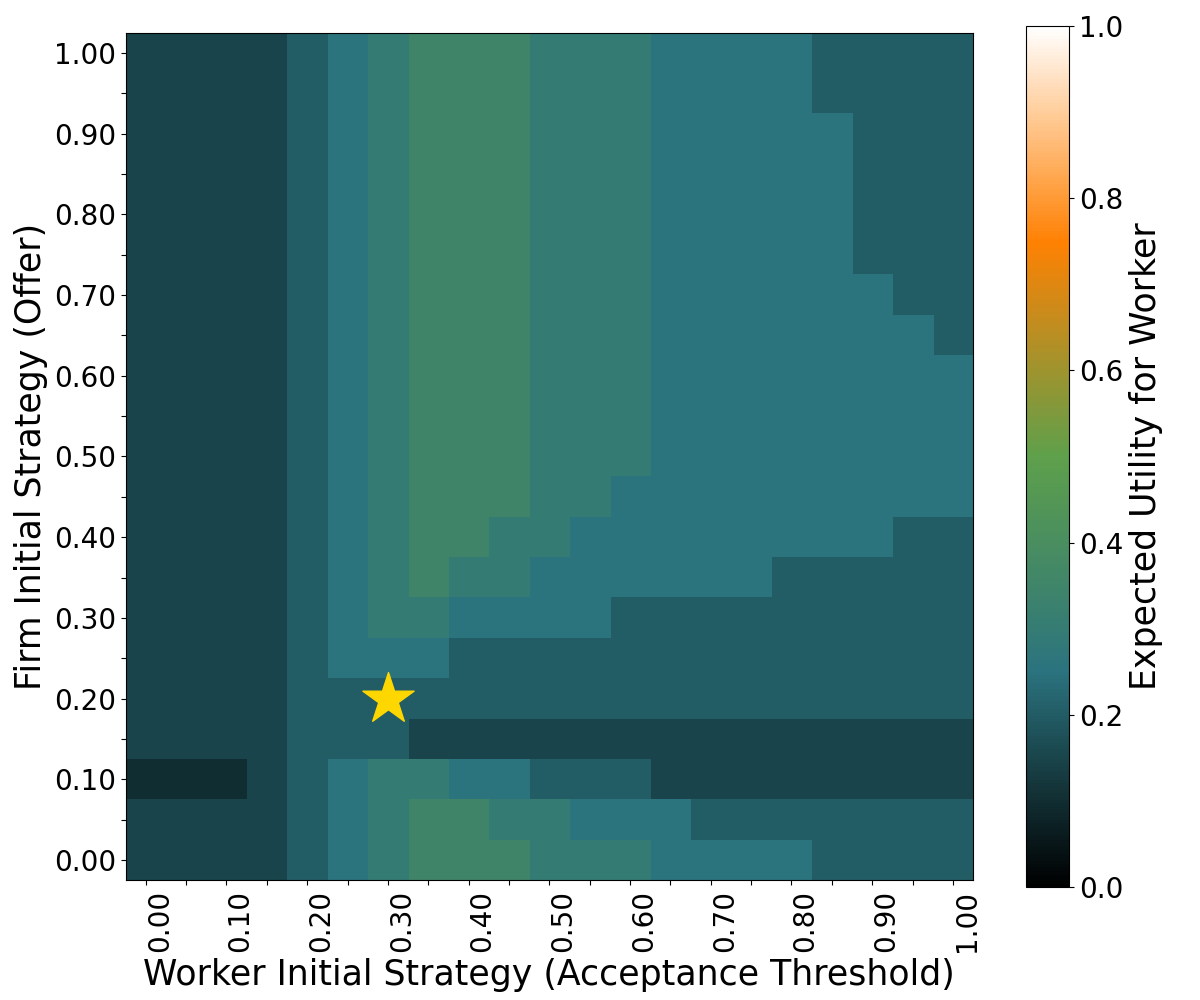}
        \caption{$D=20,\eta =0.8$,$\alpha_f=\alpha_w=\bf{0}$.}
        \label{fig:enter-label}
    \end{subfigure}\hfill
    \begin{subfigure}{0.32\textwidth}
        \includegraphics[width=\linewidth]{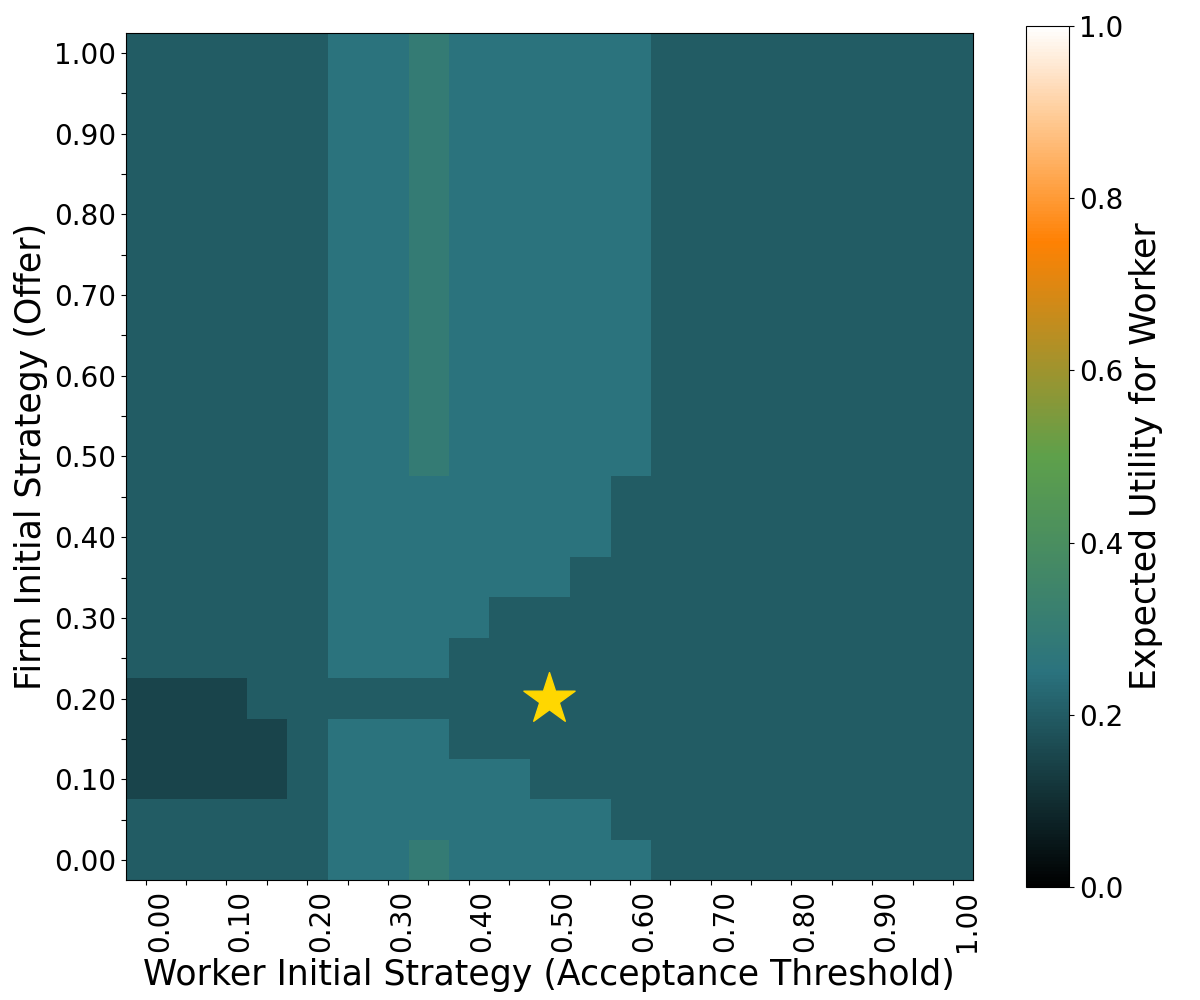}
        \caption{$D=20,\eta = 0.25$,$\alpha_f=\alpha_w=\bf{0}$.}
        \label{fig:enter-label}
    \end{subfigure}\\
    \begin{subfigure}{0.32\textwidth}
        \includegraphics[width=\linewidth]{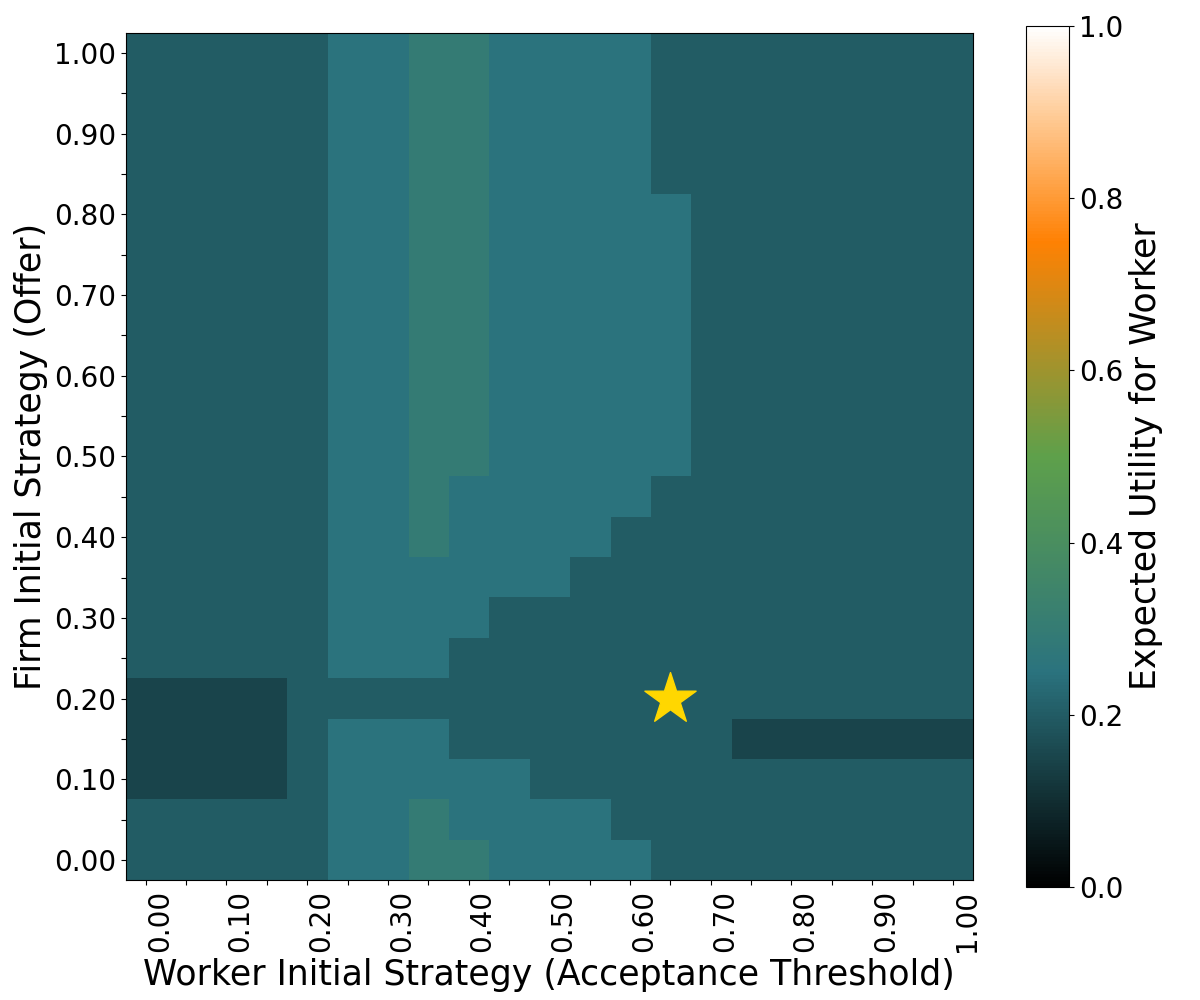}
        \caption{$D=20,\eta = 0.29$,$\alpha_f=\alpha_w=\bf{0}$.}
        \label{fig:enter-label}
    \end{subfigure}\hfill
    \begin{subfigure}{0.32\textwidth}
        \includegraphics[width=\linewidth]{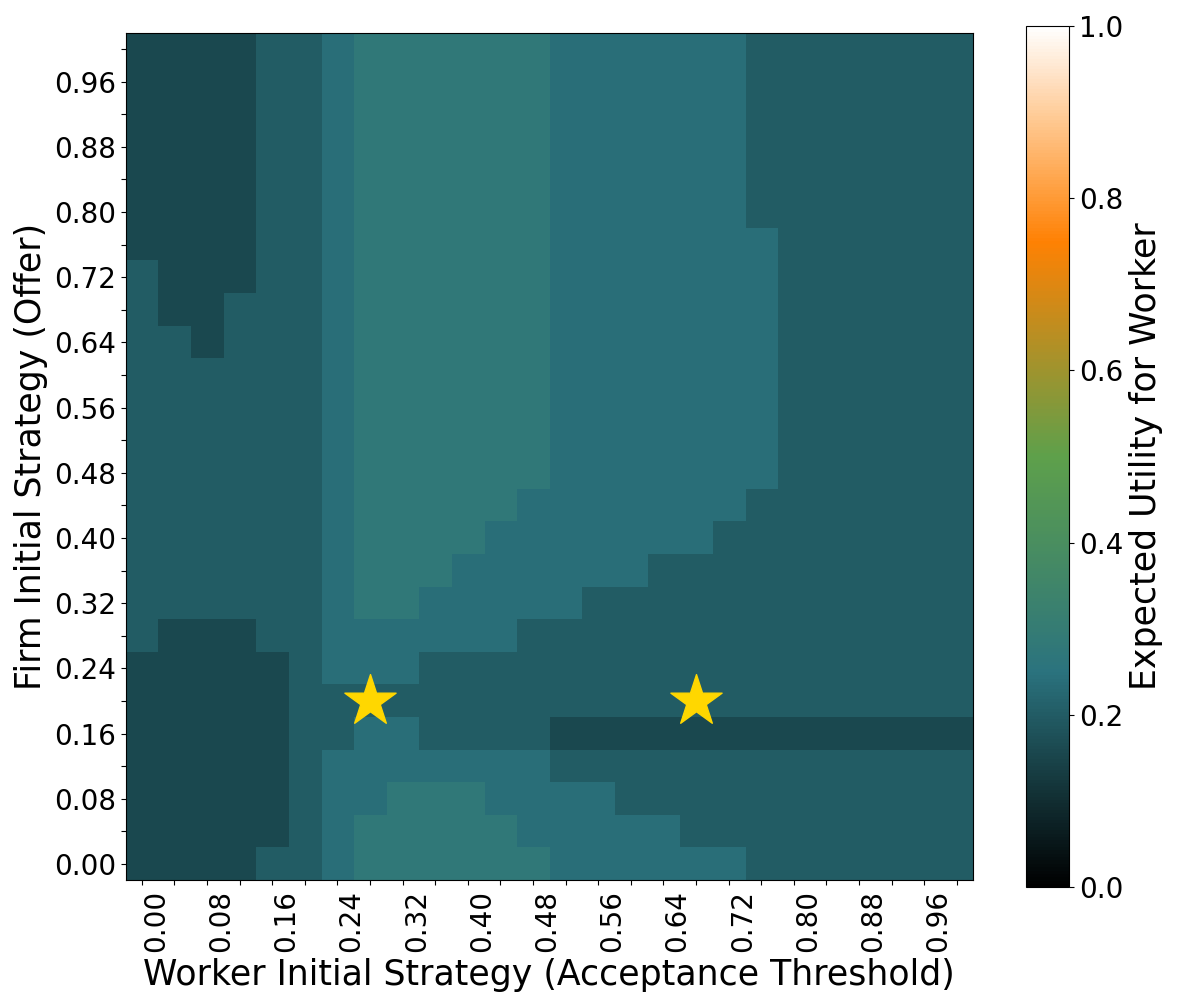}
        \caption{$D=25,\eta = 0.29$,$\alpha_f=\alpha_w=\bf{0}$.}
        \label{fig:enter-label}
    \end{subfigure}\hfill
    \begin{subfigure}{0.32\textwidth}
        \includegraphics[width=\linewidth]{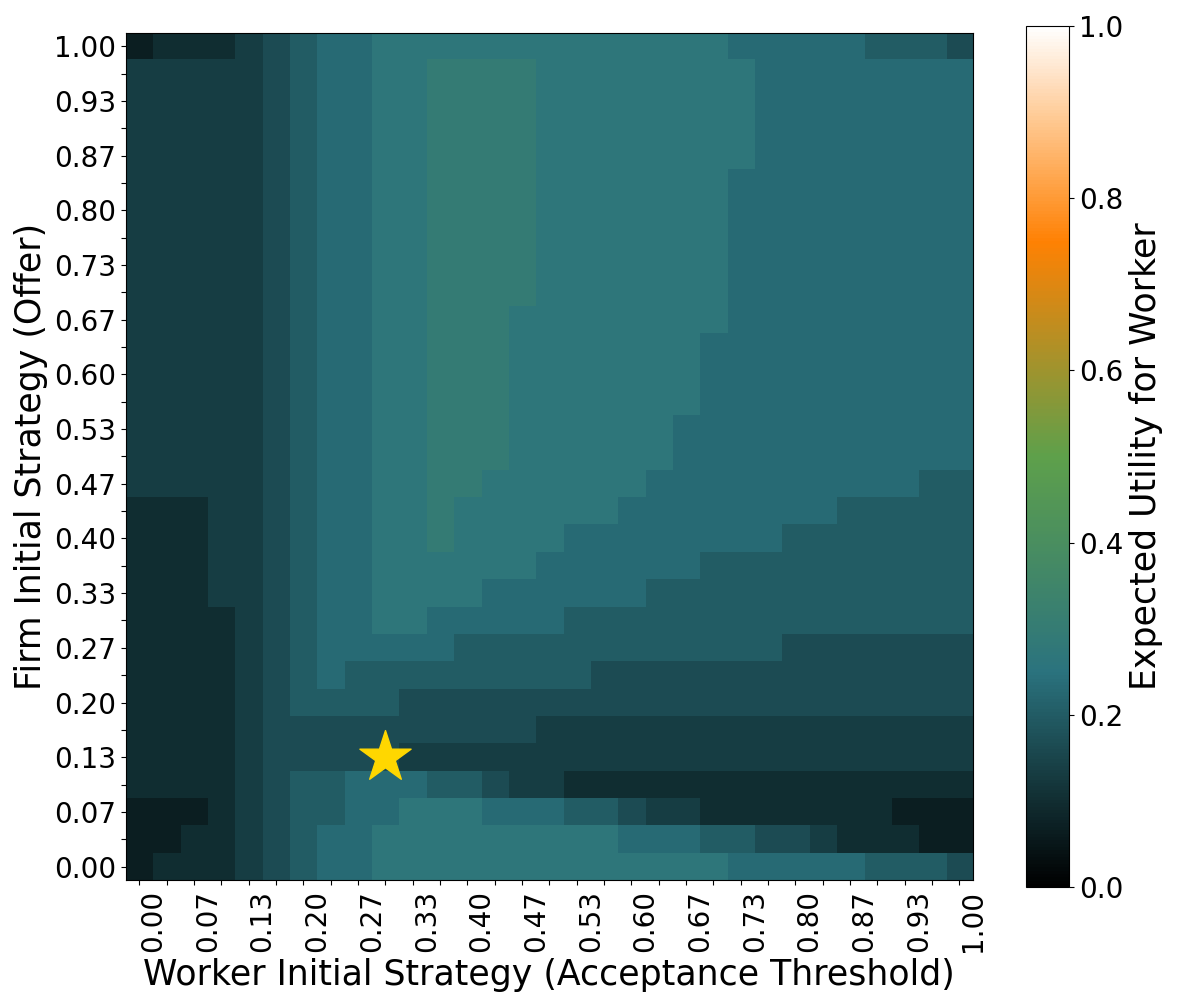}
        \caption{$D=30,\eta=0.29$,$\alpha_f=\alpha_w=1$.}
        \label{fig:enter-label}
    \end{subfigure}
    % \begin{subfigure}{0.4\textwidth}
    %     \includegraphics[width=\linewidth]{Experimental_Results/appendix-1-round/1_round_imshow_D_25_M_0.25_T_5000_del_0.5_eps_1e-07_rounds_1.png}
    %     \caption{Caption}
    %     \label{fig:enter-label}
    % \end{subfigure}
    \caption{Convergence outcomes for agents learning strategies for $\mathcal{G}^{(1)}$ via Algorithm~\ref{alg:ftrl} with $\alpha_f=\alpha_w=\bf{0}$ or $\alpha_f=\alpha_w=1$ and a variety of $D$ and $\eta$ values.}
    \label{fig:referenece_point_0}
\end{figure}

\begin{figure}[hp]
    \centering
    \begin{subfigure}{0.4\textwidth}
        \includegraphics[width=\linewidth]{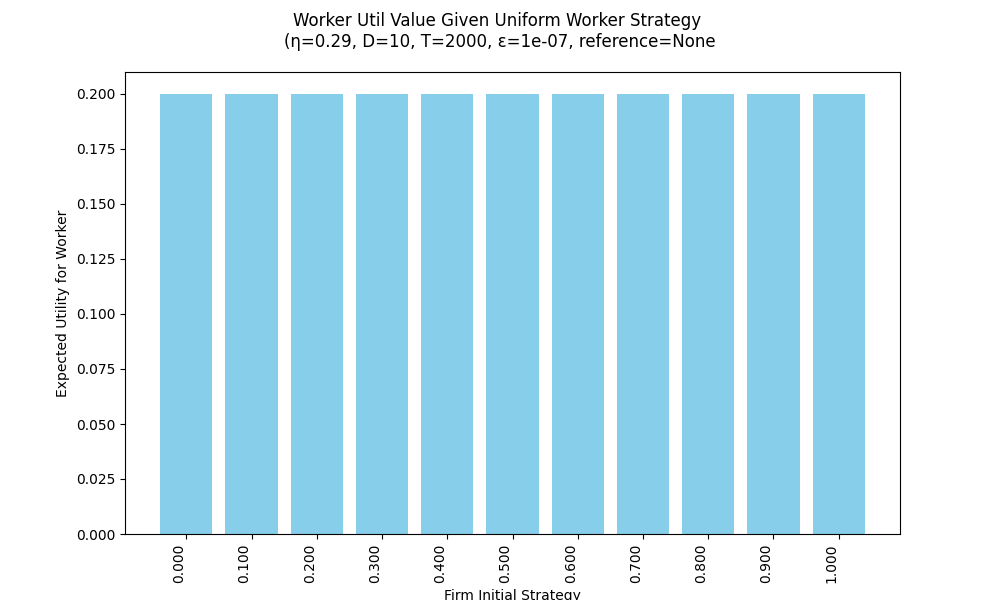}
        \caption{$D=10$.}
        \label{fig:enter-label}
    \end{subfigure}\hfill
    \begin{subfigure}{0.4\textwidth}
        \includegraphics[width=\linewidth]{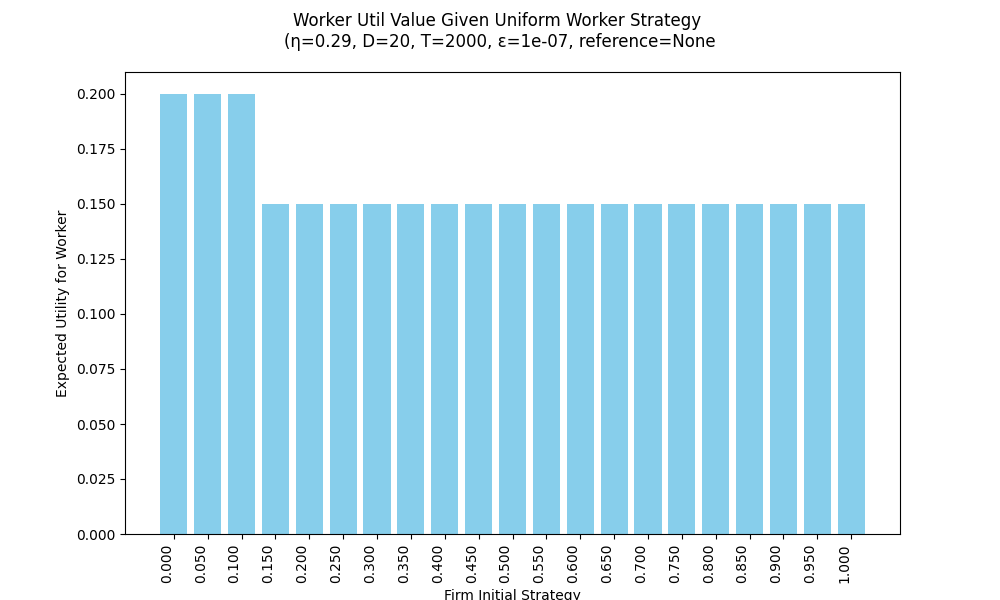}
        \caption{$D=20$.}
        \label{fig:enter-label}
    \end{subfigure}\\
    \begin{subfigure}{0.4\textwidth}
        \includegraphics[width=\linewidth]{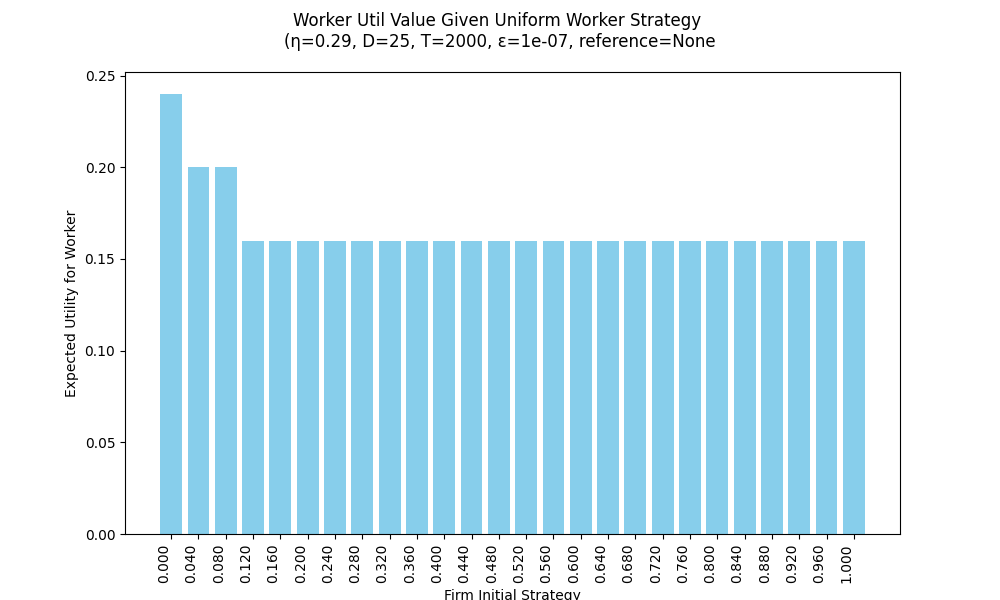}
        \caption{$D=25$.}
        \label{fig:enter-label}
    \end{subfigure}\hfill
    \begin{subfigure}{0.4\textwidth}
        \includegraphics[width=\linewidth]{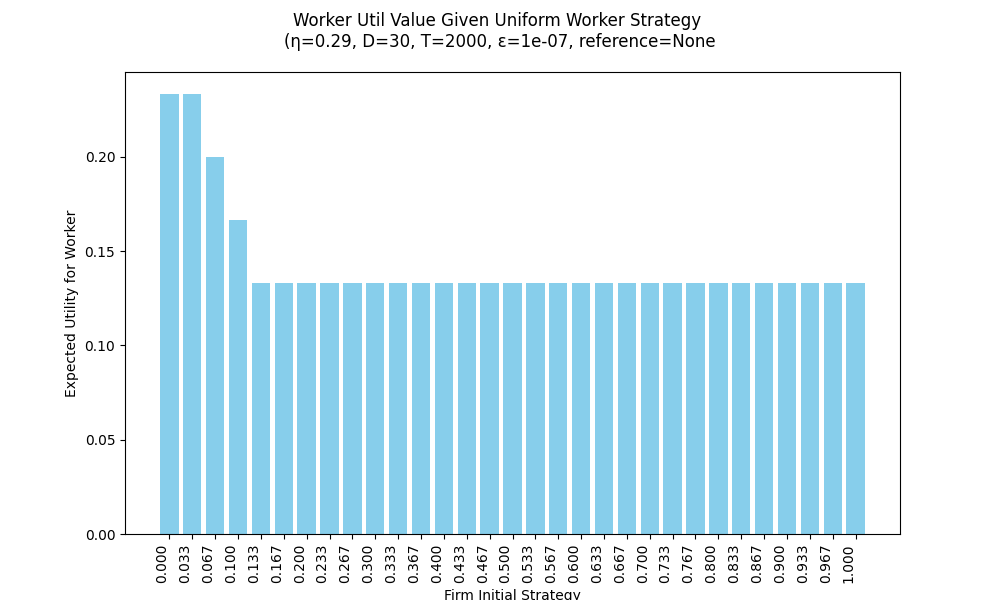}
        \caption{$D=30$.}
        \label{fig:enter-label}
    \end{subfigure}
    \caption{Worker payoff at convergence when the worker sets $x_w^{(1)}$ as the uniform mixed strategy and the firm sweeps over pure initial strategies. Experiments fix $\eta = 0.29$ and vary $D$ values.}
    \label{fig:worker-uniform-mixed}
\end{figure}

\begin{figure}[hp]
    \centering
    \begin{subfigure}{0.4\textwidth}
        \includegraphics[width=\linewidth]{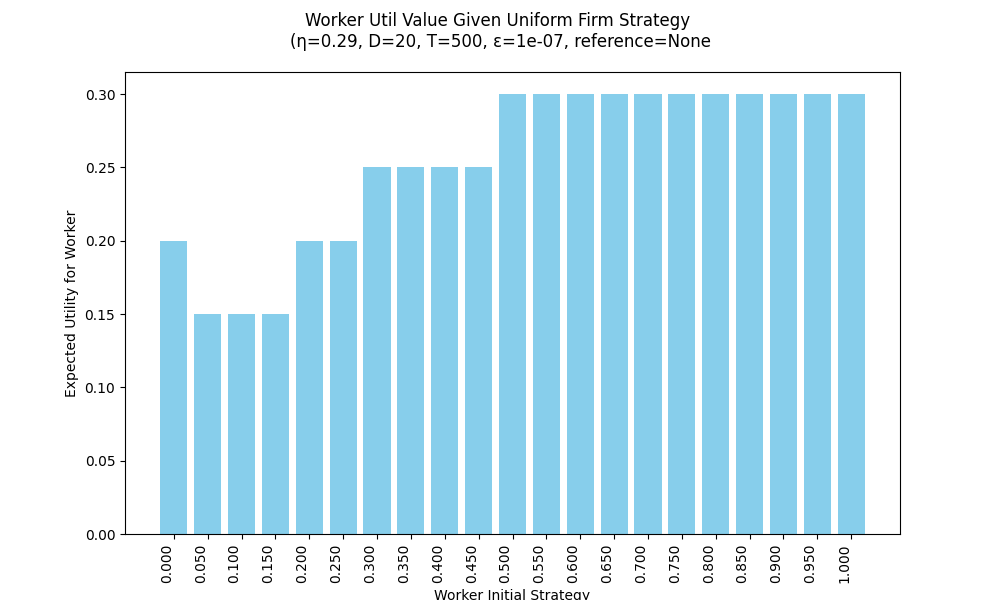}
        \caption{$D=20$.}
        \label{fig:enter-label}
    \end{subfigure}\hfill
    \begin{subfigure}{0.4\textwidth}
        \includegraphics[width=\linewidth]{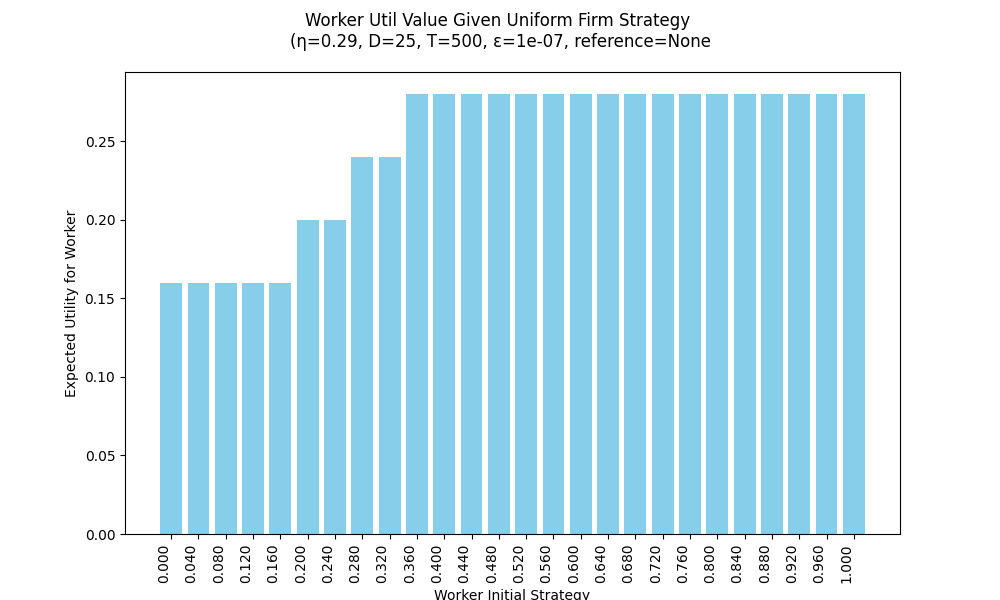}
        \caption{$D=25$.}
        \label{fig:enter-label}
    \end{subfigure}\\
    \begin{subfigure}{0.4\textwidth}
        \includegraphics[width=\linewidth]{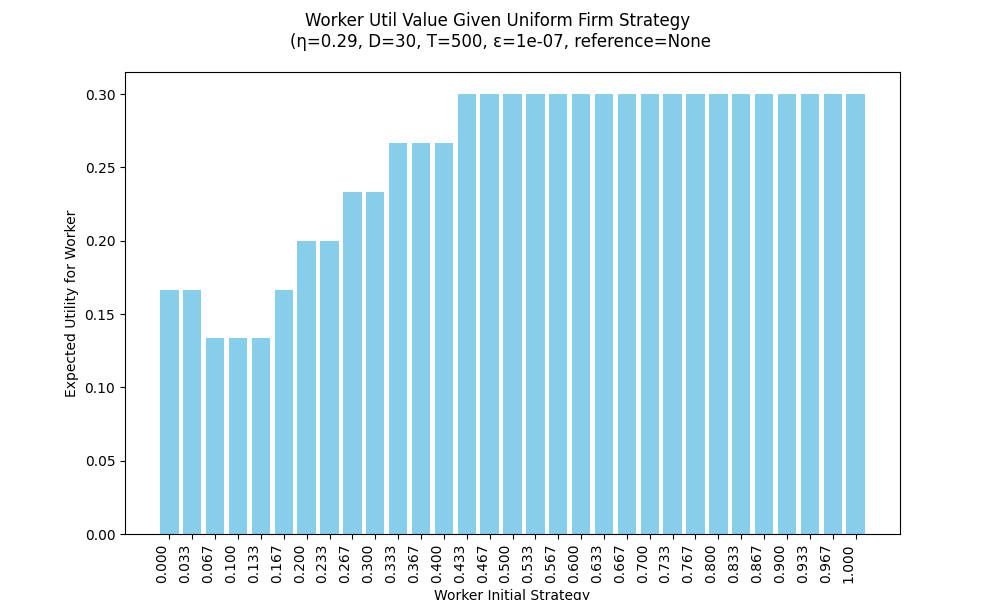}
        \caption{$D=30$.}
        \label{fig:enter-label}
    \end{subfigure}\hfill
    \begin{subfigure}{0.4\textwidth}
        \includegraphics[width=\linewidth]{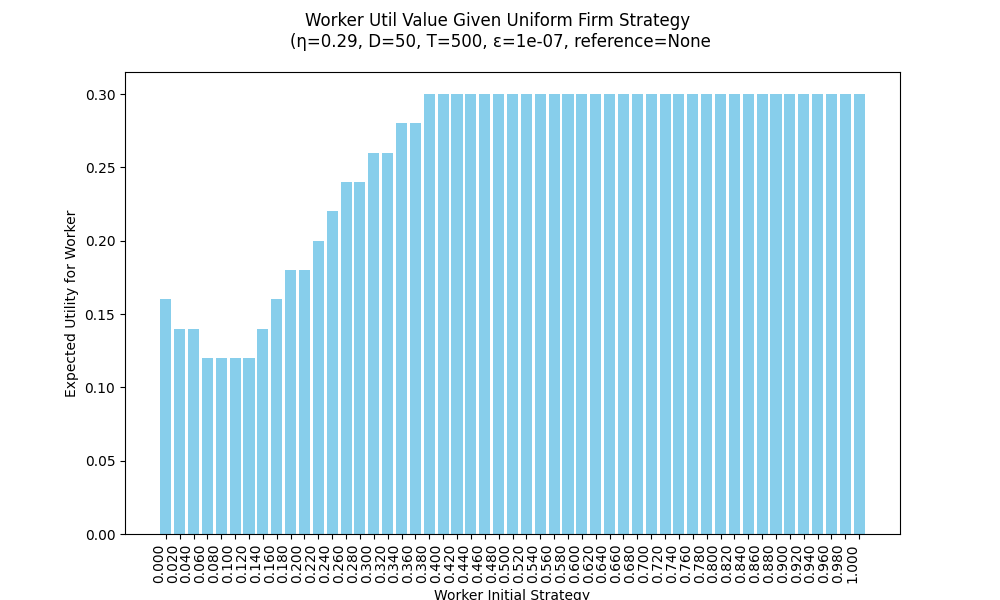}
        \caption{$D=50$.}
        \label{fig:enter-label}
    \end{subfigure}
    \caption{Worker payoff at convergence when the firm sets $x_f^{(1)}$ as the uniform mixed strategy and the worker sweeps over pure initial strategies. Experiments fix $\eta = 0.29$ and vary $D$ values.}
    \label{fig:firm_mixed_uniform}
\end{figure}

We also experimented with the worker and firm each starting with a uniform mixed initial strategy in Figures~\ref{fig:worker-uniform-mixed} and~\ref{fig:firm_mixed_uniform}, respectively. We fixed $\eta = 0.29$ and $\alpha_f=\alpha_w =\bf{0}$ in these experiments and varied $D$ from 10 to 50. Compared with Figure~\ref{fig:referenece_point_0}, the minimum value of $u_w$ only slightly decreases in both cases, however, in both cases, the proportion of pure initial strategies where the worker gets the worst outcome increases as $D$ increases.

\begin{figure}[h]
    \centering
    \begin{subfigure}{0.4\textwidth}
        \includegraphics[width=\linewidth]{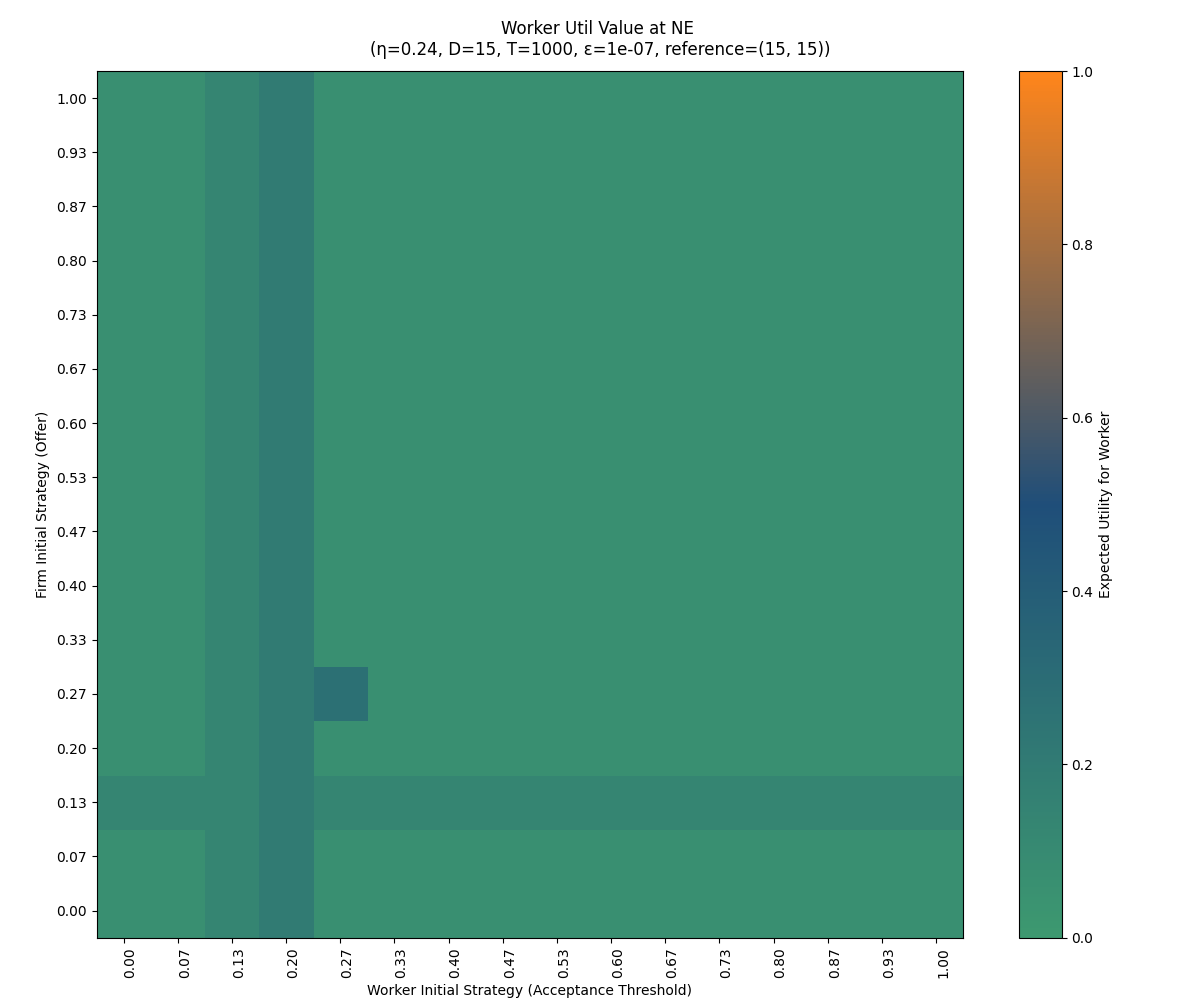}
        \caption{$D=15,\eta=0.24,\alpha_f=1,\alpha_w=1$.}
        \label{fig:enter-label}
    \end{subfigure}\hfill
    \begin{subfigure}{0.4\textwidth}
        \includegraphics[width=\linewidth]{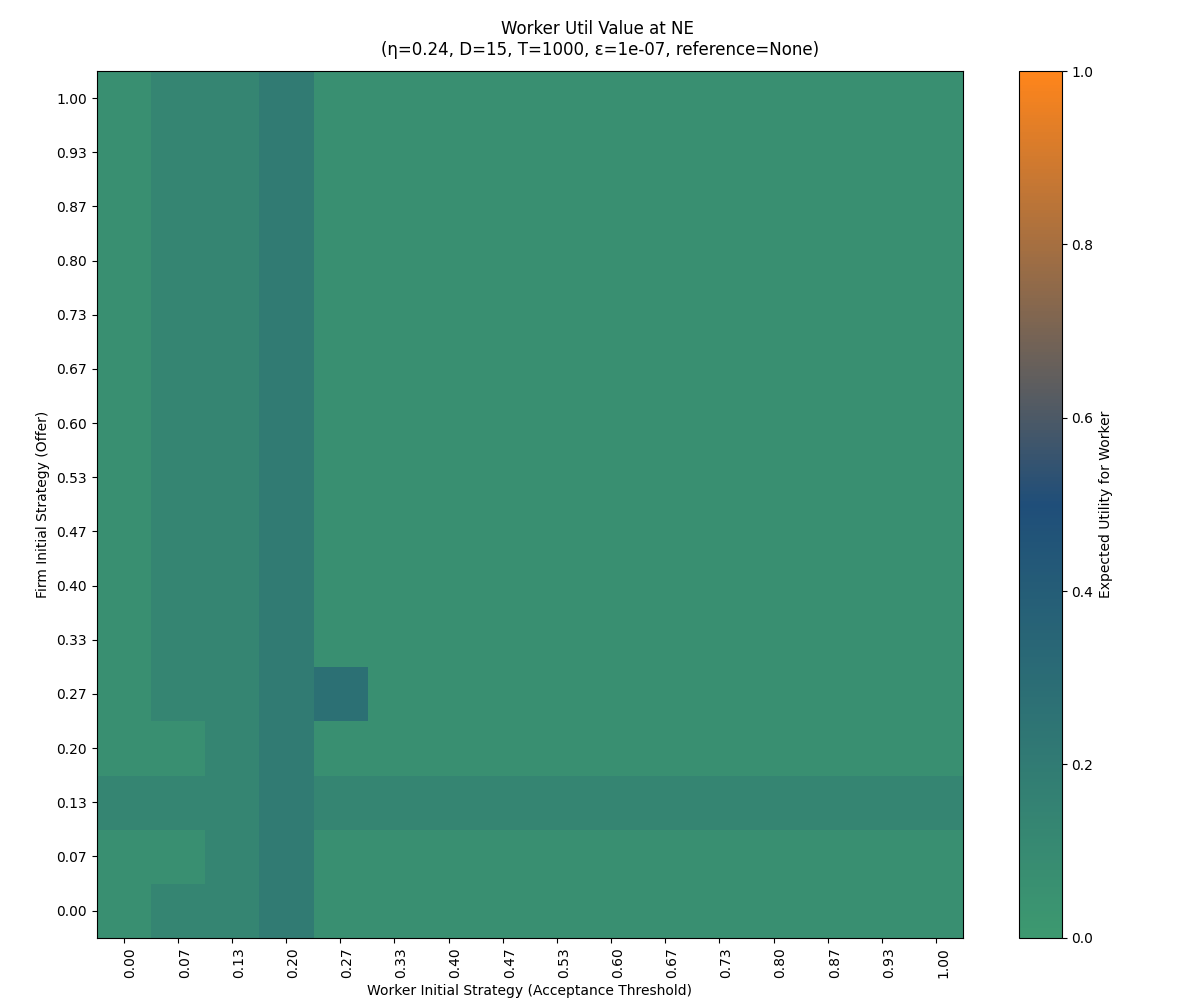}
        \caption{$D=15,\eta=0.24,\alpha_f=\alpha_w=\bf{0}$.}
        \label{fig:enter-label}
    \end{subfigure}\\
    \begin{subfigure}{0.4\textwidth}
        \includegraphics[width=\linewidth]{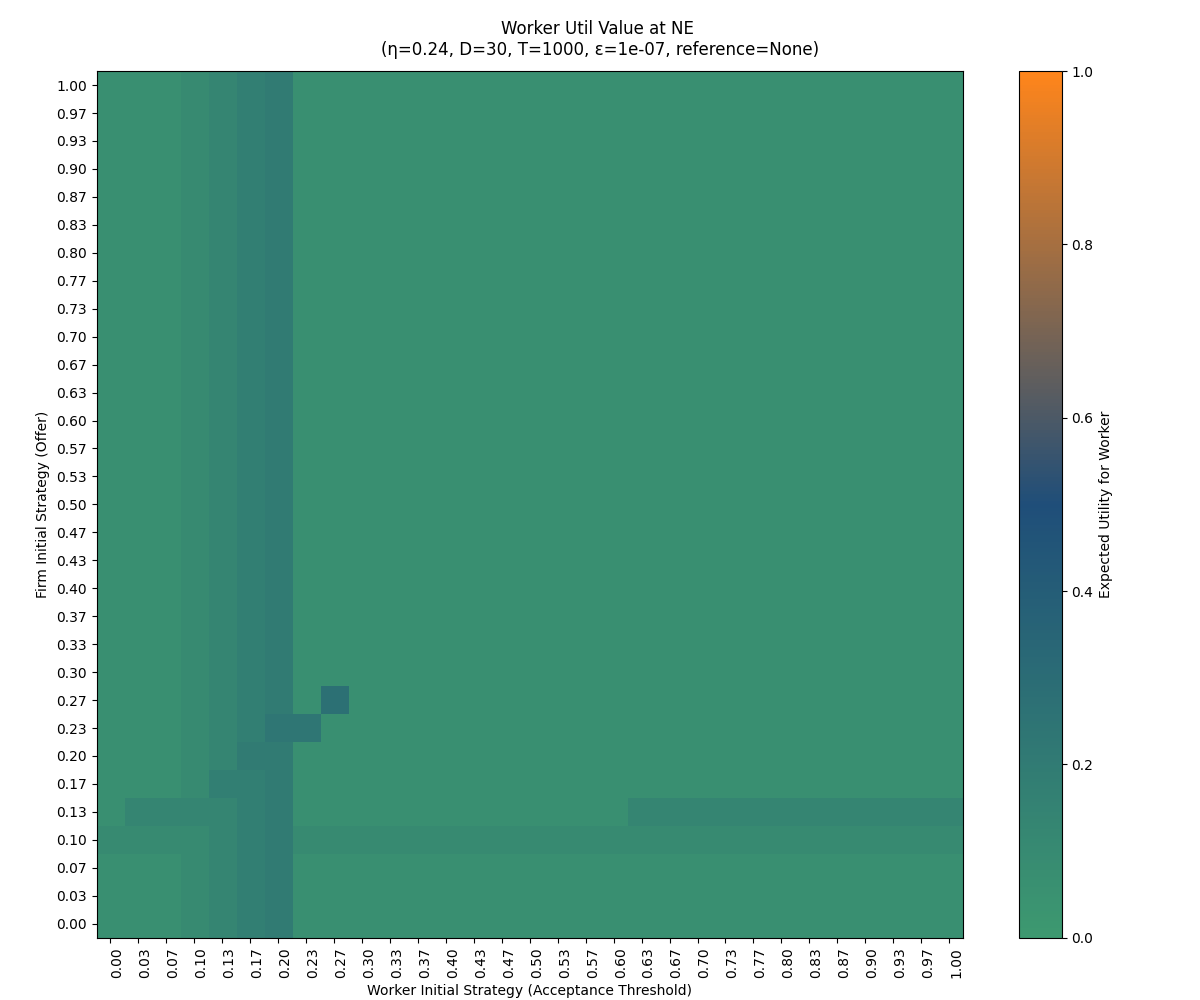}
        \caption{$D=30,\eta=0.24,\alpha_f=\alpha_w=\bf{0}$.}
        \label{fig:enter-label}
    \end{subfigure}\hfill
    \begin{subfigure}{0.4\textwidth}
        \includegraphics[width=\linewidth]{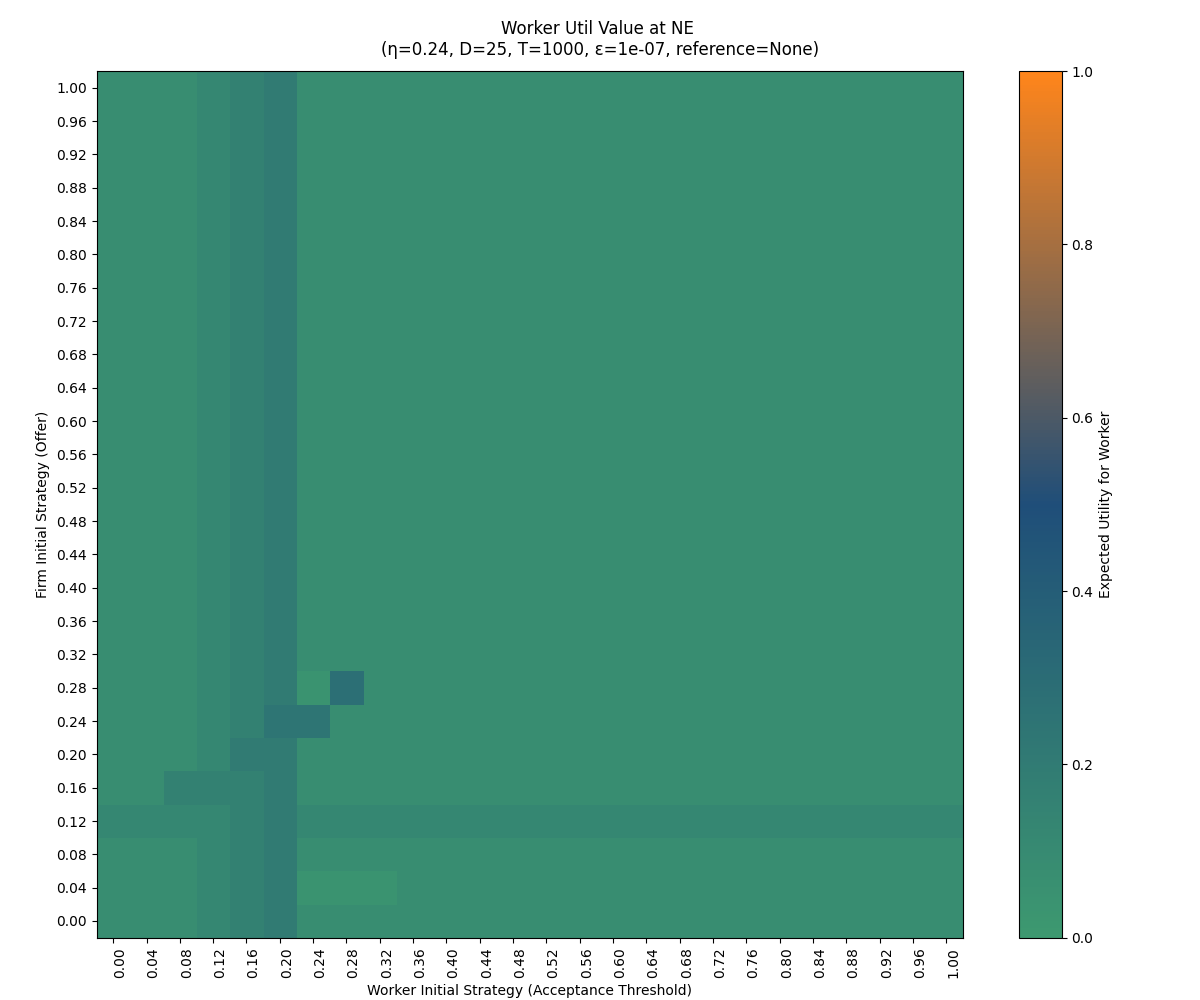}
        \caption{$D=25,\eta=0.24,\alpha_f=\alpha_w=\bf{0}$.}
        \label{fig:enter-label}
    \end{subfigure}\\
    \begin{subfigure}{0.4\textwidth}
        \includegraphics[width=\linewidth]{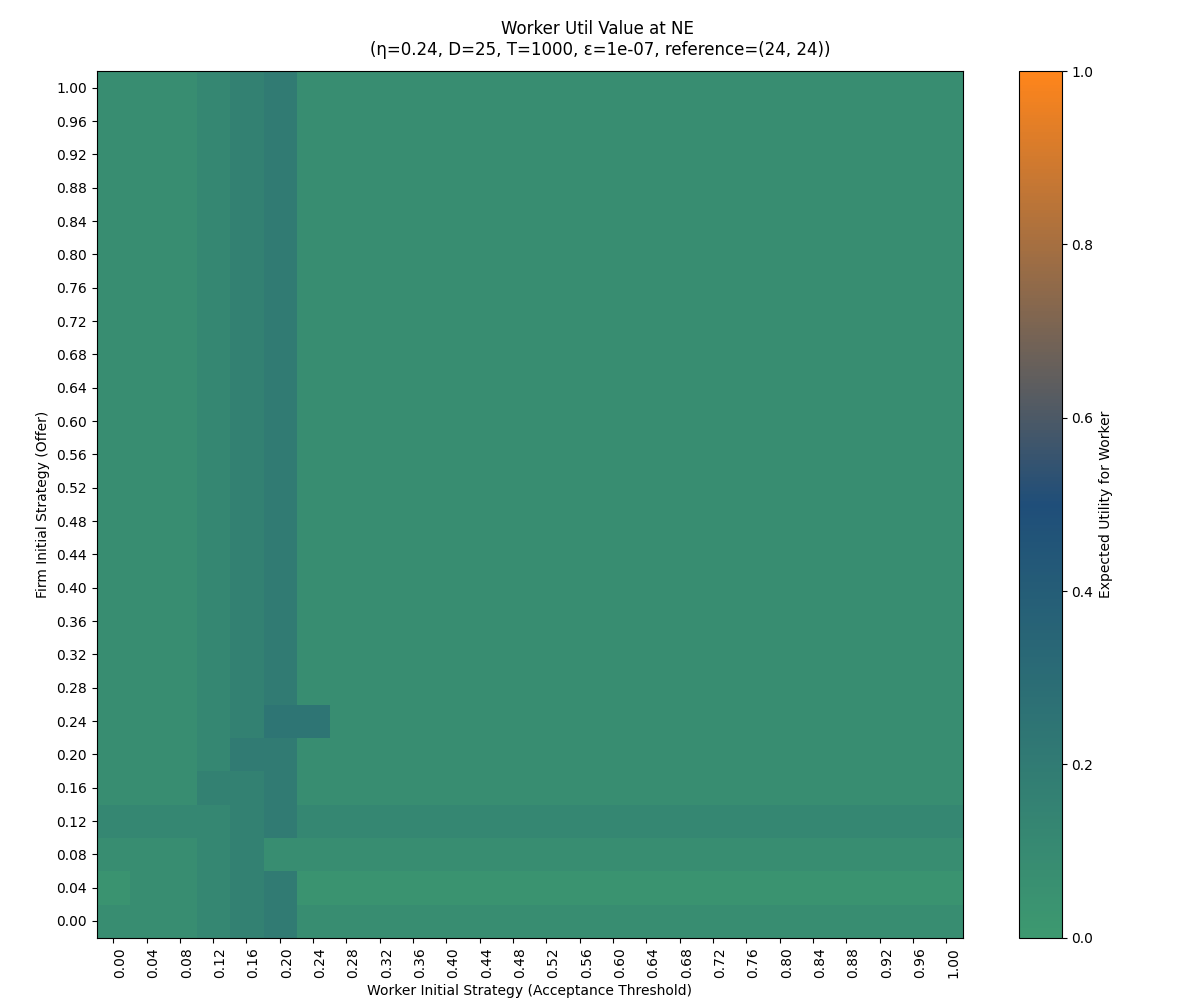}
    \caption{$D=25,\eta=0.24,\alpha_f=\frac{24}{25},\alpha_w=\frac{24}{25}$.}
        \label{fig:enter-label}
    \end{subfigure}\hfill
    \begin{subfigure}{0.4\textwidth}
        \includegraphics[width=\linewidth]{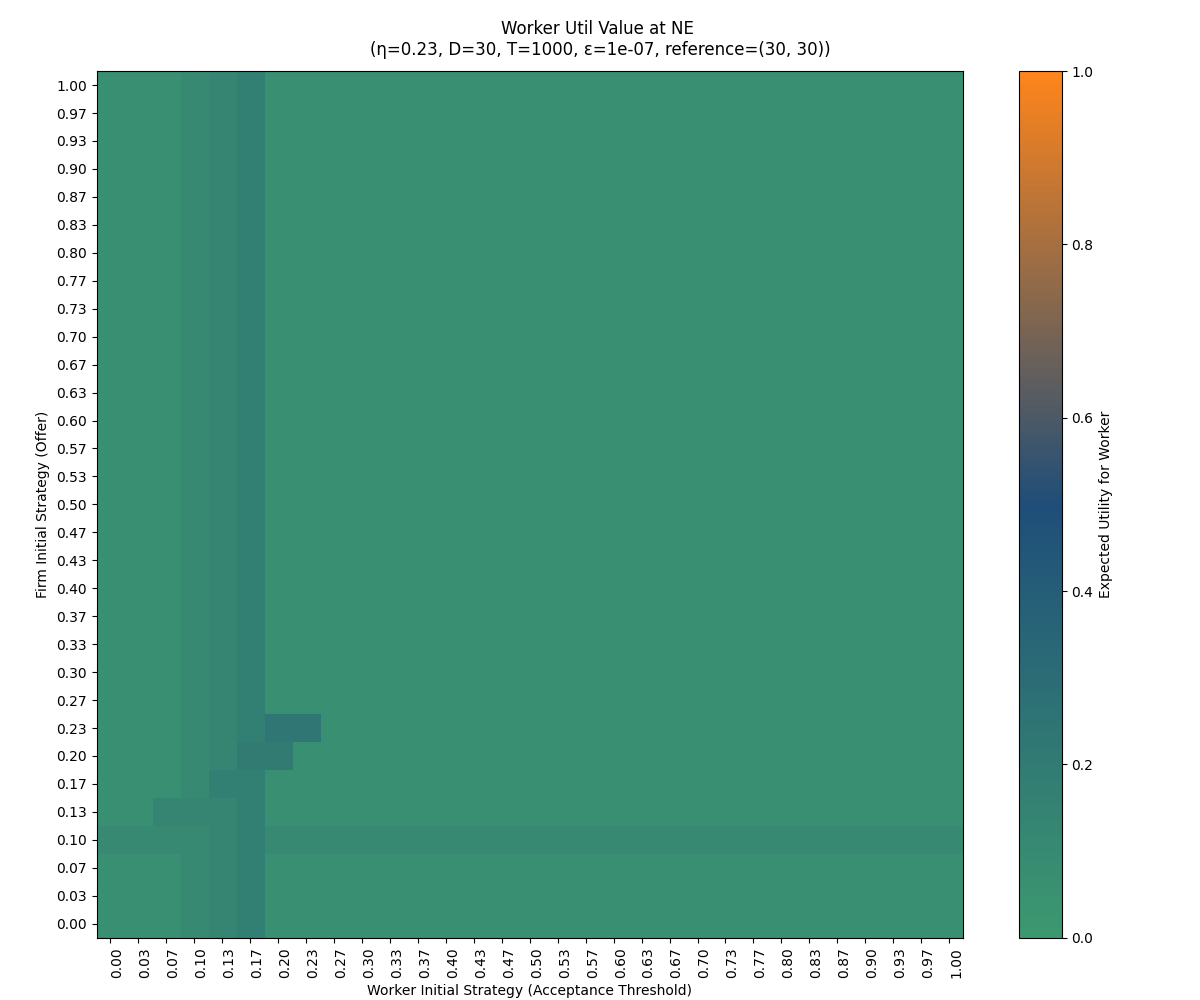}
        \caption{$D=30,\eta=0.23,\alpha_f=1,\alpha_w=1$.}
        \label{fig:enter-label}
    \end{subfigure}
    \caption{Convergence outcomes for agents learning strategies for the extensive form version of the ultimatum game via Algorithm~\ref{alg:ftrl} under a variety of parameter settings.}
    \label{fig:extensive-form-1-round}
\end{figure}

The convergence outcome plots for agents learning strategies for the ultimatum game in the extensive form are shown in Figure~\ref{fig:extensive-form-1-round}. In these experiments, the worker generally receives a lower equilibrium payoff than when agents learn strategies for the ultimatum game in the normal form, i.e., $\mathcal{G}^{(1)}$. However, we still see a variety of NE outcomes at convergence which indicates another non-trivial relationship between the initial conditions and NE outcomes at convergence indicating that it is possible to use FTRL to learn a variety of bargaining strategies in extensive form games, even in the simplest 1 round bargaining game.

\begin{figure}[h]
    \centering
    \begin{subfigure}{0.32\textwidth}
        \includegraphics[width=\linewidth]{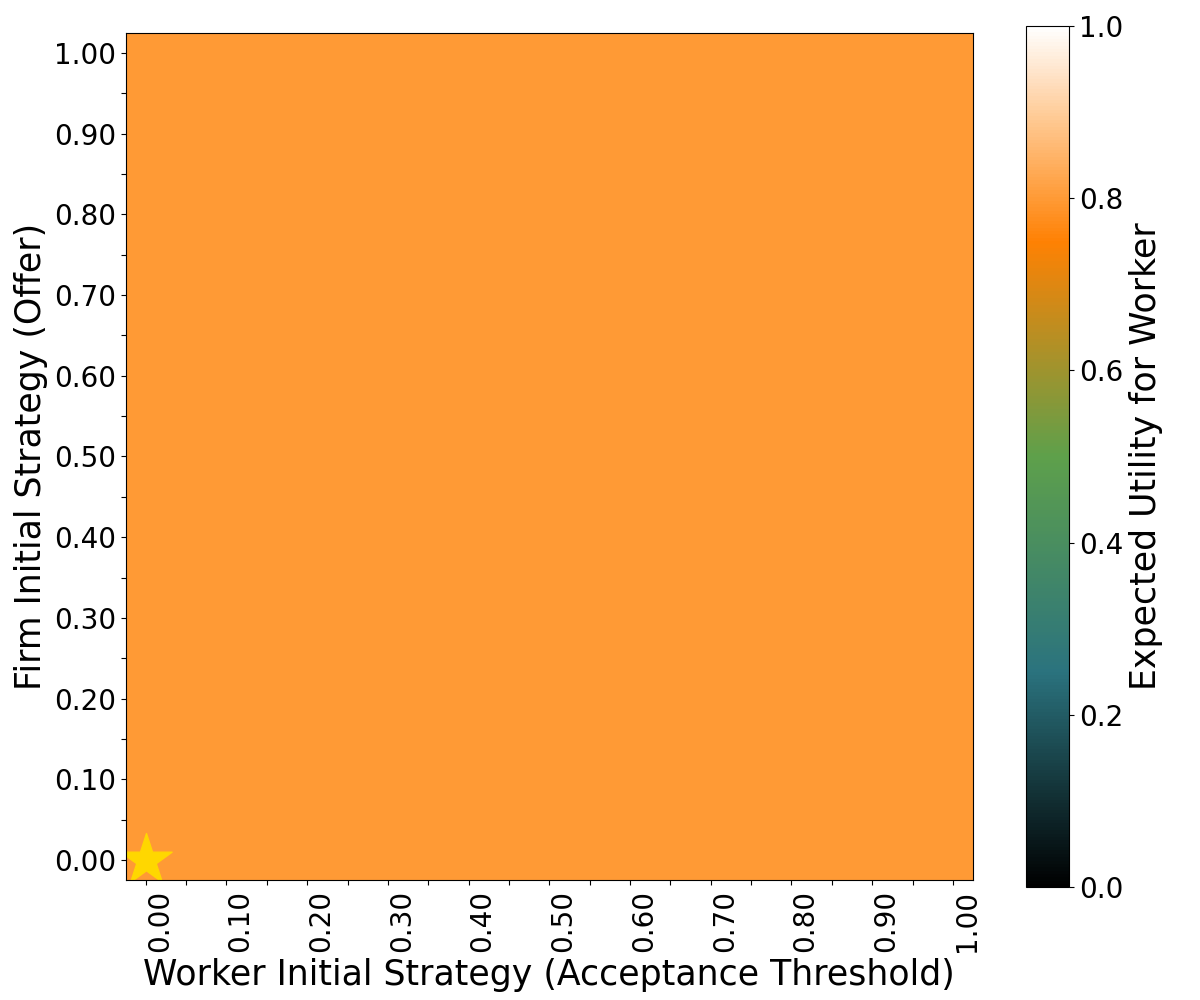}
        \caption{$\eta =0.1$.}
        \label{fig:enter-label}
    \end{subfigure}\hfill
    \begin{subfigure}{0.32\textwidth}
        \includegraphics[width=\linewidth]{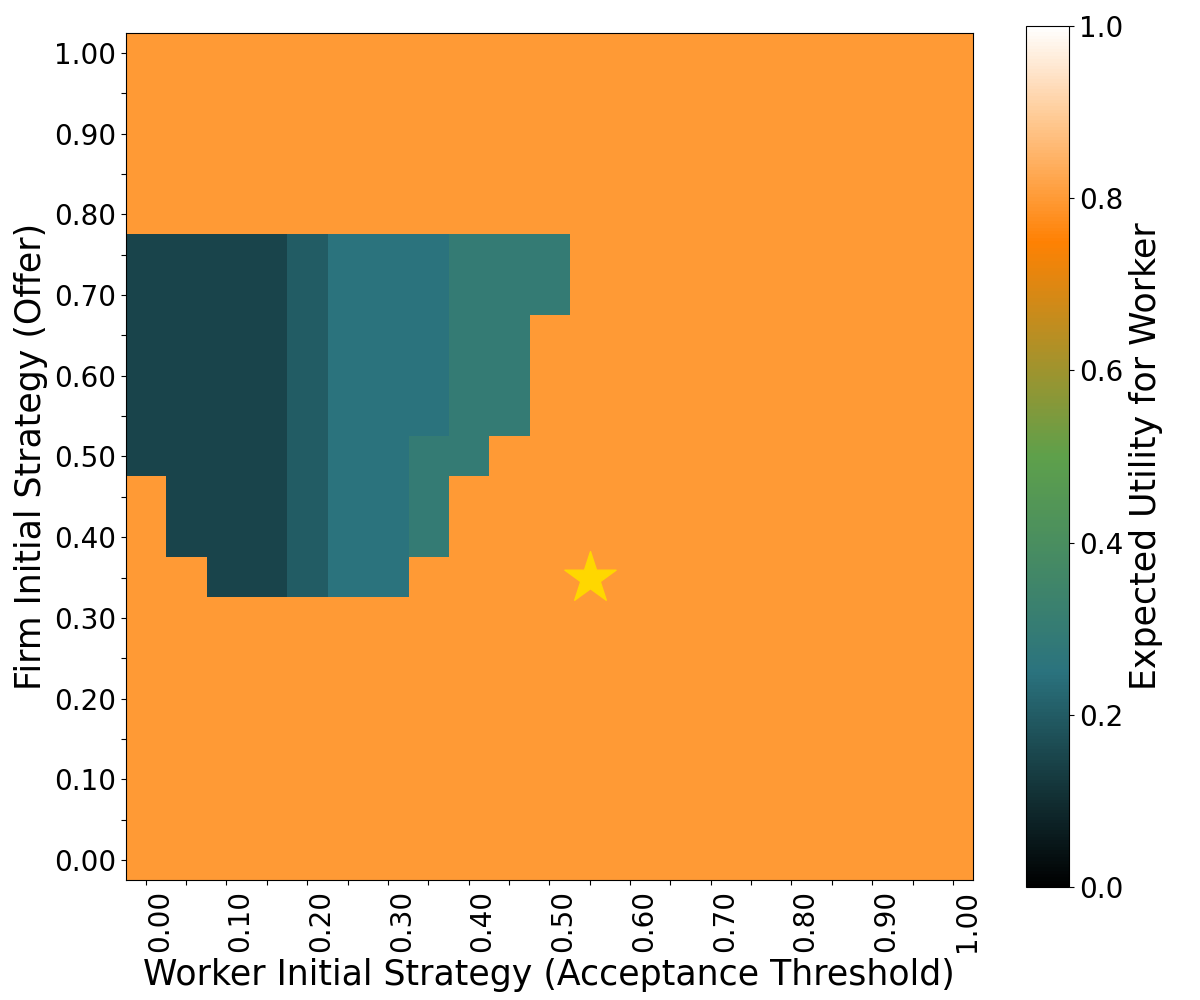}
        \caption{$\eta = 0.25$.}
        \label{fig:enter-label}
    \end{subfigure}\hfill
    \begin{subfigure}{0.32\textwidth}
        \includegraphics[width=\linewidth]{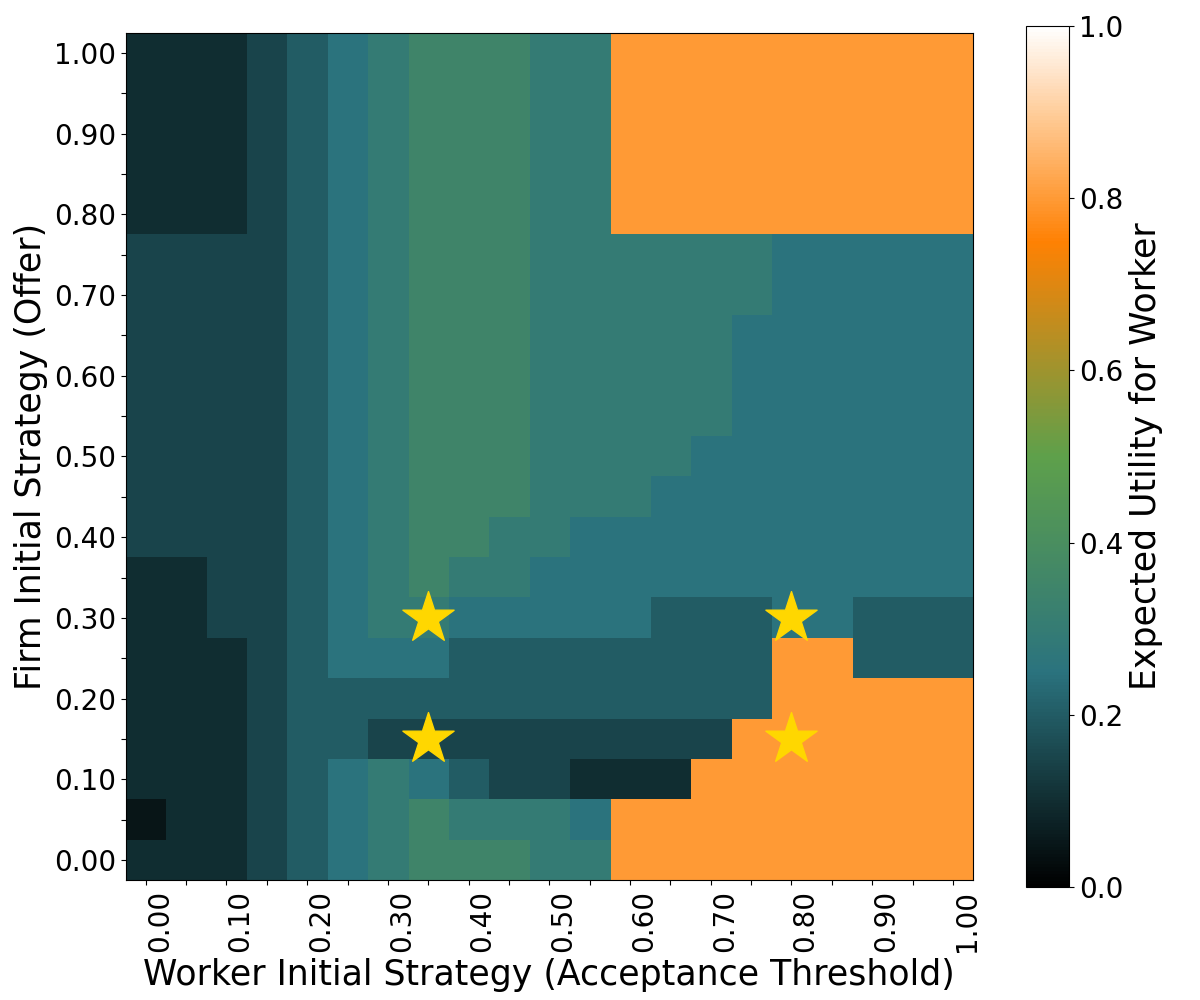}
        \caption{$\eta = 0.8$.}
        \label{fig:enter-label}
    \end{subfigure}
    % \begin{subfigure}{0.4\textwidth}
    %     \includegraphics[width=\linewidth]{Experimental_Results/appendix-1-round/1_round_imshow_D_25_M_0.25_T_5000_del_0.5_eps_1e-07_rounds_1_ref_20,_16_.png}
    %     \caption{Caption}
    %     \label{fig:nfg-plots-20-16-7}
    % \end{subfigure}
    \caption{Convergence outcomes for agents learning strategies for $\mathcal{G}^{(1)}$ via Algorithm~\ref{alg:ftrl} with $\alpha_f=1, \alpha_w=\frac{4}{5}$ and $D=20$ and a variety of $\eta$ values.}
    \label{fig:referenece_point__20,_16__20}
\end{figure}
\begin{figure}[h]
    \centering
      \begin{subfigure}{0.32\textwidth}
        \includegraphics[width=\linewidth]{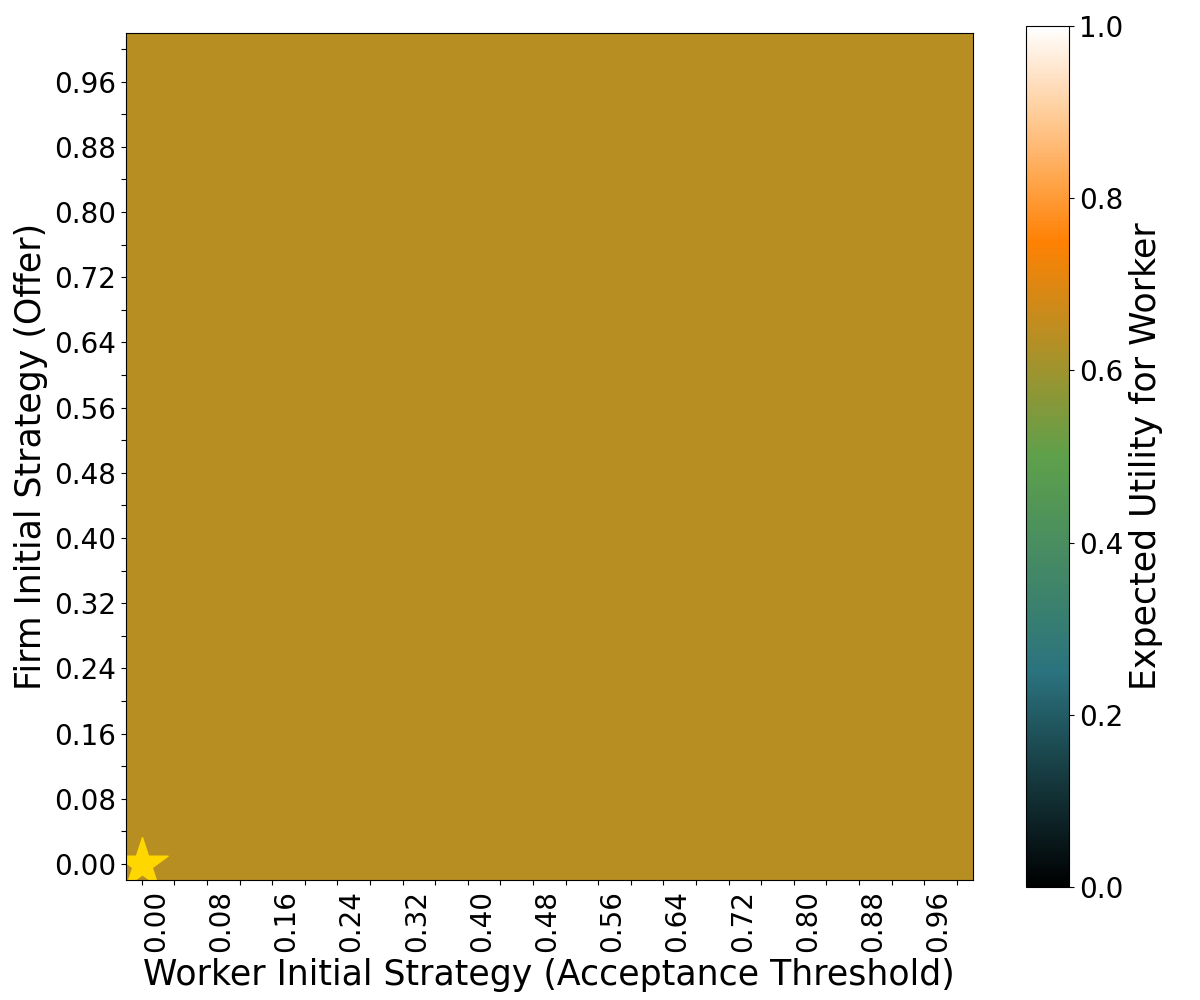}
        \caption{$\eta=0.1$.}
        \label{fig:enter-label}
    \end{subfigure}\hfill
    \begin{subfigure}{0.32\textwidth}
        \includegraphics[width=\linewidth]{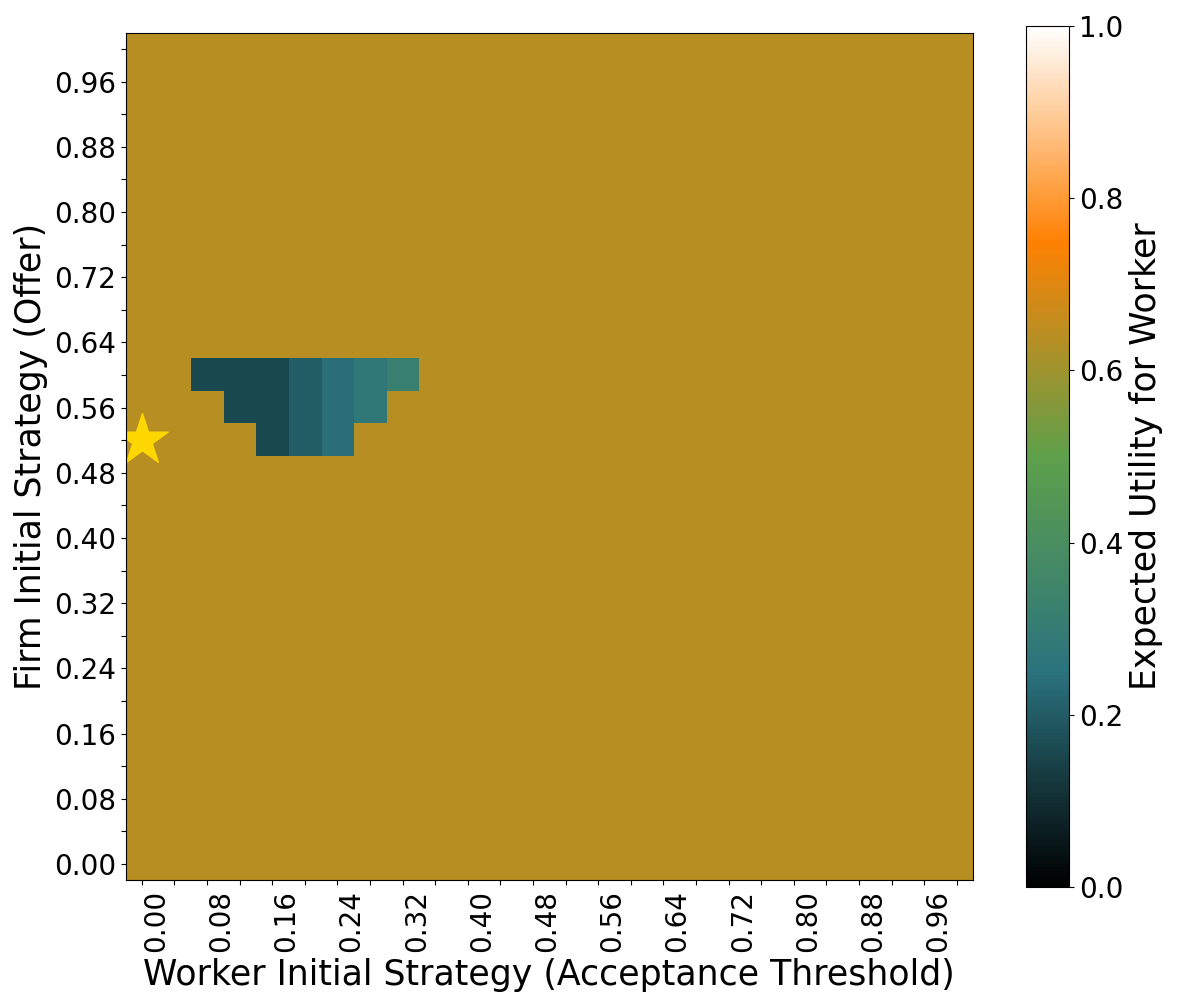}
        \caption{$\eta = 0.6$.}
        \label{fig:enter-label}
    \end{subfigure}\hfill
    \begin{subfigure}{0.32\textwidth}
        \includegraphics[width=\linewidth]{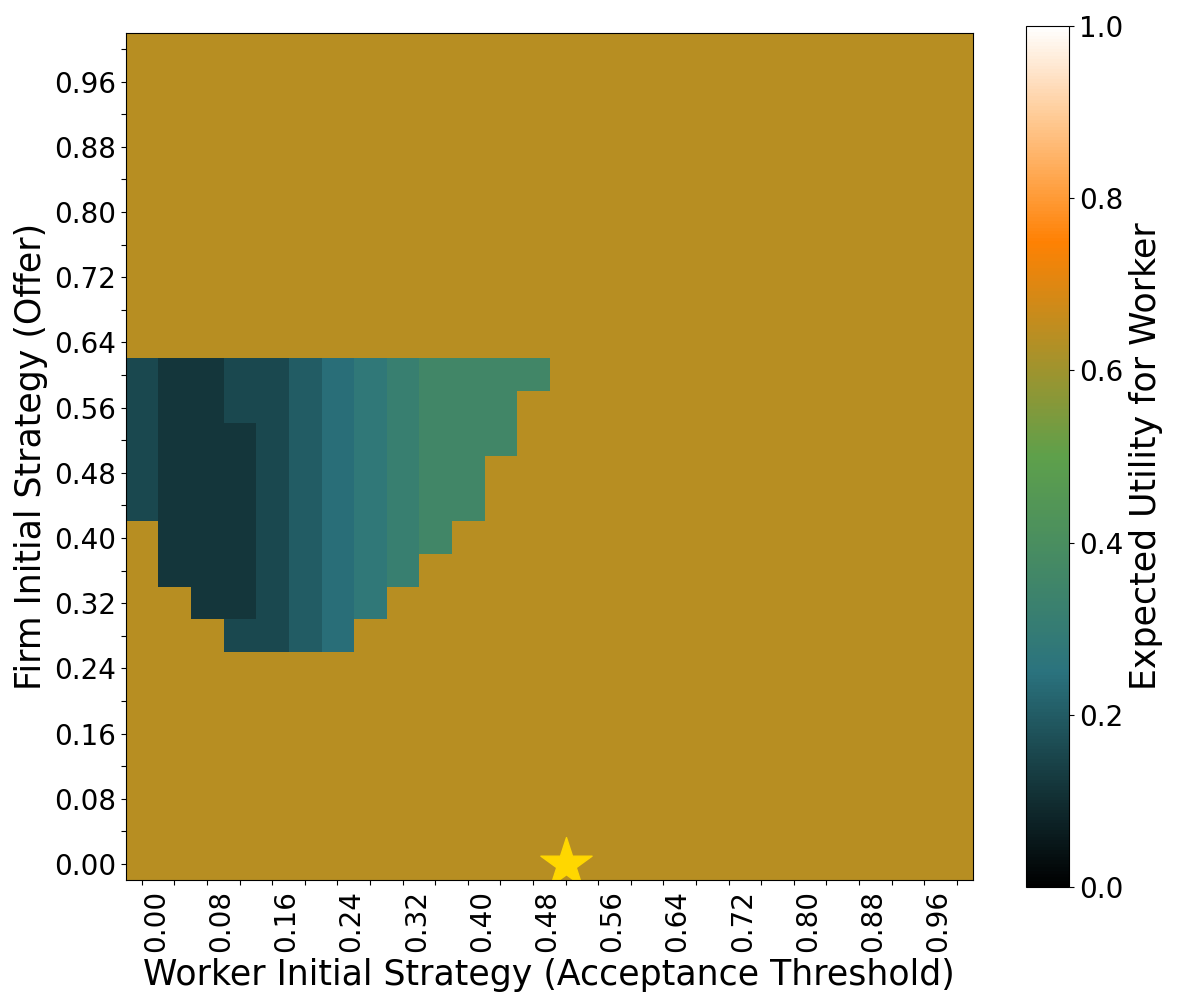}
        \caption{$\eta = 0.8$.}
        \label{fig:enter-label}
    \end{subfigure}
    \caption{Convergence outcomes for agents learning strategies for $\mathcal{G}^{(1)}$ via Algorithm~\ref{alg:ftrl} with $\alpha_f=\frac{4}{5}, \alpha_w=\frac{16}{25}$ and $D=25$ under a variety of $\eta$ values.}
    \label{fig:ref_point_20_16_25}
\end{figure}
\begin{figure}[h]
    \centering
    \begin{subfigure}{0.32\textwidth}
        \includegraphics[width=\linewidth]{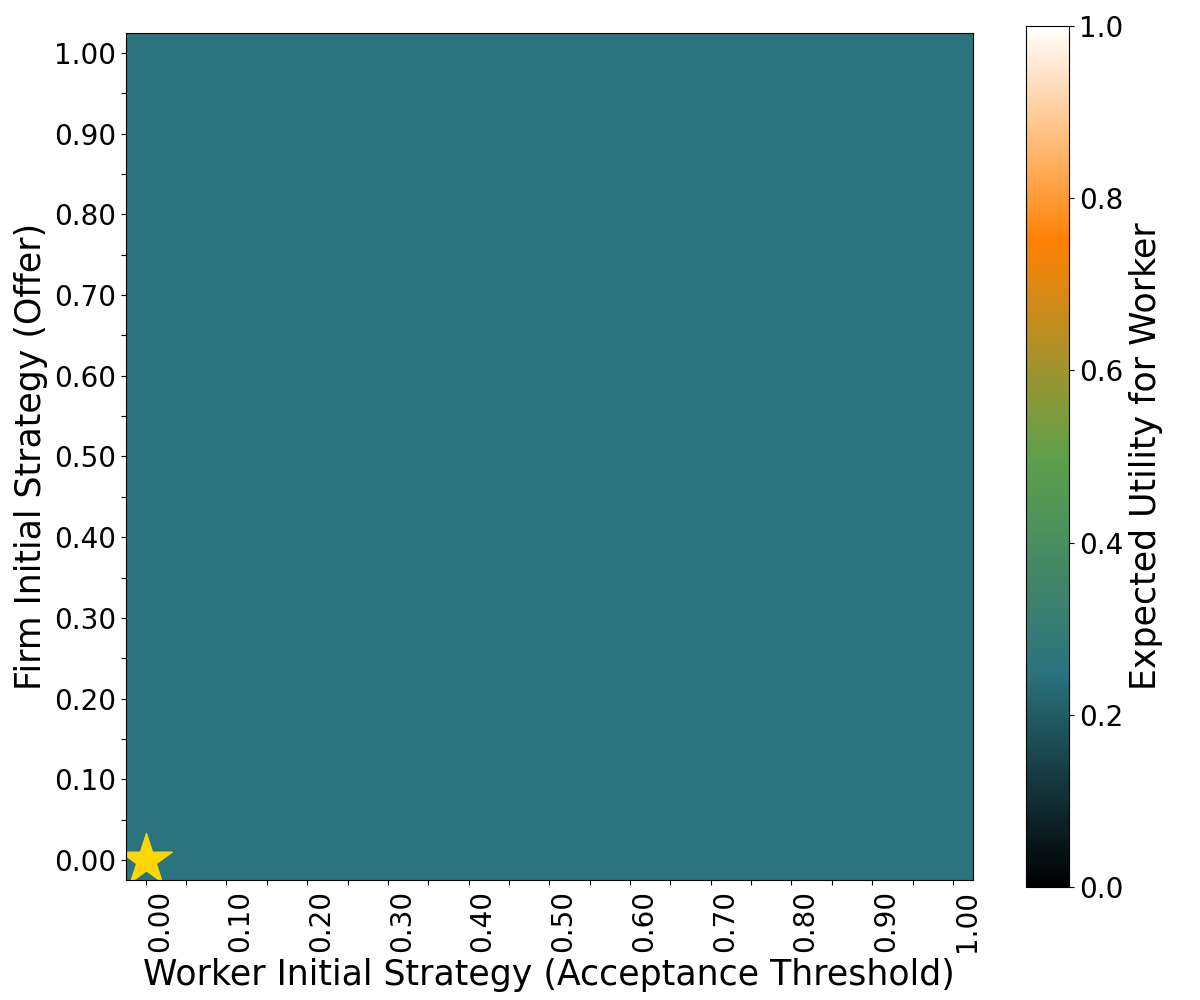}
        \caption{$\eta = 0.1$}
        \label{fig:enter-label}
    \end{subfigure}\hfill
    \begin{subfigure}{0.32\textwidth}
        \includegraphics[width=\linewidth]{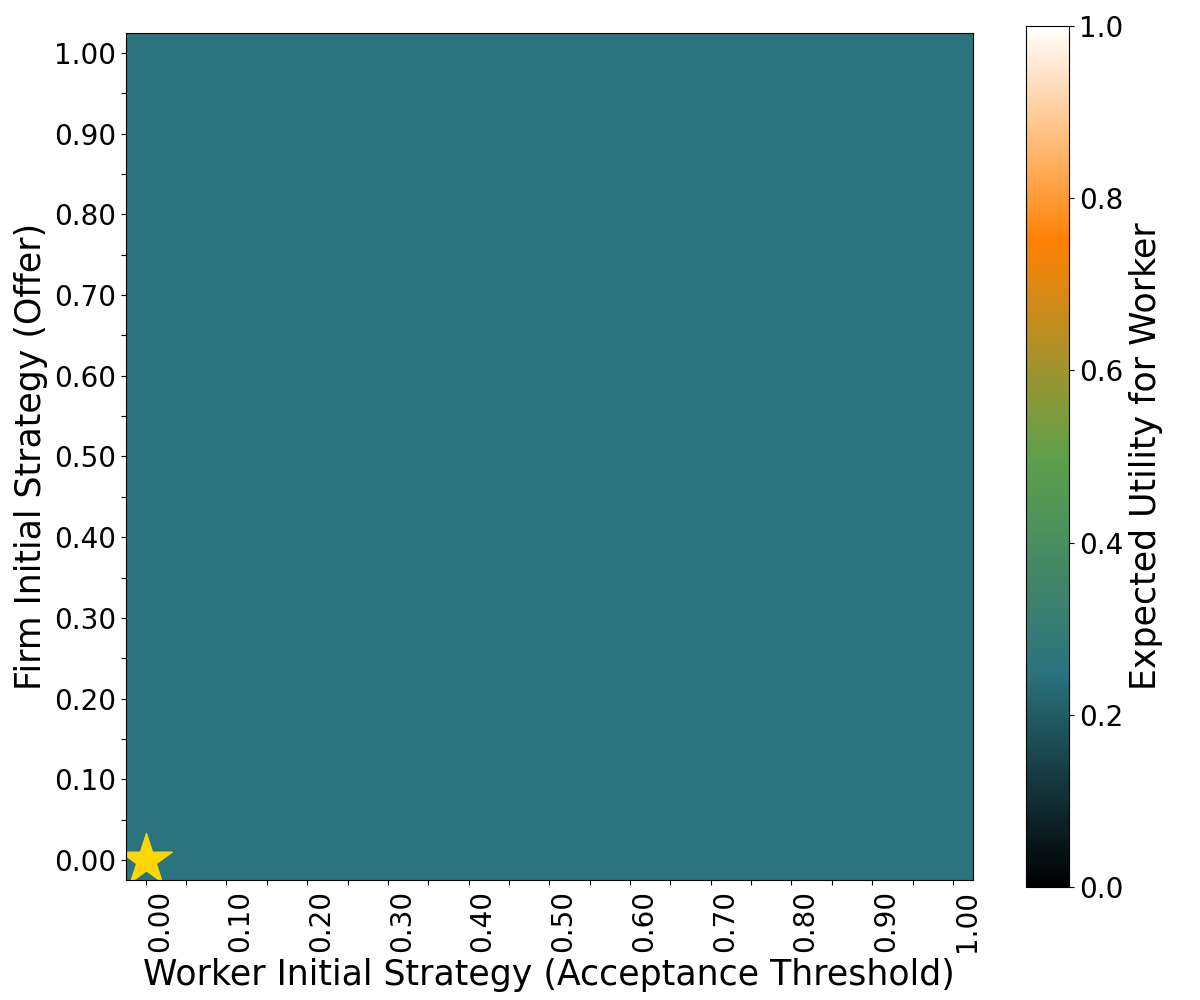}
        \caption{$\eta=0.25$.}
        \label{fig:enter-label}
    \end{subfigure}\hfill
    \begin{subfigure}{0.32\textwidth}
        \includegraphics[width=\linewidth]{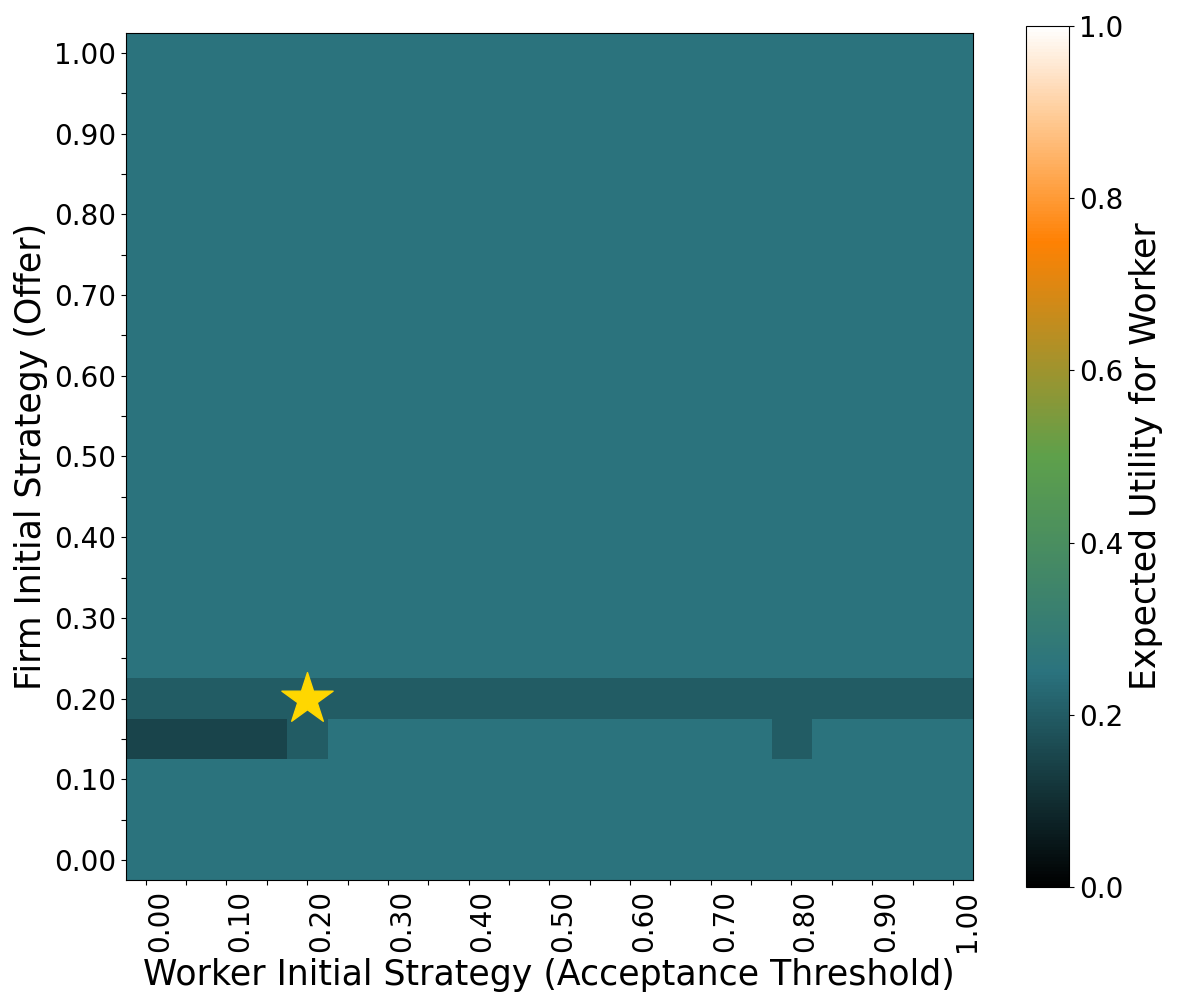}
        \caption{$\eta=0.8$.}
        \label{fig:enter-label}
    \end{subfigure}
    % \begin{subfigure}{0.4\textwidth}
    %     \includegraphics[width=\linewidth]{Experimental_Results/appendix-1-round/1_round_imshow_D_25_M_0.25_T_5000_del_0.5_eps_1e-07_rounds_1_ref_5,_15_.png}
    %     \caption{Caption}
    %     \label{fig:enter-label}
    % \end{subfigure}
    \caption{Convergence outcomes for agents learning strategies for $\mathcal{G}^{(1)}$ via Algorithm~\ref{alg:ftrl} with $\alpha_f=\frac{1}{4}, \alpha_w=\frac{3}{4}$ and $D=20$ under a variety of $\eta$ values.}
    \label{fig:referenece_point__5,_15__20}
\end{figure}

\begin{figure}[hp]
    \centering
    \begin{subfigure}{0.32\textwidth}
        \includegraphics[width=\linewidth]{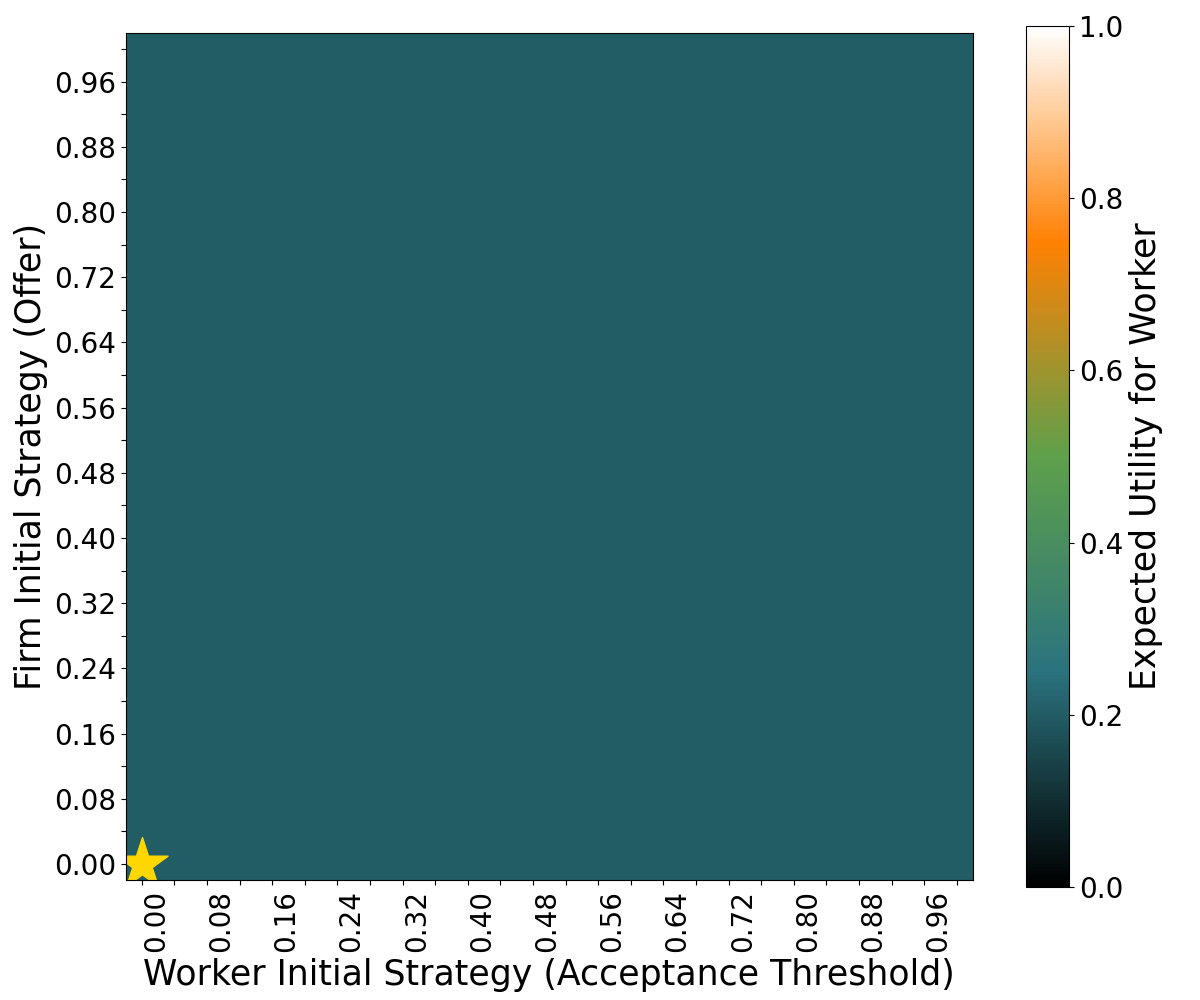}
        \caption{$\eta = 0.1$.}
        \label{fig:enter-label}
    \end{subfigure}\hfill
    \begin{subfigure}{0.32\textwidth}
        \includegraphics[width=\linewidth]{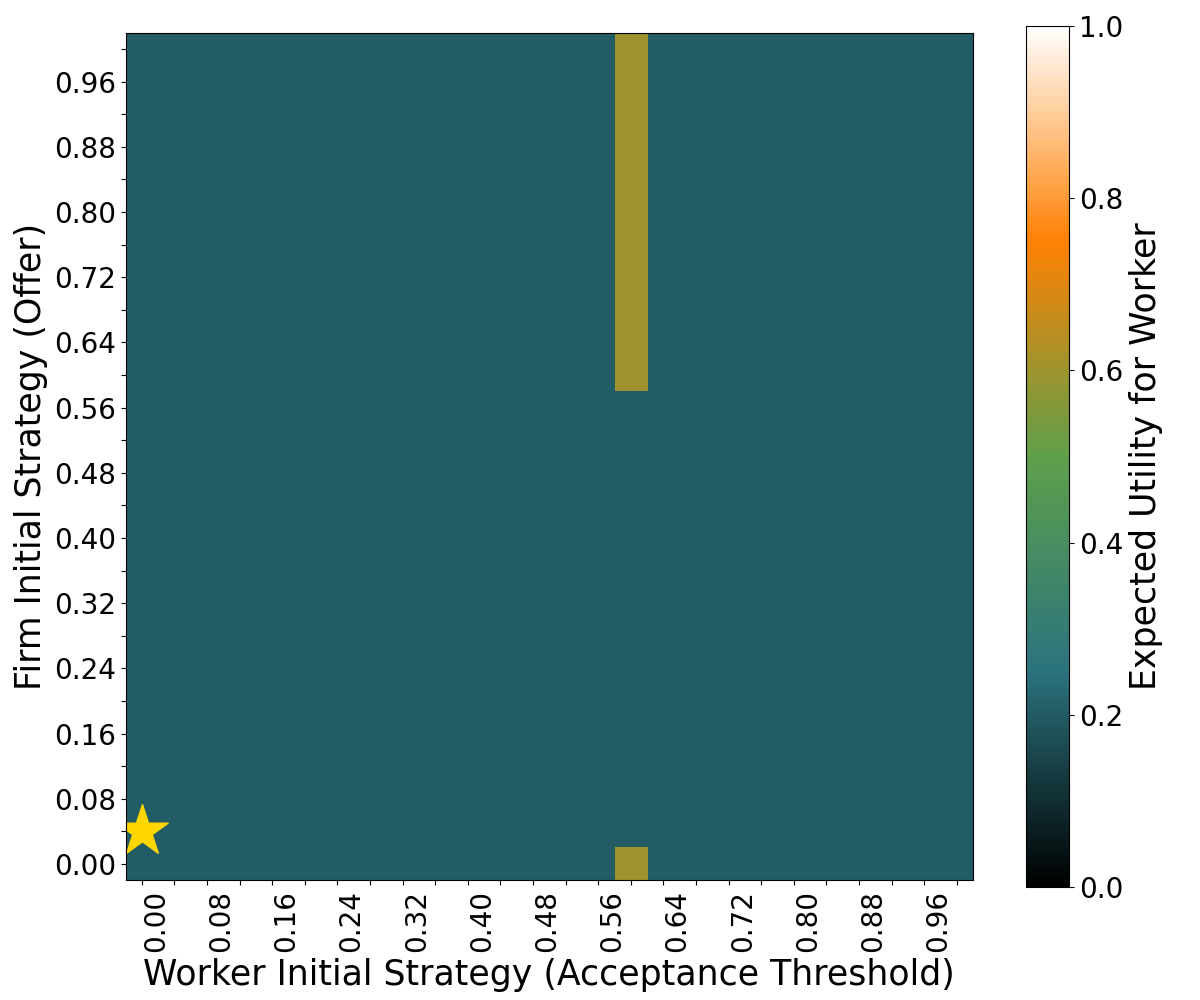}
        \caption{$\eta = 0.6$.}
        \label{fig:enter-label}
    \end{subfigure}\hfill
    \begin{subfigure}{0.32\textwidth}
        \includegraphics[width=\linewidth]{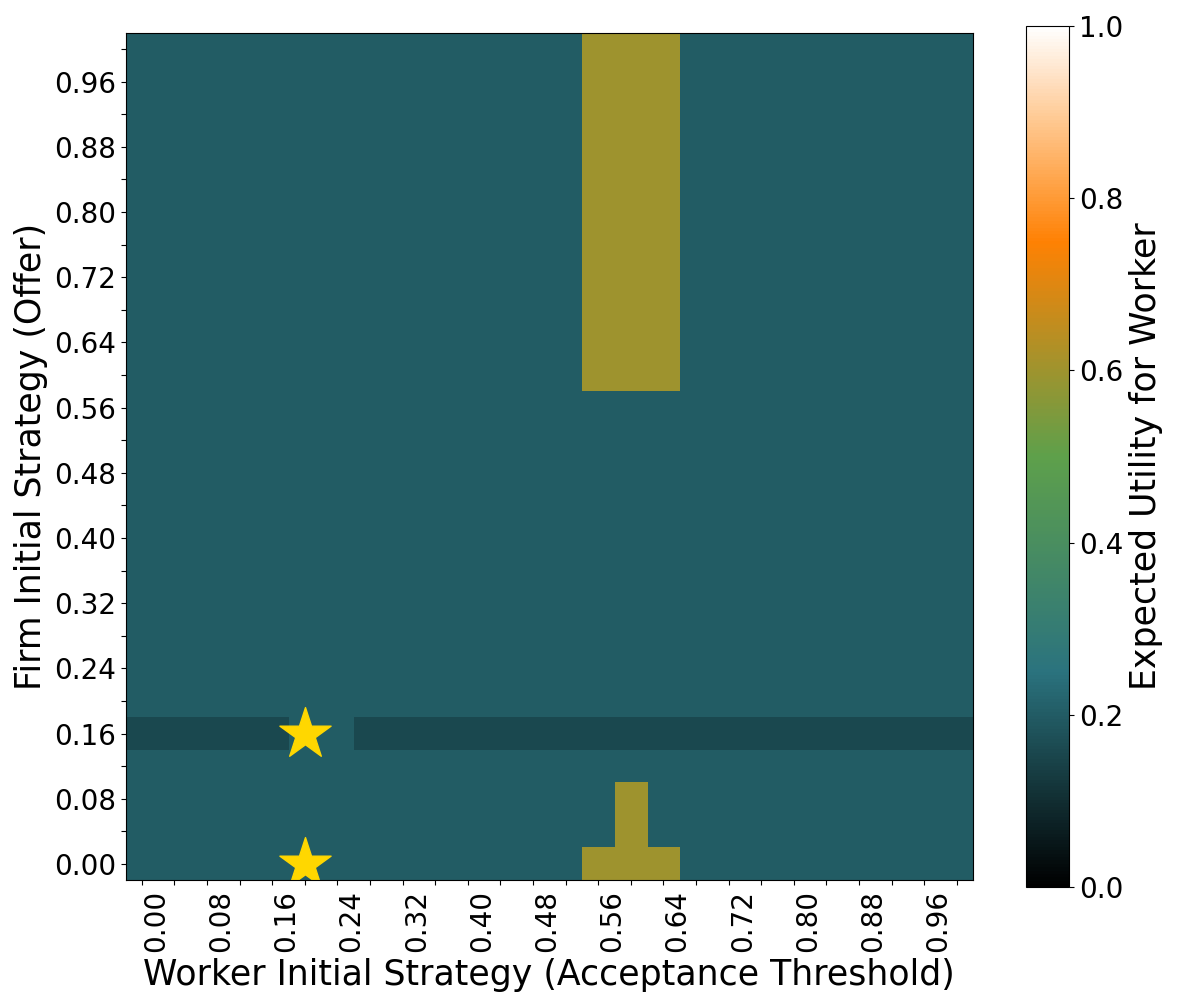}
        \caption{$\eta= 0.8$.}
        \label{fig:enter-label}
    \end{subfigure}
    \caption{Convergence outcomes for agents learning strategies for $\mathcal{G}^{(1)}$ via Algorithm~\ref{alg:ftrl} with $\alpha_f=\frac{1}{5}, \alpha_w=\frac{3}{5}$ and $D=25$ under a variety of $\eta$ values.}
    \label{fig:ref_point_5_15_25}
\end{figure}

\begin{table}[hp]
    \centering
    \begin{tabular}{|c|c|c|c|c|c|c|}
    \hline
    $D$ & $\eta$ & $(\alpha_f,\alpha_w)$ & Prop. $u_w \ge x_w^{(1)}$ & Prop. $u_w \ge \alpha_w$ & Min. $u_w$ & Max. $u_w$\\\hline
       25 & $0.1$ & ($\frac{1}{5}$, $\frac{3}{5}$) & $0.1923$ & $0.0$ & $0.2$ & $0.2$\\
25 & $0.1$ & ($\frac{4}{5}$, $\frac{16}{25}$) & $0.6538$ & $1.0$ & $0.64$ & $0.64$\\
25 & $0.25$ & ($\frac{1}{5}$, $\frac{3}{5}$) & $0.1923$ & $0.0$ & $0.2$ & $0.2$\\
25 & $0.25$ & ($\frac{4}{5}$, $\frac{16}{25}$) & $0.6538$ & $1.0$ & $0.64$ & $0.64$\\
25 & $0.6$ & ($\frac{1}{5}$, $\frac{3}{5}$) & $0.2101$ & $0.0178$ & $0.2$ & $0.6$\\
25 & $0.6$ & ($\frac{4}{5}$, $\frac{16}{25}$) & $0.6479$ & $0.9778$ & $0.16$ & $0.64$\\
25 & $0.8$ & ($\frac{1}{5}$, $\frac{3}{5}$) & $0.2308$ & $0.0562$ & $0.16$ & $0.6$\\
25 & $0.8$ & ($\frac{4}{5}$, $\frac{16}{25}$) & $0.6169$ & $0.8728$ & $0.12$ & $0.64$\\
30 & $0.29$ & ($\frac{1}{6}$, $\frac{1}{2}$) & $0.5068$ & $0.8959$ & $0.1667$ & $0.5$\\
20 & $0.29$ & ($\bf{0}$, $\bf{0}$) & $0.2358$ & N/A & $0.15$ & $0.3$\\
20 & $0.1$ & ($\bf{0}$, $\bf{0}$) & $0.2268$ & N/A & $0.2$ & $0.25$\\
20 & $0.1$ & ($\frac{1}{4}$, $\frac{3}{4}$) & $0.2857$ & $0.0$ & $0.25$ & $0.25$\\
20 & $0.1$ & ($1$, $\frac{4}{5}$) & $0.8095$ & $1.0$ & $0.8$ & $0.8$\\
20 & $0.25$ & ($\bf{0}$, $\bf{0}$) & $0.2358$ & N/A & $0.15$ & $0.3$\\
20 & $0.25$ & ($\frac{1}{4}$, $\frac{3}{4}$) & $0.2857$ & $0.0$ & $0.25$ & $0.25$\\
20 & $0.25$ & ($1$, $\frac{4}{5}$) & $0.7211$ & $0.8186$ & $0.15$ & $0.8$\\
20 & $0.5$ & ($\bf{0}$, $\bf{0}$) & $0.0272$ & N/A & $0.15$ & $0.3$\\
20 & $0.5$ & ($\frac{1}{4}$, $\frac{3}{4}$) & $0.0023$ & $0.0$ & $0.25$ & $0.25$\\
20 & $0.8$ & ($\bf{0}$, $\bf{0}$) & $0.3084$ & N/A & $0.1$ & $0.35$\\
20 & $0.8$ & ($\frac{1}{4}$, $\frac{3}{4}$) & $0.2789$ & $0.0$ & $0.15$ & $0.25$\\
20 & $0.8$ & ($1$, $\frac{4}{5}$) & $0.381$ & $0.1882$ & $0.05$ & $0.8$\\
25 & $0.29$ & ($\bf{0}$, $\bf{0}$) & $0.2574$ & N/A & $0.16$ & $0.28$\\
30 & $0.29$ & ($1$, $1$) & $0.1165$ & $0.0$ & $0.0667$ & $0.3$\\
30 & $0.25$ & ($\frac{1}{6}$, $\frac{1}{2}$) & $0.5088$ & $0.9313$ & $0.1667$ & $0.5$\\
30 & $0.29$ & ($\frac{1}{2}$, $\frac{29}{30}$) & $0.3226$ & $0.0$ & $0.1667$ & $0.5$ \\\hline
    \end{tabular}
    \caption{Attributes of $u_w$ for each plot in Figures~\ref{fig:referenece_point__20,_16__20},~\ref{fig:ref_point_20_16_25},~\ref{fig:referenece_point__5,_15__20},~\ref{fig:ref_point_5_15_25}, and~\ref{fig:referenece_point_0}. The fourth column gives the proportion of outcomes where $u_w \ge x_w^{(1)}$, the worker's initial acceptance threshold. The fifth column gives the proportion of outcomes where $u_w \ge \alpha_w$, the worker's reference point. The last two columns give the minimum and maximum $u_w$ value in each plot.}
    \label{tab:summary-statistics-2}
\end{table}

\section{Experimental Results for $\mathcal{G}^{(2)}$}
\label{appendix:efg-graphs}
In this section, we include additional graphs from our numerical experiments for agents learning strategies for $\mathcal{G}^{(2)}$ via Algorithm~\ref{alg:ftrl}. In each plot, the worker payoff at the converged equilibrium is displayed. Each algorithm run converged according to the convergence criteria with a threshold of $10^{-6}$ in Section~\ref{sec:experimental} within 15000 time steps and each convergence outcome was verified to be a $\epsilon$-Nash equilibrium for $\epsilon=10^{-7}$. The code we used to run these experiments can be accessed anonymously at the following link: \href{https://anonymous.4open.science/r/learning_bargaining_strategies-D6B7/README.md}{https://anonymous.4open.science/r/learning\_bargaining\_strategies-D6B7/README.md}. All of our experiments were run on standard laptops without any extra compute. The goal of our experiments was not to optimize run time or implement an algorithm for bargaining strategies in general, but rather to explore the space of convergence outcomes that were not covered by our theoretical results.

\begin{figure}[h]
    \centering
    \begin{subfigure}{0.45\textwidth}
        \includegraphics[width=\linewidth]{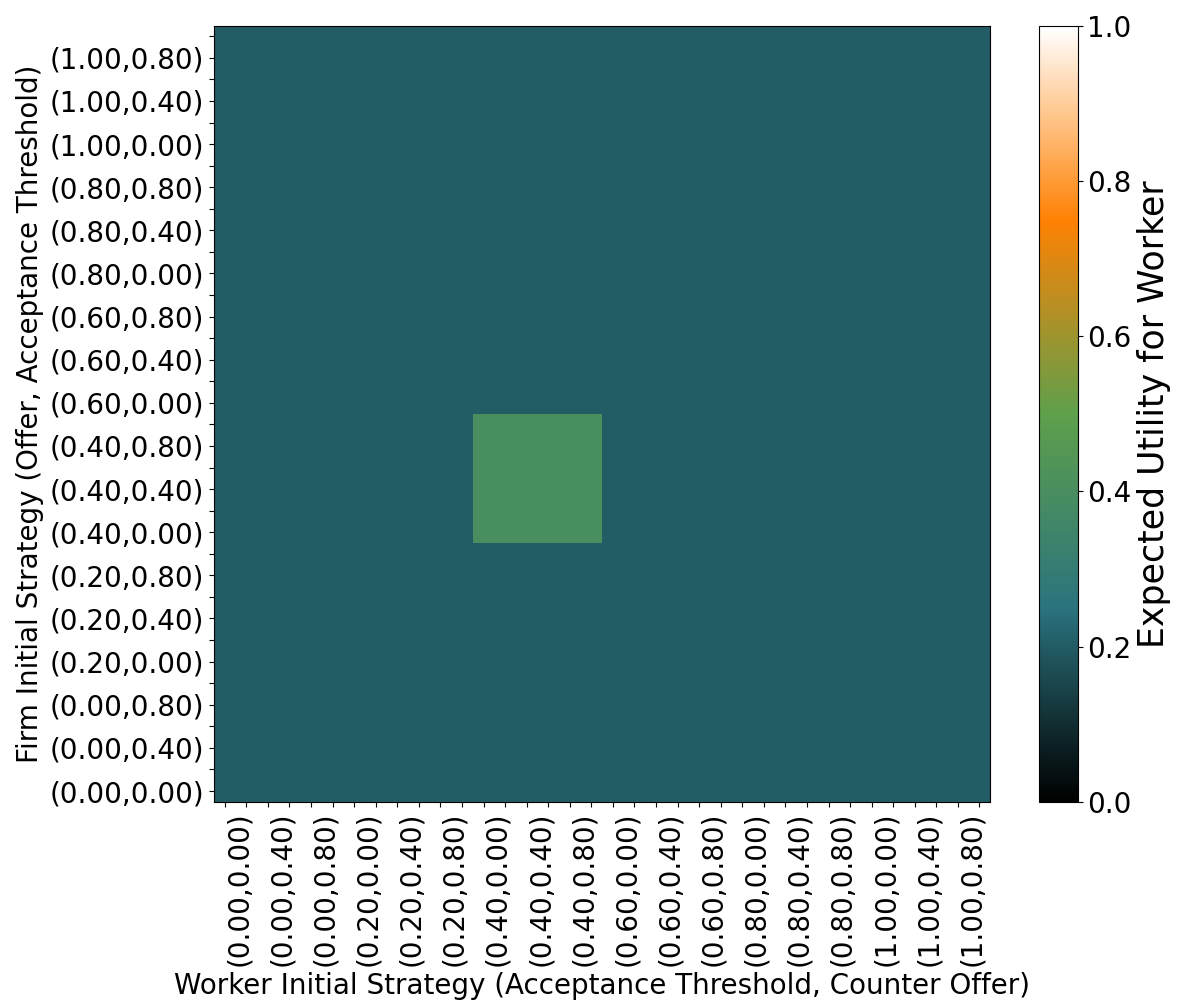}
        \caption{$\eta=0.25,\delta=0.1$.}
        \label{fig:enter-label}
    \end{subfigure}\hfill
    \begin{subfigure}{0.45\textwidth}
        \includegraphics[width=\linewidth]{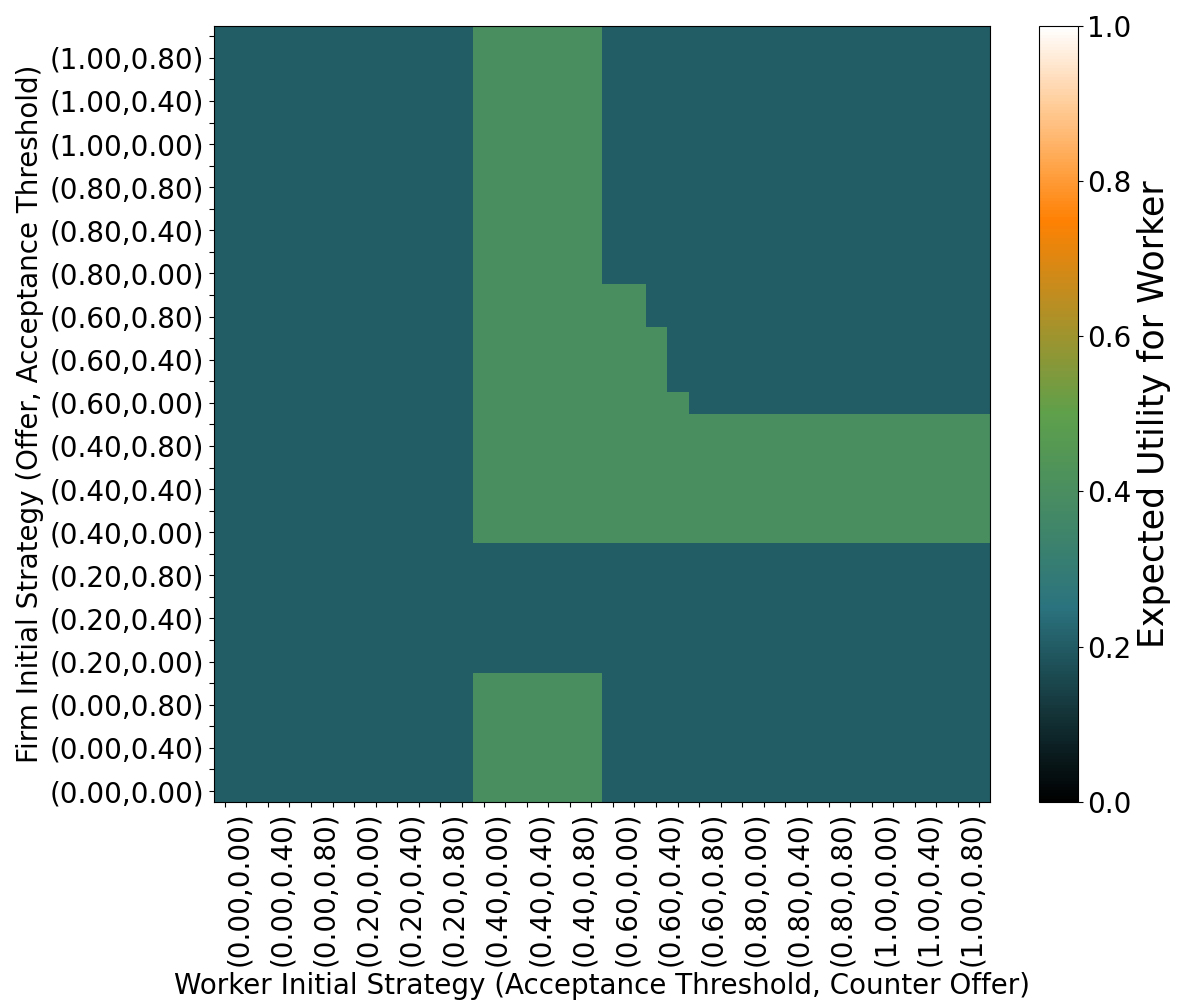}        
        \caption{$\eta=0.8,\delta=0.1$.}
        \label{fig:enter-label}
    \end{subfigure}\\
    \begin{subfigure}{0.45\textwidth}
        \includegraphics[width=\linewidth]{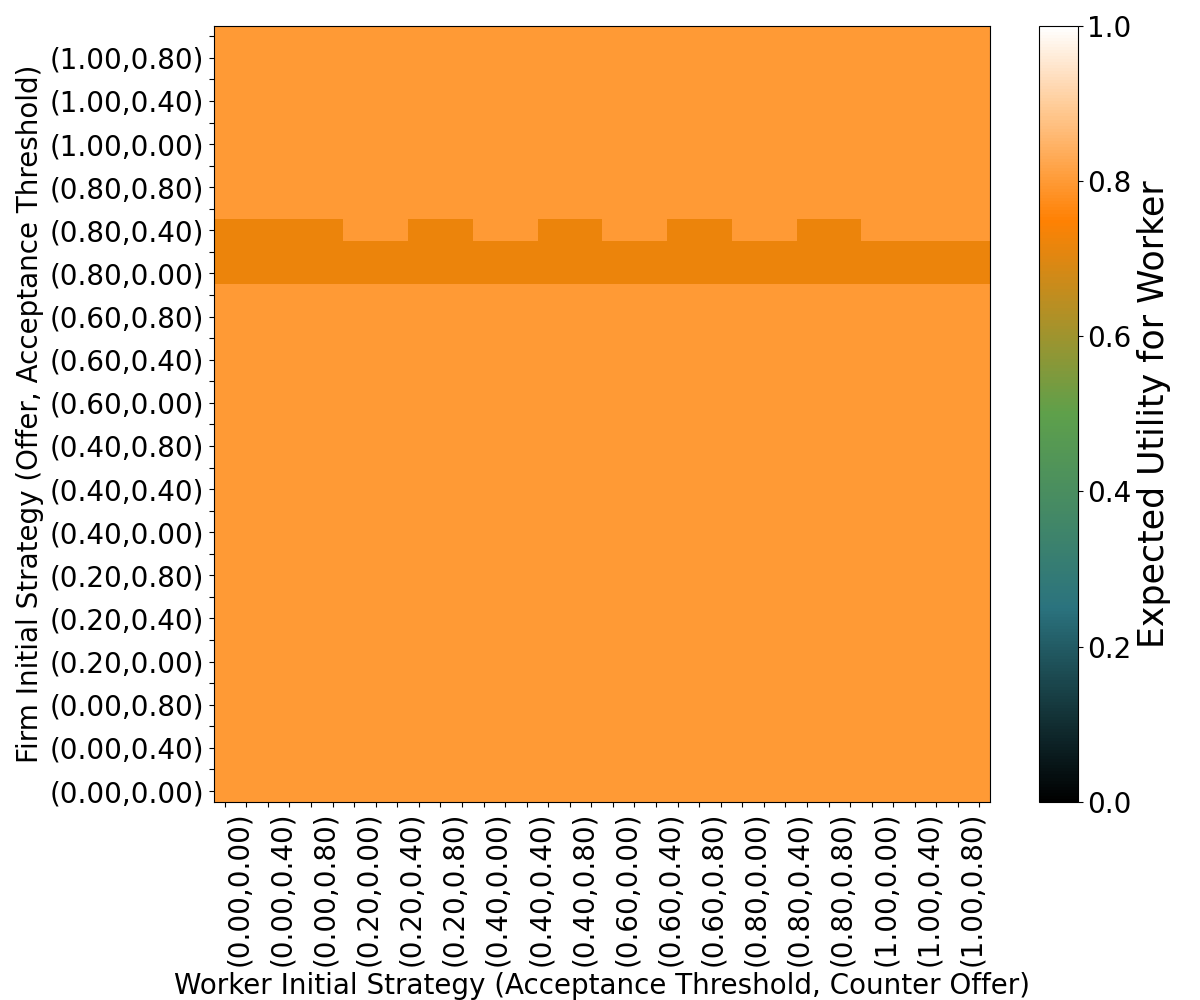}
        \caption{$\eta=0.25,\delta=0.9$.}
        \label{fig:enter-label}
    \end{subfigure}\hfill
    \begin{subfigure}{0.45\textwidth}
    \includegraphics[width=\linewidth]{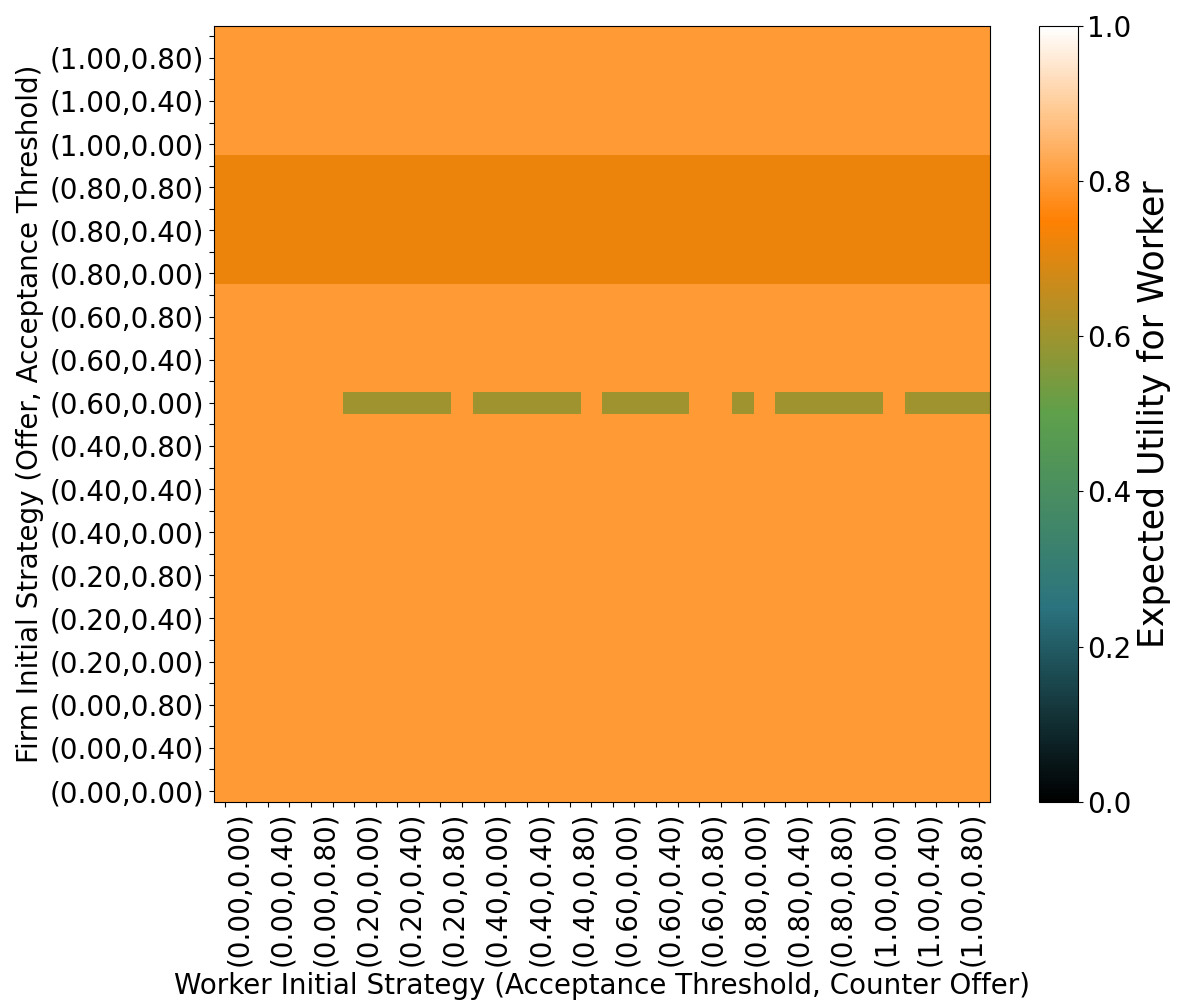}
        \caption{$\eta = 0.8,\delta=0.9$.}
        \label{fig:enter-label}
    \end{subfigure}
    \caption{Convergence outcomes for agents learning strategies for $\mathcal{G}^{(2)}$ via Algorithm~\ref{alg:ftrl} with $\alpha_f=\alpha_w=\bf{0}$ and $D=5$ and a variety of other parameter settings.}
    \label{fig:2-round-convergence-outcomes}
\end{figure}

\begin{figure}[h]
    \centering
    \begin{subfigure}[t]{0.45\textwidth}
         \resizebox{\textwidth}{4cm}{
\begin{tikzpicture}

    % \node[rectangle, draw,text width=3.cm,align=center] (title) at (0,1.5) {2-Round Alternating Bargaining Game};
    \node[circle, draw,color=blue,text width=1.5cm,align=center] (w) at (0,0) {Worker, response to $0.6$};
    \node[text width=2cm,align=center] (wA) at (-7,0) {$(0.6,0.4)$};
    \node[circle, draw,color=blue,text width=1cm,align=center] (wR0) at (-6,-2) {Firm};
    \node[circle, draw,color=blue,text width=1cm,align=center] (wR02) at (-4,-6) {Firm};
    \node[circle, draw,color=red,text width=1cm,align=center] (wR04) at (0,-8) {Firm};
    \node[circle, draw,color=red,text width=1cm,align=center] (wR06) at (4,-6) {Firm};
    \node[circle, draw,color=red,text width=1cm,align=center] (wR08) at (6,-2) {Firm};
    \node[circle, draw,color=red,text width=1cm,align=center] (wR1) at (7,0) {Firm};

    \node[text width=2cm,align=center] (fA0) at (-7,-4) {$(0,0.9)$};
    \node[text width=2cm,align=center] (fR0) at (-5,-4) {$(0,0)$};
    \node[text width=2cm,align=center] (fA02) at (-5,-8) {$(0.18,0.72)$};
    \node[text width=2cm,align=center] (fR02) at (-3,-8) {$(0,0)$};
    \node[text width=2cm,align=center] (fA04) at (-1,-10) {$(0.36,0.54)$};
    \node[text width=2cm,align=center] (fR04) at (1,-10) {$(0,0)$};
    \node[text width=2cm,align=center] (fA06) at (3,-8) {$(0.54,0.36)$};
    \node[text width=2cm,align=center] (fR06) at (5,-8) {$(0,0)$};
    \node[text width=2cm,align=center] (fA08) at (5,-4) {$(0.72,0.18)$};
    \node[text width=2cm,align=center] (fR08) at (7,-4) {$(0,0)$};
    \node[text width=2cm,align=center] (fA1) at (9,-2) {$(0.9,0)$};
    \node[text width=2cm,align=center] (fR1) at (11,0) {$(0,0)$};

    \draw[->,solid,black] (w) to node[midway,above] {Accept}(wA);
    \draw[->,solid,black] (w) to node[midway,above,sloped] {Counter 0}(wR0);
    \draw[->,solid,black] (w) to node[midway,above,sloped] {Counter $0.2$}(wR02);
    \draw[->,solid,black] (w) to node[midway,below,sloped] {\textit{1}}(wR02);
    \draw[->,solid,black] (w) to node[midway,sloped,above] {Counter $0.4$}(wR04);
    \draw[->,solid,black] (w) to node[midway,above,sloped] {Counter $0.6$}(wR06);
    \draw[->,solid,black] (w) to node[midway,above,sloped] {Counter $0.8$}(wR08);
    \draw[->,solid,black] (w) to node[midway,above,sloped] {Counter 1}(wR1);

    \draw[->,solid,black] (wR0) to node[midway,left] {Accept}(fA0);
    \draw[->,solid,black] (wR0) to node[midway,right] {\emph{0.5}}(fA0);
    \draw[->,solid,black] (wR0) to node[midway,right] {Reject}(fR0);
    \draw[->,solid,black] (wR0) to node[midway,left] {\emph{0.5}}(fR0);
    \draw[->,solid,black] (wR02) to node[midway,left] {Accept}(fA02);
    \draw[->,solid,black] (wR02) to node[midway,right] {\textit{1}}(fA02);
    \draw[->,solid,black] (wR02) to node[midway,right] {Reject}(fR02);
    \draw[->,solid,black] (wR04) to node[midway,left] {Accept}(fA04);
    \draw[->,solid,black] (wR04) to node[midway,right] {\emph{0.5}}(fA04);
    \draw[->,solid,black] (wR04) to node[midway,right] {Reject}(fR04);
    \draw[->,solid,black] (wR04) to node[midway,left] {\emph{0.5}}(fR04);
    \draw[->,solid,black] (wR06) to node[midway,left] {Accept}(fA06);
    \draw[->,solid,black] (wR06) to node[midway,right] {\emph{0.5}}(fA06);
    \draw[->,solid,black] (wR06) to node[midway,right] {Reject}(fR06);
    \draw[->,solid,black] (wR06) to node[midway,left] {\emph{0.5}}(fR06);
    \draw[->,solid,black] (wR08) to node[midway,left] {Accept}(fA08);
    \draw[->,solid,black] (wR08) to node[midway,right] {\emph{0.5}}(fA08);
    \draw[->,solid,black] (wR08) to node[midway,right] {Reject}(fR08);
    \draw[->,solid,black] (wR08) to node[midway,left] {\emph{0.5}}(fR08);
    \draw[->,solid,black] (wR1) to node[midway,right] {Accept}(fA1);
    \draw[->,solid,black] (wR1) to node[midway,above] {Reject}(fR1);
    \draw[->,solid,black] (wR1) to node[midway,left] {\emph{0.5}}(fA1);
    \draw[->,solid,black] (wR1) to node[midway,below] {\emph{0.5}}(fR1);

    \end{tikzpicture}
    }
    \caption{Off-equilibrium subtree of worker's response to a firm first round offer of $0.6$ for initial strategy choices $a_{f,r} = a_{f,p}=a_{w,r}=a_{w,p} = 0$.}
    \label{threat:credible-worker}
    \end{subfigure}\hfill
    \begin{subfigure}[t]{0.45\textwidth}
        \resizebox{\textwidth}{4cm}{
    \begin{tikzpicture}

    % \node[rectangle, draw,text width=3.cm,align=center] (title) at (0,1.5) {2-Round Alternating Bargaining Game};
    \node[circle, draw,color=black,text width=1.5cm,align=center] (w) at (0,0) {Worker, response to $0.6$};
    \node[text width=2cm,align=center] (wA) at (-7,0) {$(0.4,0.6)$};
    \node[circle, draw,color=blue,text width=1cm,align=center] (wR0) at (-6,-2) {Firm};
    \node[circle, draw,color=red,text width=1cm,align=center] (wR02) at (-4,-6) {Firm};
    \node[circle, draw,color=red,text width=1cm,align=center] (wR04) at (0,-8) {Firm};
    \node[circle, draw,color=red,text width=1cm,align=center] (wR06) at (4,-6) {Firm};
    \node[circle, draw,color=red,text width=1cm,align=center] (wR08) at (6,-2) {Firm};
    \node[circle, draw,color=red,text width=1cm,align=center] (wR1) at (7,0) {Firm};

    \node[text width=2cm,align=center] (fA0) at (-7,-4) {$(0,0.9)$};
    \node[text width=2cm,align=center] (fR0) at (-5,-4) {$(0,0)$};
    \node[text width=2cm,align=center] (fA02) at (-5,-8) {$(0.18,0.72)$};
    \node[text width=2cm,align=center] (fR02) at (-3,-8) {$(0,0)$};
    \node[text width=2cm,align=center] (fA04) at (-1,-10) {$(0.36,0.54)$};
    \node[text width=2cm,align=center] (fR04) at (1,-10) {$(0,0)$};
    \node[text width=2cm,align=center] (fA06) at (3,-8) {$(0.54,0.36)$};
    \node[text width=2cm,align=center] (fR06) at (5,-8) {$(0,0)$};
    \node[text width=2cm,align=center] (fA08) at (5,-4) {$(0.72,0.18)$};
    \node[text width=2cm,align=center] (fR08) at (7,-4) {$(0,0)$};
    \node[text width=2cm,align=center] (fA1) at (9,-2) {$(0.9,0)$};
    \node[text width=2cm,align=center] (fR1) at (11,0) {$(0,0)$};

    \draw[->,solid,black] (w) to node[midway,above] {Accept}(wA);
    \draw[->,solid,black] (w) to node[midway,below] {\emph{1}}(wA);
    \draw[->,solid,black] (w) to node[midway,above,sloped] {Counter 0}(wR0);
    \draw[->,solid,black] (w) to node[midway,above,sloped] {Counter $0.2$}(wR02);
    \draw[->,solid,black] (w) to node[midway,sloped,above] {Counter $0.4$}(wR04);
    \draw[->,solid,black] (w) to node[midway,above,sloped] {Counter $0.6$}(wR06);
    \draw[->,solid,black] (w) to node[midway,above,sloped] {Counter $0.8$}(wR08);
    \draw[->,solid,black] (w) to node[midway,above,sloped] {Counter 1}(wR1);

    \draw[->,solid,black] (wR0) to node[midway,left] {Accept}(fA0);
    \draw[->,solid,black] (wR0) to node[midway,right] {\emph{0.5}}(fA0);
    \draw[->,solid,black] (wR0) to node[midway,right] {Reject}(fR0);
    \draw[->,solid,black] (wR0) to node[midway,left] {\emph{0.5}}(fR0);
    \draw[->,solid,black] (wR02) to node[midway,left] {Accept}(fA02);
    \draw[->,solid,black] (wR02) to node[midway,right] {\emph{0.5}}(fA02);
    \draw[->,solid,black] (wR02) to node[midway,right] {Reject}(fR02);
    \draw[->,solid,black] (wR02) to node[midway,left] {\emph{0.5}}(fR02);
    \draw[->,solid,black] (wR04) to node[midway,left] {Accept}(fA04);
    \draw[->,solid,black] (wR04) to node[midway,right] {\emph{0.5}}(fA04);
    \draw[->,solid,black] (wR04) to node[midway,right] {Reject}(fR04);
    \draw[->,solid,black] (wR04) to node[midway,left] {\emph{0.5}}(fR04);
    \draw[->,solid,black] (wR06) to node[midway,left] {Accept}(fA06);
    \draw[->,solid,black] (wR06) to node[midway,right] {\emph{0.5}}(fA06);
    \draw[->,solid,black] (wR06) to node[midway,right] {Reject}(fR06);
    \draw[->,solid,black] (wR06) to node[midway,left] {\emph{0.5}}(fR06);
    \draw[->,solid,black] (wR08) to node[midway,left] {Accept}(fA08);
    \draw[->,solid,black] (wR08) to node[midway,right] {\emph{0.5}}(fA08);
    \draw[->,solid,black] (wR08) to node[midway,right] {Reject}(fR08);
    \draw[->,solid,black] (wR08) to node[midway,left] {\emph{0.5}}(fR08);
    \draw[->,solid,black] (wR1) to node[midway,right] {Accept}(fA1);
    \draw[->,solid,black] (wR1) to node[midway,above] {Reject}(fR1);
    \draw[->,solid,black] (wR1) to node[midway,left] {\emph{0.5}}(fA1);
    \draw[->,solid,black] (wR1) to node[midway,below] {\emph{0.5}}(fR1);

    \end{tikzpicture}
    }
    \caption{Equilibrium path subtree of worker's response to a firm first round offer of $0.6$ for initial strategy choices $a_{f,r} = 0, a_{f,p}= 0.6, a_{w,r}= 0.6, a_{w,p} = 0.2$.} 
    \label{threat:non-credible-firm}
    \end{subfigure}
    
    \caption{Subtrees of $\mathcal{G}^{(2)}$ starting with the worker's response node to a first offer of 0.6 after Algorithm~\ref{alg:ftrl} converges with the parameter settings $D=5,\eta=0.5,\delta=0.9$ for two sets of initial strategy choices. Only non-zero probability values are displayed and they are italicized on the opposite side of the branch of their corresponding action. Red nodes indicate \textit{non-credible} threats by the firm to reject with non-zero probability in the second round. Blue nodes indicate \textit{credible} threats to take the specified action. Finally, black nodes are on the equilibrium path, so they are not threats.}
    \label{fig:threat-trees}
\end{figure}
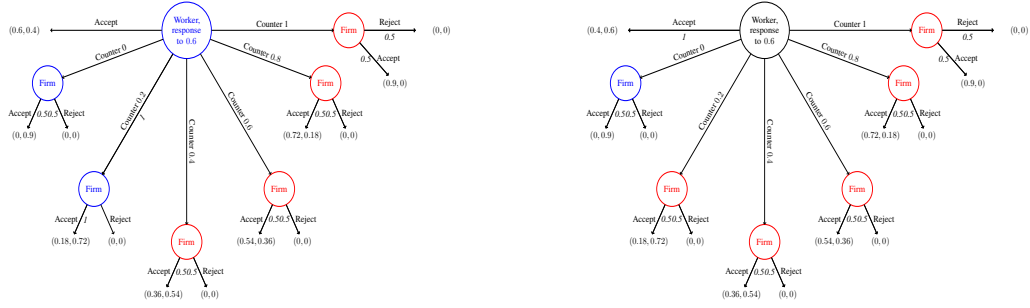

The additional results are shown in Figure~\ref{fig:2-round-convergence-outcomes} show the value of $u_w$ at convergence, but they do not include threat outcomes. We turn to Figure~\ref{fig:threat-trees} to discuss threats in more detail.

\paragraph{Threats}
To expand on our discussion of threats, we highlight two cases of threat emergence under the parameter settings $D=5,\eta=0.5,\delta=0.9$. Consider, two initial strategy profile choices: 1) $a_{f,r} = a_{f,p}=a_{w,r}=a_{w,p} = 0$ and 2) $a_{f,r} = 0, a_{f,p} = 0.6,a_{w,r}=0.6, a_{f,p} = 0.2$. The equilibrium outcome of 1) involves the firm offering $0.8$ to the worker which the worker accepts while the equilibrium outcome of 2) involves the firm offering $0.6$ to the worker which the worker accepts. Thus, the worker's response to $0.6$ is \textit{off} the equilibrium path in outcome 1) while it is \textit{on} the equilibrium path in outcome 2). We show these worker response subtrees of $\mathcal{G}^{(2)}$ in Figures~\ref{threat:credible-worker} and~\ref{threat:non-credible-firm}, respectively.
Here, we use numerical approximations of their converged strategy probabilities for illustrative purposes. 
% compare our results to what theory would predict and highlight why this is interesting
% credible threat intuition
In subtree~\ref{threat:credible-worker}, the fact that the firm is not  non-credibly threatening to reject $0.2$ in the second round means the worker can credibly threaten to reject the offer of $0.6$. The firm can see this strategy when updating, and, since the worker's threat at an offer of $0.6$ would leave the firm with $0.18$, the worker's credible threat causes the firm to prefer offering $0.8$ in the first round which leaves the firm with $0.2$ instead.
% non-credible threat intuition
Next, in subtree~\ref{threat:non-credible-firm}, the fact that the firm is now non-credibly threatening to reject $0.2$ after making a first round offer of $0.6$ to the worker. Because the worker updates their strategy based on this non-credible threat, their best response is to accept the offer of $0.6$ as opposed to rejecting and trying to get a better deal in the second round. In FTRL, each agent updates the probability mass on an action proportionally to the cumulative expected utility that action gets, relative to other actions the agent could have taken at a decision node. So, the reason both of these outcomes are possible is that the cumulative utility for the firm accepting a second round deal grows proportional to the amount of probability mass the worker puts on that second round action. So, depending on the initial conditions, the worker will either explore second round counter offers sufficiently long enough for the firm to converge to accepting in the second round or the worker will converge to accepting a first round deal before this point, leaving the firm with non-credible threats in the second round at convergence.

\end{document}